\documentclass[zpreprint,zbstdefault]{zeus_paper}

\usepackage[english]{babel}

\chardef\usc=95
\chardef\til=126
\catcode`\@=11 
\DeclareRobustCommand\xdotspace{\futurelet\@let@token\@xdotspace}
\def\@xdotspace{%
  \ifx\@let@token.\else
  \ifx\@let@token\bgroup.\else
  \ifx\@let@token\egroup.\else
  \ifx\@let@token\/.\else
  \ifx\@let@token\ .\else
  \ifx\@let@token~.\else
  \ifx\@let@token!.\else
  \ifx\@let@token,.\else
  \ifx\@let@token:.\else
  \ifx\@let@token;.\else
  \ifx\@let@token?.\else
  \ifx\@let@token/.\else
  \ifx\@let@token'.\else
  \ifx\@let@token).\else
  \ifx\@let@token-.\else
  \ifx\@let@token\@xobeysp.\else
  \ifx\@let@token\space.\else
  \ifx\@let@token\@sptoken.\else
   .\space
   \fi\fi\fi\fi\fi\fi\fi\fi\fi\fi\fi\fi\fi\fi\fi\fi\fi\fi}
\catcode`\@=12 

\newcommand{\stru}[2]{%
   \relax\ifmmode\hbox{\vrule height#1 depth#2 width0pt}%
   \else\vrule height#1 depth#2 width0pt\fi}

\newcommand{\Ronum}[1]{\uppercase\expandafter{\romannumeral#1}}
\newcommand{\ronum}[1]{\expandafter{\romannumeral#1}}
\DeclareRobustCommand{\LaTeXZ}{%
  \LaTeX\kern-.05em4\kern-.1em
  {\raisebox{-0.2ex}{$\scriptstyle\text{ZEUS}$}}\xspace}

\DeclareMathAlphabet{\mathbf}{OT1}{cmr}{bx}{sl}
\newcommand{\eVdist}{\kern-0.06667em}

\newcommand{\slashfrac}[2]{%
  \raisebox{0.5ex}{\ensuremath #1}\kern-0.12em/\kern-0.08em
  \raisebox{-.8ex}{\ensuremath #2}}

\newcommand{\sqr}[3]{%
    {\vcenter{\hrule height.#3ex\hbox{\vrule width.#2ex height#1ex
     \kern#1ex\vrule width.#3ex}\hrule height.#2ex}}}

\catcode`\@=11 
\newcommand{\parenbar}{\mathpalette\p@renb@r}
\def\p@renb@r#1#2{\vbox{%
  \ifx#1\scriptscriptstyle \dimen@.7em\dimen@ii.2em\else
  \ifx#1\scriptstyle \dimen@.8em\dimen@ii.25em\else
  \dimen@1em\dimen@ii.4em\fi\fi \offinterlineskip
  \ialign{\hfill##\hfill\cr
    \vbox{\hrule width\dimen@ii}\cr
    \noalign{\vskip-.3ex}%
    \hbox to\dimen@{$\mathchar300\hfil\mathchar301$}\cr
    \noalign{\vskip-.3ex}%
    $#1#2$\cr}}}
\catcode`\@=12 

\newcommand{\IP}{{\rm I$\kern-0.01667em$P}\xspace}

\mathchardef\qsm=63
\mathchardef\pls=43
\mathchardef\mns=512
\mathchardef\plm=518
\mathchardef\eql=61
\mathchardef\smallleft=300
\mathchardef\smallright=301
\mathchardef\les=316
\mathchardef\gre=318
\mathchardef\leq=532
\mathchardef\grq=533

\catcode`\@=11 
\newcounter{pict@width}
\newcounter{pict@height}
\newlength{\pict@scale}
\newcommand{\psfigadd}[4]{%
\setcounter{pict@width}{1*\ratio{#2+\pict@scale/2}{\pict@scale}}
\setcounter{pict@height}{1*\ratio{#3+\pict@scale/2}{\pict@scale}}
\setlength{\unitlength}{\pict@scale}
\hbox to #2{\hspace{-\fill}\begin{picture}(\thepict@width,\thepict@height)
\put(0,0){\psfig{figure=#1,width=#2,height=#3,clip=}}
\SetScale{0.283466457}
\SetWidth{1.763889}
{#4}
\end{picture}}
}
\newcounter{pict@widthfst}
\newcounter{pict@widthscd}
\newcounter{pict@widthtot}
\newcommand{\psfigaddtwo}[7]{%
\setcounter{pict@widthfst}{1*\ratio{#2+\pict@scale/2}{\pict@scale}}
\setcounter{pict@widthscd}{1*\ratio{#2+#4+\pict@scale/2}{\pict@scale}}
\setcounter{pict@widthtot}{1*\ratio{#2+#4+#6+\pict@scale/2}{\pict@scale}}
\setcounter{pict@height}{1*\ratio{#3+\pict@scale/2}{\pict@scale}}
\setlength{\unitlength}{\pict@scale}
\hbox{\hspace{-\fill}\begin{picture}(\thepict@widthtot,\thepict@height)
\put(0,0){\psfig{figure=#1,width=#2,height=#3,clip=}}
\put(\thepict@widthscd,0){\psfig{figure=#5,width=#6,height=#3,clip=}}
\SetScale{0.283466457}
\SetWidth{1.763889}
{#7}
\end{picture}}
}
\newcommand{\psfigror}[4]{%
\setcounter{pict@width}{1*\ratio{#2+\pict@scale/2}{\pict@scale}}
\setcounter{pict@height}{1*\ratio{#3+\pict@scale/2}{\pict@scale}}
\setlength{\unitlength}{\pict@scale}
\hbox{\begin{picture}(\thepict@width,\thepict@height)
\put(0,\thepict@height){\psfig{figure=#1,width=#3,height=#2,clip=,angle=270}}
\SetScale{0.283466457}
\SetWidth{1.763889}
{#4}
\end{picture}}
}
\newcommand{\psfigrol}[4]{%
\setcounter{pict@width}{1*\ratio{#2+\pict@scale/2}{\pict@scale}}
\setcounter{pict@height}{1*\ratio{#3+\pict@scale/2}{\pict@scale}}
\setlength{\unitlength}{\pict@scale}
\hbox{\begin{picture}(\thepict@width,\thepict@height)
\put(0,0){\psfig{figure=#1,width=#3,height=#2,clip=,angle=90}}
\SetScale{0.283466457}
\SetWidth{1.763889}
{#4}
\end{picture}}
}
\catcode`\@=12 
\newlength\listtextwidth

\catcode`\@=11 
\newlength{\@tabfninsert}
\newlength{\@tabfnwidth}
\newcommand{\tabfootnote}[2]{%
  \setlength{\@tabfninsert}{0.8em}
  \setlength{\@tabfnwidth}{\textwidth}
  \addtolength{\@tabfnwidth}{-\@tabfninsert}
  \addtolength{\@tabfnwidth}{-0.4em}
  \noindent\makebox[\@tabfninsert][r]{\footnotesize$^{#1}$\hfil}\hfill%
  \parbox[t]{\@tabfnwidth}{\footnotesize #2\hfill}}
\catcode`\@=12 

\newcommand {\Pom} {I\hspace{-0.2em}P}

\newcommand {\alphapom} {\mbox{$\alpha_{\Pom}$}}

\newcommand {\alphappom} {\mbox{$\alpha^\prime_{\Pom}$}}
\newcommand{\be}{\begin{equation}}
\newcommand{\ee}{\end{equation}}
\newcommand{\bea}{\begin{eqnarray}}
\newcommand{\eea}{\end{eqnarray}}

\newcommand {\eps} {\epsilon}
\newcommand {\R} {{\rm Re}}
\newcommand {\I} {{\rm Im}}
\newcommand {\bm} {\boldmath}
\def\Qb{\overline{Q}}
\def\lsim{\mathrel{\rlap{\lower4pt\hbox{\hskip1pt$\sim$}}
    \raise1pt\hbox{$<$}}}         
\def\gsim{\mathrel{\rlap{\lower4pt\hbox{\hskip1pt$\sim$}}
    \raise1pt\hbox{$>$}}}         

\newcommand{\bce}{\begin{center}}
\newcommand{\ece}{\end{center}}

\newcommand{\ba}{\begin{array}}
\newcommand{\ea}{\end{array}}

\newcommand{\bDelta}{\mbox{\boldmath $\Delta$}}

\newcommand{\bPsi}{\mbox{\boldmath $\Psi$}}
\newcommand{\bPhi}{\mbox{\boldmath $\Phi$}}

\newcommand{\bkappa}{\mbox{\boldmath $\kappa$}}

\newcommand{\bfe}{{\bf e}}

\newcommand{\bb}{{\bf b}}
\newcommand{\br}{{\bf r}}

\newcommand{\bk}{{\bf k}}

\newcommand{\bp}{{\bf p}}

\newcommand{\bV}{{\bf V}}

\def\Qb{\overline{Q}}

\newcommand{\etal}{{\sl et al.~}}

\newcommand{\bfR}{\mbox{\boldmath ${\cal R}$}}

\def\gsim{\mathrel{\rlap{\lower4pt\hbox{\hskip1pt$\sim$}}
    \raise1pt\hbox{$>$}}}         

\newcommand{\dst}{\displaystyle}
\newcommand{\fr}[2]{\frac{{\dst #1}}{{\dst #2}}}

\begin{document}

\prepnum{DESY--04--243}

\title{Diffractive Vector Meson Production at HERA:\\
 from Soft to Hard QCD \thanks{To be published in Physics of Elementary
 Particles and Atomic Nucleai, JINR, Russia}}

\author{I.P. Ivanov$^{a,b,}$\thanks{E-mail: igivanov@cs.infn.it},
N.N.Nikolaev$^{c,d,}$\thanks{E-mail: n.nikolaev@fz-juelich.de},
A.A.Savin$^{e,f,}$\thanks{E-mail: savin@mail.desy.de},
\\
\makebox[8cm][c]{\normalsize
$^a$  INFN, Gruppo Collegato di Cosenza, Cosenza, Italy}\\
\makebox[8cm][c]{\normalsize
$^b$  Sobolev Institute of Mathematics, Novosibirsk, Russia}\\
\makebox[8cm][c]{\normalsize
$^c$ IKP(Theorie), Forschungszentrum J\"ulich, Germany}\\
\makebox[8cm][c]{\normalsize
$^d$L. D. Landau Institute for Theoretical Physics, Moscow, Russia}\\
\makebox[8cm][c]{\normalsize
$^e$University of Wisconsin, Madison, USA}\\
\makebox[8cm][c]{\normalsize
$^f$on leave from Skobeltsyn Institute of Nuclear Physics, Moscow State University, Moscow, Russia}}

\date{December 2004}

\abstract{
   Experimental results from HERA on diffractive
   vector meson production and their theoretical interpretation
   within microscopic QCD are reviewed with an emphasis on the BFKL color
   dipole and $k_T$-factorization approaches.
}

\makezeustitle

\newpage
\tableofcontents

\newpage
\pagenumbering{arabic} 
\pagestyle{plain}

\section{Introduction}
\label{sect1} 

\subsection{The motivation}  
\label{sect1.1}

The Deep Inelastic Scattering (DIS) of leptons off hadrons is
interpreted as a knockout of one of the charged partons of the
target by hard Rutherford scattering followed by a complete
shattering of the target nucleon or nucleus. One of major
discoveries at the electron-proton collider HERA at DESY was the
observation that the large rapidity gap events, in which the
target nucleon emerges in the final state with a loss of a very
small fraction of its energy-momentum, constitute a substantial
and approximately scaling fraction of high-energy/small-$x$ DIS of
electrons and positrons on protons ~\cite{pl:b315:481,np:b429:477}. Although
the major features of such events and their cross sections
have been correctly predicted within perturbative QCD
~\cite{NZdifDIS}, the very existence of large rapidity gap events 
for  nuclear targets is nearly paradoxical: as well known
the deposition of dozen MeV energy is already sufficient
to break up the target nucleus, still the theory predicts that for
a sufficiently heavy nucleus and for the Bjorken variable 
$x \lsim 10^{-3}$ the fraction of
rapidity-gap DIS with retention of the target nucleus in
exactly the ground state must be exactly 50 per cent \cite{NZZdiffr} and
there is a direct evidence for that from the E665 Fermilab experiment
\cite{E665gap}. The discovery of rapidity gaps at HERA has led to
a renaissance of the physics of diffractive scattering in an
entirely new domain, in which the large momentum transfer from
leptons provides a hard scale. It also vindicated the early
suggestions of Bjorken to look for hard diffraction in hadronic
interactions \cite{BjorkenGap} and stimulated a revival of the
rapidity gap physics with hard triggers --- large-$p_{\perp}$ jets,
$W^{\pm}$-bosons, excitation of heavy flavors, --- at the
proton-antiproton collider Tevatron (for the recent review see
\cite{DinoTevatron,CDFgap,FELIX} and references therein). Whether
the existence of such a hard scale makes the diffractive DIS
tractable within the perturbative QCD or not has been a subject of
intense theoretical and experimental research during the past
decade or so.  A good summary of the pre-1997 status of the vector
meson production physics is found in the monograph of Crittenden
\cite{crittenden:1997:mesons}, the pre-1999 status of theoretical
ideas on diffractive DIS was reviewed by Hebecker
\cite{ArturPhysRep}), for the general introduction into the
physics of diffractive scattering see the recent books of Barone
and Predazzi \cite{BaronePredazzi} and Forshaw and Ross
\cite{ForshawRoss}.

The subject of this review is a special case of diffractive DIS ---
the exclusive production of vector mesons. One disclaimer is in
order: we focus on the high-energy
and/or very small-$x$ regime dominated by the pQCD pomeron exchange
and, facing the size limitations, don't discuss very interesting 
low to moderate energy data from
the HERMES collaboration which are strongly affected by the
non-vacuum exchanges (for the review and references see \cite{HERMESreview}).
The past decade the topic of high-energy diffraction
has been dominated by new fundamental data coming from
the ZEUS and H1 experiments at HERA. The interest in the exclusive
electroproduction of vector mesons is multifold. From the purely
experimental point of view, the HERA experiments offer a prime
example of diffractive scattering at energies much higher than
were attainable before. Furthermore, the self-analyzing decays of
spin-1 vector mesons allow one to unravel the mechanism of
diffraction in full complexity. Specifically, the HERA experiments
for the first time gave an unequivocal proof that the $s$-channel
helicity non-conservation persists at highest available energies
\cite{epj:c12:393,epj:c13:371}. On the theoretical side, starting from
the seminal papers on the color dipole approach by Kopeliovich,
Zakharov et al.
 \cite{KZcharmonium,NNNcomments,KNNZrs,KNNZct,NNZscanVM}
and the related momentum space approach by Ryskin
\cite{RyskinJPsi} and Brodsky et al. \cite{BFGMSvm}, it has been
understood that the exclusive diffractive production of vector mesons
in DIS is a genuinely hard phenomenon, whose major features 
can be described by pQCD. This can be understood in terms of the
shrinkage of the photon with the increase of the hard scale
\cite{NZ91,KZcharmonium,NNNcomments}, and because of this shrinkage
the diffractive production probes the hadronic properties of the
photon and vector mesons at short distances. One of the direct
manifestations of this shrinkage of the photon is a decrease of the
diffraction slope with the increase of the hard scale
\cite{NZZslope,NNPZZslopeVM,*NNPZZslopeVM1}, which has for the first time been
observed at HERA \cite{pl:b356:601,np:b468:3}, for the earlier evidence
from the NMC experiment see \cite{NMCslope}. Finally, the presence
of the hard scale enables one to test the modern theoretical ideas
on the mechanism of the $t$-channel exchange with  vacuum 
quantum numbers, i.e., the QCD
Pomeron. The way the QCD Pomeron is probed in diffractive vector
meson production is similar to, but still different from, that in
the conventional inclusive DIS. For instance, large-$t$ 
diffractive production of
vector mesons probes the QCD Pomeron in a hard regime
\cite{GinzburgPanfilSerbo} inaccessible in inclusive DIS.


\subsection{From inclusive DIS to DVCS to exclusive vector meson
production} 
\label{sect1.2}

\begin{figure}[htbp]
   \centering
   \epsfig{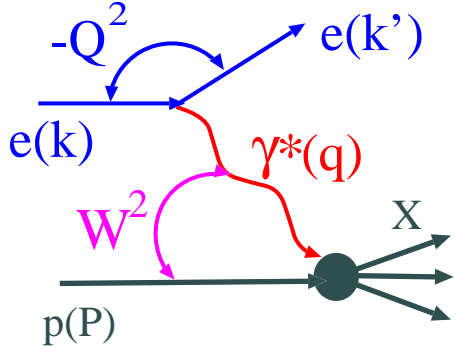}
   \caption{\it The kinematics of DIS \label{fig:DISbasic}}
\end{figure}

To this end recall the basics of inclusive DIS of leptons off
nucleons
$$
e(k)~p(P)\to e(k')~X\, .
$$
To the lowest order in QED it is treated in the one-photon
exchange approximation. The leptons serve as a source of virtual
photons of energy $\nu$ and virtuality $Q^2=-q^2$ (the scattering kinematics and
the 4-momenta are shown in Fig.~\ref{fig:DISbasic}) and the
fundamental process is the virtual photoabsorption
$$
\gamma^*(Q^2)~p(P) \to X\, .
$$

\begin{figure}[htbp]
   \centering
   \epsfig{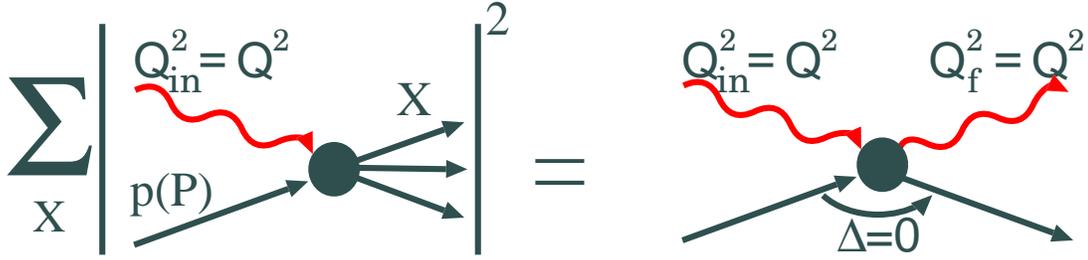}
   \caption{\it    The unitarity relation between DIS and
forward Compton scattering \label{fig:ComptonDIS}}
\end{figure}

In the fully inclusive DIS only the scattered lepton is detected
and one sums over all the hadronic final states $X$. Then the
observed inclusive DIS cross section is proportional to the
absorptive part of the forward, at vanishing momentum transfer
$\bDelta$, virtual Compton scattering amplitude
${\cal T}_{\mu\nu}(\nu,Q_f^2,Q_{in}^2,\bDelta=0)$ shown in
Fig.~\ref{fig:ComptonDIS}

\be
\gamma^*(Q^2)~p(P) \to \gamma^*(Q^2)~p(P) \label{eq:1.2.1}
\ee
and, invoking the optical theorem, can be cast in the form of the
flux of virtual transverse $(T)$ and scalar (longitudinal) $(L)$
photons times the total photoabsorption cross sections $\sigma_T$
and $\sigma_L$.

\begin{figure}[htbp]
   \centering
   \epsfig{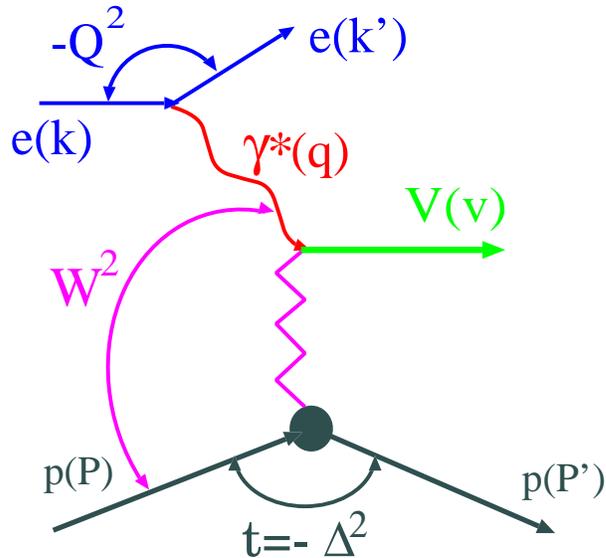}
   \caption{\it    Schematic diagram of exclusive vector-meson production in
   $ep$ interaction, $ep \rightarrow eVp$.}
   \label{fig:VMgraph}
\end{figure}

Now take a closer look at the Compton scattering amplitude as a
function of the virtuality of the incident $(in)$ and final $(f)$
state photons, $Q_{in}^2$ and $Q_f^2$, respectively. In fully
inclusive DIS this amplitude is accessible only for
$Q_{in}^2=Q_f^2 =Q^2$ and at vanishing momentum transfer
$\bDelta=0$. When continued analytically to $Q_f^2=0$ the
amplitude ${\cal T}_{\mu\nu}(\nu,0,Q^2)$ will describe the exclusive real
photon production often referred to as the Deeply Virtual Compton
Scattering (DVCS) \cite{DVCS}
\be
\gamma^*(Q^2)~p(P) \to
\gamma~p(P')\, , \label{eq:1.2.2}
\ee
while the further
continuation to $Q_f^2=-m_V^2$ gives the amplitude of the
exclusive vector meson production
\be
\gamma^*(Q^2)~p(P) \to
V(v)~p(P')\, . \label{eq:1.2.3}
\ee
Both DVCS and exclusive
vector meson production can be studied experimentally by selecting a
special final state $X=\gamma~ p(P')$ or $X=V~p(P')$,
respectively. Furthermore, both reactions can be studied at
the non-vanishing momentum transfer $\bDelta$, i.e.,
$t=-\bDelta^2 \neq 0$, for the definition of the kinematical variables see Fig.~\ref{fig:VMgraph}.

The point that inclusive DIS, DVCS, and exclusive vector meson
production are described by the same analytic function taken at
different values of $Q_f^2$ suggests from the very outset the
complementary probe of high energy pQCD in the three reactions
(\ref{eq:1.2.1}), (\ref{eq:1.2.2}), and (\ref{eq:1.2.3}). For
instance, in the forward, $\bDelta=0$, Compton scattering probed
in inclusive DIS the helicity flip amplitudes vanish for the
kinematical reason. In contrast to that, the inclusive vector
meson production at $\bDelta\neq 0$ enables one to determine  the
full set of helicity-conserving and helicity-flip amplitudes and
investigate the spin properties of hard (generalized) Compton
scattering to full complexity.


\subsection{When vector meson production is dominated by
small color dipole interactions?} 
\label{sect1.3}

\begin{figure}[htbp]
   \centering
   \epsfig{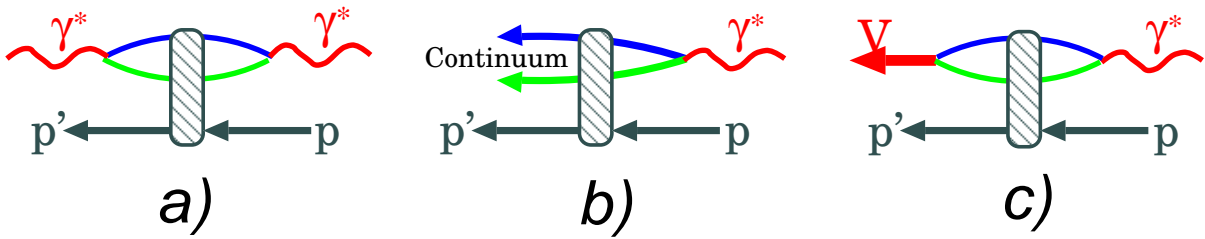}
   \caption{\it    The unified picture of Compton scattering,
diffraction excitation of the photon into hadronic continuum
states and into the diffractive vector meson
\label{fig:DISdifVMunified}}
\end{figure}

The intimate relationship between inclusive DIS, DVCS, and
exclusive vector meson production is still better seen in the
lightcone color dipole picture of small-$x$ DIS which illustrates
nicely the interplay of the scattering mechanism and the
(partonic) structure of particles. It is needless to recall the
outstanding role of the photon-matter interactions in the
conception and formation of the quantum mechanics and quantum
field theory. In the early years of the nonrelativistic quantum
mechanics the photon has been regarded as structureless and the
focus of the theory was on spectral lines, photo-effect and the
related phenomena. With the advent of the first quantum field
theory --- the Quantum Electro Dynamics (QED), --- it has become clear
that the fundamental transition
\be
\gamma \Leftrightarrow e^+e^-
\label{1.3.1}
\ee
between bare particles gives rise to a concept
of a dressed physical photon that contains all bare states to
which it couples via (\ref{1.3.1}) and still higher order QED
processes. At low energies, the virtual vacuum polarization gives
rise to the well known Uehling-Serber radiative correction to the
Coulomb potential; at higher energies the familiar Bethe-Heitler
$e^+e^-$ pair production in the Coulomb field of a nucleus can be
viewed as materialization of the $e^+e^-$ component of the
physical photon (see Bjorken, Kogut, Soper
\cite{BjorkenKogutSoper}). The Compton scattering which is behind
inclusive DIS at very small values of the Bjorken variable
$x$ can be viewed as (i) the
transition of the virtual photon to the $q\bar{q}$ pair (the color
dipole) at a large distance
\be
l \sim {1\over m_N x}\,,
\label{1.3.2}
\ee
upstream the target (here $m_N$ is the nucleon mass), 
(ii) interaction of the
color dipole with the target nucleon, and (iii) the projection of
the scattered $q\bar{q}$ onto the virtual photon 
(Fig.~\ref{fig:DISdifVMunified}a). Notice the very special choice of the
stage (iii): if one lets the scattered color dipole materialize as
hadrons, one ends up with the large rapidity gap DIS --- the
diffractive excitation $\gamma^*~p(P) \to p(P')X$. Here the
production of continuum hadronic states $X$ is modeled by the
continuum $q\bar{q}$ states (Fig.~\ref{fig:DISdifVMunified}b),
whereas the projection of the scattered $q\bar{q}$ color dipole
onto the vector meson gives the exclusive (diffractive, elastic)
vector meson production, and projection onto the real photon gives
so-called DVCS (Fig.~\ref{fig:DISdifVMunified}c).
The amplitude of the transition of the photon into the $q\bar{q}$ state, alias the
$q\bar{q}$ wave function of the photon, and the amplitude of
scattering the color dipole off the target are the universal
ingredients in all the processes. The wave function of the virtual
photon is well known \cite{NZ91}, and different processes probe the color
dipole scattering amplitude at different dipole sizes
\cite{NNNcomments}. 

For instance, irrespective of the photon's virtuality $Q^2$, the
inclusive diffractive DIS into the continuum states is controlled for the
most part by interaction of large color dipoles \cite{NZdifDIS}.
The scaling violations in the proton structure function (SF),
$\partial F_{2p}(x,Q^2)/\partial\log Q^2$,
come from small color 
dipoles of size
\be
r^2\sim {4 \over Q^2+ 4m_q^2}\,, \label{eq:1.3.3.0} 
\ee
whereas the absolute value of $F_{2p}(x,Q^2)$ receives contributions 
from large to small dipole sizes
\cite{NZglue,INdiffglue}, 
\be
{4\over Q^2+4m_q^2} < r^2 < {1\over
m_q^2}\,. 
\label{eq:1.3.4} 
\ee
In contrast to the inclusive DIS and inclusive diffractive DIS,
the amplitude of the exclusive vector meson production is
dominated by the contribution from small dipoles of size
\cite{NNNcomments,KNNZrs} 
\be
r\sim r_S \approx {6 \over \sqrt{Q^2+m_V^2}}\,, \label{eq:1.3.3} 
\ee
often referred to as the
scanning radius (formula (\ref{eq:1.3.3}) is applicable only
if $r_S$ is smaller than the typical hadron size). This, 
exclusive vector meson production offers a cleaner
environment for testing transition from soft to 
hard scattering.

The color dipole formalism is entirely equivalent to 
the BFKL formalism of the (transverse) momentum dependent
gluon distributions in the leading $\log{1\over x}$
approximation \cite{FKL75,KLF77,*KLF771,BL}. 
Within this formalism, often referred to as
the $k_{\perp}$-factorization, Eq.~(\ref{eq:1.3.3})
suggests that the vector meson production probes the gluon
density of the target at pQCD hard scale
\cite{RyskinJPsi,KNNZrs,KNNZct,NNZscanVM,Igorhardscale} 
\be
\overline{Q}^2 \approx {1\over 4}(Q^2+m_V^2) ={9 \over r_S^2},
\label{eq:1.3.5} 
\ee
which is large for heavy quarkonia ($J/\Psi,\Upsilon,...$) or for
large $Q^2$. In the hard regime of small scanning radius, the
vector meson production amplitudes will only depend on the wave
function of vector mesons at a vanishing quark-antiquark
separation in the two-dimensional transverse, or impact-parameter,
space. There still remains a certain sensitivity to the separation
of quarks in the longitudinal direction, which nonrelativistically
is conjugate to the longitudinal Fermi motion of the quark and
antiquark in the vector meson or the partition of the longitudinal
momentum of the vector meson between the quark and antiquark in
the relativistic lightcone language. As a result, the vector meson
production amplitude is not calculable from the first principles
of pQCD, still the sensitivity to the soft input can to a large
extent be constrained by the decay $V\to e^+e^-$, which proceeds
via the short-distance annihilation $q\bar{q}\to e^+e^-$. Then
Eq.~(\ref{eq:1.3.3}) suggests that, upon factoring out the
emerging $V\to e^+e^-$ decay amplitude, the vector meson production
amplitudes will depend on the hard scale $\overline{Q}$ in a
universal manner. Finally, the energy dependence of the vector
meson production amplitude offers a more local probe of the
properties of the hard pQCD Pomeron than the inclusive DIS.


\subsection{The scale for the onset of hard regime} 
\label{sect1.4}

Before opening the issue of hard production of vector mesons, one
needs to define the typical soft production. Here a brief comment
on the venerable Vector Dominance Model (VDM) is in order.
Because of the obvious dominance by the vector meson pole
contribution, the point that at $Q_f^2 = -m_V^2$ the amplitude of
the production of the timelike virtual photon $\gamma^*(Q_f^2)$
will be proportional to the appropriate vector meson production
amplitude times the $\gamma^*(-m_V^2)V$ transition amplitude is a
tautology. Experimentally, the timelike photons are produced in
the $e^+e^-$ annihilation and the $\gamma^*(-m_V^2)V$ transition
amplitudes are measured at the $e^+e^-$ colliders and, of course,
in the decay $V_{0}\to e^+e^-$. The assumption that the ground
state vector meson pole contribution dominates the photoproduction
amplitudes, and the $\gamma^*(Q_f^2)V$ transition amplitude does
not vary substantially from the vector meson pole
$Q_f^2=-m_V^2$ down to $Q_f^2=0$ is the basis of the very successful
VDM as formulated by Sakurai \cite{Sakurai}, Gell-Mann,
Zachariasen, Scharp and Wagner \cite{GellMann1961,GMSW1962} (for
the comprehensive review of foundations and tests of the VDM, see
Bauer et al. \cite{BauerRevModPhys,*BauerRevModPhys1}). 

\begin{figure}[htbp]
   \centering
   \epsfig{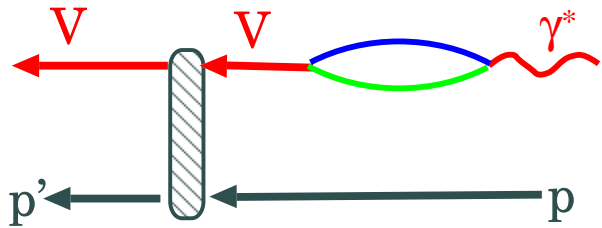}
   \caption{\it    The VDM amplitude for the vector meson photoproduction
\label{fig:VDM}}
\end{figure}

From the color dipole point
of view, the success of the VDM in real photoproduction derives
from the proximity of the distribution of color dipoles
$q_f\bar{q}_f$ in the ground state vector mesons and in the real
photon. So, the $q_f\bar{q}_f$ component of the physical
photon can be approximated by the corresponding vector meson
(quarkonium) and the amplitude of interaction of the color dipole
with the nucleon can be approximated by the vector meson-nucleon
scattering amplitude, for an illustration see Fig.~\ref{fig:VDM}.
From the naive quark model viewpoint, the $\rho$-meson is the
hyperfine partner of the $\pi$-meson and the 2-dimensional charge
radius, $R_{\pi}$, of the $\pi^{\pm}$ sets the relevant scale. 

One comes to the same conclusion from the experimentall
observation that to a very good accuracy the
$t$-dependence of elastic $\pi N$ scattering, real Compton
scattering $\gamma p \to \gamma p$ and real photoptoduction
  $\gamma p \to \rho p$ is the same
\cite{BauerRevModPhys,*BauerRevModPhys1, GammaPelast}.
Indeed, within the VDM which is a
very good approximation for real photons, the differential
cross sections of the latter two processes are proportional
to the differential cross section of elastic $\rho N$
sacttering. Then the equal $t$-dependence of the $\pi N$
and $\rho N$ elastic scattering entails an equality of t
he radii of he $\rho$-meson.

Experimentally,
the charge form factor of the pion is well described by the VDM
$\rho$-pole formula and \cite{PionRadius,CEBAFpionFF}
\be
\langle \br_{\pi}^2 \rangle^{1/2} \approx 0.55~{\rm fm}
\approx {2 \over m_{\rho}}\, . \label{eq:1.4.1}
\ee

The onset of the hard regime in diffractive vector meson production requires
that the scanning radius $r_S$ is smaller than any other hadronic scale.
First place, one needs $r_S^2 \ll
\langle \br_{\rho}^2 \rangle \approx \langle \br_{\pi}^2 \rangle$,
i.e.,
\be
Q^2 \gg  {5 \over 4} m_{\rho}^2 \approx 1~{\rm (GeV)^2}\, .
\label{eq:1.4.2}
\ee
The corrections in the small parameter $r_S/\langle \br_{V}^2 \rangle)^{1/2}$
depend on the wave function (WF) of the vector meson. With the soft, Gaussian,
wave function,  $\psi_V(r) \sim \exp(-r^2/\langle \br_{V}^2 \rangle)$,
in order for the WF-dependent corrections not to exceed $\lsim (20\div 30)\%$
one needs $Q^2 \gsim (2\div 4)~{\rm (GeV)^2}$.
For the hard, Coulomb-like, wave functions,
$\psi_V(r) \sim \exp(-r/\langle \br_{V}^2 \rangle^{1/2})$,
a still higher $Q^2 \gsim 10{\rm (GeV)^2}$ is needed for a similar
insensitivity
to the shape of the wave function,
for the related discussion of the onset of pQCD
see \cite{GinzburgIvanovHight}.
Even for the heavy $J/\Psi$ the scanning radius at $Q^{2}=0$ is large,
\be
r_s \approx {6 \over m_c} \sim \langle \br_{J/\Psi}^2 \rangle^{1/2} \approx
0.4~{\rm fm}
\label{eq:1.4.3}
\ee
(for the charmonium parameters see
\cite{NovikovPhysRep,QuiggPhysRep,NNPZdipoleVM}),
so that for the onset of the short-distance regime
insensitive to the  shape of the wave function of the $J/\Psi$ one needs
$Q^2 \gsim m_{J/\Psi}^2$. In the realistic QCD there is still another
scale - the propagation radius
for perturbative gluons which is small, $R_c\approx (0.2\div 0.3)$fm
(for the lattice QCD evaluations of $R_c$ see \cite{Meggiolaro},
for the origin of $R_c$ in the instanton models of QCD vacuum see
\cite{Shuryak}, the analysis of heavy quarkonia decays is found in
\cite{Field}).
The color dipole cross section is of true pQCD origin only for dipoles
$r \lsim R_c$, i.e., the fully pQCD description of diffractive
vector mesons requires $r_S \lsim R_c$, i.e.,
\be
Q^2 \gsim Q_{pQCD}^2 = {36\over R_c^2} \approx (20\div 30) {\rm GeV}^2.
\label{eq:1.4.4}
\ee
One must not be discouraged, though: the $r$-dependence of the 
dipole cross section does not change any dramatically from the
pure pQCD domain of $r \lsim R_c$ to the non-perturbative domain
of $r \gsim R_c$, and the fundamental concept of the scanning
radius remains viable up to $r_s \lsim 1$ fm, see also the discussion
in Sect. 3.3.2.

The large momentum transfer, $|t| \gg 1~{\rm GeV}^2$, is still
another way to probe the structure of the photon
and vector meson at short distances, $r \sim {1/\sqrt{|t|}}
 \gg 1~{\rm GeV^{-1}}$.
It is generally believed \cite{ForshawRyskinLarget,GinzburgIvanovHight}
that $|t|$ supersedes $\Qb^2$ as a hard
scale if $|t| \gsim \Qb^2$. The caveats of $t$ as large scale and
of single BFKL pomeron exchange dominance will will be
discussed to more detail in Sect. 4.11.


\subsection{The structure of the review} 
\label{sect1.5}

In this review we focus on the onset of hard pQCD regime in
exclusive vector meson production at HERA. The presentation of the
experimental data and of theoretical ideas goes in parallel,
and an intimate connection between the vector meson production 
and the inclusive DIS will be repeatedly underlined. 
For this reason the presentation of the theoretical
ideas on vector meson production will be heavily biased towards
the color dipole picture and its momentum-space counterpart --- the
so-called $k_{\perp}$-factorization.

The brief description of the H1 and ZEUS  detectors, the
kinematics of DIS and of the vector meson production,  the event
selection, the definition of major observables and of the spin
density matrix of virtual photons is presented in Section 2. The
subject of Section 3 is an overview of basic theoretical ideas
on the vector meson production. Here we discuss briefly the Regge theory
of the soft photon and hadron interactions, the QCD approach to the
vacuum exchange (the Pomeron), the flavor dependence, the
connection between the vector meson production and the leptonic
decay of vector mesons, the origin of $s$-channel helicity
non-conservation (SCHNC) and the exclusive-inclusive duality
connection between inclusive diffractive DIS  and vector meson
production. We also introduce the color dipole approach to DIS and
vector meson production and explain how the shrinkage of virtual
photons makes the vector meson production pQCD tractable. The
unified microscopic QCD approach to small-$x$ DIS and diffractive
vector meson production --- the $k_{\perp}$-factorization
approach, which is equivalent to the color dipole approach, --- is
presented to more detail in Section 4. Here we discuss both
the small-$t$ production within the diffraction cone and
major ideas on large-$t$ proton dissociative reaction.
This section can be skipped
in the first reading, but is essential for understanding the
status of theoretical calculations of the vector meson production.

In Section 5 we start the presentation of the physics results with
the helicity structure of the vector meson production. This
includes the definition of the spin observables, an introduction
into the important subject of the s-channel helicity
non-conservation (SCHNC) and the comparison of the experimental
data on the spin density matrix of produced vector mesons with the
theoretical expectations from the
color-dipole/$k_{\perp}$-factorization approach 
\footnote{Throughout this review, the 
numerical results shown for the $k_{\perp}$-factorization
are either taken from the PhD thesis \protect\cite{IgorPhD} or performed
specially for this review \protect\cite{IgorNumerics}.}
In Section 6 we discuss the $Q^2$-dependence  of the vector meson
production cross sections as well as the longitudinal-to-transverse
cross sections ratios $R_V = \sigma_L/\sigma_T$. 
We put special emphasis on the flavor dependence of cross sections,
emphasize an importance of $(Q^2+m_V^2)$ as the hard pQCD scale
and comment on the sensitivity of  $R_V = \sigma_L/\sigma_T$ to
the short distance wave function of vector mesons.
In Section 7 we review the experimental data on the energy dependence 
of the cross sections and its theoretical interpretation 
in terms of the Pomeron exchange. We show how the
the change of the energy dependence from light to heavy flavors
and from photoproduction to DIS is controlled by $(Q^2+m_V^2)$ 
as the hard pQCD scale. We comment on tricky points in comparison
of hard scales and energy dependence in inclusive DIS and 
diffractive vector meson production and on the impact of interplay
of the scanning radius $r_S$ and the position of the node of
the radial wave function for the $\Psi(2S)$ production cross 
section.  
The focus of Section 8 is on the $t$-dependence of the cross sections,
both in the low-$t$ and high-$t$ regimes. The discussion of
low-$t$ data centers on the $Q^2$, $m_V$ and $W$-dependence of 
the slope of diffraction cone. The recurrent theme is a
universality of diffraction slopes as a function of the scanning radius
and/or $(Q^2+m_V^2)$ as the hard pQCD scale. The properties of 
the Pomeron trajectory  $\alpha_{\Pom}(t)$ extracted from 
the vector meson production 
data are discussed in detail: the experimentally observed shrinkage
of the diffraction cone for the $J/\Psi$ production
gives a strong evidence for $\alpha_{\Pom}(t)$
which decreases with $t$  approximately linearly 
at $|t| \lsim 1 ~{\rm GeV^2}$, but 
then starts rising up to $\alpha_{\Pom}(t) \sim 1.3$ 
in the hard regime of large $|t|$.
Finally, in Section 9, we summarize the principal findings from
HERA experiments on diffractive vector meson production
and list open issues in the pQCD interpretation of these data.

It is important to mention here that not all currently available HERA data
are always shown in each plot where they may belong to. This is because
sometimes the published plots from H1, ZEUS and other
authors are used without any modifications. This is especially valid for
the "preliminary" H1 and ZEUS plots, that have been shown to the
conferences and are not yet submitted in form of "official" papers. Such
plots are just taken as they are. If for some compilation and figures only
very recent data are used it is explained in the correspondent caption.

\newpage


\section{The experimental overview}  
 \label{sect2}


\subsection{HERA} 
\label{sect2.1}

HERA(Hadron Electron Ring Anlage)
 is the world's first lepton-proton collider located at
Deutsches Elektronen Synchrotron (DESY) site in Hamburg, Germany 
(see e.g. ~\cite{arevns:44:413} and references therein). The HERA
ring has a circumference of about 6.3 km with two separated
synchrotron rings for electrons (positrons) and protons. It runs
10-30 m below ground level and has four experimental areas. In two
of them the beams are made to collide to provide $ep$ interactions
for the experiments H1 and ZEUS. The remaining two areas are used
by the fixed target experiments: HERMES~\cite{Ackerstaff:1998av},
which scatters longitudinally polarized electrons off stationary
polarized targets, and HERA-B~\cite{Hartouni:1995cf,H:2000ce},
which investigated $CP$-violation in the $B^0 \bar{B}^0$ system by
scattering beam halo protons off wire targets (was shut down in 2000). 
HERA was
commissioned in 1991 with the first $ep$ collision observed by H1
and ZEUS in the spring 1992. A major HERA upgrade took place
during 2000-2002 break. A significant luminosity increase should
be achieved by stronger focusing of both the electron and the
proton beams, see Tab.~\ref{tab:HERA} where the design and
achieved HERA values, as well the values of HERA after upgrade, are
summarized. Further information about HERA luminosity upgrade can
be found in ~\cite{Schneekloth:1998lu,Seidel:2000lu}.
\begin{table}[htbp!]
\begin{center}
\begin{tabular}{||c|c|c|c||}
\hline
HERA Parameters & Design & 2000 & Design after upgrade \\
\hline\hline
 p/e beam energy (GeV) & 820/30 & 920/27.5 & 920/30 \\
\hline
 p/e beam current (mA) & 160/58 & $>$100/$>$50  & 140/58 \\
\hline
Number of bunches proton/electron & 210 & 180/189 &  180/189  \\
\hline
Time between crossings (ns) & \multicolumn{3}{c||}{96}  \\
\hline
Proton $\beta$-function $x/y$ (m) & 10/1 & 7/0.5 &  2.45/0.18  \\
\hline
Electron $\beta$-function $x/y$ (m) & 2/0.7 & 1./0.7 &  0.63/0.26  \\
\hline Specific luminosity ($cm^{-2} s^{-1} mA^{-2}$) &
3.4x10$^{29}$ & 8x10$^{29}$ & 1.6x10$^{30}$ \\
\hline
Luminosity ($cm^{-2} s^{-1}$) & 1.5x10$^{31}$ & 2x10$^{31}$ & 7x10$^{31}$  \\
\hline
\end{tabular}
\caption{\it    HERA parameters. \label{tab:HERA}}
\end{center}
\end{table}


\subsection{The detectors H1 and ZEUS} 
\label{sect2.2}

\begin{figure}[htbp]
   \centering
   \epsfig{file=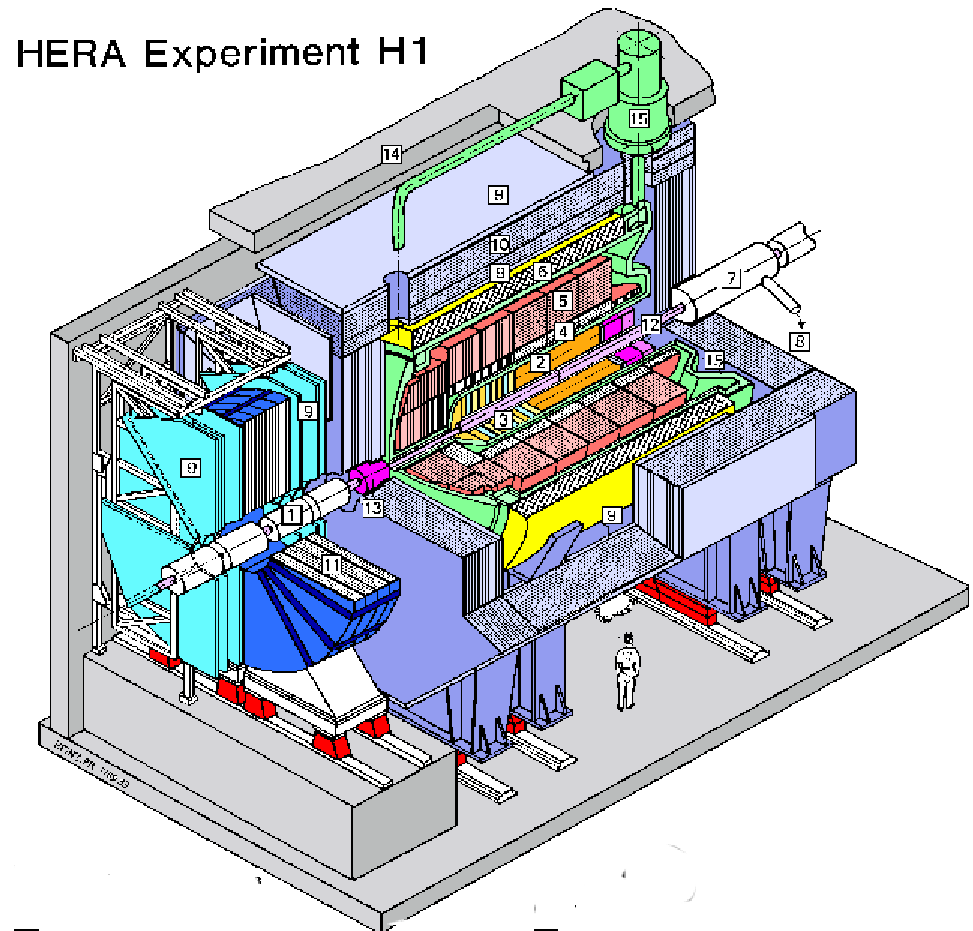,width=160mm}
   \caption{\it    The H1 detector \newline
   The main components are: \newline
   1,7,11 - Beam, Compensating and Muon toroid magnets ; \newline
   2 - Central tracking detector ; \newline
   3 - Forward tracking and Transition radiators; \newline
   4,5 - Liquid Argon Calorimeter; \newline
   6- Superconducting coil; \newline
   9 - Muon chambers ;  \newline
   12,13 - Warm electromagnetic and Plug calorimeters. \label{fig:H1}}
\end{figure}

\begin{figure}[htbp]
   \centering
   \epsfig{file=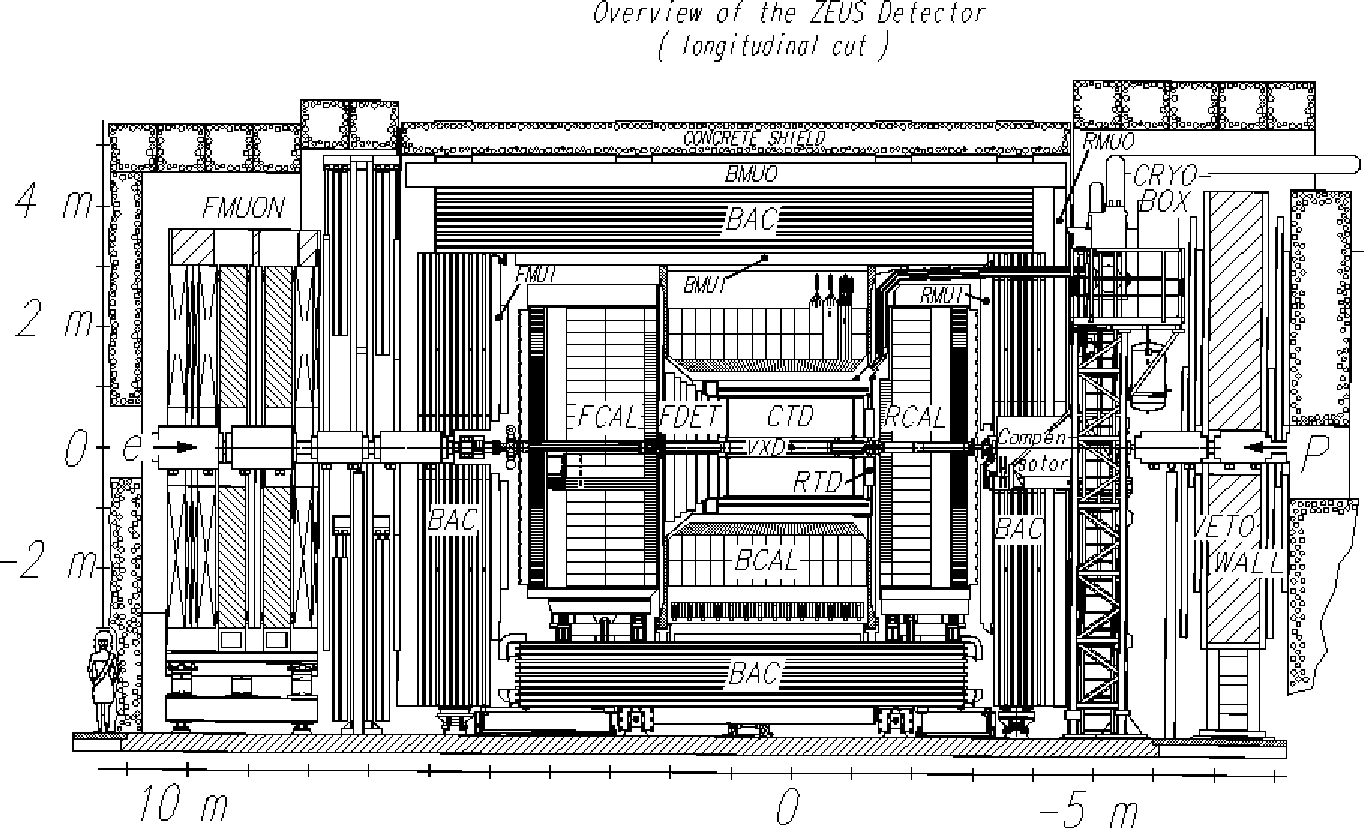,width=100mm,angle=-90}
   \caption{\it    The ZEUS detector \newline
   The main components are: \newline
   VXD - Vertex Detector,
   after 2000 upgrade Silicon Microvertex Detector ; \newline
   CTD - Central Tracking Detector ; \newline
   FDET - Forward Detector ; \newline
   RTD - Rear Tracking Detector ; \newline
   F/RMUON - Forward/Rear Muon Chambers; \newline
   BMUOI/O - Barrel Muon Inner/Outer Chambers; \newline
   F/B/RCAL - Forward/Barrel/Real Calorimeters; \newline
   BAC - Backing Calorimeter. \label{fig:ZEUS}}
\end{figure}

The H1 and ZEUS are general purpose detectors with nearly hermetic
calorimetric coverage and a large forward-backward asymmetry to
accommodate the boost of the $ep$ center-of-mass in the direction
of the proton beam.

The H1 and ZEUS detectors are described in details elsewhere
~\cite{nim:a386:310,zeus:1993:bluebook}.
The detectors are shown in Figs.~\ref{fig:H1},~\ref{fig:ZEUS}. The
main difference between the H1 and ZEUS detectors is the choice of
the calorimetry. In the H1 case the main liquid argon calorimeter
with different tracking detectors inside is surrounded by a large
diameter superconducting solenoid thus minimizing the amount of
inactive material in the path of the particles between the
interaction point and the calorimeter. In the ZEUS case only
tracking chambers are placed inside a superconducting solenoidal
magnet, surrounded by a uranium-scintillator sampling calorimeter
with equal response to the electromagnetic and hadronic
components. Both detectors are surrounded by muon chambers. Some
of the components most relevant for the vector meson analysis are
outlined below.


\subsubsection{Tracking detectors} 
\label{sect2.2.1}

Charged particles are measured both for H1 and ZEUS by the central
tracking detectors operating in magnetic field of 1.15 T and 1.43
T respectively. Both trackers are build mainly of drift, jet and
proportional chambers. The part closest to the beam pipe in H1
case uses silicon detectors (Central Silicon Tracker). During
2000-2001 shutdown ZEUS has also installed a Silicon Micro Vertex
Detector that should significantly improve the resolution of the
tracking system and the vertex reconstruction.

The polar angle coverage is $15^\circ < \theta < 164(165)^\circ$
for H1(ZEUS) correspondingly. The relative transverse-momentum
resolution is $\sigma(p_T)/p_T \approx 0.006p_T$ with $p_T$ in GeV
for both experiments. Charged particles in the forward direction
are detected in the forward tracking detector covering the polar
angle range $7^\circ<\theta<25^\circ$ and
$7^\circ<\theta<28^\circ$ for the H1 and ZEUS respectively, the
backward part ($172^\circ<\theta<176^\circ$) is covered by the
backward silicon tracker, BST in the H1, and by the Small Rear
Tracking Detector in the ZEUS cases.

Charged tracks measured by the tracking system are used to
reconstruct the interaction vertex for each event.


\subsubsection{Calorimetry} 
\label{sect2.2.2}

The tracking detectors of H1 are surrounded by a liquid argon
calorimeter (LAr, $4^\circ<\theta<154^\circ$, $\sigma/E:
0.12/\sqrt{E}$ and $0.50/\sqrt{E}$
 for
electromagnetic and hadronic showers correspondingly, E in GeV)
and a scintillating fiber calorimeter (spaghetti calorimeter,
SpaCal, $153^\circ<\theta<178^\circ$, $0.075/\sqrt{E}$ for
electromagnetic showers).

The central tracking detector of ZEUS is placed inside of a thin
super-conducting coil. Surrounding the solenoid is the
high-resolution uranium-scintillator calorimeter (CAL) which
covers the angular range $2.6^\circ < \theta < 176.2^\circ$ with
equal response to the electromagnetic and hadronic components and
with energy resolution of $0.18/\sqrt{E}$ and $0.35/\sqrt{E}$ for
the electromagnetic and hadronic components correspondingly.

In 1998-2000 a Forward Plug Calorimeter (FPC, lead-scintillator
sandwich calorimeter~\cite{nim:a450:235}) was installed in the
20x20 cm$^2$ beam hole of the forward part of the CAL with only a
small hole of radius 3.15 cm in the center to accommodate the
beam pipe. It extended the pseudorapidity coverage of the forward
calorimeter from $\eta < 4.0$ to $\eta <5.0$. A similar device ---
Beam Pipe Calorimeter (BPC) --- was installed in the rear region of
the ZEUS detector, 294 cm away from the nominal $ep$ interaction
point, mainly to measure the lepton scattered at very small angle.
Both these calorimeters were removed during 2000-2001 shutdown
because of the changed beam-pipe geometry for the HERA luminosity
upgrade.


\subsubsection{Muon detectors} 
\label{sect2.2.3}

The H1 muon system consists of an instrumented iron return yoke
(Central Muon Detector, CMD, $4^\circ<\theta<171^\circ$) and a
Forward Muon Detector (FMD, $3^\circ<\theta<17^\circ$).

The ZEUS muon system covers the polar angles between
$10^\circ<\theta<171^\circ$, in additional the forward  part has
additional drift chambers for high-momentum muon reconstruction
for polar angles between 6$^\circ$ and 30$^\circ$.


\subsubsection{Forward detectors and proton taggers} 
\label{sect2.2.4}

 Both H1 and ZEUS have very forward detectors, placed along the beam-line in the
 direction of the proton beam, 20-90 m away from the nominal interaction point.

 Forward Proton Spectrometer (FPS in H1) and Leading
 Proton Spectrometer (LPS in ZEUS) consist of movable "Roman Pots" forming
 together with the magnets of HERA a kind of magnetic spectrometer. Scattered
 protons with a different energy and/or angle compared to the nominal beam
 protons are separated from the beam and are detected at appropriate positions.

 Also in the proton direction experiments have placed simple scintillation
 counters (5.15 m and 23-24 m from the nominal interaction point in the
 ZEUS case, five stations between 9 and 92 m in the H1 case)
 that are used as Proton Remnant Taggers (PRT).
 These taggers cover very high region of pseudorapidity
 (e.g. $4.3 < \eta < 5.8$ for ZEUS) and are
 used to tag the events where the proton dissociate.


\subsubsection{Luminosity detectors and electron taggers} 
\label{sect2.2.5}

The luminosity is determined from the rate of the Bethe-Heitler
bremsstrahlung process $ep \rightarrow e\gamma p$, where the
high-energy photon is detected in a lead-scintillator calorimeter
(LUMI) located at $Z=-107$ m in the HERA tunnel in  the ZEUS case or
by a crystal Cherenkov calorimeter (PD) located at $Z=-103$ m in the
H1 case.

 In the lepton direction the experiments have Photoproduction Taggers
 (PT) at 8
 and 44 m from the nominal interaction point for ZEUS, Electron Taggers (ET)
 in the H1 case.
 They detect leptons scattered
 under very small angle (less than few mrads). The leptons measured in the PT
 (ET) are used to tag photoproduction events, thus significantly reducing the
 background.


\subsection{Kinematics and cross sections} 
\label{sect2.3}


\subsubsection{Kinematics of DIS} 
\label{sect2.3.1}

Because of the small electromagnetic coupling $\alpha_{em}\approx
1/137$, the deep inelastic scattering of leptons off protons is
treated in the one-photon exchange approximation. The generic diagram
for DIS   $ e(k)~p(P) ~\rightarrow ~e(k')~X$ is shown in 
Fig.~\ref{fig:DISbasic}. The relevant kinematic variables  are:
\begin{itemize}
\item $Q^2 = -q^2=-(k-k')^2$, the negative squared four-momentum
of the virtual photon; 
\item $W^2 = (q+P)^2 =2m_p \nu +m_p^2-Q^2$, the squared
center-of-mass energy of the
  photon-proton system; 
\item $y = (P \cdot q)/(P \cdot k)$, the fraction of the positron
  energy transferred to the photon in the proton rest frame.
\item $x=Q^2/2(P \cdot q)$, the Bjorken variable, which in the
parton model interpretation of DIS has a meaning of the fraction
of the proton's lightcone momentum carried by the struck charged
parton.
\end{itemize}


\subsubsection{The flux and polarization of photons} 
\label{sect2.3.2}

The amplitude of DIS equals 
\be
T(e(k)~p(P) ~\rightarrow ~e(k')~X)
= \frac{4\pi \alpha_{em}}{Q^2}\langle
e(k')|l_{\mu}|e(k)\rangle\cdot g_{\mu\nu}\cdot \langle X|
J_{\nu}|p(P)\rangle\, ,
\label{eq:2.3.2.1} 
\ee
where $l_{\mu}$ and $J_{\mu}$ stand for the electromagnetic current 
of leptons and hadrons. The leptons serve as a source of photons 
and the physical process is the virtual photoabsorption
\[
\gamma^*(q)p(P) ~\rightarrow ~X\,.
\]
The virtual photons have three polarization states: the two
spacelike transverse ones with helicities $\lambda_{\gamma^*}=\pm
1$,
\be
\bfe(\pm) = -\frac{1}{\sqrt{2}}(\pm \bfe_{x}+i\bfe_{y})
\cdot e^{\pm i \Phi}
\label{ephoton}
\ee
and the timelike {\it scalar}  state (often misnamed the {\it
longitudinal} one, hereafter we follow this tradition)
\[
e_\mu(L)= -\sqrt{\frac{Q^2}{(P\cdot q)^2 +P^2Q^2}}\left[P_\mu+ \frac{(P\cdot
q)}{Q^2}q_\mu\right]\, .
\]
For the purpose of future convenience, here we choose the $z$-axis along 
the photon's 3-momentum, the $x$-axis 
in the $\gamma p \to V p$ reaction plane, and $\Phi$ is the azimuthal 
angle between the reaction and the $(e,e')$ scattering planes
(for more details see below section 5.1).
The complete set includes still another spacelike vector
\[
e_\mu(S) = \frac{1}{\sqrt{Q^2}}\;q_\mu\, .
\]
Making use of the expansion
\[
g_{\mu\nu} = e^*_{\mu}(L)e_{\nu}(L) -
\bfe^*_{\mu}(+)\bfe_{\nu}(+) - \bfe^*_{\mu}(-)\bfe_{\nu}(-)
-e^*_{\mu}(S)e_{\nu}(S)\, ,
\]
and of the current conservation, $(e(S)\cdot J)=0$,
one can write down 
\bea
\langle e(k')|l_{\mu}|e(k)
\rangle\cdot g_{\mu\nu}\cdot \langle X| J_{\nu}|p(P)\rangle &=&
\langle e(k')|l_{\mu}|e(k)\rangle e^*_{\mu}(L)\cdot
\langle X| J_{\nu}|p(P)\rangle e_{\nu}(L)\nonumber\\
 & -&
\langle e(k')|l_{\mu}|e(k)\rangle\bfe^*_{\mu}(+) \cdot
\langle X| J_{\nu}|p(P)\rangle \bfe_{\nu}(+)
\nonumber\\
&-& \langle e(k')|l_{\mu}|e(k)\rangle \bfe^*_{\mu}(-) \cdot
\langle X| J_{\nu}|p(P)\rangle \bfe_{\nu}(-)\, .
\label{eq:2.3.2.2} 
\eea 
Now notice that 
\be
T(\gamma^*(\lambda_{\gamma*};q)p(P) ~\rightarrow ~X)=\sqrt{4\pi
\alpha_{em}}\cdot \langle X| J_{\nu}|p(P)\rangle
e_{\nu}(\lambda_{\gamma*}) 
\label{eq:2.3.2.3} 
\ee
is precisely an
amplitude of the photoabsorption for the photon of polarization
$\lambda_{\gamma*}$, and $\langle e(k')|l_{\mu}|e(k)\rangle e^*_{\mu}(L),
-\langle e(k')|l_{\mu}|e(k)\rangle e^*_{\mu}(+), -\langle
e(k')|l_{\mu}|e(k)\rangle e^*_{\mu}(-)$ define the emission by leptons of
photons of appropriate polarization, which is quantified by 
the spin density matrix of the photon $\rho_{\lambda'\lambda}$. 
Then, making use of the
expansion (\ref{eq:2.3.2.2}) the differential cross section for
the leptoproduction of the specific final state $X$ can be
expressed through the photoabsorption cross sections as 
\bea
\frac{d\sigma(ep\to e'X)}{dQ^2dy} d\tau_X = \Gamma_T(Q^2,y)
\sum_{\lambda',\lambda=+,-,L}\rho_{\lambda'\lambda}
d\sigma_{\lambda'\lambda}(\gamma^* p ~\to~ X)\, ,
\label{eq:2.3.2.4} 
\eea 
where $d\tau_X$ is the element of the
appropriate phase space, 
\bea
d\sigma_{\lambda'\lambda}(\gamma^* p
~\to~ X) =\frac{1}{4\sqrt{(p\cdot q)^2+Q^2m_p^2}}
{\cal T}^*(\gamma^*_{\lambda'}p ~\rightarrow ~X)
{\cal T}(\gamma^*_{\lambda} p ~\rightarrow ~X)d\tau_X
\label{eq:2.3.2.5} 
\eea 
and 
\bea
\Gamma_T(Q^2,y)=\frac{\alpha_{em}}{\pi Q^2
y}\cdot\left(1-y+\frac{1}{2}y^2\right) 
\label{eq:2.3.2.6} 
\eea 
is the flux of
transverse photons. With this normalization the spin density
matrix of the photon equals 
\bea
\left(\begin{matrix}
\rho_{++} & \rho_{+-} &\rho_{+L}\\
\rho_{-+} & \rho_{--} &\rho_{-L}\\
\rho_{L+} &\rho_{L-} & \rho_{LL}\\
\end{matrix}\right)
=\left(\begin{matrix}
\frac{1}{2} & -\frac{1}{2}\epsilon\, e^{2i\Phi} & \frac{1}{2}\sqrt{\epsilon(1+\epsilon)}\, e^{i\Phi}\\
 -\frac{1}{2}\epsilon\, e^{-2i\Phi}  & \frac{1}{2} &-\frac{1}{2}\sqrt{\epsilon(1+\epsilon)}\, e^{-i\Phi}\\
 \frac{1}{2}\sqrt{\epsilon(1+\epsilon)}\, e^{-i\Phi} &-\frac{1}{2}\sqrt{\epsilon(1+\epsilon)}\, e^{i\Phi}& \epsilon \\
\end{matrix}\right)\, ,
\label{eq:2.3.2.7} 
\eea 
where 
\be
\epsilon = \frac{1-y
-y^2\frac{Q^2}{4\nu^2}}{1-y +\frac{1}{2}y^2
+y^2\frac{Q^2}{4\nu^2}} \approx \frac{2(1-y)}{(1-y)^2 +1}
\label{eq:2.3.2.8} 
\ee
is the longitudinal polarization of the
virtual photon. We also indicated the small-$x$ approximation for
$\varepsilon$ which is appropriate for DIS at HERA. 

Notice, that
because of the current conservation one can define the
longitudinal photon interaction amplitude in terms of the current
component $J_z$, which is customary in electronuclear physics, for
instance, see \cite{DombeyRevModPhys,BoffiPhysRep,SchillingWolf}.
It does not affect the observed cross section (\ref{eq:2.3.2.4})
because the different normalization of the amplitude
(\ref{eq:2.3.2.2}) for longitudinal photons is compensated for by
the change of the relevant components of the spin density matrix
of the photon.


\subsubsection{The transverse and longitudinal cross sections for DIS}
\label{sect2.3.3}

In the fully inclusive DIS one integrates over the whole phase
space of the state $X$ and sums over all states $X$. Then by
virtue of the optical theorem one can relate the photoabsorption
cross section to the absorptive part of the Compton forward
scattering amplitude 

\be
 \sum_{X}
\sigma_{\lambda'\lambda}(\gamma^* p ~\to~ X)
=\frac{1}{2\sqrt{(p\cdot q)^2+Q^2m_p^2}} {\rm Im}
{\cal T}_{\lambda'\lambda}(\gamma^* p \to \gamma^* p). \label{eq:2.3.3.1}
\ee
The crucial point is that for the unpolarized target the
helicity-flip, $\lambda'\neq \lambda$, amplitudes vanish in the
forward scattering. Then the virtual photon-proton cross section,
$\sigma^{\gamma{^*} p}$, can be determined from the measured
positron-proton cross section: 
\be
\sigma^{\gamma^{*} p} =
\sigma_{\rm T}^{\gamma^{*} p}+ \ \epsilon \sigma_{\rm
  L}^{\gamma^{*} p} = \frac{1}{{\it \Gamma}_{\rm T}(Q^2,y)}
\cdot \frac{d^2 \sigma^{ep}}{dQ^2 dy}, 
\label{eq:2.3.3.2} 
\ee
where
$\sigma_{\rm T}^{\gamma^{*} p}=\sigma_{++}=\sigma_{--}$ and
$\sigma_{\rm L}^{\gamma^{*} p}$ are the transverse and the
longitudinal virtual photoproduction cross sections, respectively.
The often discussed total inclusive cross section, $\sigma_{\rm
tot}^{\gamma^{*} p} = \sigma_{\rm T}^{\gamma^{*} p}+ \ \sigma_{\rm
  L}^{\gamma^{*} p}$ can be determined from $\sigma^{\gamma^{*} p}$
through the relation: 
\be
\sigma_{\rm tot}^{\gamma^{*} p} = \
\frac{1+R_{DIS}}{1+\epsilon R_{DIS}} \sigma^{\gamma^{*} p},
\label{eq:2.3.3.3}
\ee
where 
\be
R_{DIS}=\frac{\sigma_{\rm L}^{\gamma^{*} p}} { \sigma_{\rm
  T}^{\gamma^{*} p}}\, .
\label{eq:2.3.3.4} 
\ee
(Because $R$ is heavily used for different ratios, we supply it
by the subscript DIS.)
In the kinematic range of most of the
discussed measurements, the value of $\epsilon$ is close to unity,
and  because $R_{DIS}$ is small, $\sigma^{\gamma^{*} p}$ differs from
$\sigma_{\rm tot}^{\gamma^{*} p}$ by less than one percent.


\subsection{Kinematics of diffractive vector meson production} 

Diffractive vector meson production corresponds to the special
two-body final state which contains  only the vector meson and
scattered proton
\[
e(k)~p(P) ~\rightarrow ~e(k')~V(v)~p(P'),
\]
where $V$=\{$\omega$,$\rho$,$\phi$,$J/\psi$,$\Psi$',$\Upsilon$\}
and $k$, $k'$, $P$, $P'$, and $v$ are the four-momenta of the
incident lepton (positron or electron), scattered lepton, incident
proton, scattered proton and vector meson, respectively, see
Fig.~\ref{fig:VMgraph}.

The new kinematic variable is $t = (P - P')^2 = (v - q) ^2=
-\bDelta^2+t_{min}$, the
squared four-momentum-transfer at the proton vertex. 
At high energies the longitudinal momentum transfer 
$\Delta_L=m_p(Q^2+m_V^2)/W^2$ is small, $t_{min}=-\Delta_L^2$
can be neglected, and $t\approx t'\equiv -\bDelta^2$. Besides 
$t$, the new important variables are the orientation of the
production plane with respect to the electron scattering plane and
the appropriately defined polar and azimuthal angles of the decay
pions, which will be discussed in Section 5.

The major background process is the proton-dissociative reaction 
$e\, p\,\rightarrow\, e\, V\, Y$, and in addition to the above 
quantities, $M_Y$, the mass of the diffractive excitation of
the proton, is used.


\subsection{The event reconstruction}  

For the photoproduction events, $Q^2 \approx 0$, $Q^2$ ranged from
the kinematic minimum, $Q^2_{min} = M^2_e y^2/(1-y) \approx
10^{-12}$ GeV$^2$, where $M_e$ is the positron mass, up to
$Q^2_{max} \approx 1 $ GeV$^2$, the value at which the scattered
positron starts to be observed in the calorimeter, with a median
$Q^2$ of approximately $5 \cdot 10^{-5}$ GeV$^2$ (differs slightly
for ZEUS and H1 and from year to year with modifications in
calorimeter geometry). Since the typical $Q^2$ is small, it can be
neglected in the reconstruction of the other kinematic variables.


For the DIS events the kinematic variables are reconstructed using
the momenta of the decay particles and the polar and azimuthal
angles of the measured scattered lepton. Neglecting the transverse
momentum of the outgoing proton with respect to its incoming
momentum, the energy of the scattered positron can be expressed
as:
\[
E_{{e}^{\prime}} \simeq [2E_{e} -(E_{V} -
p^Z_{V})]/(1-\cos{\theta_{{e}^{\prime}}}),
\]
where $E_{e}$ is the energy of the incident lepton, $E_{V}$ and
$p^Z_{V}$ are the energy and longitudinal momentum of the vector
meson $V$, and $\theta_{{e}^{\prime}}$ is the polar angle of the
scattered lepton.  The value of $Q^2$ was calculated from:
\[
Q^2 = 2E_{{e}^{\prime}} E_{{e}} (1 + \cos{\theta_{{e}^{\prime}}})\,.
\]

The photon-proton center-of-mass energy, $W$, can be expressed as
$W^2 \approx 2E_p (E - p_{Z})_{V}+Q^2$, where $E_p$ is the
laboratory energy of the incoming proton and $(E - p_{Z})_{V}$  is
the difference between the energy and the longitudinal momentum of
the vector meson. The fraction of the positron momentum carried
by the photon is calculated from $y = (E - p_{Z})_{V}/2E_e$.

The squared four-momentum transfer  at the proton vertex is given
by
$|t|=(p_{{e}^{\prime}}+p_V)^2_X+(p_{{e}^{\prime}}+p_V)^2_Y$.



\subsection{Data samples and event selection} 

The kinematic region for each particular data sample can be found
in Tabs.~\ref{tab:allVMPHP},~\ref{tab:allVMDIS}. The tables
summarize all the recent data discussed in this paper, for the
overview of the pre-1997 experimental data, see~\cite{crittenden:1997:mesons}.

\begin{table}[htbp!]
\begin{sideways}
\begin{minipage}[b]
{\textheight} \vspace{-0.5cm}
\begin{center}
\begin{tabular}{||c|c|c|c|c|c|c|c|c|c||}
\hline VM & Mode & Decay Ch. & $Q^2$, GeV$^2$ & $W$, GeV
& $t$, GeV$^2$ & Year & Lumin.,$pb^-1$ & Exp. & Ref.\\
\hline\hline
\multicolumn{10}{||c||}{PHOTOPRODUCTION WITHIN THE DIFFRACTION CONE} \\
\hline
 $\rho^0$ & Elastic, P.-Diss. & $\pi^+ \pi^-$ & $4 \cdot 10^{-6}$ & 50-100  &
 0.-0.5 & 1994
         & 2.17 & ZEUS & ~\cite{epj:c2:247}\\
\hline
 $\rho^0$ & Elastic & $\pi^+ \pi^-$ & $10^{-4}$ & 25-70  &
 0.073-0.45 & part of 1999
         & 3.0 & H1 & ~\protect\cite{cpaper:ichep2002:991}\\
\hline
 $J/\psi$ & Elastic & $e^+e^-$, $\mu^+\mu^-$ & 0.05 & 26-285  & 0-1.2 & 1996-97
         & 20.5 & H1 & ~\protect\cite{pl:b483:23}\\
\hline
 $J/\psi$ & Elastic & $\mu^+\mu^-$ & $5 \cdot 10^{-5}$ & 20-170  & 0-1.8 & 1996-97
         & 38.0 & ZEUS & ~\protect\cite{epj:c24:345}\\
\hline
 $J/\psi$ & Elastic & $e^+e^-$ &  $5 \cdot 10^{-5}$ & 20-290  & 0-1.2 & 1998-2000
         & 55.2 & ZEUS & ~\protect\cite{epj:c24:345}\\
\hline

 $\Psi (2S)$ & Elastic & $l^+ l^-$, $J/\psi \pi^+ \pi^-$ & $10^{-4}$ & 40-160  &
 n.d. & 1993-94
         & 6.3 & H1 & ~\protect\cite{pl:b421:385}\\
\hline
 $\Psi (2S)$ & Elastic, P-Diss & $l^+ l^-$, $J/\psi \pi^+ \pi^-$ & 0.055 &
 40-150  &
 0-5 & 1996-2000
         & 77.0 & H1 & ~\protect\cite{pl:b541:251}\\
\hline
 $\Psi (2S)$ & Elastic & $e^+ e^-$ & $5 \cdot 10^{-5}$ &
 50-125  &
 n.d. & 1999-2000
         & 55.3 & ZEUS & ~\protect\cite{cpaper:epc2001:562}\\
\hline
 $\Upsilon$ & Elastic & $\mu^+\mu^-$ & 0.11 & 70-250  & 0-1.2 & 1994-97
         & 27.5 & H1 & ~\protect\cite{pl:b483:23}\\
\hline
 $\Upsilon$ & Elastic & $\mu^+\mu^-$ & $5 \cdot 10^{-5}$ & 80-160  & n.d. & 1995-97
         & 43.2 & ZEUS & ~\protect\cite{pl:b437:432}\\
\hline
\multicolumn{10}{||c||}{PHOTOPRODUCTION AT HIGH $t$} \\
\hline
 $\rho^0$, $\phi$,  & Elastic, P.-Diss. & $\pi^+\pi^-$, $K^+K^-$ & $7 \cdot 10^{-6}$
 & 85-105  & 0-3 & 1995 & 2.0 & ZEUS & ~\protect\cite{epj:c14:213}\\
 $J/\psi$ &  & $\mu^+\mu^-$, $e^+e^-$ & & & & & & & \\
\hline
 $\rho^0$, $\phi$, & P.-Diss. & $\pi^+\pi^-$, $K^+K^-$ & $7 \cdot 10^{-6}$
 & 80-120  & 0-12 & 1996-97 & 25.0 & ZEUS & ~\protect\cite{epj:c26:389}\\
 $J/\psi$ &  & $\mu^+\mu^-$, $e^+e^-$ & & & & & & & \\
\hline
 $J/\psi$ & P.-Diss. & $\mu^+\mu^-$ & n.d. & 50-160  & 1-21 & 1999
         & 19.1 & H1 & ~\protect\cite{cpaper:ichep2002:993}\\

\hline
 $J/\psi$ & P.-Diss. & $\mu^+\mu^-$ & 0.06 & 50-200  & 2-30 & 1996-2000
         & 78.0 & H1 & ~\protect\cite{pl:b568:205}\\

\hline \hline
\end{tabular}

\caption{\it    The recent H1 and ZEUS photoproduction 
measurements discussed throughout
this paper. An overview of older data can be found in
~\protect\cite{crittenden:1997:mesons}. Information that is not
available is labeled by "n.d.". The determination
of the diffraction slope is based on the data at $|t| \lsim 1$ GeV$^{2}$. 
\label{tab:allVMPHP} }

\end{center}
\end{minipage}
\end{sideways}
\end{table}

\begin{table}[htbp!]
\begin{sideways}
\begin{minipage}[b]
{\textheight} \vspace{-0.5cm}
\begin{center}
\begin{tabular}{||c|c|c|c|c|c|c|c|c|c||}
\hline VM & Mode & Decay Ch. & $Q^2$, GeV$^2$ & $W$, GeV
& $t$, GeV$^2$ & Year & Lumin.,$pb^-1$ & Exp. & Ref.\\
\hline\hline
\multicolumn{10}{||c||}{DEEP INELASTIC SCATTERING} \\
\hline
 $\rho^0$ & P.-Diss. & $\pi^+ \pi^- $ & 7-35 & 60-180  & 0-1.5 & 1994
         & 2.8 & H1 & ~\protect\cite{zfp:c75:607}\\
\hline
 $\rho^0$ & Elastic & $\pi^+ \pi^- $ & 0.25-0.85 & 20-90  & 0-0.6 & 1995
         &  3.8 & ZEUS & ~\protect\cite{epj:c6:603}\\
\hline
 $\rho^0$ & Elastic & $\pi^+ \pi^- $ & 3-50 & 32-167  & 0-0.6 & 1995
         &  6.0 & ZEUS & ~\protect\cite{epj:c6:603}\\
\hline
 $\rho^0$ & Elastic & $\pi^+ \pi^- $ & 0.25-0.85 & 20-90  & 0-0.6 & 1995
         &  3.8 & ZEUS & ~\protect\cite{epj:c12:393}\\
\hline
 $\rho^0$ & Elastic & $\pi^+ \pi^- $ & 3-30 & 40-120  & 0-0.6 & 1995
         &  6.0 & ZEUS & ~\protect\cite{epj:c12:393}\\
\hline
 $\rho^0$ & Elastic & $\pi^+ \pi^- $ & 1-60 & 30-140  & 0-0.5 & 1995-96
         & 4.0 & H1 & ~\protect\cite{epj:c13:371}\\
\hline
 $\rho^0$ & Elastic & $\pi^+ \pi^- $ & 2.5-60 & 40-120  & 0-3.0 & 1997
         & 6.0 & H1 & ~\protect\cite{pl:b539:25}\\
\hline
 $\rho^0$ & Elastic & $\pi^+ \pi^- $ & 2-80 & 32-160  & 0-0.6 & 1996-97
         & 38.0 & ZEUS & ~\protect\cite{cpaper:epc2001:594}\\
\hline
 $\rho^0$ & Elastic and P.-Diss. & $\pi^+ \pi^- $ & 2-80 & 50-140  & 0-2
  & 1996-97 & 38.0 & ZEUS & ~\protect\cite{cpaper:ichep2002:818}\\
\hline
 $\rho^0$ & Elastic & $\pi^+ \pi^- $ & 8-60 & 40-180  & 0-0.5 & 2000
         & 42.4 & H1 & ~\protect\cite{cpaper:ichep2002:989}\\
\hline
 $\omega$ & Elastic & $\pi^+ \pi^- \pi^0$ & 3-20 & 40-120  & 0-0.6 & 1996-97
         & 37.7 & ZEUS & ~\protect\cite{pl:b487:273}\\
\hline
 $\phi$ & Elastic & $K^+ K^- $ & 6-20 & 42-134  & 0-0.6 & 1994
         & 2.8 & H1 & ~\protect\cite{zfp:c75:607}\\
\hline
 $\phi$ & Elastic & $K^+ K^- $ & 1-15 & 40-130  & 0-0.5 & 1995-96
         & 3.1 & H1 & ~\protect\cite{pl:b483:360}\\
\hline
 $\phi$ & Elastic & $K^+ K^- $ & 2-70 & 35-145  & 0-0.6 & 1998-2000
         & 66.4 & ZEUS & ~\protect\cite{ZEUS2004_phi_preliminary}\\
\hline
 $J/\psi$ & Elastic & $\mu^+\mu^-$, $e^+ e^-$ & 2-40 & 50-150  & n.d. & 1995
         &  6.0 & ZEUS & ~\protect\cite{epj:c6:603}\\
\hline
 $J/\psi$ & Elastic & $\mu^+\mu^-$, $e^+ e^-$ & 2-80 & 25-180  & n.d. & 1995-97
         & 27.3 & H1 & ~\protect\cite{epj:c10:373}\\
\hline
 $J/\psi$ & Elastic & $\mu^+\mu^-$, $e^+ e^-$ & 0.15-0.8; ~2-100 & 30-220  \
         & 0-1 & 1998-2000
         & 69.0;~83.0 & ZEUS & ~\protect\cite{np:b695:3}\\
\hline $\Psi$(2S) & Elastic & $J/\psi \pi^+ \pi^-$ & $ 1-80 $ &
40-180  & n.d. & 1995-97
         & 27.3 & H1 & ~\protect\cite{epj:c10:373}\\
\hline \hline \hline
\end{tabular}
\caption{\it    The recent H1 and ZEUS electroproduction measurements discussed throughout
this paper. Overview of older data can be found in
~\protect\cite{crittenden:1997:mesons}. Notation is the same as 
in Tab.~\ref{tab:allVMPHP}. \label{tab:allVMDIS} }

\end{center}
\end{minipage}
\end{sideways}
\end{table}
\pagebreak


\section{An overview of theoretical approaches to
diffractive scattering}


\subsection{The rudiments of the Regge theory of
strong interactions}\label{sect3.1}

As Bjorken emphasized, the foundations of the Regge theory are as
solid as QCD itself \cite{BjorkenEilat}. Because the physics of
diffractive scattering is permeated by ideas and concepts from the
Regge theory of strong interactions, a brief introduction into
this subject is in order. For the more rigorous treatment and for
technicalities one must consult the textbooks
\cite{ReggeTextbook,BaronePredazzi,ForshawRoss}, the review papers
\cite{KaidalovPhysRep,AlberiPhysRep,IrvingPhysRep} and the
collection of reprints \cite{Caneschi}.


\subsubsection{The $s$-channel asymptotics from the $t$-channel exchanges:
spin and energy dependence}  

There is a deep connection between the high-energy behavior of
a binary reaction $a~b~\to~c~d$ and the spin, $J$, of the {\it
elementary} particle with mass $M$ exchanged in the $t$-channel:
\bea
A_{ab \to cd}(W^2,t) = {\frac{g_{ac}(t) g_{bd}(t)}{t-M^2}}(W^2)^J
\label{eq:3.1.1.1}\\
\frac{d\sigma(a~b \to c~d)}{dt} \propto{\frac{g_{ac}^2(t)
g_{bd}^2(t)} {(t-M^2)^2}} W^{4(J-1)}. \label{eq:3.1.1.2}
\eea
Although it follows in a straightforward manner from the analysis
of Feynman diagrams, it is instructive to look at
(\ref{eq:3.1.1.1}) from the $t$-channel point of view. In the
crossed channel
$$
a~\bar{c} \to \bar{b}~d
$$
the total c.m.s. energy squared is $t=(p_a+p_{\bar{c}})^2 =
(p_a-p_c)^2=M^2$, the momentum transfer squared is
$(p_a-p_{\bar{b}})^2=s=(p_a+p_b)^2$, and the exchanged particle
emerges as a resonance at $t=M^2$ in the partial wave $J$.
The angular dependence of this contribution to the scattering
amplitude is given entirely by the Legendre polinomial
\be
A_{ab
\to cd}^{\sigma_t}(W^2,t) = A_J(t) P_{J}(\cos\theta_t) =
{\frac{G_{a\bar{c}}(t) G_{\bar{b}d}(t)}{t-M^2}} [\sigma_t+
(-1)^J]P_{J}(-\cos\theta_t)\, , \label{eq:3.1.1.3}
\ee
where (for the sake of simplicity we take 
$m_a=m_b=m_c=m_d=\mu$)
\be
\cos\theta_t = 1 + \frac{2W^2}{t-4\mu^2}\, . \label{eq:3.1.1.4}
\ee
The so-called signature $\sigma_t=\pm 1$ separates the
crossing-even and crossing-odd amplitudes; for instance, in the
crossing-even $\pi^0\pi^0$ scattering $\sigma_t=+1$ and the
contribution from the odd-partial waves to (\ref{eq:3.1.1.3})
vanishes identically.

The amplitude (\ref{eq:3.1.1.3}) depends on $W^2$ only through the
Legendre polinomial and can readily be continued analytically into
the high energy domain of $W^2\gg M^2,\mu^2,|t|$, which amounts to
$-\cos\theta_t \gg 1$ and $P_J(-\cos\theta_t) \propto
(-\cos\theta_t)^J \propto (W^2)^J$, i.e., we derived the
asymptotics (\ref{eq:3.1.1.1}) by analytic continuation from the
$t$-channel to $s$-channel scattering.


\subsubsection{The Regge trajectories} 

On the one hand, the existence of high-spin resonances is an
ultimate truth of the physics of strong interactions; on the other
hand, the exchange by an elementary particles of spin $J>1$ would
conflict the fundamental Froissart bound \cite{FroissartBound}
\be
A(W^2,t) < W^2 \log^2 W\,. \label{eq:3.1.2.1}
\ee
The sole known
way out of this trouble is offered by the Regge theory: one must
improve upon the above mock-up analytic continuation going from
the sum over integer (or half integer) partial waves to the
Sommerfeld-Watson integral over the complex angular momentum $J$
with which the analytic continuation to $-\cos\theta_t \gg 1$ must
be complemented with the appropriate deformation of the
integration contour on the complex-$J$ plane
\cite{GribovComplexJ,*GribovComplexJ1}. The key point is that the asymptotic
behaviour of the $s$-channel amplitude will be controlled by
singularities of the partial wave $A_{J}(t)$ in the complex-$J$
plane. If the singularity is the (Regge) pole,
\be
A_J(t) \propto
\frac{1}{J-\alpha_R(t)} \label{eq:3.1.2.2}
\ee
then one obtains
precisely the amplitude of the form (\ref{eq:3.1.1.1}) 
with $J=\alpha_R(t)$.
The $t$-channel unitarity dictates
\cite{GribovComplexJ,*GribovComplexJ1,GribovShrinkage,*GribovShrinkage1} that the Regge pole must be
a moving one, i.e. it must have a finite slope $\alpha_R'(t)$.
Experimentally, the Regge trajectory $\alpha_R(t)$ for the
$s$-channel scattering at $t<0$ can be extracted from the energy
dependence of the differential cross sections, and can be linked
to the resonance mass spectrum by extrapolation of the
mass-dependence of the spin of $t$-channel resonances,
$J_n=\alpha_R(M_n^2)$. Such Chew-Frautschi plots are well
approximated by straight lines,
\be
\alpha_R(M_n^2) \approx
\alpha_R(0) +\alpha'_R t\,. \label{eq:3.1.2.3}
\ee
For instance,
for the $\rho,\omega,A_2,f_2$ families of resonances with
non-vacuum quantum numbers such extrapolations suggest the
intercept $\alpha_R(0) \approx 0.45$, in very good agreement with
the results from the scattering experiments. To cite few examples,
the $\rho$-trajectory is best studied in the charge-exchange
$\pi^-p\to \pi^0 n$, the $A_2$-trajectory is probed in $\pi^- p \to
\eta n$, the $\omega$-trajectory is probed in the regeneration
$K_L \to K_S$ on the isoscalar target, the $\pi$-trajectory is
probed in the charge-exchange $np\to pn$, etc. For classic
reviews on the Regge trajectories see \cite{IrvingPhysRep}, a
more recent discussion of the Chew-Frautschi plots is found in
\cite{AnisovichRegge}. The high-lying Pomeron, $\rho,\omega,A_2,f_2$
are the natural spin-parity exchanges, i.e., the spin $J$ and 
parity $P$ of particles lying on the corresponding Regge 
trajectory are related by $P=(-1)^J$. The un-natural spin-parity
$\pi,A_1$ exchnages, $P=-(-1)^{J}$, have much lower intercepts, 
$\alpha_{\pi}(0)
\approx \alpha_{A_1}(0)\approx 0$.


\subsubsection{The universality aspects of the Regge exchange} 
\label{sect3.1.3}

The Ansatz (\ref{eq:3.1.1.1}) bears all the salient features of
the realistic reggeon-exchange amplitude:
\begin{enumerate}

\item The trajectory $J=\alpha_R(t)$ is universal for all
beam and target particles, it only depends on the $t$-channel
quantum numbers.

\item Dependence on the initial and final state particles has
a factorized form.

\item If one parameterizes the $t$-dependence of the near forward
differential cross section by the so-called slope parameter $B$,
\be
\frac{d\sigma}{dt} \propto \exp(-B|t|)\,,
\label{eq:3.1.3.1}
\ee
then the factorization property entails 
\be
B(a~b\to c~d) =
B_{ac}+B_{bd}+B_R, \label{eq:3.1.3.2} 
\ee
where $B_{ac}$ and
$B_{bd}$ come from the form factors of the $a\to c$ and $b\to d$
transitions, and $B_{R}$ characterizes the exchanged reggeon.

\item Notice that
\be
(W^2)^{\alpha_R(t)} =
(W^2)^{\alpha_R(0)}\cdot (W^2)^{\alpha_R't} =
(W^2)^{\alpha_R(0)}\cdot \exp[-\alpha_R'|t|\log(W^2)],
\label{eq:3.1.3.3}
\ee
what entails Gribov's  growth of the slope
parameter with energy, alias the shrinkage of the diffraction cone
\cite{GribovShrinkage,*GribovShrinkage1}: 
\be
B_R = 2\alpha_R'\log\left(\frac{W^2}{s_0}\right). 
\label{eq:3.1.3.4} 
\ee

The slope of all the non-vacuum Regge trajectories is about the same,
\be
\alpha_R'\approx \frac{1}{2m_\rho^2}\approx 0.9~ {\rm
GeV}^{-2}\, , \label{eq:3.1.3.5}
\ee
for the recent summary see \cite{AnisovichRegge}.

\item The phase of the reggeon exchange amplitude is uniquely
fixed by the analytic continuation of the signature factor
$\eta(\sigma_t,t)= \sigma_t
+(-1)^{\alpha_R(t)}=\sigma_t-\exp[-i\pi(\alpha_R(t)-1)]$:
\be
\frac{Re A(W^2,t)}{Im A(W^2,t)} =
\begin{cases}
\tan[{\frac{\pi}{2}}(\alpha_R(t)-1)], & \text{if $\sigma_t=+1$},\\[2mm]
\cot[{\frac{\pi}{2}}(\alpha_R(t)-1)], & \text{if $\sigma_t=-1$}.\\
\end{cases}
\label{eq:3.1.3.6}
\ee
\end{enumerate}


\subsubsection{The vacuum exchange: the Pomeron trajectory
from hadronic scattering} \label{sect3.1.4}

Elastic scattering is driven via unitarity by strongly
absorptive inelastic multiproduction processes, which is nicely
illustrated by the impact parameter representation --- the high
energy version of the partial wave expansion. In high energy
elastic scattering the momentum transfer, $\bDelta$, is the
two-dimensional vector transverse to the beam momentum. The 
elastic scattering amplitude can be cast in the form of
the Fourier transform
\be
{1\over W^2}A(W^2,\bDelta)=2i\int d^2\bb \left[1-S(\bb)\right]
\exp(-i\bb\bDelta), \label{eq:3.1.4.1}
\ee
where
$S(\bb)=\exp(2i\delta(\bb))$ is the S-matrix for elastic
scattering at an impact parameter $\bb$ and the angular momentum
$l=|\bp|\cdot |\bb|$. The total elastic and inelastic cross
sections equal
\bea
\sigma_{el}=\int d^2\bb
\left|1-S(\bb)\right|^2\, ,\nonumber\\
\sigma_{in}=\int d^2\bb \left[1-|S(\bb)|^2\right]\,.
\label{eq:3.1.4.2}
\eea 
Strong absorption implies the
predominantly imaginary scattering phase. One often uses the
so-called profile function $\Gamma(\bb)= 1-S(\bb)$. The small
momentum transfer expansion in (\ref{eq:3.1.4.1}) gives
\be
{1\over W^2}A(W^2,\bDelta)=2i\int d^2\bb
\Gamma(\bb)\cdot [1-{1\over 2}(\bb\bDelta)^2] ={1\over
W^2}A(W^2,0)\cdot \left(1 -{1\over 2} B\bDelta^2\right)\, ,
\label{eq:3.1.4.3}
\ee
so that the diffraction slope $B$ is
determined by the mean impact parameter squared
\be
B = {1\over 2}
\langle \bb^2\rangle = {1\over 2}\cdot \frac{\int d^2\bb\,
\bb^2\Gamma(\bb)}{\int d^2\bb\,\Gamma(\bb)}\,. \label{eq:3.1.4.4}
\ee
The extreme case is the scattering on the absorbing black disc
of radius $R$ for which $|S(\bb)|= \theta(R-|\bb|)$, which is a
good approximation for the scattering of nucleons off heavy
nuclei. Then 
\be
\sigma_{el} =\sigma_{in} = \frac{1}{2}\sigma_{tot}=\pi R^2\, 
\label{eq:3.1.4.5}
\ee
and the diffraction slope equals
\be
B_{el}=\frac{1}{4}R^2\,. \label{eq:3.1.4.6}
\ee
Such a flat,
energy independent, elastic scattering
 must be contrasted
to the two-body reactions with the non-vacuum exchange which
constitute a tiny fraction of high energy inelastic collisions of 
hadrons and have cross sections that vanish at high energy,
\be
\sigma(ab\to cd) \propto \frac{1}{W^{4(1-\alpha_R(0))}} <
\frac{1}{W^2}\,. \label{eq:3.1.4.7}
\ee

The importance of strong absorption for high-energy hadron
interactions
is evident form the proximity of central partial waves
of $pp$ scattering to the unitarity limit, $\Gamma(\bb)\leq 1$,
\cite{AmaldiSchubert}, although the periphery of the nucleon is
still gray, and for all the hadrons $\sigma_{el}$ is still
substantially smaller than $\sigma_{in}$, see the plots in the
Review of Particle Physics \cite{PDG2002}. As emphasized first by
Pomeranchuk, the particle-antiparticle cross section differences
vanish at high energy, see \cite{PDG2002}, and from the
$t$-channel viewpoint the elastic scattering is dominated by the
vacuum exchange. In 1961 Chew and Frautschi conjectured that the
vacuum channel too can be described by the reggeon --- first dubbed
the Pomeranchukon, later shortened to Pomeron, --- exchange with an
appropriate spin-2, $C$-even, isoscalar, positive parity resonance
lying on the Pomeron trajectory (the early history of the Pomeron
is found in \cite{Caneschi}.

If the Pomeron were a simple Regge pole, it would have been
utterly distinct from the non-vacuum reggeons:
\begin{itemize}

\item For all hadrons and real photons the total cross sections
rise with energy and the phenomenological Pomeron trajectory has
$\alpha_{\Pom}(0)=1+\Delta_{\Pom} \approx  1.1 > 1$ (notice that from now on the $\Delta$ is still  used for the
four-momentum exchanged, but
$\Delta_{index}$ is used to define the variation of the intercept of
the coorrespondent {\it index} trajectory from unity).
Such a rise
of the vacuum component of the total cross section,
\be
\sigma_{vac} =\sigma_{Pom} \propto (W^2)^{\Delta_{\Pom}}
\label{eq:3.1.4.8}
\ee
can not go forever, though. At  asymptotic
energies it would conflict the Froissart bound. Furthermore, the
partial waves of elastic scattering would overshoot the unitarity
bound. Indeed, in the often used exponential approximation,
${1\over W^2}A(W^2,\bDelta) \propto \exp(-{1\over 2}B\bDelta^2)$,
and neglecting the small real part of the small-angle scattering
amplitude, one finds
\be
\Gamma(\bb)= {\sigma_{tot} \over 4\pi
B}\cdot \exp(-{\bb^2 \over 2B}) \label{eq:3.1.4.9}
\ee
and with
the unlimited growth of $\sigma_{tot}$ one would run into
$\Gamma(\bb)>1$. The unitarity (absorption, multipomeron
exchange,...) corrections, which must eventually tame such a growth
of $\Gamma(\bb)$ and of $\sigma_{tot}$ with energy, were shown to
be substantial already at moderate energies
\cite{Capella1,Capella2,ITEPPomeron,*ITEPPomeron1}. The multipomeron absorption
affects substantially the determination of $\Delta_{\Pom}$: the
first estimate
\[
\Delta_{\Pom}\sim 0.13
\]
with the perturbative treatment of absorption 
based on Gribov's reggeon field theory \cite{GribovRFT,*GribovRFT1,AGK,*AGK1}
goes back to the
1974-75 papers by Capella, Tran Thahn Van and Kaplan
\cite{Capella1,Capella2}. Within a more realistic model for
absorption, the ITEP group \cite{ITEPPomeron,*ITEPPomeron1} found the equally
good description of the hadronic cross section data with
substantially larger $\Delta_{\Pom} \approx 0.23$. If one follows
the Donnachie-Landshoff suggestion \cite{DLsigtot} to ignore the
absorption corrections altogether and stick to the simplified pole
terms, then the Particle Data Group finds $\Delta_{\Pom}=0.095$
\cite{PDG1998}. However, according to the 2002 edition of the Review of
Particle Properties \cite{PDG2002}, à still better fit the the
experimental data is provided by the parameterization
\cite{CudellFroissart}
\be
\sigma_{vac}(ab)=\sigma_{vac}(\bar{a}b)=Z_{ab} +2B\log^2(W/W_0),
\label{eq:3.1.4.10}
\ee
 which is consistent with the Froissart
bound and from the Regge theory viewpoint corresponds to the
triple-pole singularity at $j=1$, i.e., $\Delta_{\Pom}\equiv 0$!

\item The shrinkage of the diffraction cone in elastic scattering
suggests very small slope of the Pomeron trajectory
$\alpha_{\Pom}(t)$: the combined analysis of the experimental data
on elastic $pp,\bar{p}p,\pi^{\pm}p,K^{\pm}p$ scattering at the
CERN SPS/FNAL and CERN ISR energies gave $\alpha_{\Pom}'\approx
0.13\pm 0.025$ GeV$^{-2}$ (\cite{Burq}, for the review see
\cite{GoulianosPhysRep}). The extrapolation of these fits
under-predicts the $\bar{p}p$ diffraction slope at the Tevatron,
which call for $\alpha_{\Pom}' \approx 0.25$ GeV$^{-2}$.
Incidentally, the last value of $\alpha_{\Pom}'$ has been used by
theorists ever since 1974-75 \cite{Capella1,Capella2}, but it must
be taken with the grain of salt: the observed growth of the
diffraction cone can to a large extent be due to the
unitarity/absorption driven correlation, cf. Eqs.
(\ref{eq:3.1.4.5}) and  (\ref{eq:3.1.4.6}), between the total
cross section and the diffraction slope so that the Tevatron data
can well be reproduced with the still smaller values of
$\alpha_{\Pom}'$ \cite{KNPsigtot}.

\end{itemize}

To summarize, the Donnachie-Landshoff (DL) parameterization
\cite{DLsigtot} 
\be
\alpha_{soft}(t)=1.1+0.25 {\rm GeV^{-2}}\cdot t
\label{eq:3.1.4.11}
\ee
must only
be regarded as a convenient short hand description of the local, $
W< 1$ TeV, energy dependence of the vacuum component of the
elastic scattering of hadrons.


\subsubsection{The diffraction slope: variations from elastic
scattering to single to double diffraction 
excitation } \label{sect3.1.5}

The variation of the diffraction slope (\ref{eq:3.1.3.2})
from elastic scattering to single (SD) to double (DD) diffraction  
excitation exhibits certain universal features
\cite{HoltmannDD,AlberiPhysRep,KaidalovPhysRep}. An excellent
guidance is provided by a comparison of elastic proton-nucleus, $pA \to pA$,
to quasielastic, $pA\to p'A^*$, scattering. The latter reaction, 
in which
one sums over all excitations and breakup of the target nucleus without
production of secondary particle,  
must be regarded as diffraction excitation of the target nucleus.

The crucial point is that at a sufficiently large $(p,p')$ 
momentum transfer such that the recoil energy exceeds the  
typical nuclear binding energy, which can viewed as hard scattering,
the $t$-distribution of 
scattered protons in quasielastic (nucleus-dissociative)
$pA\to p'A^*$ is the same 
as in elastic $pp$ scattering, $B_{diss}(pA\to p'A^*) \approx B_{pp}$
\cite{Glauber,Czyz}. The quasielastic $pA\to p'A^*$ becomes sort of
a deep inelastic scattering with quasifree bound nucleons behaving
as partons of a nucleus and quasifree  $pN \to p'N$ scattering 
being a counterpart of the Rutherford scattering of leptons off charged
partons in DIS off the proton.  The summation over breakup of a
nucleus into all continuum excitations is
important, for excitation of the specific discrete state $A^*$ of
a target nucleus, $pA\to p' A^*$, the diffraction slope will still
be large, 
\be
B_{AA*}\sim B_{el} \approx {1\over 4} R_A^2.
\label{eq:3.1.5.1*}
\ee

Now define the ratio of differential cross sections
\be
Ratio(diss/el)(t)= \frac{d\sigma_{diss} ({pA \rightarrow  p'A^*})}{dt} \Bigg/
\frac{d\sigma_{el} ({ pA \rightarrow p'A})}{dt}. 
\label{eq:3.1.5.1}
\ee

\begin{figure}[htbp]

   \centering

   \epsfig{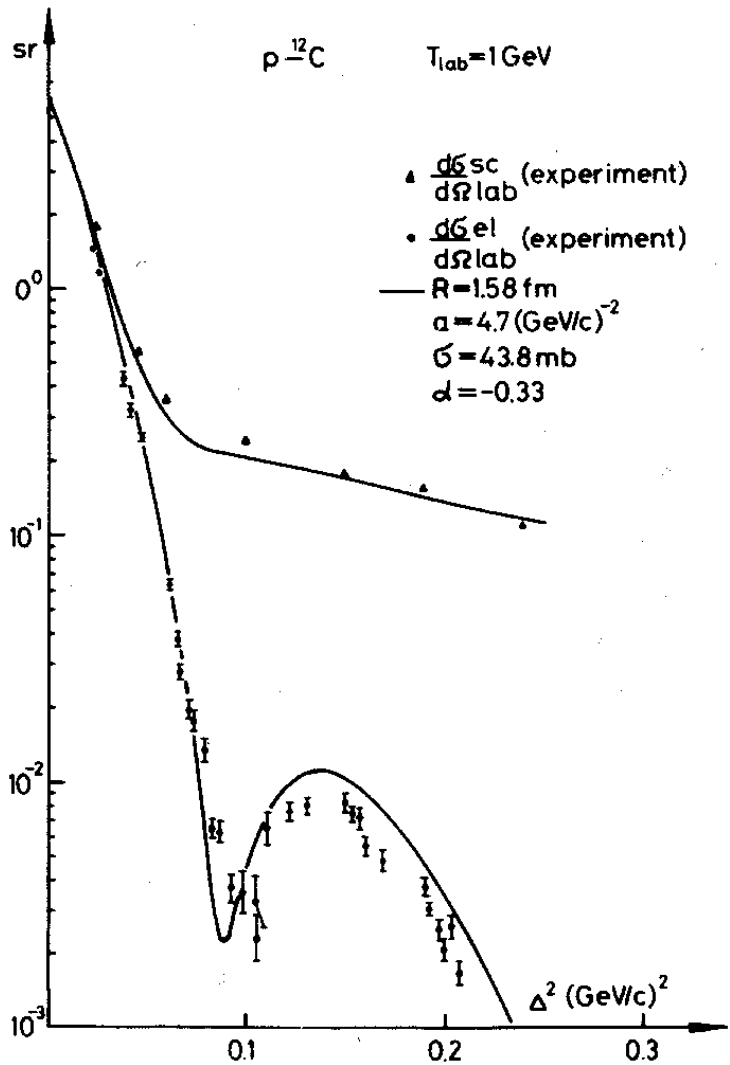}

   \caption{A comparison of elastic (points, the lower set of data
points and the lower curve) and
combined elastic plus nucleus-dissociative (triangles,the upper set 
of data points and the upper curve) $p^{12}C$ scattering data
\protect\cite{Friedes_pC12,Palevsky_pC12}. The theoretical
calculations are from Czyz et al. \protect\cite{Czyz}.}

   \label{fig:pC12elasticdissociative}

\end{figure}

Elastic 
scattering: 
$Ratio(diss/el)(t) \ll 1$ is 
the dominant process within the diffraction cone, $R_A^2|t|\ll 1$. 
However, elastic
scattering dies out rapidly for $R_A^2|t|\gsim 1$, where quasielastic
scattering takes over: $ Ratio(diss/el)(t) \gg 1$.
This point is clearly illustrated by the experimental data 
\cite{Friedes_pC12,Palevsky_pC12} on elastic
and nucleus-dissociative  $p ^{12}C$ scattering shown in 
Fig.~\ref{fig:pC12elasticdissociative}. Notice the diffractive dip-bump 
structure, familiar from optical diffraction, in the differential
cross section of  pure elastic scattering. For a sufficiently
hard scattering, $|t|=\Delta^2 \gsim 0.06~{\rm (GeV)}^2$, the sum of 
the elastic and nucleus-dissociative
cross sections, $d\sigma_{sc}=d\sigma_{el}+d\sigma_{diss}$ is
clearly dominated by the nucleus-dissociative $d\sigma_{diss}$.

In the regime of strong absorption the integrated cross 
section of quasielastic or nucleus-dissociative scattering is small
\cite{Glauber},
\be
 \sigma_{diss}{(pA \rightarrow  p'A^*})\ll \sigma_{el}({ pA \rightarrow p'A})
 \approx {1\over 2} \sigma_{tot}^{pA}.
\label{eq:3.1.5.2}
\ee

Exactly the  same considerations apply to elastic scattering and 
diffraction excitation of hadrons and real photons, $a=p,\pi,K,\gamma$
on the free nucleon target, $b=p$. Let $B_N$ be the contribution to the
diffraction slope of electric $pp$ scattering from the Pomeron-proton-proton
vertex, so that 
\be
B_{el}^{pp} = 2B_N+B_{\Pom}.
\label{eq:3.1.5.3}
\ee
In the single or target-dissociative (SD) reaction, $pp \to pY$, 
and double dissociation (DD), $pp \to XY$, one must distinguish
the low-mass (LM = resonances, low-mass continuum states, ...) and high-mass 
(HM) states $X,Y$. 
The boundary
between the low-mass (exclusive low mass states, resonances, ...) 
and high-mass
continuum excitations is $M_{X,Y}\sim  2\,GeV$. 
The case of small-mass excitation
is an exact counterpart of excitation of discrete nuclear states in
$pA \rightarrow  p'A^*$. Then (\ref{eq:3.1.5.1*}) suggests that 
the contribution to the diffraction slope from the $pY$ transition
$B_{pY}\approx B_N$, so that 
in SD and DD into low-mass states
\be
B_{SD}(LM) = B_{pY}+ B_N+B_{\Pom} \approx B_{DD}(LM,LM)= B_{pY}+ B_{pX}+B_{\Pom} 
\approx B_{el}\, ,
\label{eq:3.1.5.4}
\ee
in good agreement with the experimental data from in the
CERN ISR and FNAL experiments \cite{A76,B78,W75,C75,C80}. 
The SD into high-mass (HM) continuum, $pp \to pY(HM)$, corresponds 
to the complete breakup of the target proton and the reaction can be
viewed as elastic scattering of the beam proton on one of the
constituents of the target. Consequently, the dependence on the size 
of the target proton vanishes, $B_{pY} \approx 0$, and
in SD into high-mass states (often referred to as
the triple-Pomeron region) and mixed low\&high mass DD 
\be
B_{SD}(HM) \approx B_DD(LM,HM) \approx B_{N} + B_{\Pom}
\approx {1\over 2} B_{el} \approx 6~ {\rm GeV}^{-2}\, .
\label{eq:3.1.5.6}
\ee
In DD $pp \to X(HM)Y(HM)$ with excitation of high-mass states from both the
target and beam $B_{pX} \approx B_{pY} \approx 0$ and 
only the $t$-channel exchange $B_{\Pom}$ contributes to
diffraction slope. Experimentally, this component 
is abnormally small \cite{C75,C80}
\be
B_{DD}(HM,HM) \approx B_{\Pom} \sim (1-2)~ {\rm GeV}^{-2}\, .
\label{eq:3.1.5.5}
\ee

Finally, although in $\pi p, Kp,pp$ scattering only the central
partial waves are close to the strong absorption limit, and
the ratios $\sigma_{el}/\sigma_{tot} \sim (0.15\div 0.25)$ are
still substantially smaller than ${1\over 2}$ for the strongly
absorbing nuclear target, the strong inequality
$\sigma_{dis}(pp\to p'Y) \ll \sigma_{el}(pp\to pp)$ holds
in close similarity to (\ref{eq:3.1.5.2}). Typically, in 
$pp$ interactions $R_{pp}(diss/el)=\sigma_{SD}/\sigma_{el}
\lsim 0.3$, for the review
see \cite{AlberiPhysRep,KaidalovPhysRep,GoulianosPhysRep}.


\subsection{The Regge theory and QCD} \label{sect3.2}

\begin{figure}[htbp]

   \centering

   \epsfig{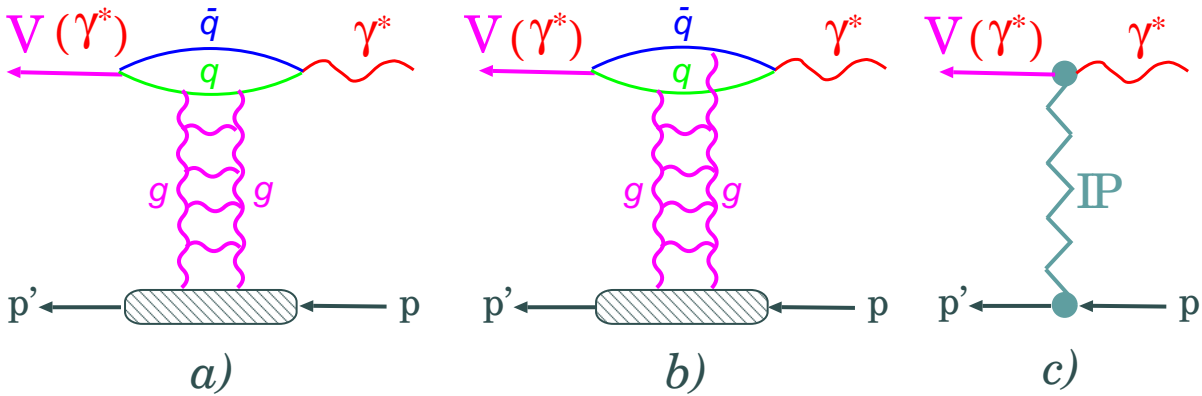}

   \caption{\it    (a,b) The subset of two-gluon tower pQCD diagrams for
the Pomeron exchange contribution (c) to the Compton scattering
(DIS) and diffractive vector meson production. Not shown are two
more diagrams with $q\leftrightarrow \bar{q}$.}

   \label{fig:Twogluontower}

\end{figure}

In the realm of DIS the high energy limit amounts to the small-$x$
limit. The SF's of small-$x$ DIS are related
to the total cross sections as
\[
F_{T,L}(x,Q^2)=\frac{Q^2}{4\pi^2\alpha_{em}}\sigma_{T,L}(x,Q^2)\,
.
\]
Instead of the transverse SF one usually discusses
$F_2(x,Q^2)=F_{T}(x,Q^2)+ F_{L}(x,Q^2)$. The QCD parton model
decomposition of the proton SF into the valence and sea quark
contributions
\bea
F_{2}(x,Q^2) &=& x \sum_{f} e_f^2
[q_f(x,Q^2)+\bar{q}_f(x,Q^2)]
\nonumber\\
&=&
  \frac{4}{9}x\cdot u_v(x,Q^2)+ \frac{1}{9}x\cdot d_v(x,Q^2)
+2x\sum_{f} e_f^2\bar{q}_f(x,Q^2) \label{eq:3.2.1}
\eea must be
viewed as a decomposition of the photoabsorption cross section
into the non-vacuum (non-single) and
vacuum (singlet) components. From the viewpoint of the QCD
evolution, the valence component corresponds to slowing down of
the valence quarks to $x\ll 1$ and depends on the target. At
small $x$, the sea evolves from glue and 
will be the same for the proton and
neutron as well as antinucleon targets, i.e. it
must be associated with the Pomeron exchange. The density of
small-$x$ gluons exceeds greatly the density of charged partons,
which entails that (i) one can model high energy inelastic
interactions by production of the multigluon final states  and
(ii) to the so-called leading-$\log\frac{1}{x}$ the small-$x$
evolution is driven by the splitting of gluons into gluons, with
the splitting $g \to q\bar{q}$  only at the last stage of the
evolution. As a result, the QCD vacuum exchange is  
modeled by the
tower of color-singlet two-gluon exchange diagrams
of Fig.~\ref{fig:Twogluontower}, which is described in terms of the
so-called  unintegrated or differential gluon density
\[
{\cal F}(x,\bkappa^2)=\frac{\partial G(x,\bkappa^2)} {\partial
\log \bkappa^2}\, ,
\]
where $\bkappa$ is the gluon transverse momentum.

At not so small $x$, the $Q^2$-dependence of the parton densities
is governed by the  DGLAP 
evolution \cite{DGLAP1,*DGLAP2,DGLAP3,*DGLAP4,DGLAP5}. 
Here the evolution
goes from smaller to larger $Q^2$, so that once the boundary
condition is taken at a sufficiently large $Q_0^2$ then one stays
in the perturbative domain. However, in the language of inelastic
multiparticle states the DGLAP evolution amounts to summing only the final
states with strong ordering of transverse momentum and as such,
it accounts to only a small
part of the available transverse phase space. This restriction on
the transverse phase space becomes excessively prohibitive and
must be lifted at very small $x$. The practical method of summing
the leading-$\log\frac{1}{x}$ contributions to the unintegrated
gluon density ${\cal F}(x,\bkappa^2)$ without restrictions on the
transverse momenta of partons has been developed in 1975 by Fadin,
Kuraev and Lipatov \cite{FKL75,KLF77,*KLF771} and refined further by
Balitsky and Lipatov (\cite{BL,Lipatov86,*Lipatov861}, for the review see
\cite{LevPhysRep}). One has to pay a heavy price, though: the BFKL
evolution receives a substantial contribution from soft,
nonperturbative transverse momenta of final state partons, where
the running strong coupling $\alpha_S$ is not small and the
sensitivity to models of infrared-regularization can not be
eliminated
(\cite{NZZdipoleBFKL,NZZspectrum1,*NZZspectrum11,NZZspectrum2,Ciafalonispectrum,RossHancock} and references therein). Although the fully
satisfactory quantitative solution to this problem is as yet
lacking, many of the properties of the QCD vacuum exchange must be
regarded as well established:

\begin{itemize}

\item Discard the asymptotic freedom, i.e., make the approximation
$\alpha_S=~const$ and allow the infinite propagation range for
gluons. Such a model is free of a dimensional parameter and
possesses the scale-invariance property, which allows for an exact
solution. The $j$-plane singularity of the model is a fixed cut
(branching point) \cite{FKL75,KLF77,*KLF771,BL} at
\be
-\infty < j \leq
1+\Delta_{BFKL} = 1 + \frac{12\log 2}{\pi} \alpha_S
\label{eq:3.2.2}
\ee
with vanishing $\alpha_{BFKL}'=0$, which
is natural in view of the lack of any dimensional parameter in the
model.

\item One can cope with the asymptotic freedom within the BFKL
approach only at the expense of a certain regularization of the
infrared growth of $\alpha_S$. One only needs to account for
the finite propagation length, $R_c$, of perturbative gluons as
suggested, for instance by the lattice QCD studies
\cite{Meggiolaro,Field,Shuryak}. In their 1975 paper Fadin, Kuraev and
Lipatov remarked that in this case the branching point is
superseded by a sequence of moving Regge poles \cite{FKL75}. The
positions of the poles were estimated in 1986 by Lipatov
\cite{Lipatov86,*Lipatov861} 
\be
\Delta_n \approx \frac{\Delta_{BFKL}}{n+1}\,.
\label{eq:3.2.3} 
\ee

Herebelow, when discussing the pure Pomeron amplitudes, we shell refer to
$\Delta_{n}$ as the intercept, which must not cause a confusion.
Within the color dipole approach the poles
differ by the number of nodes in the eigen-cross section as a
function of the dipole size $r$ \cite{BFKLRegge1,*BFKLRegge2}. The rightmost
pole has a node-free eigen-cross section, the nodal structure of
the eigen-cross sections and the $n$-dependence of the intercept
of subleading vacuum poles found in \cite{BFKLRegge1,*BFKLRegge2} are very
close to the quasiclassical approximation results of Lipatov
\cite{Lipatov86,*Lipatov861}. The intercept of the rightmost pole
$\Delta_{\Pom}$, the slopes of the emerging Regge trajectories and
positions of nodes in the eigen-cross sections depend on the
infrared regularization (\cite{NZZspectrum1,*NZZspectrum11,NZZspectrum2,NZZslope}
and references therein).

\end{itemize}


\subsection{Poor man's approximations to the QCD Pomeron} 
\label{sect3.3}


\subsubsection{The $Q^2$-independence of the Pomeron intercept} 
\label{sect3.3.1}

\begin{figure}[htbp]
   \centering
   \epsfig{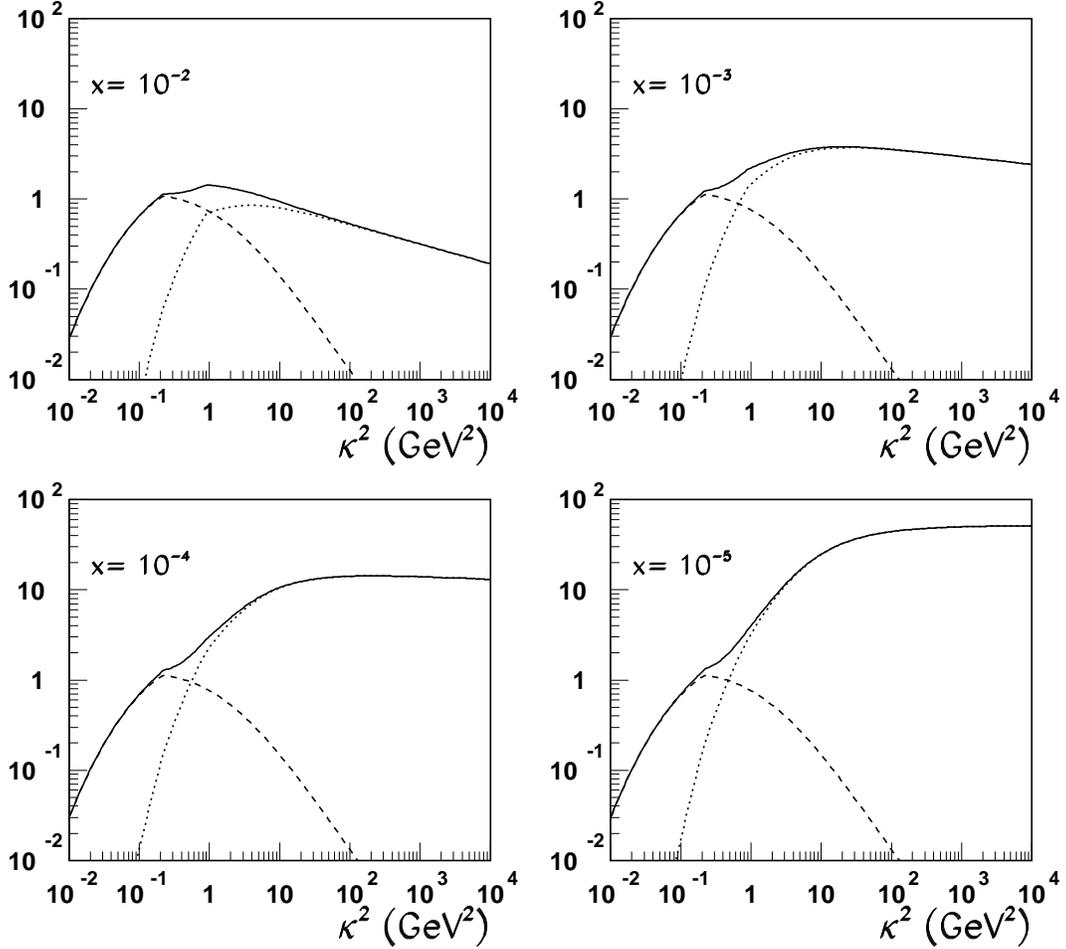}
   \caption{\it    The differential gluons structure function of
the proton determined in \protect\cite{INdiffglue,INanatomy,*INanatomy1} from
the $k_{\perp}$-factorization analysis of the experimental data on
$F_{2p}(x,Q^2)$. Notice the transition from the $x$-independent
soft component at small $\bkappa^2 < 1$ GeV$^2$ (shown by the
dashed curve) to the hard component (the dotted curve), which
converges to the derivative $\partial G_{DGLAP}(x,\bkappa^2)/\partial
\log\bkappa^2 $ of the integrated gluon density determined from the
LO DGLAP fit to $F_{2p}(x,Q^2)$. This particular example is for
the GRV LO parameterization \protect\cite{GRV}, very similar
results are found for the MRS \protect\cite{MRS} and CTEQ
\protect\cite{CTEQ} parameterizations.}
   \label{fig:DIFGlueGRV}
\end{figure}

For each and every pole the intercept does not depend on
the probe. In application to DIS that means an independence of
intercepts on $Q^2$
 \cite{NZZdipoleBFKL,NZZspectrum1,*NZZspectrum11,NZZspectrum2,NZHERA},
only the residues can depend on $Q^2$, so that the $x$-dependence
of structure functions will be of the form
\be
F_{2}(x,Q^2) = \sum_{n=0}
F^{(n)}(Q^2)\left(\frac{1}{x}\right)^{\Delta_n}
+F_{2}^{soft}(Q^2)\, . \label{eq:3.3.1.1}
\ee
Examples of such a
BFKL-Regge expansion for the proton and photon SF's with energy
independent soft contribution $F_{2}^{soft}(Q^2)$, i.e.,
$\Delta_{soft}=0$, are found in \cite{BFKLRegge1,*BFKLRegge2,NSZpion,NZcharm,NSZgg}.
If one reinterprets the soft contribution in terms of the soft,
nonperturbative, unintegrated gluon density, then similar Regge-BFKL
expansions hold for the integrated gluon density, $G(x,Q^2)$, and the
unintegrated gluon density
\[
{\cal F}(x,\bkappa^2)=\frac{\partial G(x,\bkappa^2)} {\partial
\log \bkappa^2}\, .
\]
An example of the decomposition of ${\cal F}(x,\bkappa^2)$ into
the soft and hard components is found in
\cite{INdiffglue,INanatomy,*INanatomy1} and is shown in
Fig.~\ref{fig:DIFGlueGRV}.

\begin{figure}[htbp]
   \centering
   \epsfig{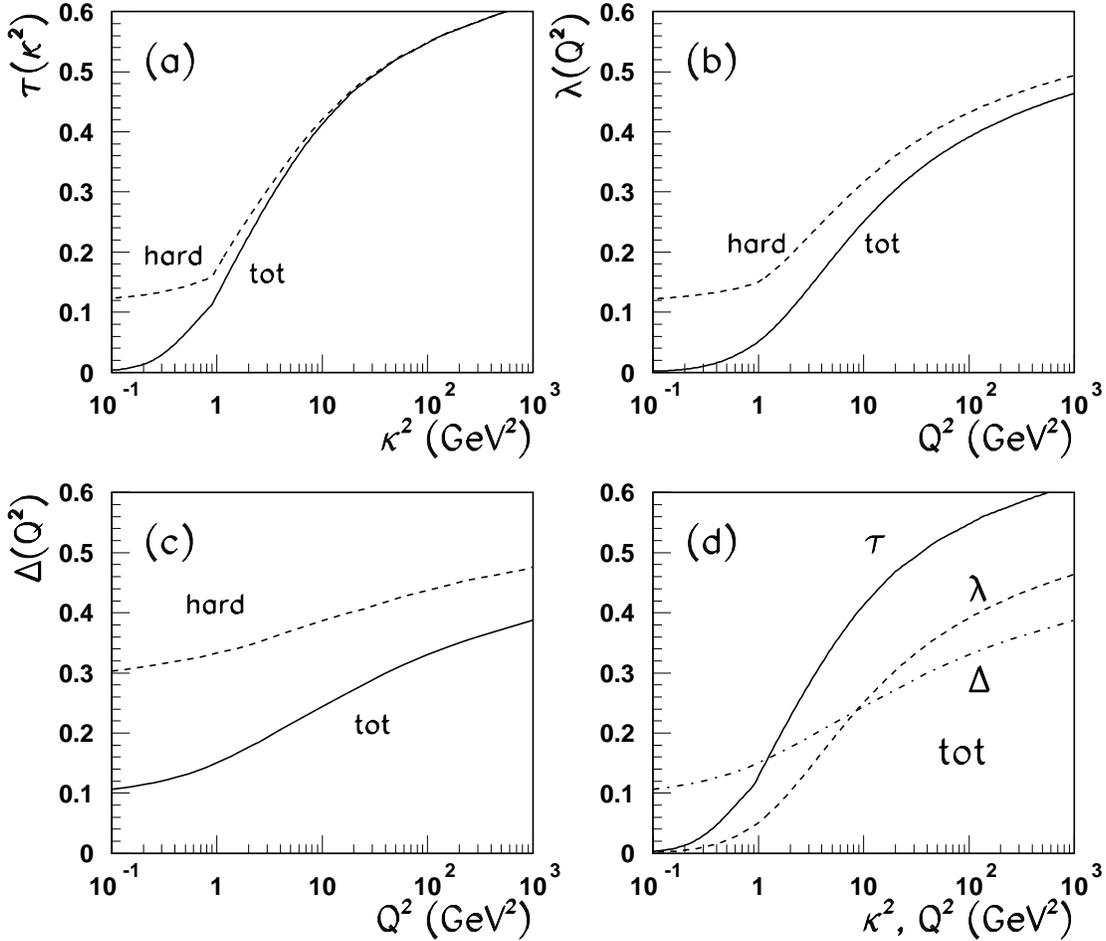}
   \caption{\it    The effective  intercepts $\tau(\bkappa^2),
\lambda(Q^2),~\Delta(Q^2)$ of the local $x$-dependence,
$10^{-3} < x < 10^{-5}$, of ${\cal
F}(x,\bkappa^2),G(x,Q^2),F_{2}(x,Q^2)$, respectively for the
$k_{\perp}$-factorization analysis
\protect\cite{INdiffglue,INanatomy,*INanatomy1} with the large-$\bkappa^2$
behavior of ${\cal F}(x,\bkappa^2)$ tuned to the GRV LO
parameterization \protect\cite{GRV} as described in the text. In
boxes (a)-(c) the dashed lines are for the hard components, the
solid lines are found if the soft components are included. The box
(d) shows how the intercepts change from ${\cal F}(x,\bkappa^2)$
to $G(x,Q^2)$ to $F_{2}(x,Q^2)$. The very close results are found
for intercepts of parameterizations
tuned to converge at large $Q^2$ to the MRS LO \protect\cite{MRS}
and CTEQ LO \protect\cite{CTEQ}.}
   \label{fig:intercepts_grv}
\end{figure}

From the viewpoint of the energy dependence, the Regge cut also can
be viewed as an infinite sequence of Regge poles. One can 
approximate the local
$x$-dependence of the BFKL-Regge expansion (\ref{eq:3.3.1.1}) by
\be
F_{2}(x,Q^2) = F(Q^2) \left(\frac{1}{x}\right)^{\Delta(Q^2)}\,
, \label{eq:3.3.1.2}
\ee
which must not be interpreted that the
Pomeron is a Regge pole with $Q^2$-dependent intercept, for such
an warning see, for instance, Bjorken \cite{BjorkenEilat}. 
An example of how the effective intercept $\Delta(Q^2)$ changes 
with the range of $x$ is found in \cite{NZBFKL,NZHERA,NZpomintercept},
the variations of the effective intercept from the
unintegrated gluon density ${\cal F}(x,\bkappa^2) \propto
\left(\frac{1}{x}\right)^{\tau(\bkappa^2)}$ to the integrated
gluon density 
$G(x,Q^2)\propto\left(\frac{1}{x}\right)^{\lambda(Q^2)}$ and to
the proton SF  $F_{2}(x,Q^2)$ is found in
\cite{INdiffglue,INanatomy,*INanatomy1}, see Fig.~\ref{fig:intercepts_grv},%
where we show separately the intercept for the hard components of ${\cal
F}(x,\bkappa^2),G(x,Q^2),F_{2}(x,Q^2)$ and for the same quantities
with the soft contributions included. These intercepts
parameterize the local $x$-dependence for $10^{-3} < x < 10^{-5}$.
The striking finding is that while
$\tau_{hard}(\bkappa^2)$ and $\lambda_{hard}(Q^2)$ exhibit a very
strong scale-dependence, i.e., the contributions form the
subleading BFKL poles are large, the $\Delta_{hard}(Q^2)$ is about
$Q^2$ independent one, $\Delta_{hard}(Q^2)\sim 0.35$-0.45.




\subsubsection{The contributions from the soft region beyond
pQCD} 

Here one faces three major questions: (i) is the rise of soft hadronic
cross sections driven by small dipoles in hadrons, (ii) what is
the mechanism of interaction of non-perturbative large dipoles and
(iii) is the soft contribution relevant to the large-$Q^2$ DIS?

The first question can be answered in the affirmative: the
somewhat model-dependent estimates suggest strongly that the rise
of the hadronic and real photoabsorption cross sections receive a
large if not a predominant contribution from the interaction of
small-size color dipoles in hadrons \cite{NSZpion,INdiffglue}.
This suggests a weak energy dependence of the genuine soft vacuum
exchange: $\Delta_{soft}\approx 0$. The discussion of the
potential importance of hard contributions to hadronic cross
section was initiated in \cite{KNPsigtot}, for the recent work
along these lines see \cite{Cudelltwopole}.

From the color dipole viewpoint, the pure pQCD considerations stop
at the dipole size $r \gsim R_c\sim (0.2\div 0.3)$ fm and can not describe
the bulk of the hadronic cross sections. It is plausible 
that at such large dipole sizes the color
dipoles spanned between the constituent quarks do still remain the
important degrees of freedom, but the corresponding soft dipole
cross section remains a model-dependent phenomenological quantity,
for which we only have constraints from soft hadronic diffractive
scattering or from real or moderate-$Q^2$ photoabsorption
\cite{NZ91,NSZpion,NSZgg,INdiffglue}. Such a soft dipole cross
section can be modeled either by the non-perturbative two-gluon
exchange \cite{NZ91,NZHERA,NSZpion,INdiffglue} or within the
closely related model of the stochastic QCD vacuum suggested by
the Heidelberg group \cite{DoschSoftSigma}.
Purely phenomenological attempts to guess the shape of this 
soft cross section and its continuation 
into the hard region \cite{GolecBiernat} 
should not be disregarded as well. 

From the practical point of view, the available models for the dipole
cross section suggest a smooth $r$-dependence of the dipole cross 
across $r\sim R_c$ up to $r\sim 1$ fm. Because the pQCD BFKL component
of the dipole cross section rises with energy much faster than 
the energy-independent soft dipole cross section, at higher
energies the dominance
of the pQCD component of the dipole cross section
will extend beyond $r\sim R_c$, for which reason the 
lower boundary for the pQCD dominance will be lower
than given by eq. ~(\ref{eq:1.4.4}).
One can come to the same conclusions 
from the smooth $\bkappa^2$-dependence of the unintegrated gluon
density from soft to hard region and the dominance of the
hard component at large ${1\over x}$ which is clearly seen in
Fig.~\ref{fig:DIFGlueGRV}.

Regarding the question (iii), even at very large $Q^2$ the virtual
photons contain the hadronic size $q\bar{q}$ components and the
SF's receive a non-vanishing, even substantial at $x\sim
10^{-2}$, contribution from the interaction of soft dipole.
Within the more familiar DGLAP approach such a contribution is
hidden in the input parton densities; the sensitivity of the DGLAP
evolution to the input partons is an old news, although eventually
the rising perturbative QCD component would take over at very
large $Q^2$ \cite{NZHERA,BFKLRegge1,*BFKLRegge2,NSZpion}. Recently there were
many suggestions to start with the Regge parameterization of
photoabsorption at small to moderate $Q^2 < Q_b^2$ and take
$F_2^{(Regge)}(x,Q_b^2)$ as a boundary condition at $Q^2=Q_b^2$
for the DGLAP evolution at large $Q_b^2$
(\cite{LaszloReggeDIS1,LaszloReggeDIS2,KaidalovReggeDIS} and
references therein).


\subsubsection{The two-Pomeron approximation}\label{sect3.3.3}


The transition from  the unintegrated gluon density, 
${\cal F}(x,\bkappa^2)$,
to the conventional, integrated one, $G(x,Q^2)$, involves an
integration, $G(x,Q^2)= \int^{Q^2}(d\bkappa^2/\bkappa^2) {\cal
F}(x,\bkappa^2)$. Similarly, to the DGLAP approximation the
small-$x$ SF involves an integration,
$F_{2}(x,Q^2) \propto
\int^{Q^2}(d\bkappa^2/\bkappa^2)G(x,\bkappa^2)$. Each integration
shifts the nodes to larger value of $Q^2$ and, furthermore,
enhances the relative contribution from the node-free rightmost
eigen-function. The model-dependent estimates within the color dipole model
show that the QCD vacuum exchange contribution to DIS is
numerically dominated by the rightmost Pomeron pole plus the
energy-independent soft exchange contributions \footnote{To this
end it is instructive to recall the early doubts in
the necessity of the hard Pomeron contribution for
description of the observed
cross sections \cite{DLnohardPom,PredazzinohardPom,CudellNoHardPole}} 
because the
subleading Pomeron pole contributions have a node in the
practically important region of $Q^2 \sim 10\div 40$ GeV$^2$
\cite{NZZspectrum1,*NZZspectrum11,NZHERA,NZpomintercept,NSZpion,NSZgg}. This is
the reason behind the remarkable flat $Q^2$ dependence of
$\Delta_{hard}(Q^2)$ shown in
Fig.~\ref{fig:intercepts_grv}. Consequently, within the kinematical
range of HERA, the hard contribution to the proton SF  can be well
approximated by a simple Regge-pole formula with the intercept
$\Delta_{hard} \sim 0.35$-0.45 \cite{INdiffglue,INanatomy,*INanatomy1}. This
finding is a dynamical justification of the two-pole approximation
\cite{DLtwopole,DLtwopoles2,Cudelltwopole}. The specific models
\cite{NSZpion,NSZgg,INdiffglue} give the concrete
$Q^2$-dependence of the residues; on general grounds there are
no reasons for decoupling of the effective hard Pomeron from soft
amplitudes, including the real photoproduction. We emphasize that
the two-Pomeron parameterization only holds in the limited range
of $x$ and should not be extrapolated far beyond the kinematical
range of HERA.


\subsection{The basics of the theory of diffractive
vector meson production} 

Here we comment briefly on properties of diffractive vector meson
production starting with the nonrelativistic quark model in
conjunction with the vector dominance model. It offers a useful
insight into such fundamental issues as the flavor dependence, the
relation between the vector meson production and $V^0 \to
e^+e^-$ decay and the way the short distance wave function of vector
mesons is probed in vector meson production. Then we qualify those
properties in the color dipole approach.


\subsubsection{The flavor dependence, the relation to
the decay $V^0 \to e^+e^-$ and VDM} 

\begin{figure}[htbp]
   \centering
   \epsfig{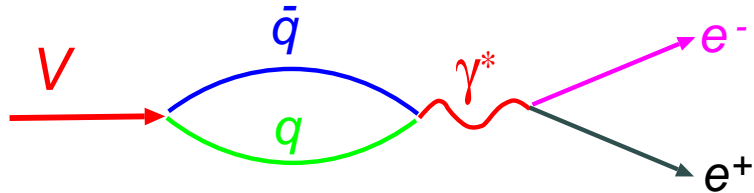}
   \caption{\it    \it  The decay of the vector meson into the lepton pair
via annihilation $q\bar{q}\to e^+e^-$.}
   \label{fig:VMdecay}
\end{figure}

On the one hand, the $V^0\to e^+e^-$ decay amplitude can be
parameterized in terms of the matrix element of the
electromagnetic current
\be
\langle 0|J_{\mu}|V\rangle =
-\sqrt{4\pi\alpha_{em}} g_V c_V V_{\mu}\, , \label{eq:3.4.1.1}
\ee
where $V_{\mu}$ is the vector meson polarization vector,
so that the decay width equals
\be
\Gamma(V^{0}\to e^+e^-)=
\frac{4\pi \alpha_{em}^2 g_V^2 c_V^2}{3m_V^3}. \label{eq:3.4.1.2}
\ee
Here the charge-isospin factors $c_V$ are
$c_{\rho}=\frac{1}{\sqrt{2}}(e_u-e_d) = \frac{1}{\sqrt{2}},~
c_{\omega}=\frac{1}{\sqrt{2}}(e_u+e_d) = \frac{1}{3\sqrt{2}},~
c_{\phi}=e_s =  -\frac{1}{3}, ~ c_{J/\Psi}=e_c=\frac{2}{3},
~c_{\Upsilon}=e_{b}= -\frac{1}{3}$. One also often uses the 
parameter
$$
{1\over f_V} = {g_V c_V \over m_V^2}.
$$
On the other hand, in the
nonrelativistic quark model the vector meson is the weakly bound
spin-triplet, $S$-wave $q\bar{q}$ state, and the decay $V^{0}\to
e^+e^-$ proceeds via annihilation $q\bar{q}\to e^+e^-$,
\be
\Gamma(V^{0}\to e^+e^-)= |R_V(0)|^2 \langle
v_{q\bar{q}}\sigma(q\bar{q}\to e^+e^-)\rangle = \frac{4
\alpha_{em}^2 c_V^2}{m_V^2}|R_V(0)|^2 \,, \label{eq:3.4.1.3}
\ee
where $v_{q\bar{q}}$ is the relative velocity of the quark and
antiquark in the vector meson and $R_V(0)$ is the 
radial wave function at the origin \cite{VanRoyen,*VanRoyen1}.
This gives a useful relationship
\be
g_V = R_V(0)\sqrt{\frac{3m_V}{\pi}}, \label{eq:3.4.1.4}
\ee
which amounts to the nonrelativistic calculation of the Feynman
diagram of Fig.~\ref{fig:VMdecay}.

\begin{figure}[htbp]
   \centering
   \epsfig{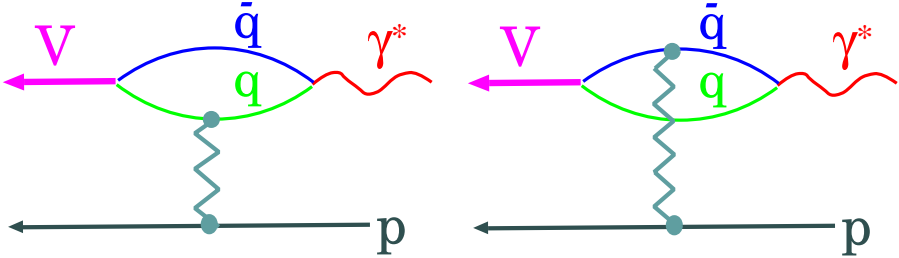}
   \caption{\it  The Additive Quark Model approximation for the
vector meson production amplitude.}
   \label{fig:AQMamplitude}
\end{figure}

Consequently, in the simplified VDM approximation, for transverse photons,
\bea
{\cal T}(\gamma^*p\to Vp)& = &\frac{\sqrt{4\pi\alpha_{em}} g_V
c_V}{Q^2+m_V^2} {\cal T}(Vp\to Vp)
\label{eq:3.4.1.5}\\
&=& \sqrt{\frac{3\Gamma(V^{0}\to e^+e^-)}{m_V\alpha_{em}}} \cdot
\frac{m_V^2}{Q^2+m_V^2} {\cal T}(Vp\to Vp)
\label{eq:3.4.1.6}\\
&=&\frac{ c_V R_V(0)\sqrt{12\alpha_{em} m_V} }{Q^2+m_V^2} {\cal T}(Vp\to
Vp). \label{eq:3.4.1.7}
\eea Precisely the same result is found if
one computes the vector meson production amplitude through the
diagrams of Fig.~\ref{fig:AQMamplitude}
and applies the additive quark model,
\bea
{\cal T}(Vp\to Vp) = {\cal T}(qp\to
qp)+ {\cal T}(\bar{q}p\to \bar{q}p)\,. \label{eq:3.4.1.8}
\eea In the
case of the $\rho p$ and $\omega p$ final states a very good
parameter free description of the E401-FNAL measurements of the
differential cross section of photoproduction is found if one
takes isoscalar elastic $\pi N$ scattering amplitudes for ${\cal T}(Vp\to
Vp)$ (\cite{GammaPelast,BusenitzVDM} and references therein). The
$sp,\bar{s}p$ amplitudes needed for the $\phi p$ state can be
extracted from the $\pi N,KN,\bar{K}N$ elastic scattering
amplitudes
\bea
{\cal T}(\phi p\to \phi p) = {\cal T}(K^+p \to K^+p)+ {\cal T}(K^-p \to
K^-p)- {\cal T}(\pi^-p \to \pi^-p)\, , \label{eq:3.4.1.9}
\eea which
gives a perfect description of the $t$-dependence of the E401-FNAL
data on photoproduction of $\phi p$ \cite{BusenitzVDM}.
Specifically, (\ref{eq:3.4.1.9}) correctly reproduces the
experimentally observed change of the diffraction slope from
$B(\gamma p \to \omega p) = 12.6\pm 2.3$ GeV$^{-2}$ to $B(\gamma p
\to \phi p) =6.8 \pm 0.8$ GeV$^{-2}$. In terms of the discussion in
Section ~\ref{sect3.1.3}, see Eq.~(\ref{eq:3.1.3.2}), this inequality of
diffraction slopes suggest that the spatial size of the $\phi$
made of the heavier strange quarks is substantially smaller than
the spatial size of the $\omega$ made of the light $u,d$ quarks.
However, the observed differential cross section is only a half of
what is predicted by (\ref{eq:3.4.1.6}) and (\ref{eq:3.4.1.9}).
Within the color dipole approach the culprit is the oversimplified VDM
approximation (\ref{eq:3.4.1.9}): the interaction of the 
quarkonium is controlled by not the number and flavor of quarks 
in the state but rather its size \cite{PumplinSubtractive}.


\subsubsection{Vector meson production in the color dipole
approach}   \label{sect3.4.2}

In the color dipole approach, thanks to Lorentz dilation of time
at high energies, the partonic fluctuation (to the lowest
 order, $q\bar q$ pair)
of the incident photon is frozen in transverse (impact parameter)
space during the interaction with the target. This allows one to
cast the photoproduction amplitude in a quantum-mechanical form
\cite{KZcharmonium,KNNZrs,NNZscanVM}
\be
{\cal T} = \langle \Psi_V|\hat \sigma_{dip}|\Psi_\gamma\rangle =
\int dz\, d^2 \br \;\Psi_V^*(\br)\;\sigma_{dip}(x,\br)\;
\Psi_\gamma(\br)\,, \label{eq:3.4.2.1}
\ee
where $z$ and $(1-z)$ are fractions of the photon's lightcone
momentum carried by the quark and antiquark, respectively. The
basic quantity here, the cross section of the color dipole
interaction with the target $\sigma_{dip}(\br)$, can be calculated
for the forward scattering case through the unintegrated gluon
distribution,
\be
\sigma_{dip}(x,\br) = {4\pi \over 3}\int
{d^2\bkappa \over \kappa^4} {\cal
F}(x,\bkappa)\,\alpha_s[\mbox{max}(\bkappa^2,A/\br^2)]\,
[1-\exp(i\bkappa\br)]\,, \label{eq:3.4.2.2}
\ee
where 
\be
A \sim 9\div10
\label{eq:3.4.2.201}
\ee 
follows from the properties of Bessel functions 
\cite{NZglue} . Eqs.~(\ref{eq:3.4.2.1}) and (\ref{eq:3.4.2.2}) sum
to the leading log${1\over x}$ the towers of two gluon exchange diagrams of
Fig.~\ref{fig:Twogluontower}, as manifested by the unintegrated
glue ${\cal F}(x,\bkappa)$ in the integrand of
(\ref{eq:3.4.2.1}). The $x$-dependence of the dipole cross section
is governed by the color dipole BFKL equation 
(\cite{NZZdipoleBFKL,NZ3Pom,NZpomdiffr,*NZpomdiffr1}, see also \cite{MuellerBFKL,MuellerPatel}),
for the discussion of the choice
\be
x=x_g\approx 0.4\cdot {Q^2+m_V^2 \over W^2}
\label{eq:3.4.2.202}
\ee 
see below Section 4.6.

In due turn, the unintegrated glue of the proton can be extracted
from the experimental data on the proton structure function
\cite{INdiffglue,INanatomy,*INanatomy1}, so that there is a microscopic QCD
link between inclusive DIS and vector meson production, if the vector meson 
is treated in the $q\bar{q}$ Fock-state approximation. For small dipoles
there is a useful relationship to the integrated gluon structure
function of the proton \cite{BGNPZunit,NZglue}
\be
\sigma_{dip}(x,\br) =
\frac{\pi^2}{3}r^2\alpha_S\left(\frac{A}{r^2}\right)
G\left(x,\frac{A}{r^2}\right).\label{eq:3.4.2.3}
\ee
Now comes the crucial point: the lightcone
wave function of the virtual photon shrinks with $Q^2$, namely,
$\Psi_\gamma(\br) \propto \exp(-\epsilon r)$, where
\cite{KZcharmonium,NNNcomments,KNNZrs}
\be
\varepsilon^2 =
z(1-z)Q^2 +m_f^2\, , \label{eq:3.4.2.4}
\ee
where $m_f$ is mass of the quark of the flavor $f$.
Then, for a
sufficiently large $Q^2$, the dominant contribution to the virtual
photoproduction amplitude will come from $r\sim r_S =3/\epsilon$,
so that 
\be
{\cal T} \propto c_V  r_S^2 \sigma_{dip}(x_g,r_S)
\Psi^*_V(z,r_S) \propto r_S^4 \alpha_S\left(\frac{A}{r_S^2}\right)
G\left(x_g,\frac{A}{r_S^2}\right),
\label{eq:3.4.2.5} 
\ee
where in the integrand of
(\ref{eq:3.4.2.5}) the $z$-dependent factors coming from the
photon wave function have been suppressed.

Note that the "quark mass" term $m_f^2$ here must not be omitted
even for the light flavors. This "quark mass" serves as an
effective parameter that bounds from above the transverse size of the
$q\bar q$ state in a real photon. One can discuss the large-size 
properties of the photon only under certain assumptions on the
color-dipole cross section for large dipoles or the unintegrated
gluon density for nonperturbative soft gluon momenta:
the early choice has been  $m_{u,d} \approx 0.15$ GeV 
\cite{NZHERA,BFKLRegge1,*BFKLRegge2}, the more recent $k_{\perp}$-factorization
analysis  
\cite{INdiffglue} of the low-$Q^2$ $F_{2p}$ data 
suggests $m_{u,d} \approx 0.22$ GeV.

The result (\ref{eq:3.4.2.5}) has all the properties of the
amplitude (\ref{eq:3.4.1.7}) subject to important QCD
modifications:
\begin{itemize}

\item The color-dipole cross section is flavor independent, and
the charge-isospin factors are precisely the same as in the VDM.

\item For $r_S \ll R_V$ the vector meson production is obviously
short distance dominated and tractable within pQCD
(\cite{KZcharmonium,NNNcomments,KNNZrs,RyskinJPsi,BFGMSvm}, for
refinements on the applicability of pQCD see Collins
\cite{collinstheorem}). The amplitude is proportional to the
vector meson wave function at vanishing transverse $q\bar{q}$
separation, $\Psi^*_V(z,0)$, which is closely related to the
so-called vector meson distribution amplitude
\cite{BrodskyLepage,ChernyakPhysRep}.

\item To the nonrelativistic approximation, $z\sim 1/2$ and
$m_V\approx 2m_q$, one has $\epsilon^2 \approx \frac{1}{4}
(Q^2+m_V^2)$, and the factor $r_S^2 \propto 1/(Q^2+m_V^2)$
reproduces the $Q^2$ dependence dictated by the vector meson
propagator.

\item However, $\sigma_{tot}(Vp\to Vp)$
which enters (\ref{eq:3.4.1.7}), is substituted for by 
\be
\sigma_{dip}(x_g,r_S) \approx  \frac{3\pi^2}{\overline{Q}^2}
\alpha_S({\overline Q}^2)G(x_g,{\overline Q}^2)\, ,
\label{eq:3.4.2.6} 
\ee
where we used (\ref{eq:3.4.2.201}) and (\ref{eq:1.3.5}) by which 
$A/r_S^2 \approx \Qb^2$. For large dipoles, $r_S \sim R_V$,
which dominate in real photoproduction,  $\sigma_{dip}(R_V)
\approx \sigma_{tot}(Vp\to Vp)$, but for small dipoles, $r_S \ll R_V$,
which dominate electroproduction, $\sigma_{dip}(x_g,r_S)\ll
\sigma_{tot}(Vp\to Vp)$ and the simplified VDM is bound to fail.

\item 
For small scanning radii, $r_S \ll R_V$, such that $\Psi_V(z,r_S) 
\approx ~const$, the dependence on $Q^2$ and
the mass of the vector meson $m_V$ only enters through the
scanning radius $r_S$. Hence the fundamental prediction
\cite{NNZscanVM} that cross sections for
different vector mesons taken at the same value of $r_S$, i.e.,
the same value of $(Q^2+m_V^2)$, must exhibit similar
dependence on energy and $(Q^2+m_V^2)$.    

\item Vector meson production probes the integrated gluon SF of
the target proton at hard scale $\overline{Q}^2$ given by
(\ref{eq:1.3.3}) (\cite{RyskinJPsi,KNNZct,NNZscanVM}, for a more
accurate definition of $\overline{Q}^2$ for light vector mesons
see \cite{Igorhardscale}).

\item Notice an inapplicability of the simplified VDM to heavy
quarkonia, for which by virtue of small $\alpha_S$  the Bohr radius
$$
R_V=a_B \approx \frac{4}{m_V\alpha_S} \gg r_S\, .
$$

\item Finally, as far as the $t$-dependence is concerned, $r_S$
can be regarded as the transverse size of the $\gamma^*\to V$
transition vertex, so that for the fixed value of $x$ 
the diffraction slope is predicted
\cite{NZZslope,NNPZZslopeVM,*NNPZZslopeVM1} to decrease with $(Q^2+m_V^2)$: 
\be
B(Q^2) \sim B_N + Cr_S^2 \approx B_N + {{\rm const} \over
Q^2+m_V^2}. \label{eq:3.4.2.7} 
\ee

\end{itemize}

Because the color dipole cross section and the unintegrated gluon
SF are related by the Fourier transform, all the
above results can be rederived in the momentum space
representation, often referred to as the
$k_{\perp}$-factorization or impact factor representation. 
The relevant formalism goes back to the
1978 seminal paper by Balitsky and Lipatov \cite{BL}, although
the term "$k_{\perp}$-factorization" has been coined much later on
by several groups \cite{Levinktfact,*Levinktfact1,Cataniktfact,Collinsktfact}.
The detailed application of the $k_{\perp}$-factorization to the
vector meson production is found in
\cite{KNZspinflip,*KNZspinflip1,INSDwave,IgorKtfact,IgorPhD} and will be reviewed in the
following section 4, the first momentum space derivation of the leading
$\log \overline{Q}^2$ approximation is due to Ryskin
\cite{RyskinJPsi} and Brodsky et al. \cite{BFGMSvm}, some
corrections to the leading
 $\log \overline{Q}^2$ approximation were discussed
by Levin et al. \cite{LevinKfactVM}. Referring to Section 4 for
a detailed discussion of the helicity amplitudes within
$k_{\perp}$-factorization, here we only cite the gross features of
the longitudinal and transverse cross sections: 
\bea
\sigma_T
\propto \frac{1}{(Q^2+m_V^2)^4}\left[\alpha_S(\overline{Q}^2)
G(x_g,\overline{Q}^2)\right]^2\, ,\label{eq:3.4.2.8}\\
\sigma_L \propto \frac{Q^2}{m_V^2}\cdot
\frac{1}{(Q^2+m_V^2)^4}\left[\alpha_S(\overline{Q}^2)
G(x_g,\overline{Q}^2)\right]^2\, . \label{eq:3.4.2.9} 
\eea 
Here the factor
$\sim Q^2/m_V^2$ in the $\sigma_L$ is a generic consequence of the
electromagnetic gauge invariance, as has been understood in early
70's \cite{SakuraiGVDM,SchildknechtGVDM}.

The Heidelberg group \cite{DoschVMstochastic}
starts with the soft color dipole cross
section evaluated within the stochastic QCD vacuum model
\cite{DoschSoftSigma}. It shares with other color dipole models
the predictions for the $Q^2$ dependence, but the energy
dependence does not follow from the first principles of the model
and needs to be introduced by hand \cite{doschenergy}.


\subsubsection{Production of excited vector mesons}\label{section2s}

The $\rho^0,\omega^0,\phi^0$ and $J/\psi$ are the ground state
vector mesons. The $\Psi'(3686)$ is the well established 
radial excitation $2S$-state, the $\Psi''(3770)$ is a solid
candidate for the orbital excitation $D$-wave state 
\cite{NovikovPhysRep,QuiggPhysRep}, the radial vs. orbital 
excitation assignment in the $\rho,\omega,\phi$ family is not
definitive yet \cite{PDG2002}. 

The salient feature of the $2S$ radial excitations is a node
of the radial wave function, $\Psi_{2S}(z,r)$, at $r = r_{node} \sim 
R(1S)=R_V$,
which suppresses the $V'(2S)$ production amplitude in comparison to 
the corresponding $V(1S)$ production  amplitude
\cite{KZcharmonium,NNNcomments,NNZrhoprim,NNPZdipoleVM}.
The strength of the node effect depends on the proximity of 
the scanning radius $r_S$ to the 
node position $r_{node}$. At $r_S \ll r_{node}$ 
(in the under-compensation regime),
which can take place at high $Q^2$ or for very heavy mesons,
the contribution from $r> r_{node}$ is small and suppression is weak.
The under-compensation regime is relevant to the $\Psi'(2S)$ 
production where the color dipole model predicts the 
rise of
the ratio $\sigma(\Psi'(2S))/\sigma(J/\psi(1S))$ with 
rising $Q^2$.  
For light vector mesons at small $Q^2$ the over-compensation 
scenario of $r_S \gsim r_{node}$ and strong cancellation is 
not excluded \cite{NNZrhoprim,NNPZdipoleVM,NNPZZslopeVM,*NNPZZslopeVM1}. 
In this scenario the $V'(2S)$ and $V(1S)$ production amplitudes 
will be of the opposite sign, which can be tested 
experimentally via the S\"oding-Pumplin effect \cite{soeding,Pumplin},
and the differential cross sections $d\sigma(V'(2S))/dt$ may
exhibit a sharp forward dip \cite{NNPZdipoleVM,NNPZZslopeVM,*NNPZZslopeVM1,IgorPhD}. 
In such a regime
even a small shift of $Q^2$ would strongly alter the cancellation 
pattern, giving rise to an anomalous $Q^2$ dependence of
the ratio $\sigma(V'(2S))/\sigma(V(1S))$, of the $t$-dependence of  
$d\sigma(V'(2S))/dt$ and of the ratio $\sigma_L/\sigma_T$ for the
$V'(2S)$ - the latter effect is due to a slightly different impact
of the node effect on different helicity amplitudes. 
A subsequent discussion of sensitivity of the node effect to 
the wave function of vector mesons is found in
\cite{DoschRhoRhoprim,NemchikRhoprim,HoyerPsiprim,HufnerPsiprim}, 
the change of numerical results for the $\Psi'(2S)$ 
from one model to another must be regarded as marginal.

The case of the orbital excitation $V''(D)$ is quite different
\cite{INSDwave}:
here the radial wave function vanishes at the origin, and the 
$Q^2$ dependence of the $V''(D)$ production will be smooth.
There are some subtle changes in the helicity amplitudes: in both
the $V(1S)$ and $V''(D)$ the $q\bar{q}$ pair is in the spin-triplet
state, but the total spin of the pair is along in $V(1S)$, and opposite 
to in $V''(D)$, the spin of the meson. 

The node effects echoes in the hard scale for the $V'(2S)$ production.
In the under-compensation regime of relevance to the $\Psi'(2S)$ 
the contribution to the production amplitude from large color 
dipoles, $r > r_{node}$, is canceled by the contribution
from small dipoles, $r < r_{node}$. As a result, the $\Psi'(2S)$ 
production amplitude is dominated by color dipoles of smaller size 
than it is the case for the $J/\Psi(1S)$ and color dipoles 
models predict the hierarchy of hard scales
$\Qb^2(\Psi'(2S)) >\Qb^2(J/\Psi(1S))$. Consequently, the $\Psi'(2S)$
production amplitude must grow with energy faster than the
$J/\Psi(1S)$ production amplitude \cite{NNPZdipoleVM}. 
Furthermore, 
the (negative valued) contribution to the production amplitude
from large dipoles,  $r> r_{node}$, has a steeper $t$-dependence than
the (positive valued) contribution from small dipoles, $r> r_{node}$.
As a result, the diffraction slope in the $\Psi'(2S)$ production
is predicted to be smaller than in the $J/\Psi(1S)$ production 
\cite{NNPZdipoleVM,NNPZZslopeVM,*NNPZZslopeVM1}. 


\subsubsection{Unitarity and saturation in the color dipole
language}

The unitarization of rising scattering amplitudes in QCD remains
one of the hot and as yet unsolved issues. As emphasized in
Section 3.1.4, the unlimited growth of the model partial waves
must be tamed and the unitarity bound $\Gamma(\bb) \leq 1$ must be
met in a consistent treatment of high energy scattering. The
theory is still in the formative stage, though. Some of the early
works on unitarization have been mentioned in Section 3.1.4, the
problem of unitarity is most acute for interactions with nuclei,
in which case the impulse approximation partial waves
$\Gamma_0(\bb) \propto A^{1/3}$. For the nuclear targets the
presence of a new large parameter --- the optical thickness of a
nucleus --- leads to certain simplifications like the applicability
of the eikonal approximation for the color dipole-nucleus
scattering
\cite{NZ91PhysLett,NZ91,SlavaNucleus,SlavaNucleus1,MuellerSaturation,NonlinearKt,*NonlinearKt1}.
The recent development in imposing the unitarity on nuclear
amplitudes, often referred to as the color glass condensate, is
summarized in \cite{RajuReview,Gatchina,IancuMueller,Leonidov},
for a review of the early works see \cite{GLRPhysRep,LRPhysRep}.
A review of the enormous literature on the subject goes beyond
the scope of this review, we
rather present a brief introduction into major ideas.

\begin{figure}[htbp]

   \centering

   \epsfig{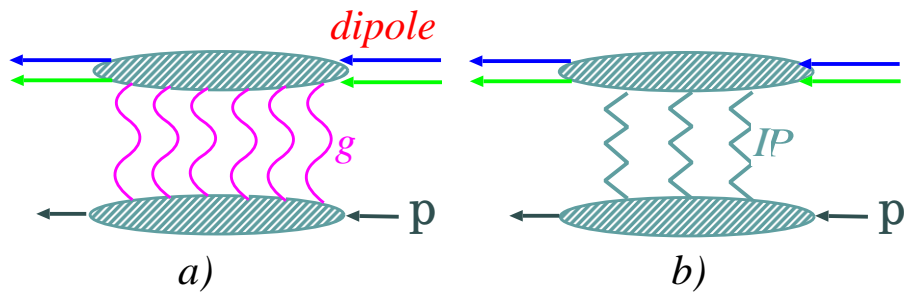}

   \caption{\it  (a) The multigluon $t$-channel exchange diagram contribution
to the color dipole scattering amplitude  and (b) its
approximation by multiple exchange by the two-gluon Pomerons.}

   \label{fig:MultiPomeron}

\end{figure}

Let $\Gamma_0(\br,\bb)$ be the profile function for the color
dipole-nucleon scattering evaluated in the single-Pomeron exchange
approximation of Fig.~\ref{fig:Twogluontower}. The Gaussian
approximation (\ref{eq:3.1.4.9}) is not imperative but convenient
for the sake of illustration. In the full fledged QCD one needs to
sum all multigluon $t$-channel exchanges between the color dipole
and nucleon, including interactions between all exchanged not
shown in Fig.~\ref{fig:MultiPomeron}a. A poor man's
approximation to this as yet unsolved problem
(\cite{BartelsRyskin4gluon,BartelsRyskinUnitarization} and
references therein) is the multiple exchange by bare Pomerons,
in general case the interactions between gluons from different 
Pomerons must be included. When the multipomeron
exchanges are evaluated in the eikonal approximation, one obtains the
``unitarized" profile function \cite{BGNPZunit}
\be
\Gamma(\br,\bb) = 1 -\exp[-\Gamma_0(\br,\bb)]\, ,
\label{eq:3.4.4.1}
\ee
whereas the so-called $K$-matrix
unitarization gives \cite{BGNPZunit}
\be
\Gamma(\br,\bb) =
{\Gamma_0(\br,\bb) \over 1 +\Gamma_0(\br,\bb)}\, .
\label{eq:3.4.4.2}
\ee
The latter has been suggested also
from the consideration of the so-called fan diagrams
(\cite{KaidalovUnitarization}, similar results are found
from different approximate non-linear evolution equations
\cite{Balitsky,MuellerAntiKovchegov}), for the so-called $U$-matrix
approach see \cite{TroshinUmatrix}.

The principal point is that the unitarized partial waves do always
respect the unitarity bound $\Gamma_0(\br,\bb) \leq 1$. The
partial waves saturate at the black-disc limit,
$\Gamma(\br,\bb)\approx 1$, for all impact parameters such that
$\Gamma_0(\br,\bb) \gg 1$, i.e.,
\be
\bb^2 \lsim
2B(\br)\log\Gamma_{0}(\br,\bb=0)\,. \label{eq:3.4.4.3}
\ee
The two
unitarized forms (\ref{eq:3.4.4.1}) and (\ref{eq:3.4.4.2}) only
differ by the rate of approach to the black disc limit, the
$K$-matrix unitarized dipole cross section takes a particular
simple form  \cite{BGNPZunit}
\be
\sigma(x,\br) = 4\pi B(\br)
\log\left(1+ {\sigma_0(x,\br)\over 4B(\br)}\right)\,,
\label{eq:3.4.4.4}
\ee
which shows clearly how the power like
small-$x$ growth of the bare Pomeron cross section
$\sigma_0(x,\br)\propto x^{-\Delta_{\Pom}}$ is superseded by the
$\propto \log{1\over x}$ behavior, or $\propto \log^2{1\over x}$
if one allows the Regge growth of the diffraction slope $B(x,\br)$
\cite{Dubovikov}.

Finally, the unitarization alters dramatically the
$\br$-dependence of the dipole cross section from
(\ref{eq:3.4.2.3}). At asymptotically small $x$ the unitarization
is at work already for small dipoles, where the $\br$-dependence
of diffraction slope $B(x,\br)$ can be neglected, see
(\ref{eq:3.4.2.7}). so that the dipole cross section would
saturate, $\sigma(x,\br) \approx 4\pi B$, for dipoles
\be
\br^2
\gsim r_{\rm sat}^2 = {12 B \over \pi \alpha_S(r)G(x,q^2=A/r^2)}\,. 
\label{eq:3.4.4.5}
\ee
The smaller is $x$, the larger is the
gluon SF in the numerator in the r.h.s. of (\ref{eq:3.4.4.5}) and
the smaller is the saturation scale $r_{\rm sat}^2$. 
Recently, specific
parameterizations for the saturating dipole cross section without
an explicit reference to the unitarity properties of partial waves
have been proposed \cite{GolecBiernat,BartelsSaturation}. (In
principle, the saturation rate and the saturated cross section
must be adjusted to describe the diffractive hadronic scattering
and real photoproduction \cite{NZ91,NZdifDIS}, 
which has not been done in the model
\cite{GolecBiernat,BartelsSaturation}.) However, with the
realistic dipole cross sections the  unitarization effects for DIS
\cite{BGNPZunit} and for vector meson production \cite{NNZscanVM}
were found to be marginal. The extraction of the $S$-matrix for the
color dipole scattering from the vector meson production data 
by Munier, Mueller and Stasto also
shows that the dipole-nucleon scattering is not yet close to the
strong absorption regime \cite{StastoUnitarization}. Similar
conclusion follows from the impact parameter extension
\cite{KowalskiSaturation} of the saturation model
\cite{GolecBiernat,BartelsSaturation}. Those findings are not
surprising, though: as shown in \cite{NZZdiffr} in the limit of
strong saturation the diffractive rapidity gap DIS must make
precisely 50 per cent of the total DIS cross section, whereas
experimentally the fraction of diffractive DIS is about 10 per
cent \cite{pl:b315:481,np:b429:477}.

To summarize, as soon as impact parameter dipole model has been
adjusted to fit the experimental data on DIS structure functions
and the total cross section and the $t$-dependence of diffractive
vector meson production, it is expected to have partial waves
consistent with the unitarity constraints in the energy and $Q^2$
range in which the experimental data are available. The same must
be true of the unintegrated gluon SF of the proton extracted in
\cite{INdiffglue} from the DIS data. Applying unitarity
corrections to the vector meson production amplitudes evaluated
with such an unintegrated gluon SF would be the double counting.


\subsubsection{Color dipole model and Generalized VDM}

\begin{figure}[htbp]

   \centering

   \epsfig{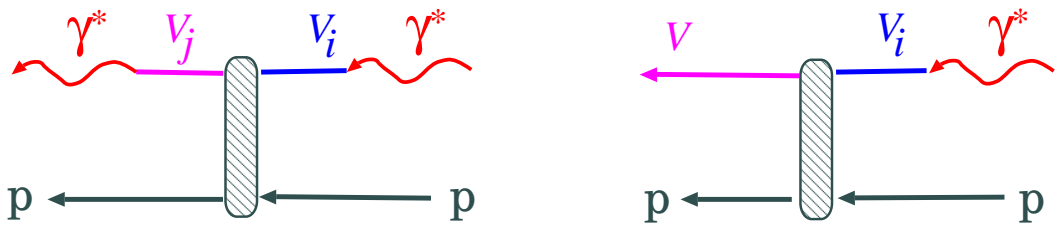}

   \caption{\it  The Generalized Vector Dominance Model diagrams
for Compton scattering (DIS) and diffractive vector meson
production.}

   \label{fig:GVDM}

\end{figure}

The simplified VDM must be regarded as the leading term of the
mass-dispersion relation calculation of the $Q^2$ dependence of
the virtual photoproduction amplitude. The importance of
contributions from the more distant singularities --- the higher
vector states and the continuum, which we denote generically as
$V_i$, --- rises with $Q^2$
\cite{GribovGVDM,*GribovGVDM1,SakuraiGVDM,SchildknechtGVDM}. Within the
resulting Generalized VDM (GVDM) for DIS the calculation of
$\gamma^*p\to \gamma^*p$ must allow for transitions of photons to
all higher vector states, $\gamma^*\to V_i$, followed by the diagonal
and off-diagonal scattering $V_i p \to V_j p$ and the transition
$V_j \to \gamma^*$, see Fig.~\ref{fig:GVDM}.

Similarly, the transitions $\gamma^*\to V_i$ followed by the
off-diagonal scattering
 $V_i p \to V p$ would contribute to the vector meson production
$\gamma^*p\to V p$. If viewed as the mass dispersion relation,
the GVDM can not fail, but the practical application requires the
knowledge of all the diagonal and off-diagonal amplitudes $V_i p
\to V_j p$ and of the $Q^2$ dependence of transitions $\gamma^*\to
V_i$. The color dipole model provides the QCD input for the GVDM
analysis \cite{NNNcomments,BFNZgvdm,*BFNZgvdm1}, the equivalence of the two
approaches emphasized in \cite{NZ91} has been elaborated by
Schildknecht et al.
\cite{SchildknechtCD,*Kuroda:2003np,*Cvetic:1999fi,*Cvetic:2001ie}.


\subsubsection{The s-channel helicity non-conservation (SCHNC)}\label{SCHNC}


The scattering of the $q\bar{q}$ dipole on the target via exchange
of the two-gluon tower exactly conserves the $s$-channel helicity of the
quark and antiquark (for the QED case see 
\cite{Yennie,LandauLifshitz1}). This does not imply the 
conservation of the helicity of photons in the off-forward 
Compton scattering. As a spin-1 particle, photon is similar
to the deuteron. In the
non-relativistic case the pure $S$-wave deuteron with spin up
consists of the spin-up proton and spin-up neutron, the 
longitudinal deuteron consists of the spin-up proton 
and spin-down neutron and vice versa. 

The perturbative QED transition of the photon to the $q\bar{q}$ pair is 
described by the familiar vertex $e_f\bar{q}\gamma_{\mu} q A_{\mu}$.
The longitudinal (scalar) virtual photon with helicity $\lambda_{\gamma}=0$
consists of the $q\bar{q}$ Fock state with $\lambda +\overline{\lambda}=\lambda_{\gamma}=0$,
in close similarity to the $S$-wave deuteron.  
The crucial point is that 
the transverse photon with helicity $\lambda_{\gamma}=\pm 1$
besides the $q\bar{q}$ state with
$\lambda+\overline{\lambda}=\lambda_{\gamma}=\pm 1$ contains the
state with $\lambda + \overline{\lambda}=0\neq \lambda_{\gamma}$, 
in which the helicity
of the photon is carried by the orbital angular momentum in the
$q\bar{q}$ system (see \cite{SlavaSpinFlip} for an early
discussion of this mechanism in application to the spin-flip in
the nucleon scattering). Furthermore, it is precisely the  state
chiral-even state with antiparallel helicities,
$\lambda + \overline{\lambda}=0$, which gives the dominant
contribution to the absorption of transverse photons and the
proton SF $F_{2p}(x,Q^2)$ in the Bjorken limit.
The perturbative transition of transverse photons to the chiral-odd 
state with parallel helicities, 
$\lambda+\overline{\lambda}=\lambda_{\gamma}=\pm 1$,
vanishes in the massless quark limit.
 
The helicity structure of vector mesons is about the same.
From the point of view of the vector meson production, it is
important that the transverse and longitudinal $\gamma^*$ and $V$
share the intermediate $q\bar{q}$ state with
$\lambda + \overline{\lambda} =0$, which allows the s-channel
helicity non-conserving (SCHNC) transitions between the transverse
(longitudinal) $\gamma^*$ and longitudinal(transverse) vector
meson V \cite{KNZspinflip,*KNZspinflip1,IKspinflip}. This mechanism of SCHNC
does not require an applicability of pQCD.

Hereafter we only discuss the experimental data from HERA 
taken with unpolarized protons, hence proton can be treated as
a spinless particle, see, however, a brief discussion
in Section 5.5. Depending on the spin-parity
of the $t$-channel exchange, the helicity amplitudes 
satisfy \cite{NaturalUnnatural}
\be
{\cal T}_{-\lambda_V -\lambda_\gamma} = \pm (-1)^{\lambda_V- \lambda_\gamma}
{\cal T}_{\lambda_V \lambda_\gamma}\,,
\label{eq:3.4.6.1}
\ee
where the $+(-)$ sign applies to natural (unnatural) parity exchange.
As discussed in Section 3.1.2, Pomeron and all the highest lying subleading
reggeons have the natural spin-parity. Under the dominance of 
the natural spin-parity exchange, the number
of independent helicity amplitudes is reduced to five:
\begin{eqnarray}
&&L \to L\,;\quad T \to T\  (\lambda_\gamma = \lambda_V)\nonumber\\
&&T\to L\,;\quad L\to T\nonumber\\
&&T \to T'\  (\lambda_\gamma = -\lambda_V).\label{eq:3.4.6.2}
\end{eqnarray}
The first line contains helicity-conserving amplitudes. They are
predicted and found to be the dominant ones. They do not vanish
for the forward production, $\bDelta=0$. The second line in
(\ref{eq:3.4.6.2}) contains two single helicity-flip amplitudes. They
must be proportional to $|\bDelta|$ in the combination 
$(\bfe \cdot \bDelta)$ or $(\bV^* \cdot \bDelta)$, since there is no other
transverse vector at our disposal. The last line contains the double
helicity-flip amplitude, which must be proportional to 
$(\bfe \cdot \bDelta)(\bV^* \cdot \bDelta)$.

One can thus predict that the $s$-channel helicity conserving
amplitudes will dominate in the almost forward production of
mesons. As $t$ increases, the relative importance of helicity-flip
amplitudes will grow, and, at high enough $t$, they might become
competitive to the helicity conserving amplitudes.

\subsubsection{Diffractive vector meson production from
extended Bloom-Gilman duality.}


The Bloom-Gilman inclusive-exclusive duality relates the $x\to 1$
behavior of DIS to elastic $ep$ scattering \cite{BloomGilman}.
Roughly speaking, if one stretches the $x$-dependence of the DIS
cross section determined for the continuum masses $W$ to the
elastic limit $W\to m_p$, then the DIS cross section
integrated over the interval
\be
0 <1-x < \frac{W_0^2-m_p^2}{Q^2}
\label{3.4.7.1}
\ee
will, with the judicious choice of the duality
interval $[m-p,W_0]$, be equal to the elastic $ep$ cross section
(for the recent active discussion of duality in DIS in connection
with the JLab data see \cite{JlabDuality,*Armstrong:2001xj,*Liuti:2001qk}). Genovese et al. argued
\cite{GNZlong} that similar parton-duality relationship must hold
between the diffraction excitation of the small mass continuum
\[
ep \to e'Xp'
\]
and exclusive vector meson production. In terms of the diffractive
Bjorken variable
\[
\beta = \frac{Q^2}{Q^2+M_X^2}
\]
the diffractive structure functions for the transverse and
longitudinal photons have the large-$\beta$ behavior
\cite{GNZlong,GNZcharm,NPZschnc}
\bea
F_T(x,\beta,Q^2) \propto (1-\beta)^2 G^2(x,q_T^2)\,, \label{eq:3.4.7.2}\\
F_T(x,\beta,Q^2) \propto \frac{1}{Q^2} G^2(x,q_L^2)\,. \label{eq:3.4.7.3}
\eea 
The relevant hard scales equal
\cite{GNZcharm,GNZlong,Bartelshardscale,*Bartelshardscale1}
\bea
q_T^2 \sim
\frac{m_q^2} {M_X^2} (Q^2+M_X^2)\, , \label{eq:3.4.7.4} \quad q_L^2 \sim
\frac{1}{4} (Q^2+M_X^2) \, . \label{eq:3.4.7.5}
\eea 
The
integration over the duality interval $[M_{min}\sim 2m_q,M_{T,L}]$, i.e.,
$1-\beta_{T,L} < M_{T,L}^2/Q^2$, yields the correct large-$Q^2$
dependence of $\sigma_{L,T}$. Furthermore, both hard scales
$q_{T,L}^2$ tend to the scale $\overline{Q}^2$ of Eq.~(\ref{eq:1.3.3})
so that Eqs.~(\ref{eq:3.4.7.2}),~(\ref{eq:3.4.7.3}) yield
precisely the same dependence on the gluon structure function as
in (\ref{eq:3.4.2.7}),~(\ref{eq:3.4.2.8}). Motivated by this
observation \cite{GNZlong}, Martin et al. suggested to 
evaluate the vector meson
production cross sections $\sigma_{T,L}$ from the duality integral
\cite{martinduality,martinduality2}. This way one encounters 
a very strong sensitivity of such evaluations of $\sigma_{T,L}$
to the duality interval,
\bea
\sigma_T \propto (M_T^6 -M_{min}^6)\, , \\
\label{eq:3.4.7.6} 
\sigma_{L} \propto (M_L^2-M_{min}^2)\, ,
\label{eq:3.4.7.7} 
\eea
which is especially strong in the case of $\sigma_T$.

Similar in spirit 
to the duality is the unorthodox color evaporation model
(CEM). In its original formulation \cite{BuchmullerHebeckerCEM}
it simply states that the color of the $q\bar{q}$ pair 
produced in $\gamma^* g \to q\bar{q}$ subprocess 
happens to be bleached by soft 
final-state interactions leading to the rapidity gap events with the
probability $1/9$. Within CEM the charmonium production 
is described by the formation of colored open charm $c\bar{c}$ states
which masses $M_{c\bar{c}}\leq 2m_D$, where $m_D$ is the mass of the
$D$-meson \cite{HalzenCEM}:
\be
\sigma_{onium}= {1\over 9}\int_{2m_c}^{2m_D} dM_{c\bar{c}}
{d\sigma_{c\bar{c}}\over dM_{c\bar{c}}}\,,
\label{eq:3.4.7.8} 
\ee
where $1/9$ is the color bleaching probability. Assuming that
about 50\% of the onium goes into the $J/\Psi$, Amundson et al.
are able to describe the photo- and hadroproduction of the $J/\Psi$ 
\cite{HalzenCEM2,HalzenCEM3}. Gay Ducati et al. find similar agreement with
the total cross section of elastic charmonium photoproduction
\cite{GayDucatiCEM,GayDucatiCEM2}. 
Here we only notice that in the near-threshold 
process $\gamma^* g \to c\bar{c}$ produces the spin-singlet
S-wave $c\bar{c}$ pair. Arguably, the color bleaching can 
not flip the spin of nonrelativistic heavy quarks and
that must lead to strong suppression factor in the
estimate (\ref{eq:3.4.7.8}). In contrast to that, in 
hadronic collisions the near-threshold open charm
can be produced in the spin-triplet state via
$q\bar{q},gg \to g \to c\bar{c}$ and the spin dynamics
of heavy nonrelativistic quarks does not prohibit the
formation of $J/\Psi$ by color bleaching.  


\subsubsection{Models which respect the Froissart bound}
\label{sect3.4.8}

Only a limited range of energy, $x$, and $Q^2$ is spanned by the
available experimental data. We already mentioned of an equally
good description of the available soft cross section data by the
soft Pomeron pole exchange and logarithmic parameterizations, 
see Section 3.1.4.
We also recall an observation by Buchm\"uller and Haidt
\cite{BuchmullerHaidt} that gross features of the small-$x$ proton
structure function as measured at HERA are reasonably well
reproduced by a very simple parameterization
\be
F_2(x,Q^2) = a +
m \log\frac{x_{0}}{x}\log \frac{Q^2}{Q_{0}^2}\, .
\label{eq:3.4.8.1}
\ee
From the Regge theory viewpoint this
corresponds to the dipole singularity at $j=1$. The dipole
singularity model for virtual photoproduction of vector mesons has
been proposed by Fiore et al. \cite{FioreDoublePolePomeron,*Fiore:2001bg,LaszloHeavyVM}, in
this specific example the nonlinear Pomeron trajectory with the
branching point singularity at the two-pion threshold in the
$t$-channel is used. For each and every vector meson a good
description of the vector meson production cross sections is found
at the expense of five free parameters. A very closely related model
was proposed by Martynov et al. \cite{dipolepom,dipolePomeron2}.
Troshin and Tyurin suggested a parameterization of vector meson
production amplitudes in which the high energy growth is tamed by
the $U$-matrix unitarity constraints \cite{TroshinUmatrix}. 
Haackman et al. \cite{KaidalovSoftVM}
start with the soft Pomeron with $\Delta_{\Pom}>0$
and impose the unitarization by reggeon field theory methods
as mentioned in Section 3.1.4. The
drawback of such models is that the $Q^2$ dependence of the vector
meson production is parameterized rather than predicted from the
microscopic QCD.

\newpage


\section{The $k_{\perp}$-factorization: unified microscopic QCD
description of DIS and vector meson production}


\subsection{The leading $\log\frac{1}{x}$ and
$q\bar{q}$ Fock state approximations}

The color dipole and $k_t$-factorization approaches to small-$x$
DIS are conjugate to each other, the technical correspondence is
given by Eq.~(\ref{eq:3.4.2.2}). The advantage of the former is in
its simple quantum-mechanical representation, still some 
technical issues such as the definition of the lightcone wave
functions, the separation of the $S$-wave and $D$-wave states of
vector mesons, and the r\^ole of the so-called skewed, or
off-diagonal, gluon distribution functions are more transparent in
the momentum-space representation.

The starting point is the BFKL diagram for small-$x$ DIS,
Fig.~\ref{fig:Twogluontower} and the reference reaction is the
non-forward Compton scattering $\gamma^*p \to
\gamma^*(\bDelta)p(-\bDelta)$. The vector meson production is
obtained from the Compton diagram replacing the outgoing {\it
pointlike} photon $\gamma^*$ by the {\it non-pointlike} vector
meson $V$. To the leading $\log\frac{1}{x}$ the
effect of perturbative higher, $q\bar{q}g,~q\bar{q}gg$ etc., 
Fock states in the
{\sl pointlike}  photon amounts to the BFKL evolution of the color
dipole cross section or of the unintegrated gluon SF
while retaining the $q\bar{q}$ Fock state approximation
\cite{FKL75,KLF77,*KLF771,BL,NZZdipoleBFKL,MuellerBFKL,MuellerPatel}.
Namely, in the DIS counterpart of (\ref{eq:3.4.2.1}) one
calculates the photoabsorption cross section as an expectation
value of the dipole cross section over the lowest $q\bar{q}$ state
of the photon:
\be
\sigma_{tot}(\gamma^*p) = \int_{0}^{1}dz \int
d^2\br \Psi_{\gamma^*}^*(z,\br)\sigma_{dip}(x,\br)
\Psi_{\gamma^*}(z,\br)\,. \label{eq:4.1.1}
\ee

\subsection{The helicity and chiral structure of the photon}


In the momentum representation the chiral structure of the 
$\bar{q}\gamma_{\mu}q A_{\mu}$ vertex is as follows. The photon
polarization vectors are described in Section 2.3.2, here we
only notice that in the Sudakov representation
\be
e_{\mu}(L) = -{1\over Q}\left(q'+{Q^2\over W^2}p'\right)
\label{eq:4.2.1}
\ee
where the two Sudakov lightcone vectors are defines as
\be
P
= p'  + \frac{m_p^2}{W^2}{q}'; \quad {q} = {q}'-xP;\quad q'^2 = p'^2
= 0; \quad x = {Q^2 \over W^2}\ll 1.
\label{eq:4.2.2}
\ee
Hereafter it will be convenient to use twice the quark and antiquark helicity, 
$\lambda,\overline{\lambda}=\pm 1$, which should not cause a 
confusion. For the transverse photons, $\lambda_{\gamma}=\pm 1$, 
in the momentum representation the perturbative QED
vertex gives the structure
\bea
\bar{q}_{\lambda}\gamma_{\mu}q_{\overline{\lambda}} e_{\mu}(\lambda_{\gamma})=
{1\over \sqrt{z (1-z)} }\left\{-\sqrt{2}m_f \delta_{\lambda_{\gamma},\lambda}
 \delta_{\lambda,\overline{\lambda}} 
+2 \delta_{\lambda,-\overline{\lambda}} [z\delta_{\lambda_{\gamma},\lambda} 
-(1-z)\delta_{\lambda_{\gamma},\overline{\lambda}}] (\bk \cdot \bfe(\lambda_{\gamma}))\right\}
\nonumber\\
\label{eq:4.2.3}
\eea
and for the longitudinal (scalar) photons
\bea
\bar{q}_{\lambda}\gamma_{\mu}q_{\overline{\lambda}} e_{\mu}(\lambda_{\gamma}=0)=
-2Q\sqrt{z (1-z)} \delta_{\lambda,-\overline{\lambda}} \,.
\label{eq:4.2.4}
\eea
Here $z$ and $(1-z)$ are the fractions of the
photon's lightcone momentum carried by the quark and
antiquark, respectively, and $\bk$ and $-\bk$ are the
corresponding transverse momenta. The perturbative chiral-odd component of the transverse photon 
with parallel helicities vanishes 
for massless quarks. The scaling contribution to the DIS structure 
function $F_2(x,Q^2)$ comes from the chiral-even component with 
antiparallel helicities.

\subsection{The lightcone helicity and chiral structure of vector mesons
and rotation invariance}


In the vector meson, the quark and antiquark are in the
spin-triplet state and either $S$- or $D$-wave. The lightcone wave
function $\Psi_V(z,\bp)$ is a probability amplitude for expansion
of the vector meson in $q\bar{q}$ states with invariant mass
\be
M^2= \frac{\bk^2+m_f^2}{z(1-z)} = 4(m_f^2+\bp^2)\, .
\label{eq:4.3.1}
\ee
One calculates first the
amplitude of production of the $q\bar{q}$ pair,
\be
\gamma^* p \to
(q\bar{q}) p'\,, \label{eq:4.3.3}
\ee
and then projects it onto the vector state 
by weighting with $\Psi_V(z,\bk)$ and
the relevant helicity factors. 
 The use of the 3-dimensional
momentum of the quark in the $q\bar q$ rest frame, 
$\bp = \left(\bk,(z-\frac{1}{2}) M\right)$, 
\be
\frac{d^3\bp}{M} = \frac{dz d^2\bk}{4z(1-z)}\, , \label{eq:4.3.2}
\ee
is helpful to see a link to the conventional
quantum-mechanical description.

The helicity/chiral structure of the vector meson for the widely 
used $Vq\bar{q}$ extension of the QED vertex of the form
\be
V_{\mu}\bar{q}_{f}\gamma_{\mu} q_f
\Gamma_V(z,\bk)\, , \label{eq:4.3.4}
\ee
is the same as for the photon subject to the substitution
$Q\to M$ for the longitudinal vector meson. The vertex (\ref{eq:4.3.4})
gives a certain admixture of the $S$ and $D$ waves. The $SD$-mixing 
is familiar
from the case of the deuteron, where it originates from the
pion-exchange tensor interaction, the presence of the 
potential-dependent $SD$-mixing
in vector mesons
is a generic feature of potential models (for the review see
\cite{NovikovPhysRep}).

The rotation-invariant lightcone description of the pure $S$ and
$D$-wave states and the corresponding vertices $S_{\mu}$ and
$D_{\mu}$ are found in \cite{INSDwave,IgorPhD}, for the related
discussion see also \cite{JausSDwave,AnisovichSDwave}. To generate
the pure $S$-wave state one needs to add the
generalized Pauli vertex. Upon applying the
Gordon identities,  the pure $S$-wave vertex can be cast in the form
\be
\Gamma_S(z,\bk)\bar{q}_{f}S_{\mu}q_f V_{\mu}= 
\Gamma_S(z,\bk)\bar{q}_{f}\left\{\gamma_{\mu}- {1\over (M+2m)}(p_f - 
p_{\bar{f}})_{\mu}\right\}
q_f V_{\mu}\,
\label{eq:4.3.5}
\ee
with the helicity/chiral structure
\bea
\bar{q}_{\lambda}S_{\mu}q_{\overline{\lambda}} V_{\mu}(\pm 1)&=&
{1\over \sqrt{z (1-z)} }\left\{-\sqrt{2} m_f\delta_{\lambda_{V},\lambda}
 \delta_{\lambda,\overline{\lambda}} + 2
\delta_{\lambda,-\overline{\lambda}} [z\delta_{\lambda_{V},\lambda} 
-(1-z)\delta_{\lambda_{V},\overline{\lambda}}](\bk \cdot \bV(\lambda_{V}))\right.\nonumber\\
&+&\left. { 2 (\bk \cdot\bV(\lambda_{V})) \over M+2m_f}
\left[m_f(1-2z)\delta_{\lambda, -\overline{\lambda}} +
\sqrt{2}(\bk \cdot\bV(-\lambda))\delta_{\lambda,\overline{\lambda}}\right] \right\}\,
\nonumber\\
\bar{q}_{\lambda}S_{\mu}q_{\overline{\lambda}} V_{\mu}(0)&=&
-2M\sqrt{z (1-z)} \delta_{\lambda,-\overline{\lambda}} \nonumber\\
 &-&{ M(1-2z) \over (M+2m_f)\sqrt{z(1-z)}}
\left[m_f(1-2z)\delta_{\lambda, -\overline{\lambda}} +
\sqrt{2}(\bk\cdot \bV(-\lambda))\delta_{\lambda,\overline{\lambda}}\right] \,.
\label{eq:4.3.6}
\eea
Note that momenta $p_f$ and $p_{\bar{f}}$ correspond to on-mass-shell fermions,
see details in \cite{IgorPhD}, which justifies
the usage of the Gordon identity.
The corresponding  vertex functions $\Gamma_{S}$ will only depend on
the "radial" variable $M^2$ and can be related to the
momentum-space radial wave functions $\psi_{S}(z,\bk)$:
\be
\Gamma_{S,D}(M^2) = \psi_{S}(z,\bk)(M^2-m_V^2). \label{eq:4.3.7}
\ee

An important part of the rotation-invariant description is
that the transversity condition must be imposed at the level of
the $q\bar{q}$ pair, which leads to the concept of the {\it
running longitudinal} polarization vector $V_{L}(M)$, which has
the Sudakov expansion
\be
V(\lambda_V=0) = {1 \over M}\left( {q}' -
\frac{M^2}{W^2}p'\right) \label{eq:4.3.8}
\ee
such that it is
orthogonal to the 4-momentum of the on-mass shell $q\bar{q}$ pair,
$(v_{q\bar{q}}\cdot V_{L}(M))=0$, where
\be
v_{q\bar{q}} = {q}' +
\frac{M^2}{W^2}p'\, \quad v_{q\bar{q}}^2=M^2\,. \label{eq:4.3.9}
\ee
This running polarization vector has been used in (\ref{eq:4.3.6}).

The lightcone extension of the considerations in Section 3.4.1
gives the $V^{0}\to e^+e^-$ decay constant for the $S$-wave state
\cite{IgorPhD}
\bea
g_V & =& N_c\int \frac{d^3\bp}{(2\pi)^3}
\psi_S(\bp) \cdot
\frac{8}{3}(M+m_f)  \nonumber\\
& = & N_c\int_{0}^{1} dz \int \frac{d^2\bk}{(2\pi)^3z(1-z)}
\psi_S(\bp) \cdot
\frac{2}{3}M(M+m_f)  \nonumber\\
&=& g_V \int_{0}^{1} dz \phi_V(z)\,, \label{eq:4.3.10}
\eea where
$\phi_V(z)$ is the so-called distribution amplitude for the pure
$S$-wave vector meson. The general phenomenology of distribution
amplitudes can be found in \cite{RadyushkinDA,*RadyushkinDA1,*Efremov:1979qk,ChernyakVMwf,*ChernyakVMwf1,ChernyakPhysRep}
and \cite{BraunVM}. 
 If the vector meson is saturated by the
$q\bar{q}$ state, then (\ref{eq:4.3.10}) is supplemented by the
normalization condition
\be
1 = \fr{N_c}{(2\pi)^3} \int d^3 {\bf
p}\ 4M |\psi^S({\bf p}^2)|^2\, = \fr{N_c}{(2\pi)^3} \int\frac{dz
d^2 {\bf k}} {z(1-z)} M^2 |\psi^S({\bf p}^2)|^2\,.
\label{eq:4.3.11}
\ee

The calculations with the fixed longitudinal
polarization vector defined for fixed $M=m_V$ break the rotation
invariance, which is often the case with parameterizations used in
the literature
\cite{NNZscanVM,NNPZZslopeVM,*NNPZZslopeVM1,Frankfurt1996,
Frankfurt1998,StastoUnitarization,ForshawColorDip}.
In technical terms, the fixed polarization vector leads to 
a mixing of the longitudinal spin-1 state and spin-0 states.
One of the drawbacks of the fixed polarization vector is that
the $V\to e^+e^-$ decay width would depend 
on the polarization state of the vector meson,
while the rotation invariant approach with the running longitudinal
polarization vector guarantees that the decay constants for the 
transverse and longitudinal
vector mesons are identical.
Quite often, in the fixed-polarization-vector approaches,
the manifestly different radial wave
functions are introduced for the transverse and longitudinal
vector mesons \cite{DoschRhoRhoprim,KowalskiSaturation}.

The principal effect of the Pauli vertex in the helicity/chiral 
expansion (\ref{eq:4.3.6}) is the chiral-odd parallel-helicity
component of the transverse vector meson which does not vanish 
for massless quark. Going back from vector mesons to
real photons, Ivanov et al. \cite{IvanovChiralOdd,*IvanovChiralOdd1} 
argued
that the related nonperturbative chiral-odd component in the
real photon is large. They relate the normalization of this 
component of the real photon wave function to the quark
condensate and its magnetic susceptibility \cite{IoffeChiralOdd}.

\subsection{The impact factor representation for the helicity
amplitudes}

\begin{figure}[htbp]

   \centering

   \epsfig{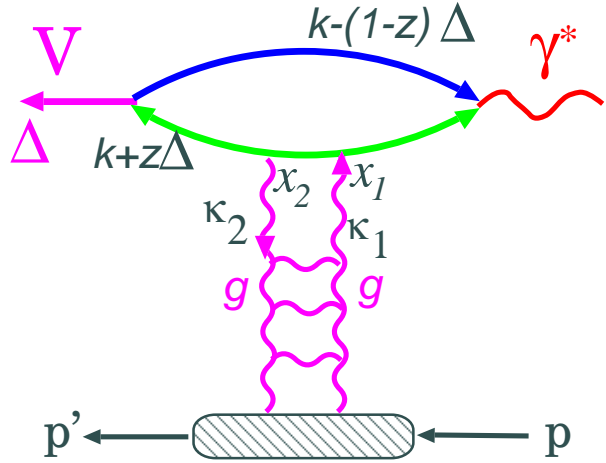}

   \caption{\it   The kinematical variables entering the $k_{\perp}$
factorization representation (\ref{eq:4.4.2}) for 
vector meson production amplitudes.}
   \label{fig:Kfactorization}
\end{figure}

The $k_\bot$-factorization, or impact factor, representation for the
vector meson production repeats closely that for the Compton
scattering amplitude in the case of DIS \cite{BL,Cataniktfact}.
The three changes are that now the momentum transfer $\bDelta\neq
0$, the vertex function for the $S$-wave vector meson is 
different from the $\gamma_{\mu}$ vertex (\ref{eq:4.3.3}) for the
photon, and the conventional unintegrated gluon SF which describes
the $t$-channel exchange is replaced by the off-forward (skewed)
unintegrated gluon structure function of the target, ${\cal
F}(x_1,x_2,\bkappa_1,\bkappa_2)$. Here  
\be
x_1 \approx {Q^2+M_1^2 \over W^2},~~ x_2 \approx {M_1^2 -m_V^2\over W^2}
\label{eq:4.4.0}
\ee
and $M_1$ is the invariant mass of the intermediate $q\bar{q}$ pair,
for the kinematical variables see Fig. \ref{fig:Kfactorization}. 

The imaginary part of the
total amplitude can be written as \cite{INSDwave,IgorPhD}
\be
\mbox{Im } {\cal T}=
{W^2}{c_{V}\sqrt{4\pi\alpha_{em}}\over 4\pi^{2}} \int {d^{2} 
\bkappa \over \bkappa^{4}} \alpha_{S}({\rm
max}\{\bkappa^2,\varepsilon^2+\bk^2\}) {\cal{F}}(x_1,x_2,
\bkappa_1,\bkappa_2) \int {dzd^2 \bk \over z(1-z)} \cdot
I(\lambda_V,\lambda_\gamma)\, , \label{eq:4.4.1} 
\ee
where for the
pure $S$-wave vector mesons the integrands
$I(\lambda_V,\lambda_\gamma)$ have the form
\bea
I^S(L,L) &=& 
 4 QM z^2 (1-z)^2 \left[ 1 + { (1-2z)^2\over 4z(1-z)} {2m_f \over
M+2m_f}\right]\Psi^*_2 \Phi_2 \,; \label{eq:4.4.2}
\\[1mm]
I^S(T,T)_{\lambda_V=\lambda_\gamma} &=& m_f^2 \Psi^*_2\Phi_2 
+ [z^2+(1-z)^2](\bPsi_1^{\,*}\cdot\bPhi_1) \label{eq:4.4.3}\\
&&+ \frac{m_f}{M+2m_f}\left[ (\bk\cdot\bPsi_1^{\,*})\Phi_2
- (2z-1)^2(\bk\cdot\bPhi_1)\Psi_2^*\right]\,; \nonumber
\\[1mm]
I^S(T,T)_{\lambda_V=-\lambda_\gamma} &=&
2z(1-z)(\Phi_{1x}\Psi_{1x}^*-\Phi_{1y}\Psi_{1y}^*) \label{eq:4.4.4}\\&&-
\frac{m_f}{M+2m_f}\left[(k_x\Psi_{1x}^*-k_y\Psi_{1y}^*)\Phi_2
- (2z-1)^2(k_x\Phi_{1x} - k_y \Phi_{1y})\Psi_2^*\right]\,; \nonumber
\\[1mm]
I^S(L,T) &=&
-2Mz(1-z)(2z-1)(\bfe\bPhi_1)\Psi^{\,*}_2
\left[ 1 + { (1-2z)^2\over 4z(1-z)} {2m_f \over M+2m_f}\right]\nonumber\\
&&+ {M m_f\over M+2m_f}(2z-1)(\bfe\bPsi_1^{\,*})\Phi_2\,;
\label{eq:4.4.5}
\\[1mm]
I^S(T,L) &=&
-2Qz(1-z)(2z-1)\left[(\bV^*\bPsi_1^{\,*})\Phi_2 
- {2m_f \over M+2m_f}(\bV^*\bk) \Psi_2^*\Phi_2\right]\,.
\label{eq:4.4.6}
\eea

Here $\br = \bk +(z-\frac{1}{2})\bDelta$ and
\bea
\Phi_2= -{1
\over (\br+\bkappa)^2 + \varepsilon^2} -{1 \over
(\br-\bkappa)^2 + \varepsilon^2} + {1 \over (\br +
\bDelta/2)^2 + \varepsilon^2} + {1 \over (\br -
\bDelta/2)^2 + \varepsilon^2} \, ,
\label{eq:4.4.7}
\eea
\bea
\bPhi_1 =
-{\br + \bkappa \over (\br+\bkappa)^2 +
\varepsilon^2} -{\br - \bkappa \over (\br-\bkappa)^2
+ \varepsilon^2} + {\br + \bDelta/2 \over (\br +
\bDelta/2)^2 + \varepsilon^2} + {\br - \bDelta/2 \over
(\br - \bDelta/2)^2 + \varepsilon^2} \,,
\label{eq:4.4.8}
\eea
for the definition of $\varepsilon^2$, see eq.~(\ref{eq:3.4.2.4}). 
Here  $1/(\bk^2+\varepsilon^2)$ and  $\Psi_2 \equiv \psi_V(z,\bk)$
describe transitions into the 
$\bar{q}$ states with the sum of helicities of
the quark and antiquark
$\lambda+\bar{\lambda}=\lambda_{\gamma^*,\lambda_V}$,
whereas $\bk/(\bk^2+\varepsilon^2)$ and
$\bPsi_1 \equiv \bk \psi_V(z,\bk)$ describe
transitions of transverse and vector meson
into the $\bar{q}$ states with 
$\lambda+\bar{\lambda}=0$, in which the
helicity of the photon and vector meson is carried 
by the orbital angular momentum
in the $q\bar{q}$ state.  

In the calculation of the
double helicity-flip amplitude (\ref{eq:4.4.4}) the $x$-axis is
chosen along the momentum transfer $\bDelta$. The point made in
Section 3.4.4 that the helicity flip proceeds via the intermediate
state with  $\lambda+\bar{\lambda}=0$ is manifest in
(\ref{eq:4.4.3})--(\ref{eq:4.4.5}). The corresponding integrands
for the $D$-wave states can be found in \cite{INSDwave}.

The $z$-dependence of the integrands 
shows that the end-point contributions ($z \ll 1$ or $1-z \ll 1$)
are suppressed in the longitudinal amplitude ${\cal T}(L,L)$ already 
in the integrands, while for the other helicity amplitudes
this suppression comes from the wave functions, 
see discussion in Section 4.7 and 4.9 below. The factor
$(2z-1)$ in the integrands of the helicity-flip
amplitudes ${\cal T}(L,T)$ and ${\cal T}(T,L)$ corresponds
to the longitudinal Fermi momentum of quarks in the vector meson,
which makes manifest the relativistic origin of helicity flip.
The expected hierarchy of the helicity flip
amplitudes is as follows \cite{KNZspinflip,*KNZspinflip1,IKspinflip}.
Roughly,
\bea
&&{|{\cal T}_{01}| \over \sqrt{|{\cal T}_{11}|^2 + |{\cal T}_{00}|^2}}
\sim {\sqrt{|t|} \over \sqrt{Q^2 + m_V^2}}\,,\label{eq:4.4.9}\\
[2mm]
&& {|{\cal T}_{10}| \over
\sqrt{|{\cal T}_{11}|^2 + |{\cal T}_{00}|^2}} \sim {\sqrt{|t|} \over \sqrt{Q^2 +
m_V^2}}
{Q m_V \over Q^2 + m_V^2}\,,\label{eq:4.4.10}\\[2mm]
&& {|{\cal T}_{1-1}| \over \sqrt{|{\cal T}_{11}|^2 + |{\cal T}_{00}|^2}}
\sim {|t| \over m_V\sqrt{Q^2 + m_V^2}}\,. \label{eq:4.4.11}
\eea
For heavy flavour
vector mesons, the helicity flip amplitudes are expected to be
further suppressed by the non-relativistic Fermi motion.

The real part of the amplitude can be reconstructed from the
imaginary part using the derivative analyticity relation
\cite{GribovDerivative,*GribovDerivative1,BronzanDerivative}: 
\be
\mbox{Re } {{\cal T}\over W^2}
=  {\pi \over 2} {\partial \over \partial \,\log W^2} \mbox{ Im
}\frac {\cal T}{W^2}. \label{eq:4.4.12} 
\ee

\subsection{The off-forward unintegrated gluon density: the
$\bDelta$-dependence within the diffraction cone
and the BFKL Pomeron trajectory}\label{sect4.5}


Thanks to a large amount of high-precision data on $F_{2p}$ both
in the soft and hard regimes, the simple, ready-to-use
parameterizations for the forward unintegrated gluon density
${\cal F}(x,\bkappa^2)$ are now available \cite{INdiffglue}.
These parameterizations can be exploited in different high-energy
reactions and bring the gluon density of the proton under control.

\begin{figure}[htbp]

   \centering

   \epsfig{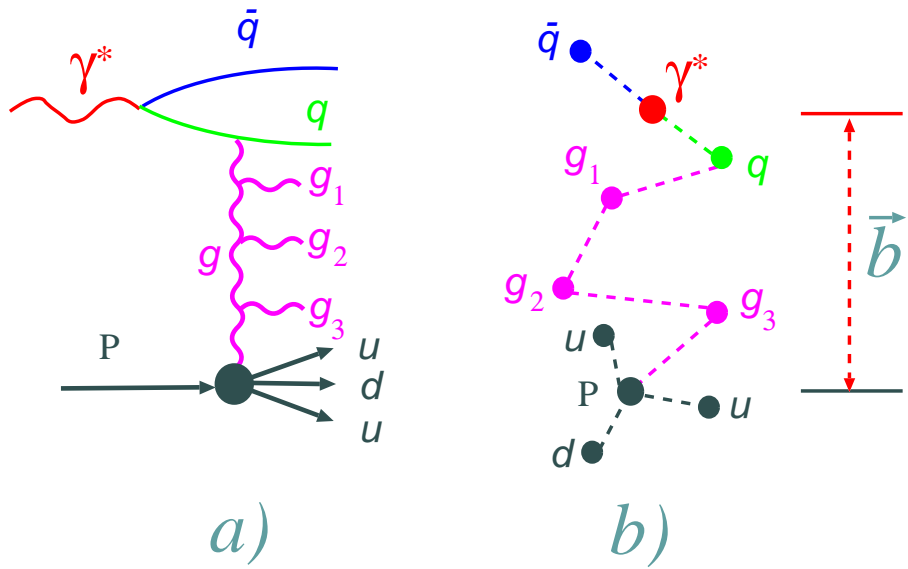}

   \caption{\it   The sequential splitting of gluons in the
Feynman diagram (a) for production of the multigluon final state
viewed as a random walk in the impact parameter space of gluons
from the $q\bar{q}$ pair of the photon $\gamma^*$ to the proton
target.}
   \label{fig:RandomWalk}
\end{figure}

For the practical application of the formalism of section 4.4, one
needs the off-forward unintegrated gluon distribution 
${\cal F}(x_1,x_2,\bkappa+{1\over 2}\bDelta, -\bkappa + {1\over 2}\bDelta)$. 
Its dependence on the momentum transfer $\bDelta$ 
comes from two courses. The first one is the soft quantity 
that can be dubbed the two-gluon form
factor of the proton. The second is the $\Delta$-dependence of the
BFKL two-gluon ladder. When viewed in the impact parameter space,
at each splitting 
of the gluon into two gluons, $g\to gg$, the hard gluon,
which carries the large longitudinal momentum of the parent
gluon, emerges at the same impact parameter as the parent gluon,
whereas the soft one, which carries small longitudinal momentum,
emerges at an impact parameter $|\Delta\bb_i| \sim {1\over
|\bkappa_{i}|}$ from the parent gluon. Consequently, 
as illustrated in Fig.~\ref{fig:RandomWalk}, 
the splitting of gluons in the process of
the $\log{1\over x}$ evolution is accompanied by the
Gribov-Feinberg-Chernavski random walk
 \cite{GribovDiffusion,FeinbergDiffusion,*FeinbergDiffusion1}
of small-$x$ gluons to larger and larger impact parameters $\bb$.
The asymptotic freedom, i.e., the running $\alpha_S$, enhances the
r\^ole of large random walks of the order of the perturbative
gluon propagation radius $R_c$. This suggests that for $\langle
\bb^2\rangle$ will rise proportionally to the number of gluon
splittings, i.e.,
\be
\langle \bb^2\rangle \propto R_c^2
\log{1\over x} \propto R_c^2 \log W^2 \, . \label{eq:4.5.1}
\ee

In conjunction with the Regge formula (\ref{eq:3.1.3.4}) and the
definition (\ref{eq:3.1.4.4}), this entails the finite slope
$\alpha_{BFKL}'$ of the Regge trajectory of the hard BFKL Pomeron.
Evidently, the dimensionfull $\alpha_{BFKL}'$ is the soft
parameter and as such it depends manifestly on the infrared
regularization of QCD. The solution of the color dipole BFKL
equation with Yukawa-type cutoff and infrared freezing of
$\alpha_S$ gave $\alpha_{BFKL}'= 0.12\div 0.15$ ~GeV$^{-2}$
\cite{NZZslopePisma,NZZslope,BFKLRegge1,*BFKLRegge2}. The results for the
shrinkage rate $\alpha_{BFKL}'$ depend on the admixture of
subleading BFKL poles and exhibit weak dependence
on $\bkappa^2$. The quoted value is found for the specific
boundary condition, which gives a good description of the HERA
results on the proton structure function
\cite{BFKLRegge1,*BFKLRegge2,NSZpion}. 

When viewed in the momentum space, the
same Gribov-Feinberg-Chernavski diffusion suggests the weakening
of the $\bkappa$-$\bDelta$ correlation with the number of
splittings. Indeed, Balitsky and Lipatov have shown that the
dependence of the off-forward gluon density 
$(\bkappa\cdot \bDelta)$ corresponds to subleading singularities
\cite{BL,Lipatov86,*Lipatov861}, and, arguably, can be neglected for HERA
energy range.

Consequently, for small momentum transfers within the diffraction
cone the
$\bDelta$-dependence can be factored out as 
\be
{\cal F}(x_1,x_2,\bkappa+{1\over 2}\bDelta, -\bkappa + {1\over
2}\bDelta)= {\cal F}(x_1,x_2,\bkappa, -\bkappa)
\exp\left(-{b_{3\Pom}\bDelta^2 \over 2}\right)\,.
\label{eq:4.5.2} 
\ee
We parameterize $b_{3\Pom}$ as
\be
b_{3\Pom} = b_{2G} + 2\alpha_{BFKL}'\log{W^2 x_{0}\over Q^2+m_V^2}
\, ,
\label{eq:4.5.3} 
\ee
where the soft parameter $b_{2G}$ can be regarded as a
slope of the form factor of the proton as probed by 
the color singlet two-gluon state. In principle, one can
determine it experimentally isolating the BFKL contribution 
to diffractive DIS into high mass states. Strictly speaking,
this parameter $b_{2G}$ as well as the Pomeron slope $\alpha'$ 
can change from the soft, non-perturbative, to hard, BFKL, 
gluon density, taking \protect\cite{IgorPhD,IgorNumerics}
the universal parameters, $b_{2G}=B_N=4$ GeV$^{-2}$ 
with $x_0 = 3.4\cdot 10^{-4}$ and the $\bkappa^2$-independent
$\alpha_{\Pom}^{soft}= \alpha_{BFKL}'=0.25$ GeV$^{-2}$ 
is the poor man's approximation.

\subsection{The off-forward unintegrated gluon density: the
dependence on skewness}


Bartels was the first to observe  \cite{BartelsSkewness} that 
two gluons enter the amplitude at $x_2 \neq x_1 \approx x$,
because the invariant mass squared $M_1^2$ of the intermediate 
$q \bar q$ system is close to $M^2\approx m_V^2$ for the final
$q\bar{q}$ state and is far from the virtuality of the incident
photon $-Q^2$. 
Such a skewed unintegrated gluon density can be, in
principle, accessed in DVCS \cite{DVCS}, but that is not yet a
practical solution. Shuvaev et al. \cite{ShuvaevSkewness} and
Radyushkin \cite{RadyushkinSkewness} argued that at small $x$ the
skewed distribution can be related to the conventional one: if ${\cal
F} \propto x^{-\lambda}$, then
\be
{\cal F}(x_1,x_2 \ll x_1,\bkappa,-\bkappa) = R_g\cdot {\cal
F}(x_1,\bkappa)\,;\quad R_g = {2^{2\lambda+3} \over \sqrt{\pi}}
{\Gamma(\lambda+{5\over 2}) \over \Gamma(\lambda+4)}\,.
\label{4.6.1}
\ee
The above factor $R_g$ can be effectively
accounted for in a form of the $x$-rescaling
\be
R_g \cdot
\left({1 \over x_1}\right)^\lambda = \left({1 \over
c(\lambda)x_1}\right)^\lambda\,, \label{eq:4.6.2}
\ee
where
$c(\lambda)$ changes from $\approx 0.435$ at $\lambda=0$ to $0.4$
at $\lambda = 1$. Given this very flat dependence, one can take
fixed $c = 0.41$, so that
\be
{\cal
F}(x,0,\bkappa,-\bkappa) \approx {\cal F}(c
x,\bkappa)\,. \label{eq:4.6.3}
\ee
Hereafter we approximate the skewed gluon density by
the forward density take at
\be
x_g = cx_1 = c{Q^2+m_V^2 \over W^2}\, .
\label{eq:4.6.4}
\ee Of course, once this rescaling
of $x$ is implemented and the Fourier transform to the color
dipole representation is performed staring from Eqs.
(\ref{eq:4.4.1})-(\ref{eq:4.4.6}), the color dipole and
$k_{\perp}$-factorization approaches will be identical to each
other.

\subsection{The Ans\"atze for the wave function}\label{sect4.7}


For the heavy quarkonia a good insight into the functional form of
the radial wave function (WF) $\psi_V({\bf p}^2)$ comes from the
potential model calculations \cite{NovikovPhysRep,QuiggPhysRep}.
Here, at least for the $\Upsilon(1S)$, the r\^ole of the QCD
Coulomb interaction is substantial. It is less so for the
charmonium, whereas the gross properties of lighter vector mesons
which have a large size are entirely controlled by the confining
interaction and here one is bound to the model 
parameterizations \cite{IgorPhD}.
The popular harmonic oscillator WF emphasizes the confinement 
property, it
decreases steeply at large $\bp^2$,
\be
  \psi_{1S}  =  c_1\exp\left(-{{\bf p}^2 a_1^2 \over 2}\right)\,;
  \quad \psi_{2S}  =  c_2 \left(\xi_{node} - {\bf p}^2 a_2^2 \right)
\exp\left(-{{\bf p}^2 a_2^2 \over 2}\right)\,. 
\label{eq:4.7.1}
\ee
which emphasizes the contrast between the non-pointlike vector
meson and pointlike photon for which $\Gamma_{\gamma^*}(z,\bk)=
\sqrt{4\pi \alpha_{em}}={\rm const}$. The position of the node,
$\xi_{node}$, is fixed from the orthogonality condition.
The attractive pQCD 
Coulomb interaction between the quark and antiquark enhances the 
WF at small $\bfR_{q\bar{q}}$ and/or
large relative momentum, the minimal relativization
of the familiar Coulomb WF suggests 
\be
\psi_{1S}({\bf p}^2) = {c_1 \over \sqrt{M}} {1 \over  (1 +
a_1^2{\bf p}^2)^2}\,; \quad \psi_{2S}({\bf p}^2) = {c_2 \over
\sqrt{M}} {(\xi_{node} - a_2^2{\bf p}^2) \over (1 + a_2^2{\bf
p}^2)^3}\,, 
\label{eq:4.7.2}
\ee
which decreases as an inverse power of $\bp^2$, much slower than
(\ref{eq:4.7.1}). The factor $1/\sqrt{M}$ in (\ref{eq:4.7.2}) is
a model-dependent suppression to make the decay constant (\ref{eq:4.3.10})
convergent. 
Arguably, those two extreme
Ans\"atze give a good idea on the model dependence of vector meson
production amplitudes. The radius $a_1$ and the normalization
$c_1$ are fixed by the $V_{0}\to e^+e^-$ decay constant
(\ref{eq:4.3.10}) and the normalization condition (\ref{eq:4.3.11}).
The hybrid model in which the short-distance QCD Coulomb
interaction in light vector mesons has been treated perturbatively
is found in \cite{NNZscanVM,NNPZdipoleVM}.


\subsection{The hard scale $\Qb^2$:  
the link to the leading $\log Q^2$-approximation
and the exponent of the $W$-dependence}
\label{sect4.8}

For soft gluons,
\be
\bkappa^2 \ll (\varepsilon^2+\bk^2) =
z(1-z)(Q^2+M^2)\, ,
\label{eq:4.8.1}
\ee
one can expand $\Phi_2$ and
$\bPhi_1$ as \cite{NZsplit} (for the sake of simplicity we
consider $\bDelta=0$)
\bea
\Phi_2 \approx \frac{2(\varepsilon^2
-\bk^2)}{(\varepsilon^2 +\bk^2)^3}\bkappa^2=
{2\over z^2(1-z)^2(Q^2+M^2)^2}\left[
1-{2\bk^2\over z(1-z)(Q^2+M^2)}\right]\bkappa^2
\, , \label{eq:4.8.2}
\\
\bPhi_1 \approx \frac{4\varepsilon^2 \bk} {(\varepsilon^2
+\bk^2)^3}\bkappa^2= {4\bk \over z^2(1-z)^2(Q^2+M^2)^2}\left[
1-{\bk^2\over z(1-z)(Q^2+M^2)}\right]\bkappa^2\,. \label{eq:4.8.3}
\eea 
A natural approximation is $M^2\approx m_V^2$. Then the factor 
$(Q^2+m_V^2)^{-2}$ which emerges in (\ref{eq:4.8.2})
and (\ref{eq:4.8.3}) corresponds to precisely the factor 
$r_S^4$ of
the color dipole approach, see Eqs. (\ref{eq:3.4.2.5})
and (\ref{eq:3.4.2.6}). The determination of the hard scale in 
the gluon SF is a bit more subtle.  
 
Expansions (\ref{eq:4.8.2}) and (\ref{eq:4.8.3}) define the
leading log$\Qb^2$ contribution with 
logarithmic integration over $\bkappa^2$:
\be
\int_{0}^{ z(1-z)(Q^2+M^2)}
\frac{d\bkappa^2}{\bkappa^2}{\cal F}(x_g,\bkappa) =G(x_g,
z(1-z)(Q^2+M^2))\, . \label{eq:4.8.4}
\ee
The emerging running hard scale depends on $z$ and $M^2$,
for the heavy quarkonia the wave function of the vector meson
is peaked at $z\sim {1\over 2}$ and one can take $M^2\approx m_V^2$,
consequently, $z(1-z)(Q^2+M^2)=\Qb^2$ of Eq.(\ref{eq:1.3.5}).
\begin{figure}[!htb]
   \centering
   \epsfig{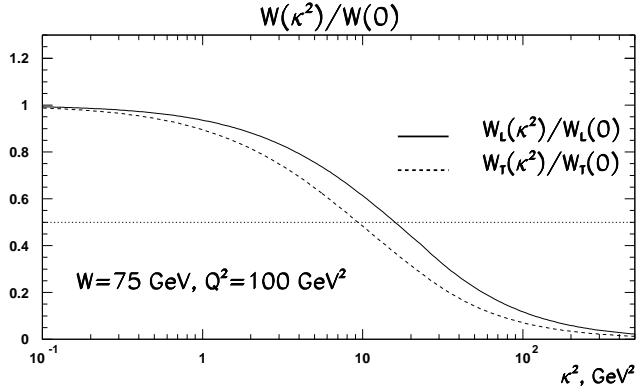}
   \caption{\em The normalized weight functions 
$W_L(Q^2,\vec\kappa^2)/W_L(0)$ and $W_T(Q^2,\vec\kappa^2)/W_T(0)$
for the $\rho$ production
calculated at $Q^2 = 100$ GeV$^2$ 
in the $k_{\perp}$-factorization approach \protect\cite{Igorhardscale}.}
   \label{fig:IgorRhoQ2weight}
\end{figure}

\begin{figure}[!htb]
   \centering
   \epsfig{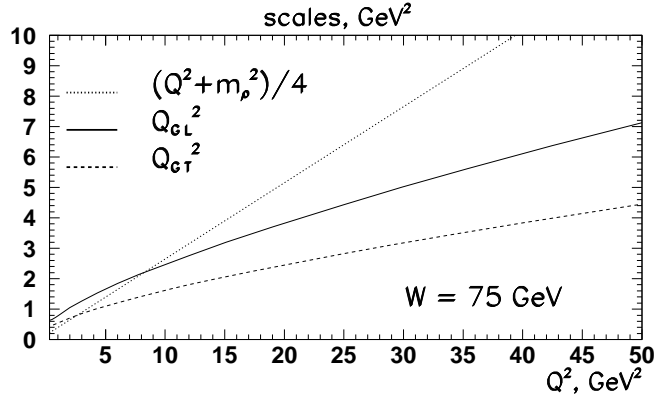}
   \caption{\em The scales $\Qb_{GL,GT}^2$ at which the
$\rho$ production maps the gluon density as a function
of $Q^2$ found  
in the $k_{\perp}$-factorization approach \protect\cite{Igorhardscale}. Shown also is the heavy flavor approximation
$\Qb_G^2=\Qb^2={1\over 4}(Q^2+m_\rho^2)$.}
   \label{fig:IgorQ2scaleRho}
\end{figure}

The contribution from small dipoles, $r< r_S$, or from hard gluons
beyond the leading $\log\Qb^2$ domain, $\bkappa^2\gsim  \Qb^2$,
is an integral part of the $k_{\perp}$-factorization approach
\cite{NNZscanVM,NZglue,INdiffglue}. In this region
\be
\Phi_2\approx {2 \over (\varepsilon^{2}+\bk^2)}\, , \quad
\bPhi_1 \approx {2 \bk \over (\varepsilon^{2}+\bk^2)}
\label{eq:4.8.5} 
\ee
and the correction to the leading  $\log \overline{Q}^2$ 
result (\ref{eq:4.8.4}) can be cast as (see also \cite{LevinKfactVM}) 
\be
 \overline{Q}^2
\int_{ \overline{Q}^2} \frac{d\bkappa^2}{(\bkappa^2)^2} {\cal
F}(x_g,\bkappa) \approx {\cal F}(x_g,\overline{Q}^2)\cdot \log C_g\,.
\label{eq:4.8.6} 
\ee
where $\log C_g\sim 1$ and depends on the exact
$\bkappa^2$-dependence of ${\cal F}(x_g,\bkappa)$. Following
\cite{NZglue,INdiffglue} one can combine (\ref{eq:4.8.4}) and
(\ref{eq:4.8.6}) as 
\be
G(x_g, \overline{Q}^2)+ {\cal F}(x_g,\overline{Q}^2)\cdot 
\log C_g \approx G(x_g, C_g\overline{Q}^2)\,.
\label{eq:4.8.7} 
\ee
Consequently, the gluon density is mapped at hard 
\be
\Qb^2_G = C_g\Qb^2\, ,
\label{eq:4.8.8} 
\ee
which is slightly different from $\Qb^2$. 
As already mentioned above, $I^S(L,L)$ of (\ref{eq:4.4.2})
is more peaked at $z\sim {1\over 2}$, whereas 
$I^S(T,T)_{\lambda_V = \lambda_\gamma}$ of (\ref{eq:4.4.3})
extends more to the end points $z \sim 0$ and $z\sim 1$. 
This leads to an inequality $\Qb^2_{GL} > \Qb^2_{GT}$
and implies that the typical dipole sizes in the $T \to T$
amplitude are somewhat larger than for the $L \to L$ amplitude.

For a more quantitative analysis the $\bkappa^2$-integrations
can be cast in the form 
\be
{1\over W^2} Im {\cal T}_{LL,TT} \equiv \int{d\vec\kappa^2 \over \vec\kappa^2}
{\cal F}(x_g,\vec\kappa) \cdot W_{L,T}(Q^2,\kappa^2)
\equiv  W_{L,T}(Q^2,0)G(x_g,\Qb_{GL,GT}^2)
\label{eq:4.8.9}
\ee
The typical behaviour of normalized        
weight functions  $W_{L,T}(Q^2,\vec\kappa^2)/W_{L,T}(Q^2,0)$ 
is shown in Fig.~\ref{fig:IgorRhoQ2weight} and for 
smooth ${\cal F}(x_g,\bkappa^2)$ they can be approximated
by the step-function $\theta(\Qb_{GL,GT}^2-\bkappa^2)$,
where $\Qb_{GL,GT}^2$ are defined by the median,
$W_{L,T}(Q^2,\Qb_{GL,GT}^2)={1 \over 2}W_{L,T}(Q^2,0)$,
the results are close to the ones found in \cite{NNZscanVM}. 
At moderately large $Q^2$ the strong scaling violations
in ${\cal F}(x_g,\bkappa^2)$ shown in Fig.~\ref{fig:DIFGlueGRV}
have a strong impact on $\Qb_{GL,GT}^2$ as shown in
Fig.~\ref{fig:IgorQ2scaleRho}. The inequality 
$\Qb_{GL}^2 > \Qb_{GT}^2$ found in \cite{NNZscanVM} is retained 
and the hierarchy of $\Qb_{GL,GT}^2$ from light to heavy
flavors is the same as of $\Qb^2$.

\begin{figure}[htbp]

   \centering

   \epsfig{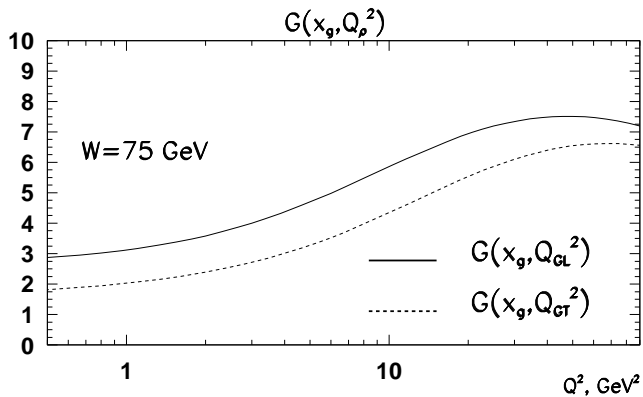}

   \caption{
The effect of different hard scales
on the integrated gluon density (\protect\ref{eq:4.8.7})
which enters the dominant SCHC amplitudes ${\cal T}_{LL}$ and  
${\cal T}_{TT}$ for the $\rho$ production as found  
in the $k_{\perp}$-factorization approach 
 \protect\cite{IgorNumerics}.}

   \label{fig:ivanov2004-g75}
\end{figure}

This effect of $\Qb_{GL}^2 > \Qb_{GT}^2$ is demonstrated 
in Fig.~\ref{fig:ivanov2004-g75}
on an example of $G(x_g,Q_{GT}^2)$ and $G(x_g,Q_{GL}^2)$ for
the $\rho$ production:
the finding of  $G(x_g,Q_{GL}^2)> G(x_g,Q_{GT}^2)$ reflects
the inequality $\Qb_{GL}^2 > \Qb_{GT}$ at equal $Q^2$.
For the same reason different helicity amplitudes 
can have a slightly different energy dependence.
The effect of different scales diminishes at larger $Q^2$
with weakening scaling violations, see Fig.~\ref{fig:DIFGlueGRV}.

A tricky point is that the small-$\bkappa$ expansion for the 
double-flip amplitude starts with the constant and it is dominated by
the soft-gluon exchange, the term $\propto \bkappa^2$ is of 
higher twist \cite{KNZspinflip,IKspinflip}.  

Of course, the above separation into the leading $\log\overline{Q}^2$
and hard regions is redundant when either the color dipole and the
full fledged $k_{\perp}$-factorization approaches are used. The
$k_{\perp}$-factorization analysis reveals somewhat better the
r\^ole of $\bk^2$ in (\ref{eq:4.8.1}) for  the determination of
the leading $\log\overline{Q}^2$ region. One may call that the
inclusion of the Fermi motion effects \cite{Frankfurt1996}, it is
automatically contained in the color dipole calculations and
should not be discussed separately. The specific form
of the Fermi motion correction suggested by Frankfurt et al.
\cite{Frankfurt1996} is not borne out by the $k_{\perp}$-factorization
analysis, though.

Finally, Eq. (\ref{eq:4.8.9}) clearly shows that the energy dependence
of the production amplitude is controlled by the $x_g$-dependence of
the integrated gluon density, i.e., by the effective intercept 
$\lambda(\Qb_{GL,GT}^2)$ defined in section 3.3.1:
\be
Im {\cal T}(W^2,t=0)_{LL,TT} \propto \left({1\over x}\right)^{1+\lambda(\Qb_{GL,GT}^2)}
\propto (W^2)^{1+\lambda(\Qb_{GL,GT}^2)}\,.
\label{eq:4.8.10}
\ee


\subsection{The production amplitude and the vector meson
distribution amplitude}

At large $Q^2$ there emerges one useful approximation usually
referred to as the collinear approximation. 
Specifically, here the explicit $\bk^2$-dependence in  $
\overline{Q}^2$ is neglected and the $d^2\bk$ integration can
be factored out. In the color dipole language this amounts to
neglecting the $r$-dependence of the vector meson wave function
and taking it at $r=0$, but keeping its $z$-dependence. The
results are best seen in the momentum representation.
For instance, the factor $z^2(1-z)^2$ in
$I^S(L,L)$, Eq.~(\ref{eq:4.4.3}) cancels the factor 
$z^{-2}(1-z)^{-2}$ in eq.~(\ref{eq:4.8.2}) for $\Phi_2$, and
the vector meson wave function
enters ${\cal T}_{LL}$ in the form of an integral
\be
\int^{\bk^2<\overline{Q}^2} d^2\bk M \psi_V^*(z,\bk)\left[ 1 + {
(1-2z)^2\over 4z(1-z)}\cdot {2m_f \over M+2m_f}\right]
\label{eq:4.9.1}
\ee
which is very similar to the distribution
amplitude defined by (\ref{eq:4.3.10}) at the factorization scale
$\overline{Q}^2$. 
We emphasize that the factor in the square brackets in
(\ref{eq:4.9.1}) depends on the $SD$-wave mixing. For the
nonrelativistic heavy quarkonia $M\approx m_V$,~ $z\approx {1\over
2}$, the proportionality of the two quantities is exact,
and one finds, cf. (\ref{eq:3.4.1.6}),
\be
{d\sigma_L \over
d|t|}\Bigg|_{t = 0} = {\pi^3 \over 12 \alpha_{em}} \cdot {Q^2
\over \Qb^8} m_V \,\Gamma(V\to e^+e^-)\,
\left[\alpha_s(\Qb^2)\cdot G(x_g,\Qb^2_G)\right]^2\,
\label{eq:4.9.2}
\ee
and
\be
R_V={\sigma_L(\gamma^*p\to Vp)\over \sigma_T(\gamma^*p\to Vp)}
\approx {Q^2 \over m_V^2}R_{LT}\,. 
\label{eq:4.9.3}
\ee
with $R_{LT}=1$ if the slight possible difference of
the $t$-dependence for the $L$ and $T$ cross sections
is neglected. 
Since the proportionality $R_{V} \propto Q^2$ at small $Q^2$
is a generic consequence of the
electromagnetic gauge invariance, the true dynamical
features of vector meson production are manifested by
the departure of $R_{LT}$ from unity. We strongly 
advocate to represent the experimental data in terms of
this parameter.


\subsection{The ratio $R=\sigma_L/\sigma_T$ and short distance
properties of vector mesons}\label{sect4.10}


One word of caution on the ratio (\ref{eq:4.9.3}) is in order
\cite{KolyaCracow}. In
the Introduction we emphasized how the vector meson production
$\gamma^*p\to Vp$ is obtained by analytic continuation from the
elastic Compton scattering $\gamma^*p\to \gamma^*p$. In the course of this
analytic continuation one changes form the pointlike
$\gamma^*q\bar{q}$ to the non-pointlike $Vq\bar{q}$ vertex.

Making use of the optical theorem for the Compton scattering, one
finds
\be
R_{Compton}= \left|{A(\gamma^*_L p\to \gamma^*_L p)
\over A(\gamma^*_T p\to \gamma^*_T p)}\right|^2=
\left({\sigma_{L}(\gamma^*p) \over \sigma_T(\gamma^*p)}\right)^2=
R_{DIS}^2 \approx 4\cdot 10^{-2}. 
\label{eq:4.10.1}
\ee
Here we used
the prediction \cite{NZHERA} for inclusive DIS $R_{DIS} =
\sigma_{L}(\gamma^*p)/ \sigma_T(\gamma^*p) \approx 0.2$, which is
consistent with the indirect experimental evaluations at HERA
\cite{pl:b393:452}. This result $R_{Compton} \ll 1$ for the elastic
scattering of pointlike photons $\gamma^*p\to \gamma^*p$ must  be
contrasted to $R_V \sim Q^2/m_V^2 \gg 1$ when the pointlike
$\gamma^*$ in the final state is swapped for the non-pointlike
vector meson. Evidently, the predictions for $R_V$ are extremely
sensitive to the presence in light vector mesons of
quasi-pointlike $q\bar{q}$ component with the WF concentrated
at short $q\bar{q}$ separation. The crude model estimates in 
\cite{NNZscanVM,NNPZdipoleVM} suggest that modifications
of the wave function by
attractive short-distance pQCD interaction do indeed
lower the theoretical results for $R_V$.

Here we just recall the pQCD radiative 
correction to the Weisskopf-Van Royen non-relativistic 
approximation
(\ref{eq:3.4.1.3}) for the leptonic decay width 
(\cite{Celmaster}, see also \cite{Barbieri}) 
\be
\Gamma(V^{0}\to e^+e^-)=  \frac{4
\alpha_{em}^2 c_V^2}{m_V^2}|\Psi_V(0)|^2 
\left(1-{8\over 3\pi}\alpha_S(m_f)\right)^2\,, 
\label{eq:4.10.2}
\ee
which even for the $J/\Psi$ suppresses the decay
width by a factor of $\sim 2$. Remarkably,
this particular NLO correction is of Abelian nature - it
derives from the Karplus-Klein QED radiative correction 
by a substitution $\alpha_{em} \to C_F \alpha_S(m_f^2)$.
In the LO approach the conservative radii of vector mesons
are fixed from
Eq.~(\ref{eq:4.3.10}) without allowance for the pQCD
correction (\ref{eq:4.10.2}). Although the 
 formula
(\ref{eq:4.10.2}) can not be directly applied to 
light vector mesons, it is reasonable to wonder what
will happen to the vector meson production phenomenology
if the Celmaster \& Barbieri et al. correction 
for the $\rho$ meson were
a factor of 3. To a crude approximation, that will enhance
the wave function at the origin by the factor 
$\approx \sqrt{3}$ and decrease the radius of the 
vector meson by the factor $\approx 3^{1/3} = 1.44$,
which is not off-scale.
We expect a substantial reduction of the predicted 
$R_V=\sigma_L/\sigma_T$ for such a squeezed $\rho$-meson.


\subsection{Perturbative QCD calculations at high-$t$}
\label{sect4.11}

Vector meson production at large $|t|$ is believed to be
dominated by small impact parameters $b \sim 1/\sqrt{|t|}$.
Simultaneously, the large $|t|$ is expected to select
the small-size configurations in the $\gamma^*V$ transition
vertex. Consequently, at $|t|$ such that $1/\sqrt{|t|} \ll
r_S$, i.e., $|t| \gg \Qb^2$, it becomes the hard pQCD scale
for the process. Arguably, in the real photoproduction
of heavy flavour mesons the correct hard scale is
$|t|+m_V^2$.

Which $|t|$ is large enough for $|t|$ or $|t|+m_V^2$ to become
the hard pQCD scale? That can be decided only a posteriori,
the answer depends on the normalization
of the hard pQCD amplitude and on how fast the
soft amplitudes do vanish at large $\bDelta$
\cite{GinzburgIvanovHight}. The large-$t$ data taken at HERA
correspond to the Regge regime of $|t|\ll W^2$. In the real
photoproduction one starts with the typical hadronic
$\gamma \to V$ transition, and our experience with large-$t$
hadronic reactions is a very discouraging one. At small $t$
within the diffraction cone the differential cross sections
have the $\exp(-B|t|)$ behaviour, but at larger $t> 1$ GeV$^2$
the slowly decreasing multiple-pomeron exchanges take over:
the $n$-pomeron exchange gives the $t$-dependence $\propto
\exp(-{B\over n}|t|)$ (for the review and references see
\cite{KaidalovPhysRep}). It is fair to say that the unequivocal
evidence for hard pQCD mechanism in high-energy elastic
proton-proton scattering is as yet missing, the soft double-pomeron
mechanism dominates for $t$ of several GeV$^2$ quite irrespective
of the specific model for the soft-pomeron amplitude (for the recent
fits to elastic $pp$ scattering see \cite{Block}).
 The model-dependent
estimates show that the rate of  decrease of the soft
amplitude slows down dramatically at large-$|t|$. For instance
in $\pi \pi$ elastic scattering at moderate energies the
dominance of the hard pQCD amplitude requires
$|t| \gsim 4~{\rm GeV}^2$ \cite{PionPionHard}.
The modern handbag mechanism for large-$t$ two-body reactions,
as well as electromagnetic form factors of nucleons and pions, 
relies on soft wave functions of hadrons (\cite{Kroll} and
references therein).

Under these circumstances the single-BFKL pomeron exchange
interpretation of the large-$t$ vector meson data is at best the
poor man's approximation. Under this very strong assumption of
two-gluon tower exchange, the $k_{\perp}$-factorization
formalism expounded in section
4.4 is perfectly applicable at large $|t|$. The approximation
of section 4.5 for the unintegrated gluon density is only
applicable within the diffraction cone, $\bDelta^2 R_c^2 \lsim 1$,
and must be modified.

The available experimental data at large-$|t|$ are for the proton
dissociate photoproduction $\gamma p \to V(\bDelta)Y$, which at 
large $t$ can be described in the equivalent parton 
approximation of Ginzburg et al. \cite{GinzburgPanfilSerbo},
\be
{d\sigma_V (\gamma^* p \to VY)\over dt dx'}=
\left({81 \over 16}g(x',|t|) + \sum_{f} [q(x',|t|)+\bar{q}(x',|t|)]\right)
{d\sigma_V (\gamma^* q \to Vq')\over dt}\, ,
\label{eq:4.11.1}
\ee
where $x'= |t|/(m_Y^2+|t|)$ is the fraction of proton's 
lightcone momentum carried by the struck parton.
It is reminiscent of the familiar collinear factorization,
but in the calculation of the hard cross section 
${d\sigma_V (\gamma^* q \to Vq')/dt}$  the 
exchange by soft gluons with the momentum $|\bkappa| \ll |\bDelta|$ 
must not be included. 
For instance, in a splitting $q \to q'g$ with the $q'-g$ relative 
transverse momentum $\bp$ the transverse size of the $q'g$ pair
is $r_{q'g}\sim 1/|\bp|$. The exchanged gluons with the wavelength
$\lambda = 1/ |\bkappa| \gg r_{q'g}$ can not resolve such a pair 
which will act as a pointlike color triplet state indistinguishable 
from the parent quark $q$. 

The first application of (\ref{eq:4.11.1}) by
Ginzburg et al. was to the process $\gamma \gamma \to V Y$ treated 
in the two-gluon exchange approximation.  In the more
advanced BFKL approach this constraint amounts to endowing the target 
partons in (\ref{eq:4.11.1}) by a $t$-dependent non-pointlike 
structure; the practical prescription has been developed by Forshaw 
and Ryskin (\cite{ForshawRyskinLarget}, see also 
\cite{MotykaMartinRyskinLarget}).

Notice that $W_{\gamma q}^2 = x'W_{\gamma p}^2$ and the Regge parameter
for $\gamma^* q \to Vq'$ is
\be
\exp(\Delta \eta)  = {x'W^2 \over m_V^2-t}\, ,
\label{eq:4.11.4}
\ee
where $\Delta \eta $ is the rapidity gap between the produced vector meson
and the hadronic debris of the proton. The lower limit of the 
$x'$-integration, $x_{min} <x' < 1$, is set by the 
experimental cuts. 

Hereafter we focus on real photoproduction. Beyond the leading 
order pQCD the only working approximations for $
{\cal F}(x_1,x_2,\bkappa+{1\over 2}\bDelta, -\bkappa + {1\over
2}\bDelta)$ in the large-$\Delta$ regime is based on the
Lipatov's solution of the leading order BFKL equation in 
the scaling approximation $\alpha_S=const$. In contrast to
the cases of DIS or diffractive vector mesons at small $\bDelta$,
where the BFKL evolution from the proton side starts from
the soft scale an becomes hard only on the virtual photon
end of the gluon ladder, at large $|t|$ the large momentum
transfer $\bDelta$ flows along the whole ladder which may
make Lipatov's scaling approximation better applicable at large 
$|t|$. Important point is that at large $\bDelta$ the
$\bDelta-\bkappa$ correlation in
$
{\cal F}(x_1,x_2,\bkappa+{1\over 2}\bDelta, -\bkappa + {1\over
2}\bDelta)$ becomes very important, technically the
dependence on the azimuthal angle between $\bDelta$ and $\bkappa$
is described by the conformal spin expansion. 
We wouldn't go into the technicalities of the formalism,
it is fair to say that for the leading helicity amplitudes
the changes from pQCD two-gluon exchange to
scaling BFKL approximation are for the most part marginal.
The sensitivity to the wave function of the vector meson 
and real photon is dramatic, though. The detailed discussion
is found in the recent paper by Poludniowski et al. 
\cite{PoludniowskiHight}, here we summarize the major points.

\begin{figure}[htbp]

   \centering

   \epsfig{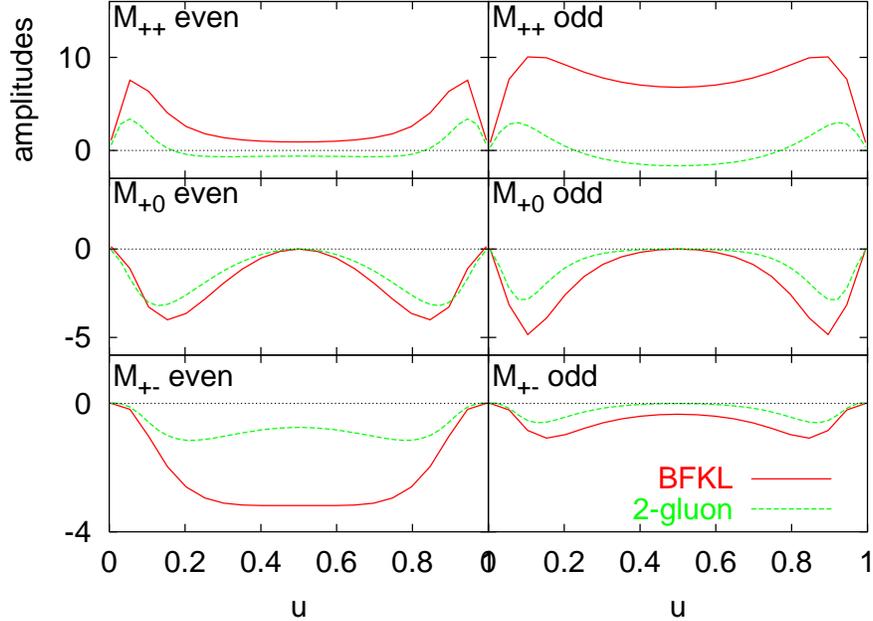}

   \caption{\it   
The helicity amplitudes differential in $z=u$ for pQCD two-gluon
and scaling BFKL approximations for $|t|=10~{\rm GeV^2}$ and 
$\alpha_S\Delta y \sim 2.4$, where $\Delta y$ is the rapidity gap
between the vector meson and debris of the proton. The pQCD two-gluon 
results have been multiplied
by a factor of 3. From Poludniowski et al.\protect\cite{PoludniowskiHight}.}

   \label{fig:Poludniowski_z_distr}

\end{figure}

The equivalent parton representation (\ref{eq:4.11.1}) makes it
obvious that the helicity properties of the target dissociative 
reaction do not depend on the target. In their analysis of 
$\gamma\gamma \to VY$ to the perturbative 
two-gluon exchange approximation
Ginzburg et al. \cite{GinzburgPanfilSerbo} 
allowed for the Fermi motion of quarks parameterized in terms
of vector meson distribution amplitudes with broad $z$-distribution
and found the dominance by the spin-flip transition $\gamma_T \to V_{L}$,
cf. the large-$t$ extension of eq.~(\ref{eq:4.4.9}). The double-flip
and non-flip amplitudes have similar $t$-dependence and
are suppressed. The found
differential cross section is of the form
\be
{d\sigma_V (\gamma_T q \to V_L q')\over dt} \propto 
\alpha_{em} m_v \Gamma(V\to e^+e^-){\alpha_S^4  \over |t|^3}.
\label{eq:4.11.5}
\ee
The flavour dependence is typical for the $\sigma_L$-dominance,
cf. eqs.~(\ref{eq:4.9.2}), (\ref{eq:6.1.1.1}). 
The sensitivity to the wave function is obvious from the 
much discussed unrealistic 
nonrelativistic limit of $z\equiv {1\over 2}$.
Here the helicity flip amplitude vanishes and , see eq. (\ref{eq:4.4.5}),
the SCHC transition dominates (\cite{GinzburgPanfilSerboJpsi}), also
see eq. (\ref{eq:4.4.5}),, the differential cross
section (\ref{eq:4.11.5}) acquires the extra factor 
$m_V^2/|t|$ and, furthermore, the predicted cross section
exhibits an accidental, artificial, zero at $|t|=m_V^2$.

Upon the radiative corrections the
electromagnetic vertex acquires the
anomalous magnetic moment (Pauli) component 
$\propto \sigma_{\mu\nu}p^{\gamma}_{\nu}/2m_f$ which, as 
we discussed in section 4.3.3, contributes to the chiral-odd 
parallel-helicity wave function of the photon.
Schwinger's  classic calculations show that for virtual photons 
the form factor of such a perturbative Pauli vertex vanishes $\propto m_f^2/Q^2$ 
(see $\S 117$ of the textbook \cite{LandauLifshitz2}). Ivanov et al
argued \cite{IvanovChiralOdd,*IvanovChiralOdd1} that for real photons the 
nonperturbative chiral-odd
vertex is substantial - they relate it to the product of the
quark condensate and the magnetic susceptibility of the vacuum 
\cite{IoffeChiralOdd} - 
and will enhance strongly the SCHC transition $\gamma_T \to V_T$.

The ideas of D.Ivanov et al. \cite{IvanovChiralOdd,*IvanovChiralOdd1}
have been extended to the scaling BFKL approximation by 
Poludniowski et al. \cite{PoludniowskiHight}, where one can 
find references to early studies. The crucial point is an
enhancement of the SCHC amplitude by the chiral-odd parallel-helicity
component in th photon. The importance of realistic
$z$-distributions in vector mesons is clearly seen from Fig.
\ref{fig:Poludniowski_z_distr} which show the helicity
amplitudes for the $\rho$ production differential in $z$ for 
chiral-even and chiral-odd photon wave function. As we stated
above, for the leading chiral-even helicity amplitude 
${\cal T}_{01}$ ($M_{+0}$
in the notations of Poludniowski et al.) the changes form the
pQCD two-gluon to scaling BFKL 
approximation are marginal. 
The BFKL approximation enhances further 
the chiral-odd contribution to the SCHC non-flip 
amplitude and makes an approximate SCHC the dominant feature
of large-$t$ vector meson production in the studied region 
of $|t| \lsim 6~{\rm GeV^2}$. The double-flip
amplitude also is enhanced.    

Finally, making use of the scaling BFKL unintegrated glue one is committed to
the BFKL intercept (\ref{eq:3.2.2}) and prediction of the steep
rise of the cross section with energy
\be
{d\sigma\over dt} \propto \left({W^2 \over m_V^2-t}\right)^{2\Delta_{BFKL}}\, ,
\label{eq:4.11.6}
\ee
which has been emphasized by Ginzburg et al. already in 1986 
\cite{GinzburgPanfilSerbo}.


\subsection{Beyond the leading $\log{1\over x}$ approximation}
\label{sect3.4.12}

The above described pQCD description of the vector meson
production is based on the manifestly leading log${1\over x}$ 
BFKL formalism. Going to the NLO $\log {1\over x}$ BFKL 
remains the major challenge to the theory. The 
principal feature of the 
leading log${1\over x}$ approximation is that adding soft
{\sl perturbative} gluons, i.e., the higher Fock states, 
to the color dipole can be reabsorbed into the 
$x$-dependence of the color dipole cross section, which is
equally true for DIS and vector meson production. 
Going to the NLO log${1\over x}$ approximation is much more 
tricky. While the nearly
decade long efforts have culminated in the derivation 
\cite{NLOBFKL} of the NLO BFKL evolution kernel, the 
matching calculations of the effect of hard gluons in the 
NLO impact factors are missing. In the case of DIS those
hard gluons are of perturbative origin, but even so, despite 
the great progress 
\cite{FadinImpactFactor,*Fadin:2001ap,*Fadin:2001ap1,BartelsImpactFactor}, the closed 
result for the impact factor of the virtual photon
is not available yet. In the case of vector mesons,
the evaluation of the NLO impact factor can not be
separated from the issue of the higher, nonperturbative,
$q\bar{q}g$,  Fock
state of the vector meson, in which the gluon carries 
a finite
fraction of the vector meson's momentum. With the reference 
to the non-relativistic intuition, one may argue that an 
admixture of such a non-perturbative $q\bar{q}g$ Fock state 
is small in the $\Upsilon$, but not the lighter quarkonia
the non-relativistic treatment of which is suspect. The
issue of nonperturbative $q\bar{q}g$ and higher Fock states
in light vector mesons remains open. 
Hence we are bound to stay within the $q\bar{q}$ Fock state
and leading log${1\over x}$ approximations.

None of the NLO corrections is expected to change substantially
the predicted $x_g$- and $Q^2$-dependences but they affect
strongly the predicted cross section. For instance, one
often includes the correction for skewness (\ref{eq:4.6.3}) 
for the reason that at small $x$ and large-$\Qb^2$ it
enhances the predicted cross section almost by a factor 2.
Similarly, the real part of the dipole amplitude 
(\ref{eq:4.4.12}) is a 
NLO correction which is substantial at large $\Qb^2$.
A consistent treatment of the potentially more important 
 Celmaster \& Barbieri et al
correction (\ref{eq:4.10.2}) and its counterpart for the
impact factor is not available yet. In their analysis of
NLO corrections Levin et al. did not consider
the Celmaster \& Barbieri et al correction but argued
that by analogy with the Drell-Yan production of lepton
pairs the cross section of near threshold diffractive
$c\bar{c}$ pairs acquires the $K$-factor \cite{LevinVM_Kfactor}
\be
K \approx 1 + {2\pi \over 3}\alpha_S\,,
\label{eq:4.12.1}
\ee
which, by virtue of duality arguments, shall propagate to
the vector meson production cross section,
while Dremin asserts that NLO Sudakov effects  
rather suppress the cross section \cite{DreminSudakov,*DreminSudakov1,*DreminSudakov2}.
The full fledged NLO $log{1\over x}$ $k_{\perp}$-factorization
analysis necessary for consistent treatment of all these 
corrections is not yet
available.
\newpage


\newpage
\section{Helicity properties of vector meson production}


\subsection{General introduction}

\begin{figure}[htbp]

   \centering

   \epsfig{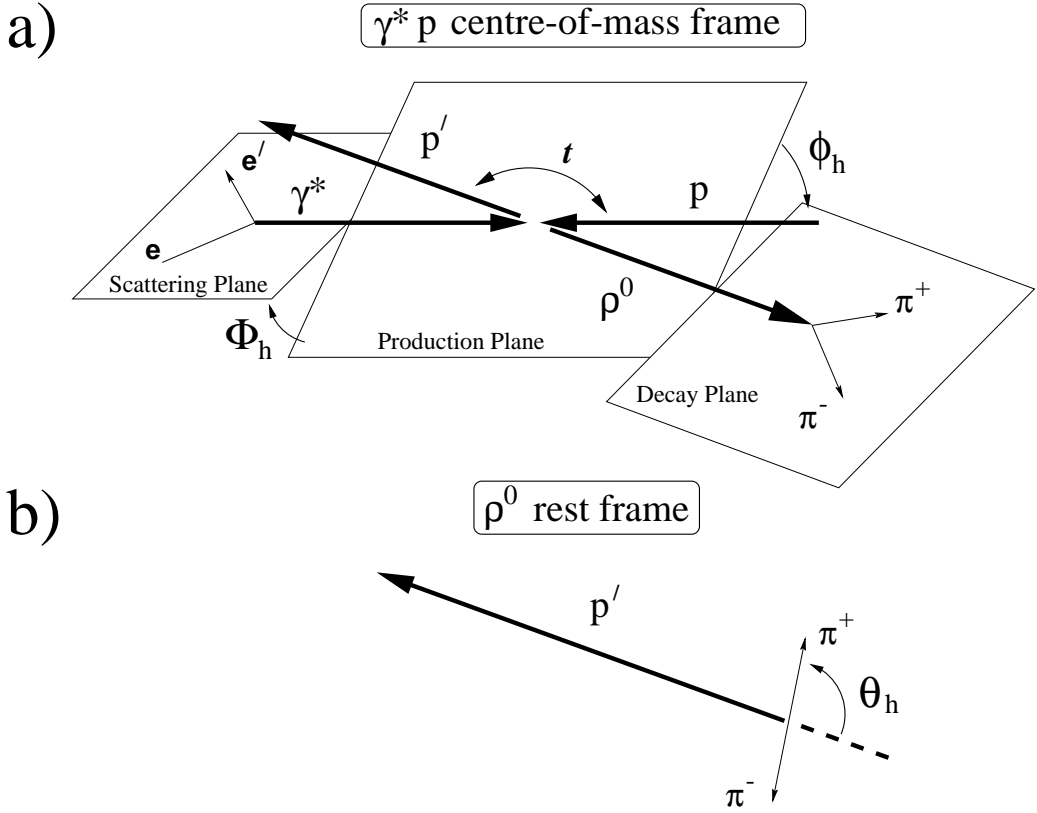}

   \caption{\it   Kinematics of the process 
$e p \rightarrow e \rho^0 p \to e \pi^+\pi^- p$.
   The angular variables used in the determination of the
helicity amplitudes from the decay distributions: $\Phi_h$, the
   azimuthal angle between the $(e,e')$ scattering plane and the 
$(\rho^0,p')$ production plane;
   $\phi_h$, the azimuthal angle between the production and decay planes; and
   $\theta_h$, the polar angle of the positively-charged decay pion defined
   with respect to the direction of the $\rho^0$ momentum in the $\gamma^* p$
   cms system. Illustration is taken from ~\protect\cite{epj:c12:393}.}

   \label{fig:HelicityDef}

\end{figure}

The angular distribution of the exclusive production of vector
mesons decaying into particle-antiparticle final state is usually
described in the so-called helicity frame 
~\protect\cite{GilmanSCHC,BauerRevModPhys,*BauerRevModPhys1}. For the
unpolarized lepton beam, such as that at HERA, the cross section
depends on three angles explained in 
Fig.~\ref{fig:HelicityDef} for the specific case of $\rho^0 \to
\pi^+\pi^-$ production.

This angular distribution can be expressed in terms of the virtual
photon density matrix $\rho_{\lambda_\gamma^\prime\lambda_\gamma}(\Phi_h)$ 
(\ref{eq:2.3.2.8}), the helicity
amplitudes of virtual photon transition with helicity
$\lambda_\gamma$ into the vector meson with helicity $\lambda_V$,
${\cal T}_{\lambda_V,\lambda_\gamma}$, and of the angular factors that
describe the vector meson decay into the final
particle-antiparticle state. For the case of decay into scalars
($\rho \to \pi^+\pi^-$ and $\phi \to K^+K^-$ decays),
these angular factors are given by spherical harmonics
$Y_{1,\lambda_V}$: 
\bea
{d\sigma \over
d\cos\theta_h\,d\phi_h\,d\Phi_h} &\equiv& \sigma\cdot
W(\cos\theta_h,
\phi_h,\Phi_h) \label{angulardef} \\
\hspace{-2cm}&=& \sum_{\lambda_\gamma,\lambda_\gamma^\prime;
\lambda_V\lambda_V^\prime} {\cal T}_{\lambda_V,\lambda_\gamma}
{\cal T}^*_{\lambda_V^\prime,\lambda_\gamma^\prime} \cdot {\rm
Y}_{1,\lambda_V}(\theta_h,\phi_h) {\rm
Y}^*_{1,\lambda_V^\prime}(\theta_h,\phi_h) \cdot
\rho_{\lambda_\gamma\lambda_\gamma^\prime}(\Phi_h)\,.
\nonumber
\label{eq:5.1.1}
\eea 
In the case of the $J/\Psi$ production one measures the angular 
distribution of leptons in the decay 
$J/\psi \to \ell^+\ell^-$, then the $D^{1/2}(\theta_h,\phi_h)$-functions 
instead of spherical harmonics will appear in (\ref{eq:5.1.1}). Since 
the photon density matrix
$\rho_{\lambda_\gamma^\prime\lambda_\gamma}$ and the angular
factors are known, the study of the angular
dependence of the cross section reveals the helicity structure of
the $\gamma^* p \to Vp$ transition.

The conservation of the $s$-channel helicity in the 
scattering of electrons in the Coulomb field is known
since 1954 (\cite{Yennie}, see also the textbook
\cite{LandauLifshitz1}).
Motivated by the early experimental data on vector meson
photo- and electroproduction, Gilman et al. suggested the
$s$-channel helicity conservation (SCHC) as the fundamental
feature of diffraction scattering (\cite{GilmanSCHC}, for the
discussion of the pre-HERA experimental situation
see \cite{BauerRevModPhys,*BauerRevModPhys1}). Within diffraction cone, 
the helicity of the vector meson $\lambda_V$ coincides
approximately with the helicity of the incident photon
$\lambda_\gamma$, see Section \ref{SCHNC}. 
It is reasonable therefore to start with the case
of strict $s$-channel helicity conservation (SCHC), $\lambda_V =
\lambda_\gamma$.

As emphasized in Section 3.1.2, the Pomeron and all higher lying secondary 
reggeons are natural parity $t$-channel exchanges. Hereafter we analyze
the helicity properties of vector meson production assuming natural
parity exchange. A good idea on why the possible contribution from
unnatural parity exchange can be neglected is 
given by the longitudinal double-spin
asymmetry. 


\subsection{Longitudinal 
double-spin asymmetry and unnatural parity exchange}

One is familiar with the helicity structure function of the proton, $g_1(x,Q^2)$,
which measures the mean helicity of partons in the longitudinally polarized
proton. It is determined experimentally in the DIS of the longitudinally 
polarized leptons off longitudinally polarized proton target, where the 
polarized leptons serve as the source of circularly polarized photons.
The measured cross section is proportional to the imaginary part of the 
helicity conserving forward Compton scattering amplitude,
\be
{\cal T}_{\lambda_{\gamma} \lambda_N, \lambda_{\gamma} \lambda_N} \propto
F_1(x,Q^2)+ \lambda_{\gamma} \lambda_N g_1(x,Q^2) 
\propto 1+A_{LL}^{(DIS)}\lambda_{\gamma} \lambda_N\, ,
\label{eq:5.2.1}
\ee
where $\lambda_{\gamma},\lambda_N$ are the helicities of the photon and
target nucleon ($\lambda_N = \pm 1$). Such a helicity dependence for
transverse photons
emerges naturally for the axial-vector ($A_1,...$) meson exchange.
The term $\propto g_1(x,Q^2)$ changes the sign when the circular polarization
of the photon is flipped and, from the t-channel exchange point of view,
corresponds to the unnatural parity exchange, see Eq.~(\ref{eq:3.4.6.1}).
As explained in Section 1.3, the vector meson production amplitude derives
from the Compton amplitude by analytic continuation in the virtuality of
the photon to the vector meson pole, which should not change dramatically
the asymmetry parameter in the amplitude. Consequently, if one  
parameterizes the unnatural parity exchange into the 
transverse vector meson production
amplitude as
\be
\left.{\cal T}_{\lambda_V \lambda_N, \lambda_{\gamma} \lambda_N}
\right|_{\lambda_V =\lambda_{\gamma}}
 \propto
 1+{1\over 2} A_{LL}^{V}\lambda_{\gamma} \lambda_N\, ,
\label{eq:5.2.2}
\ee
then the natural expectation for the transverse cross section will be
$\sigma_T$ is \cite{KolyaALL}
\be
A_{LL}^{V} \approx 2A_{LL}^{(DIS)}\,.
\label{eq:5.2.3}
\ee
There is a purposeful  difference between expansions (\ref{eq:5.2.1}) 
and (\ref{eq:5.2.2})  because in the vector meson production one measures 
the differential cross section $\propto |{\cal T}|^2$. 
In $\sigma_L$-$\sigma_T$ unseparated vector meson production the 
asymmetry is diluted for the presence of $\sigma_L$,
\be
A_1^V \approx {A_{LL}^{V} \over 1+R_V} \approx {2 A_1 \over 1+R_V}\, .
\label{eq:5.2.4}
\ee
(here we are back to the usual notation $A_1=A_{LL}^{(DIS)}$)
hence the dilution factor $(1+R_V)$ in (\ref{eq:5.2.4}) compared
to (\ref{eq:5.2.3}).  
As a matter of fact, such a relationship between the longitudinal
double-spin asymmetries for DIS and vector meson production
has been suggested already in 1976 by Fraas on the basis of
the vector dominance 
model \cite{FraasALL}.

\begin{figure}[htbp]
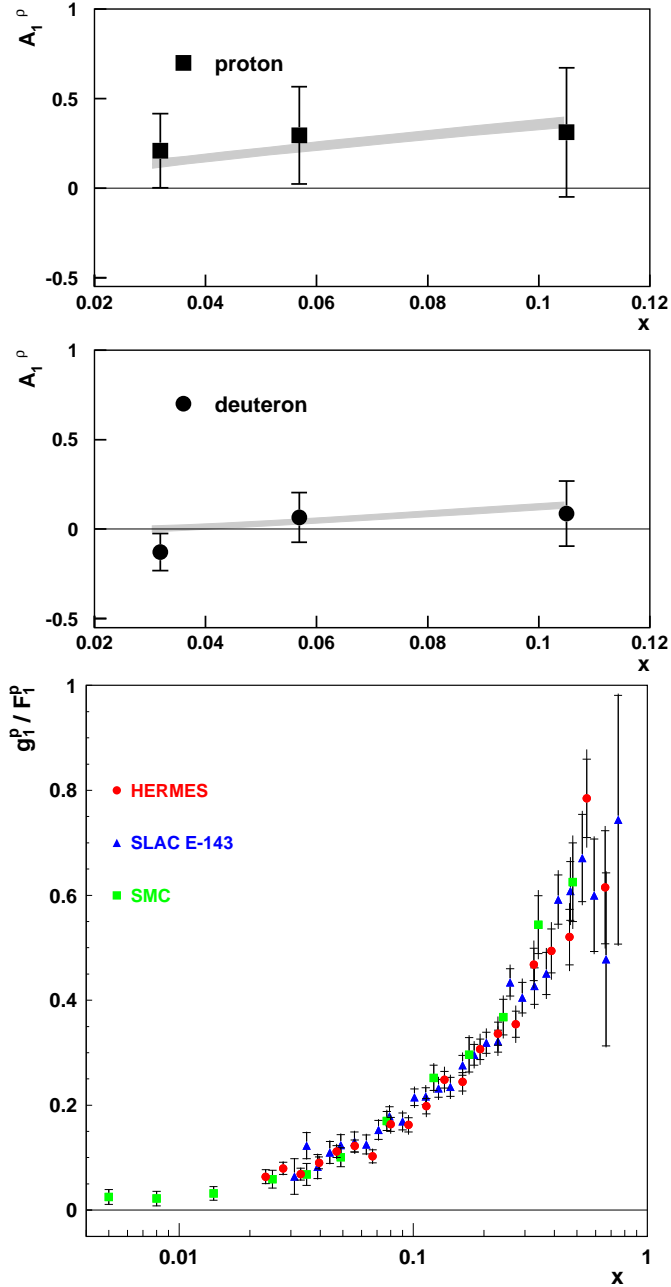

   \centering

   \hspace*{1cm}
   \epsfig{file=HermesALL.eps,width=95mm}
   \epsfig{file=A1summary.eps,width=85mm}
\caption{\it   Upper two plots show the $x$-dependence of the longitudinal double-spin
asymmetry $A_1^{\rho}$ in exclusive $\rho^0$ meson electroproduction
on the proton (top box) and deuteron (bottom box). The data 
from HERMES \protect\cite{HermesALL} are compared to 
the expectations from Eq.~(\ref{eq:5.2.4}) based on the
DIS data from the bottom plot, which shows the 
ratio of polarized to unpolarized proton structure function
from the SMC, E143 and HERMES experiments. For the review 
and references see ~\protect\cite{A1DIS}.}
\label{fig:HermesALL}
\end{figure}

The results of the first experimental determination of $A_1^{\rho}$ for
the diffractive $\rho$ production in the HERMES experiment
\cite{HermesALL} are shown in
Fig. ~(\ref{fig:HermesALL}) in comparison with the estimates from the
DIS data based on Eq.(\ref{eq:5.2.4}) but
without the dilution of $A_1^V$ by the factor $(1+R_V)$. A summary of 
the high precision experimental data on the ratio of polarized, helicity,
$g_1(x,Q^2)$,
to unpolarized, $F_1(x,Q^2)$, proton structure function is shown 
in Fig.~\ref{fig:HermesALL}, the important 
point is that the effects of unnatural parity exchange vanish at 
small $x$, for pQCD arguments in favor of that see \cite{ErmolaevG1,ErmolaevG2,ErmolaevG3}.

\subsection{The angular distribution in the SCHC approximation}

\subsubsection{Theoretical expectation: angular distributions}


Within SCHC and natural parity exchange  
we are left with two independent
helicity amplitudes, ${\cal T}_{11}$ and ${\cal T}_{00}$. Then 
the angular dependence of reaction $\gamma^* p \rightarrow V p$, 
with $V$ decaying into two scalars, is:
\bea
W(\cos\theta_h, \phi_h,\Phi_h) &=& {1 \over N} {3 \over 4\pi}
\left[\epsilon |{\cal T}_{00}|^2\cos^2\theta_h + {1\over
2}|{\cal T}_{11}|^2\sin^2\theta_h\right.
\nonumber\\[2mm]
&&\quad+\ {1\over 2}\epsilon |{\cal T}_{11}|^2\sin^2\theta_h\cos2(\Phi_h-\phi_h)\nonumber\\[2mm]
&&\quad-\ \left.\sqrt{2\epsilon(1+\epsilon)} \R({\cal T}_{11}{\cal T}^*_{00})
\sin\theta_h\cos\theta_h\cos(\Phi_h-\phi_h)\right]\,,
\label{eq:5.3.1.1} 
\eea
where $N = |{\cal T}_{11}|^2 + \epsilon |{\cal T}_{00}|^2$ and $\epsilon$ is
defined in (\ref{eq:2.3.2.8}). The first line in (\ref{eq:5.3.1.1})
is the contribution of diagonal terms of the photon density
matrix, i.e. with $\lambda_\gamma = \lambda_\gamma^\prime$; the
second line is the interference of transverse photons with
opposite helicities, $\lambda_\gamma = - \lambda_\gamma^\prime =
\pm 1$; the last line is the interference between the transverse
and longitudinal photons. Note that angular dependence
(\ref{eq:5.3.1.1}) involves only a single azimuthal angle $\psi
= \Phi_h - \phi_h$ between the $(e,e')$ and decay planes.

One can reparameterize, and analyze, 
the angular distribution (\ref{eq:5.3.1.1})
in terms $r^{\alpha}_{ij}$, which
compose the spin-density matrix of a vector meson
\cite{SchillingWolf}. The generic case and the 
involved notations for the $r^{\alpha}_{ij}$ are explained
in Section 5.3.1, here we only notice that under the assumption 
of SCHC many the elements of $r^{\alpha}_{ij}$ do vanish and 
(\ref{eq:5.3.1.1}) takes the form:

\bea
W(\cos\theta_h,\phi_h,\Phi_h) &=& {3 \over 4\pi}\left[{1
\over 2}(1-r^{04}_{00}) +
{1 \over 2}(3r^{04}_{00}-1)\cos^2\theta_h  \right.\nonumber\\[2mm]
&&+\ \epsilon \cos2\Phi_h \sqrt{2} \sin^2\theta_h\cos2\phi_h\cdot r^{1}_{1-1}\nonumber\\[2mm]
&&-\ \epsilon\sin 2\Phi_h \sin^2\theta_h \sin2\phi_h\cdot \I\{r^2_{1-1}\} \nonumber\\[2mm]
&&-\ \sqrt{2\epsilon(1+\epsilon)}\cos\Phi_h
\sin2\theta_h \cos\phi_h \cdot \sqrt{2}\R\{r^5_{10}\}\nonumber\\[2mm]
&&+\ \left. \sqrt{2\epsilon(1+\epsilon)}\sin\Phi_h \sin2\theta_h
\sin2\phi_h\cdot \sqrt{2}\I\{r^6_{10}\}
\right]\,.\label{eq:5.3.1.2}
\eea

It contains five non-zero spin-density matrix elements among which
only three are independent due to the relations
(\ref{eq:3.4.6.1}).
Their expression 
via the helicity amplitudes reads
\bea
&&r^{04}_{00}=\frac{\eps |{\cal T}_{00}|^2}{|{\cal T}_{11}|^2 + 
\epsilon |{\cal T}_{00}|^2}\,;\nonumber\\[3mm]
&&r^{1}_{1-1} = - \I\{r^2_{1-1}\} = \frac{1}{2} \frac{ |{\cal T}_{11}|^2}
{|{\cal T}_{11}|^2 + \epsilon |{\cal T}_{00}|^2}\,;\label{eq:5.3.1.3}\\[3mm]
&&\R\{r^{5}_{10}\}=- \I\{r^6_{10}\} =
\frac{1}{2\sqrt{2}}\frac{\R\{{\cal T}_{11}{\cal T}^{\star}_{00}\}}{|{\cal T}_{11}|^2 +
\epsilon |{\cal T}_{00}|^2}\,. \nonumber
\eea 
Under the  SCHC conservation, the 
ratio of the longitudinal to transverse cross sections, 
(\ref{eq:2.3.3.4}), is expressed only via matrix element $r^{04}_{00}$,
\be
R_V = \frac{1}{\epsilon}\frac{r^{04}_{00}}{1-r^{04}_{00}}\,,
\label{eq:5.3.1.4}
\ee
while the relative phase $\delta$ between the ${\cal T}_{11}$ and ${\cal T}_{00}$
amplitudes can be determined via
\be
\cos\delta = 
\fr{1+\epsilon\, R_V}{\sqrt{R_V/2}}\left(\R\{r^{5}_{10}\}-\I\{r^{6}_{10}\}\right)\,.
\label{eq:5.3.1.5}
\ee


\subsubsection{Experimental results}

\begin{figure}[htbp]

   \centering

   \epsfig{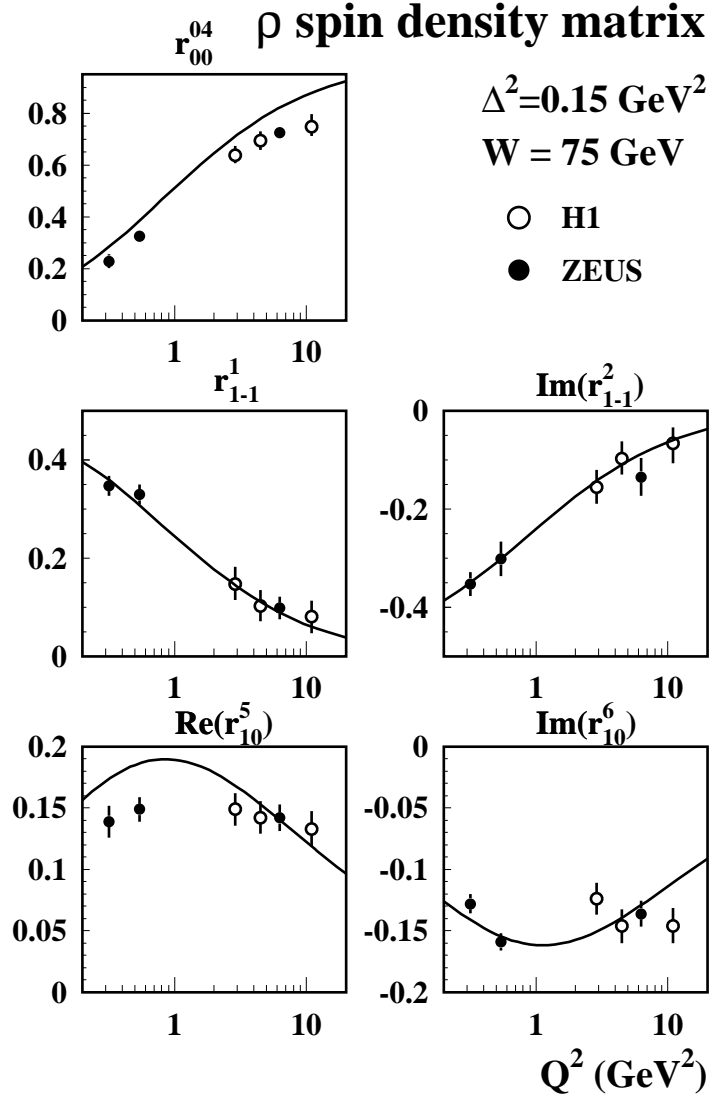}

   \caption{\it   The compilation of the ZEUS \protect\cite{epj:c12:393} 
and H1 \protect\cite{epj:c13:371}
results on the $Q^2$ behavior of the five $s$-channel
helicity conserving amplitudes for $\rho$ meson. The first matrix element
corresponds to the diagonal terms in photon density matrix; the
second line is the interference of transverse photons with
opposite helicities; the last line is the interference between
transverse and longitudinal photons. The lines show the
$k_t$-factorization predictions \protect\cite{IgorPhD,IgorNumerics}.}
   \label{fig:ivanov-rhohelicity-schc}
\end{figure}

Figure~\ref{fig:ivanov-rhohelicity-schc} shows the results 
of ZEUS  and H1 on the five "non-zero-SCHC" (hereafter
just "SCHC") matrix elements mentioned above. 
These matrix elements are placed in
three rows. The first line corresponds to diagonal terms; the
second line is the interference of transverse photons with
opposite helicities; the last line is the interference between
transverse and longitudinal photons.

\begin{figure}[htbp]
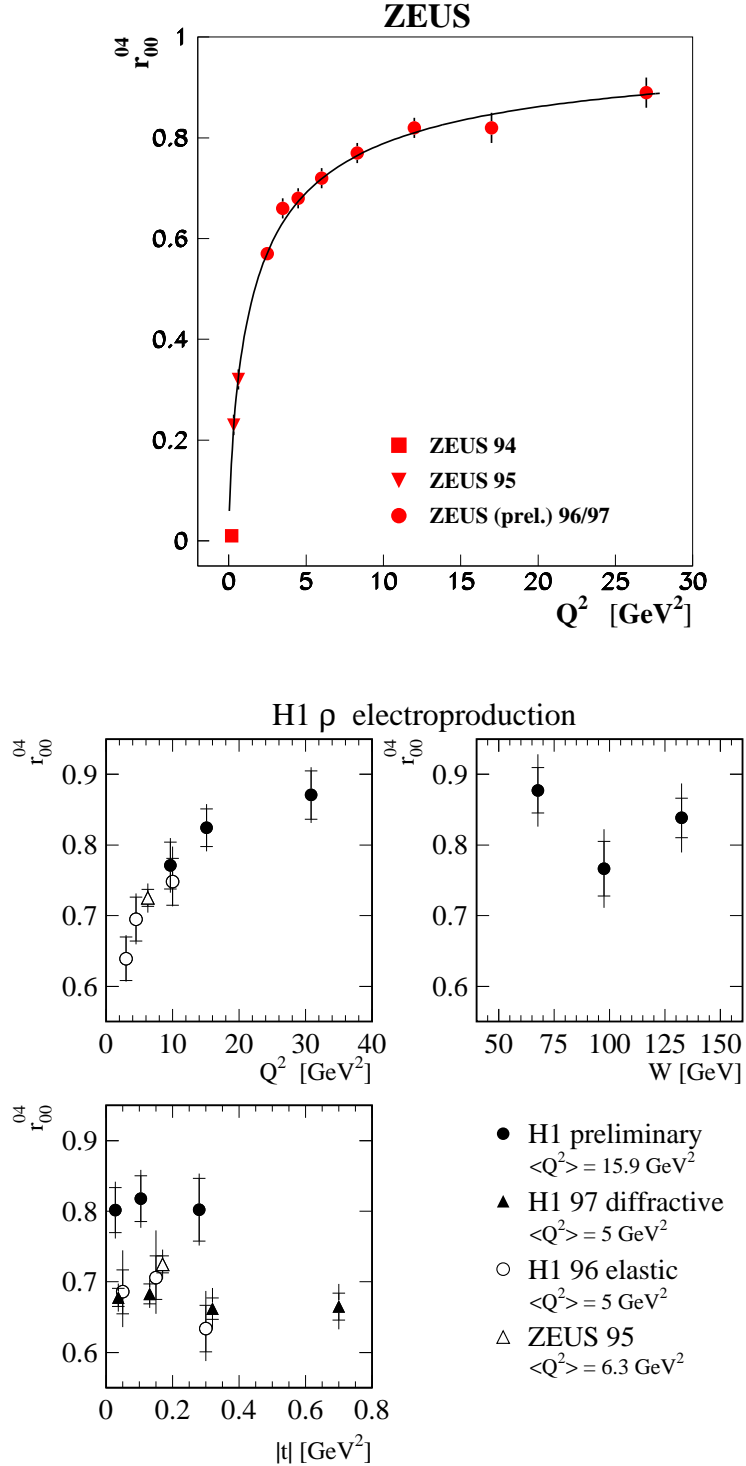

   \centering
   \epsfig{file=ZEUSDISRhoEPS01prelim_fig3.ps,width=100mm}
   \epsfig{file=H1DISRho2002prelim_fig9.eps,width=110mm}

   \caption{\it   Upper plot. The $Q^2$ behavior of the spin density matrix element $r^{04}_{00}$
   for $\rho$ production measured by ZEUS ~\protect\cite{cpaper:epc2001:594}.
The solid curve is a fit of the form 
$r^{04}_{00}=1/(1+\xi(m_{\rho}^2/Q^2)^{\kappa})$ with $\xi = 2.17\pm 0.07,
\kappa = 0.75 \pm 0.3$.
   Bottom three plots show the spin-density matrix element $r^{04}_{00}$
for $\rho$ meson electroproduction as function of $Q^2$, $W$ and
$t$ measured by H1 and ZEUS (~\protect\cite{cpaper:ichep2002:989} and
references therein). }
   \label{fig:H1ZEUSr0400}
\end{figure}

%
%
%

The matrix element $r^{04}_{00}$ is extracted from the
single-differential cross section $d\sigma/d\cos\theta_h$. The
photoproduction measurements 
~\cite{np:b463:3,epj:c2:247,pl:b377:259,epj:c24:345}
confirm that at $Q^2=0$ the matrix element $r^{04}_{00}$ is zero within
experimental uncertainties for $\rho$, $\phi$ and $J/\psi$ mesons.
This should be expected, since in the limit $Q^2 \to 0$ the
longitudinal cross section must vanish. The behavior of
$r^{04}_{00}$ as function of $Q^2$, $W$ and $t$ for $\rho$ mesons
is shown in Figs.~\ref{fig:ivanov-rhohelicity-schc},
~\ref{fig:H1ZEUSr0400}.

The steep $Q^2$ dependence of $r^{04}_{00}$ 
is mainly due to the gauge invariance driven factor $\sim Q^2/m_V^2$ present 
in the longitudinal cross section, see (\ref{eq:3.4.2.8}),
(\ref{eq:3.4.2.9}). The pattern of the $Q^2$ dependence is
very similar for all the light vector mesons and differs in 
the case of
the $J/\psi$ (not shown). The matrix element $r^{04}_{00}$ 
is the main source of the determination of 
the ratio $R_V=\sigma_L/\sigma_T$, whose $Q^2$-behavior 
will be discussed in detail
in the forthcoming Section~\ref{sect6.3}. The $t$ and $W$ dependences of
$r^{04}_{00}$ are consistent with being flat.
This indicates that the energy dependence of 
the longitudinal, ${\cal T}_{00}$,  and transverse, ${\cal T}_{11}$,
amplitudes is very close to each other in agreement 
with theoretical expectations.
The same holds also for the $t$-dependence 
of the longitudinal and transverse amplitudes.

The $Q^2$-behavior of the other SCHC matrix elements can be
read off from Eq.(\ref{eq:5.3.1.3}). 
The elements $r^1_{1-1}$ and ${\rm Im} \{ r^2_{1-1} \}$ should approach 
$\pm 1/2$, respectively, in the photoproduction limit, and are expected to decrease with
$Q^2$ increase approximately as $1/(2R)$. This tendency is well
observed in the data. The longitudinal-transverse (LT) interference driven 
elements ${\rm Re} \{ r^5_{10} \}$ and 
${\rm Im} \{ r^6_{10} \} $ should be $\propto
Q$ at small $Q^2$ and fall off with $Q^2$ growth as $1/Q$. The
experimental data from the ZEUS collaboration do follow this 
expectation, the experimental data from H1 on $r^{5,6}_{10}$ exhibit
certain departure from the theoretical expectation.

\begin{figure}[htbp]
   \centering
   
   \hspace*{2cm}
   \epsfig{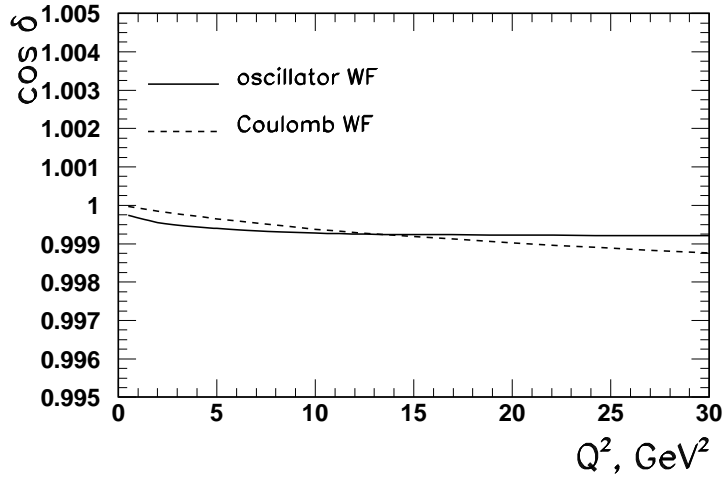}
\newline
   \epsfig{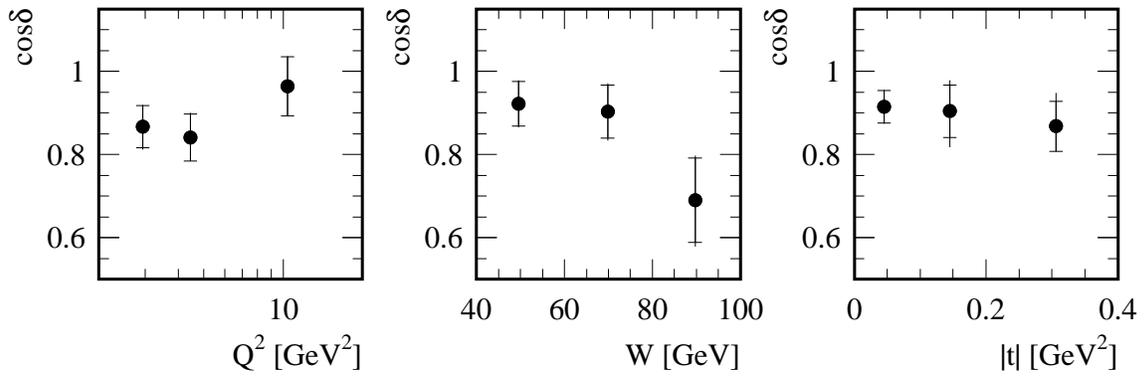}
   \caption{\it   The upper plot shows the theoretical expectation from the
$k_{\perp}$-factorization \protect\cite{IgorPhD,IgorNumerics}
for the cosine of the relative
phase $\delta$ between the transverse,
${\cal T}_{11}$, and the longitudinal, ${\cal T}_{00}$, amplitudes.
The bottom plots show
the H1 determination of $cos \delta$ 
as a function of $Q^2$, $W$ and $t$ \protect\cite{epj:c13:371}.}
   \label{fig:H1relativephaseLT}
\end{figure}

As we discussed in Section 4.8.1, because of the slightly different 
hard scales  $\Qb^2_L$ and $\Qb^2_T$ the two helicity amplitudes
 ${\cal T}_{11}$ and ${\cal T}_{00}$ 
can have a slightly different energy dependence, 
see Sections~\ref{sect3.3.1}--\ref{sect3.3.3} and
the discussion of the experimental data in Section 7. 
Then the
derivative analyticity (\ref{eq:4.4.12}) would predict
 the nonvanishing,
but very small, in the range of $1\div 3^\circ$, relative phase $\delta$ 
of the two amplitudes. The results of the H1 determination 
\cite{epj:c13:371} of $\cos \delta$
from Eq.~(\ref{eq:5.3.1.5}) are
shown in Fig.~\ref{fig:H1relativephaseLT}. The combined 
value of $\cos\delta$ suggests 
the statistically significant departure from  1,  
\be
\cos\delta =
0.925 \pm 0.022 {}^{+0.011}_{-0.022}\,,
\label{eq:5.3.2.1}
\ee
the sizeable relative phase $\delta \sim 10\div 20^\circ$ is 
difficult to accommodate in theory.
The error bars are large, though, it must be noted that ZEUS
assumes $\delta = 0$ and still obtains good fits to the angular
distribution.


\subsection{Angular dependence beyond SCHC}


\subsubsection{Theoretical expectations}

Without the assumption of $s$-channel helicity conservation,
and supposing the natural parity exchange, see Section 5.2.3,  
one is left
with five independent helicity amplitudes. The
angular dependence of the cross section (for spinless
particle-antiparticle final state such as in the decay
$\rho \to \pi^+\pi^-, \phi \to K^+K^-$) is parameterized 
in terms of 15
spin density matrix elements \cite{SchillingWolf}: 
\bea
&&{1 \over \sigma}{d\sigma
\over d\cos\theta_h\,d\phi_h\,d\Phi_h} \equiv W(\cos\theta_h,
\phi_h, \Phi_h) = \nonumber\\
&&={3 \over 4\pi}\left[{1 \over 2}(1-{\bm r^{04}_{00}}) + {1 \over
2}(3{\bm r^{04}_{00}}-1)\cos^2\theta_h - \sqrt{2}\mbox{Re }\{
r^{04}_{10}\} \sin2\theta_h
\cos\phi_h - r^{04}_{1-1} \sin^2\theta_h\cos 2\phi_h \right.\nonumber\\
&&- \epsilon \cos2\Phi_h \left(r^1_{11} \sin^2\theta_h +
r^{1}_{00} \cos^2\theta_h - \sqrt{2} \mbox{Re }\{
r^1_{10}\} \sin2\theta_h\cos\phi_h - {\boldmath r^{1}_{1-1}}
\sin^2\theta_h\cos2\phi_h\right) \nonumber\\
&&- \epsilon\sin 2\Phi_h\left(\sqrt{2}\mbox{Im}\{r^2_{10}\} 
\sin2\theta_h\sin\phi_h
+ \mbox{Im }\{ r^2_{ 1-1}\}\sin^2\theta_h \sin2\phi_h\right)\nonumber\\
&&+ \sqrt{2\epsilon(1+\epsilon)}\cos\Phi_h
\left(r^5_{11}\sin^2\theta_h + r^5_{00}\cos^2\theta_h
-\sqrt{2}\mbox{Re } \{r^5_{ 10}\}\sin2\theta_h \cos\phi_h
- r^5_{1-1}\sin^2\theta_h \cos2\phi_h\right)\nonumber\\
&&+\left. \sqrt{2\epsilon(1+\epsilon)}\sin\Phi_h
\left(\sqrt{2}\mbox{Im } \{r^6_{ 10}\}\sin2\theta_h
\sin\phi_h + \mbox{Im }\{r^6_{1-1}\} \sin^2\theta_h \sin2\phi_h\right)
\right]\,.
\label{eq:5.4.1.1} 
\eea 
One can develop an intuition when
reading this expression. The subscripts $i,k=-1,0,1$ 
of the matrix elements $r^\alpha_{ik}$ indicate
the vector meson helicities $\lambda_V$ and $\lambda_V^\prime$ of
the amplitudes interfering. Dependence on $\Phi_h$ shows the
helicities of the photon of the interfering amplitudes. As before,
the $\Phi_h$-independent terms are
diagonal in photon helicities; they are accompanied by superscript
04. Terms $\propto \cos 2\Phi_h$ and $\propto \sin2\Phi_h$ originate from the
interference of the transverse photons with opposite helicities;
these are accompanied by superscripts 1 and 2. Terms $\propto \cos
\Phi_h$ and $\propto \sin\Phi_h$ appear from the interference of the
transverse and longitudinal photons; these are accompanied by
superscripts 5 and 6. The nature of each $r^\alpha_{ik}$ can then
be understood from its indices. For example, $r^1_{11}$ comes from
interference of photons of helicities $+1$ and $-1$ and from
vector mesons with both helicities equal to $1$ or $-1$.
Therefore, $r^1_{11}$ must be proportional to the double spin-flip
amplitude.

The correspondence between the matrix elements $r^{\alpha}_{ik}$, 
the helicity amplitudes, and their SCHC and SCHNC properties is
as follows: 
\bea
\parbox{3cm}{SCHC}&&
\begin{array}{l}r^{04}_{00}=\eps |{\cal T}_{00}|^2+|{\cal T}_{01}|^2\\[3mm]
r^{1}_{1-1} = \frac{1}{2} |{\cal T}_{11}|^2
+\frac{1}{2} |{\cal T}_{1-1}|^2\\[3mm]
\I\{r^{2}_{1-1}\}=-\frac{1}{2} |{\cal T}_{11}|^2
+\frac{1}{2} |{\cal T}_{1-1}|^2\\[3mm]
\R\{r^{5}_{10}\}=\frac{1}{2\sqrt{2}}\R\{{\cal T}_{11}{\cal T}^{\star}_{00}\} +
\frac{1}{\sqrt{2}}\R\{{\cal T}_{10}{\cal T}^{\star}_{01}\}
- \frac{1}{2\sqrt{2}}\R\{{\cal T}_{1-1}{\cal T}^{\star}_{00}\}\\[3mm]
\I\{r^{6}_{10}\}=-\frac{1}{2\sqrt{2}}\R\{{\cal T}_{11}{\cal T}^{\star}_{00}\}
-\frac{1}{2\sqrt{2}}\R\{{\cal T}_{1-1}{\cal T}^{\star}_{00}\}
\end{array}\label{eq:5.4.1.2}\\[7mm]
\parbox{3cm}{strong SCHNC from single-flip 
$\gamma_T \to V_L, \propto {\cal T}_{01}$}&&
\begin{array}{l}
\R\{r^{04}_{10}\}=\frac{1}{2}\R\{{\cal T}_{11} {\cal T}^{\star}_{01}\}
+\eps\R\{{\cal T}_{10}{\cal T}^{\star}_{00}\} + \frac{1}{2}\R\{{\cal T}_{1-1}{\cal T}^{\star}_{0-1}\}\\[3mm]
\R\{r^{1}_{10}\}=\frac{1}{2}\R\{{\cal T}_{11}{\cal T}^{\star}_{0-1}\}
+\frac{1}{2} \R\{{\cal T}_{1-1}{\cal T}^{\star}_{01}\} \\[3mm]
\I\{r^{2}_{10}\}=-\frac{1}{2}\R\{{\cal T}_{11}{\cal T}^{\star}_{0-1}\}
+\frac{1}{2}\R\{{\cal T}_{1-1}{\cal T}^{\star}_{01}\}\\[3mm]
r^{5}_{00}=\sqrt{2} \R\{{\cal T}_{00}{\cal T}^{\star}_{01}\}
\end{array}\label{eq:5.4.1.3}\\[7mm]
\parbox{3cm}{weak SCHNC from single-flip $\gamma_L \to V_T$, 
 $\propto {\cal T}_{10}$}&&
\begin{array}{l}
r^{5}_{11}=\frac{1}{\sqrt{2}}\R\{{\cal T}_{10}{\cal T}^{\star}_{11}\}
-\frac{1}{\sqrt{2}}\R\{{\cal T}_{10}{\cal T}^{\star}_{1-1}\}\\[3mm]
r^{5}_{1-1}=\frac{1}{\sqrt{2}}\R\{{\cal T}_{11}{\cal T}^{\star}_{-10}\}
+\frac{1}{\sqrt{2}}\R\{{\cal T}_{10}{\cal T}^{\star}_{-11}\}\\[3mm]
\I\{r^{6}_{1-1}\}=-\frac{1}{\sqrt{2}}\R\{{\cal T}_{-10}{\cal T}^{\star}_{11}\}
+\frac{1}{\sqrt{2}}\R\{{\cal T}_{10}{\cal T}^{\star}_{-11}\}\\[3mm]
\end{array}\label{eq:5.4.1.4}\\[7mm]
\parbox{3cm}{double SCHNC from the double-flip, or two single-flips}&&
\begin{array}{l}
r^{04}_{1-1}=-\eps |{\cal T}_{10}|^2+\R\{{\cal T}_{11}{\cal T}^{\star}_{1-1}\}\\[3mm]
r^{1}_{11} = \R\{{\cal T}_{1-1}{\cal T}^{\star}_{11}\}\\[3mm]
r^{1}_{00} =-|{\cal T}_{01}|^2
\end{array}\label{eq:5.4.1.5}
\eea 
It is understood that the rhs of each and 
every line must be also divided by $\epsilon
(|{\cal T}_{00}|^2+2|{\cal T}_{10}|^2) + |{\cal T}_{11}|^2+|{\cal T}_{1-1}|^2+|{\cal T}_{01}|^2$.

The classification of each set of density matrix elements
(\ref{eq:5.4.1.2})-(\ref{eq:5.4.1.5})
in the left column of this Table
can be understood as follows: 
At large $Q^2$, the largest SCHNC amplitude is
${\cal T}_{01}$ (the transverse photon to longitudinal vector meson transition).
Therefore, the stronger SCHC violation is expected in density matrix
elements of the second group, in particular, in $r^{5}_{00}$. 
At small $Q^2$, the double-flip  ${\cal T}_{1-1}$ is
predicted to be the largest SCHNC
amplitude as it is soft-dominated and does not require the
longitudinal Fermi motion. Therefore, at small $Q^2$ the largest
non-SCHC spin density matrix elements is expected to be $r^{04}_{1-1}$ and
$r^1_{00}$. The amplitude ${\cal T}_{10}$ vanishes in the 
photoproduction limit, is of higher twist at large $Q^2$ and is expected 
to be always small.

As discussed in Section \ref{SCHNC}, the small-$t$ behavior
of various amplitudes is governed solely by the value 
of the helicity flip:
the matrix elements from the second, Eq.~(\ref{eq:5.4.1.3})  and third,
Eq.~(\ref{eq:5.4.1.4}), groups 
are expected to be $\propto \sqrt{|t'|}$, 
the elements in the last group,  Eq.~(\ref{eq:5.4.1.4}), are $\propto |t'|$.


\subsubsection{Experimental results: helicity properties at small $t$}

\begin{figure}[htbp]
   \centering
  \epsfig{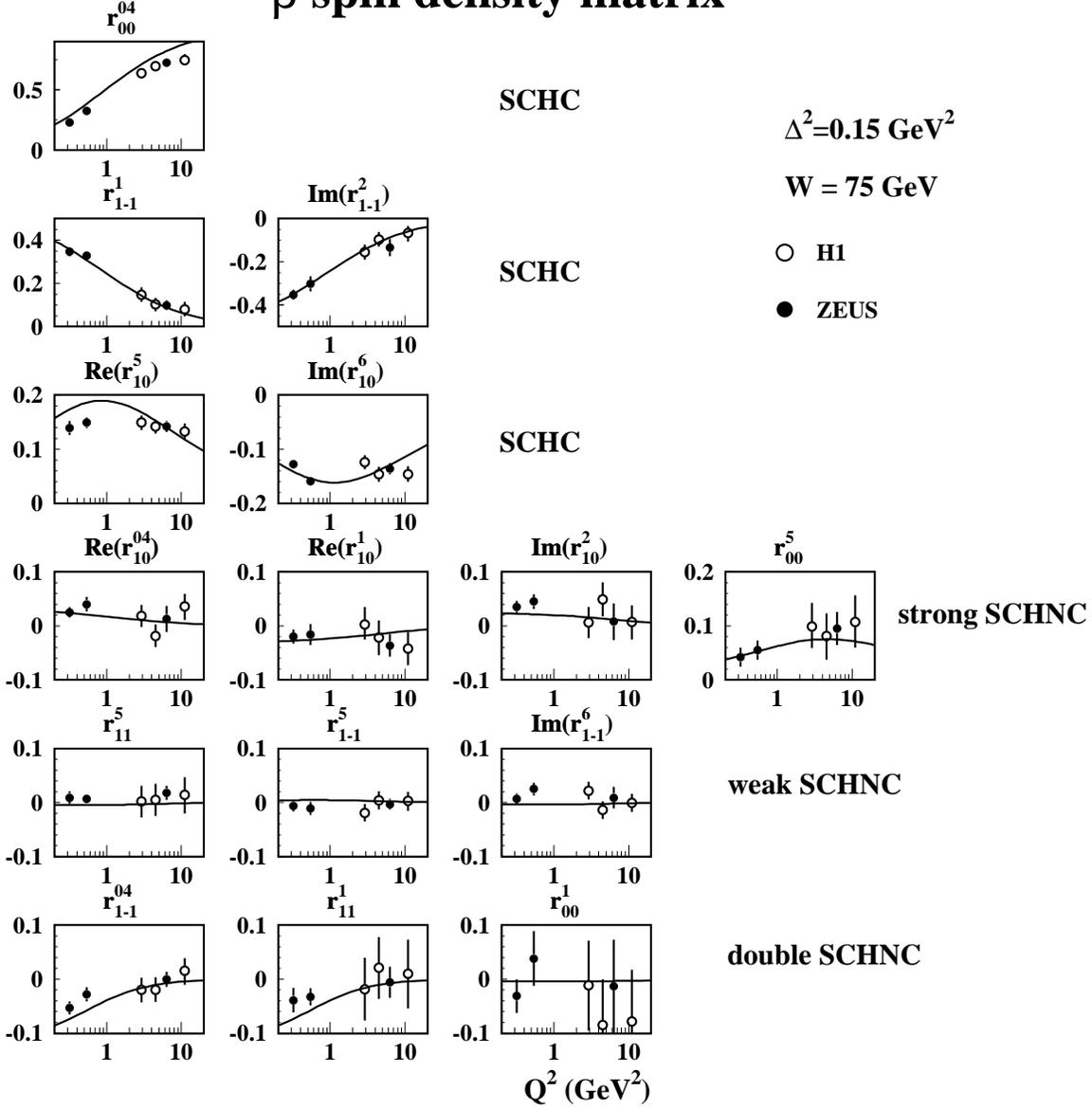}
   \caption{\it   The spin-density matrix elements measured in the reaction
$\gamma^* p \rightarrow \rho p$ as a function of $Q^2$. The
first three rows are the same as in
Fig.~\ref{fig:ivanov-rhohelicity-schc} and show the SCHC matrix
elements. 
The solid symbols present the
ZEUS~\protect\cite{epj:c12:393} and the open circules the
H1~\protect\cite{epj:c13:371} results. The curves represent the
$k_t$-factorization calculations 
\protect\cite{IgorPhD,IgorNumerics}.}
   \label{fig:ivanov-rho-density}
\end{figure}

\begin{figure}[htbp]
   \centering
   \epsfig{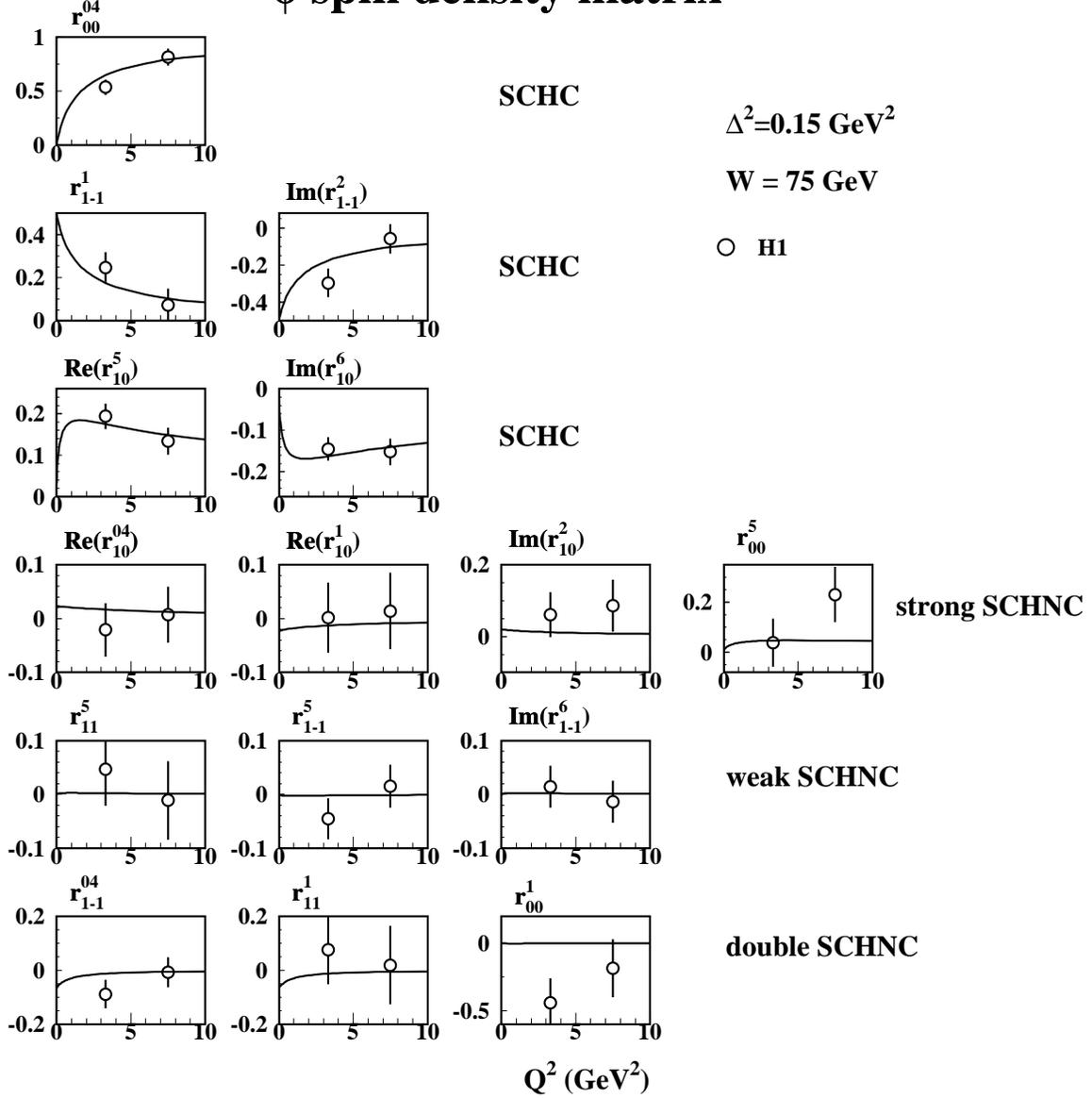}
   \caption{\it   The spin-density matrix elements measured in the reaction
$\gamma^* p \rightarrow \phi p$ as a function of $Q^2$. 
The notation is the same as in Fig.~\ref{fig:ivanov-rho-density}. 
The curves represent the $k_t$-factorization calculations
\protect\cite{IgorPhD,IgorNumerics}.}
   \label{fig:ivanov-phi-density}
\end{figure}

In experiment, the density matrix elements are obtained by
minimizing the difference between the three-dimensional
($\cos{\theta_h},\phi_h,\Phi_h$) angular distributions of the data
and those of the simulated events. 
Figure~\ref{fig:ivanov-rho-density} shows combined ZEUS 
\cite{epj:c12:393} and H1 \cite{epj:c13:371}
results of this procedure for the $\rho$ meson production. Results
for $\phi$ mesons are presented in
Fig.~\ref{fig:ivanov-phi-density}.

All the matrix elements are placed in six rows. The first three
rows, are the same as in Fig.~\ref{fig:ivanov-rhohelicity-schc} and
show the spin density matrix elements coming from SCHC transitions. 
The last three rows
represent the SCHNC matrix elements, 
which would vanish in the case of strict SCHC, for the 
definitions and classification of SCHNC as strong, weak and double
see Eqs. (\ref{eq:5.4.1.2})-(\ref{eq:5.4.1.5}) 
and discussion in Section 5.3.1 . 
The fourth row shows the four matrix amplitudes proportional
to the single-flip helicity amplitude ${\cal T}_{01}$, 
the fifth row shows the matrix elements
that are proportional to the 
single-flip helicity amplitude ${\cal T}_{10}$, 
the last row shows matrix elements 
with two helicity flips, either as a double-flip
amplitude or as a product (square) of two single-flip amplitudes.

\begin{figure}[htbp]
   \centering
   \epsfig{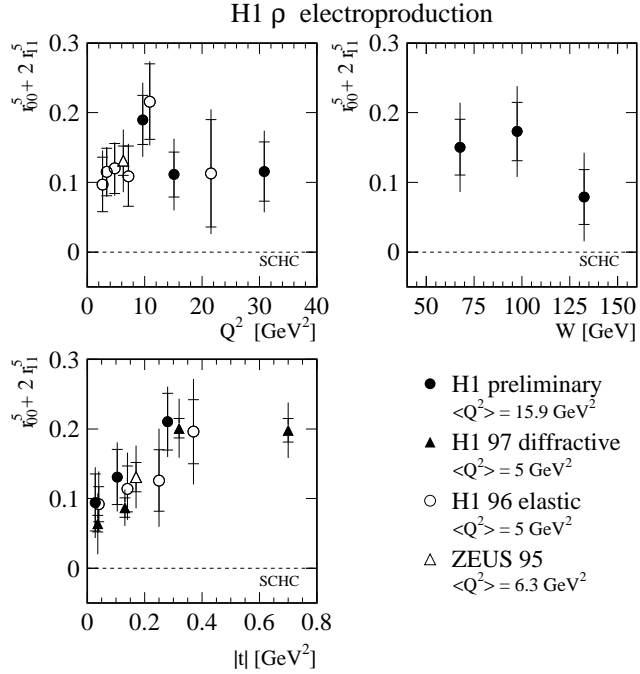}
   \epsfig{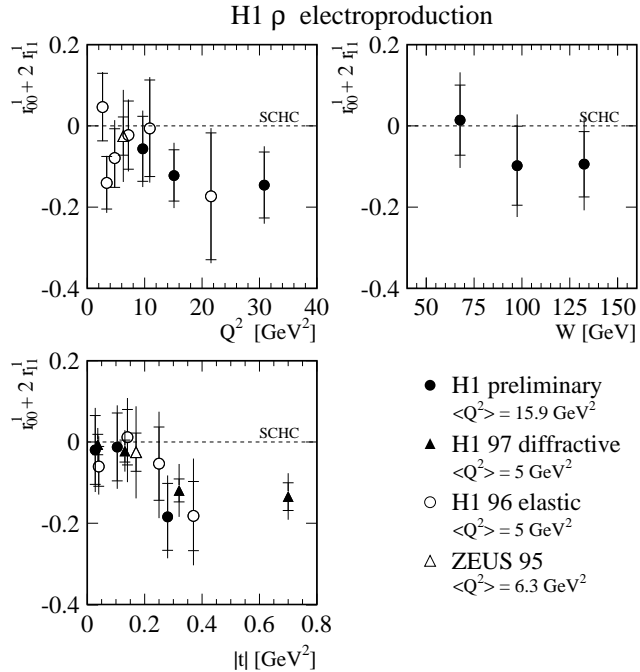}
   \caption{\it   The H1 and ZEUS measurement 
   (~\protect\cite{cpaper:ichep2002:989} and references therein)  
of the combinations
of SCHNC spin density matrix elements
 $ r^5_{00} + 2r^5_{11}$ and  $ r^1_{00} + 2r^1_{11} $ as a function of
$Q^2$, $W$, and $t$.}
   \label{fig:H1viol1}
\end{figure}


\begin{figure}[htbp]
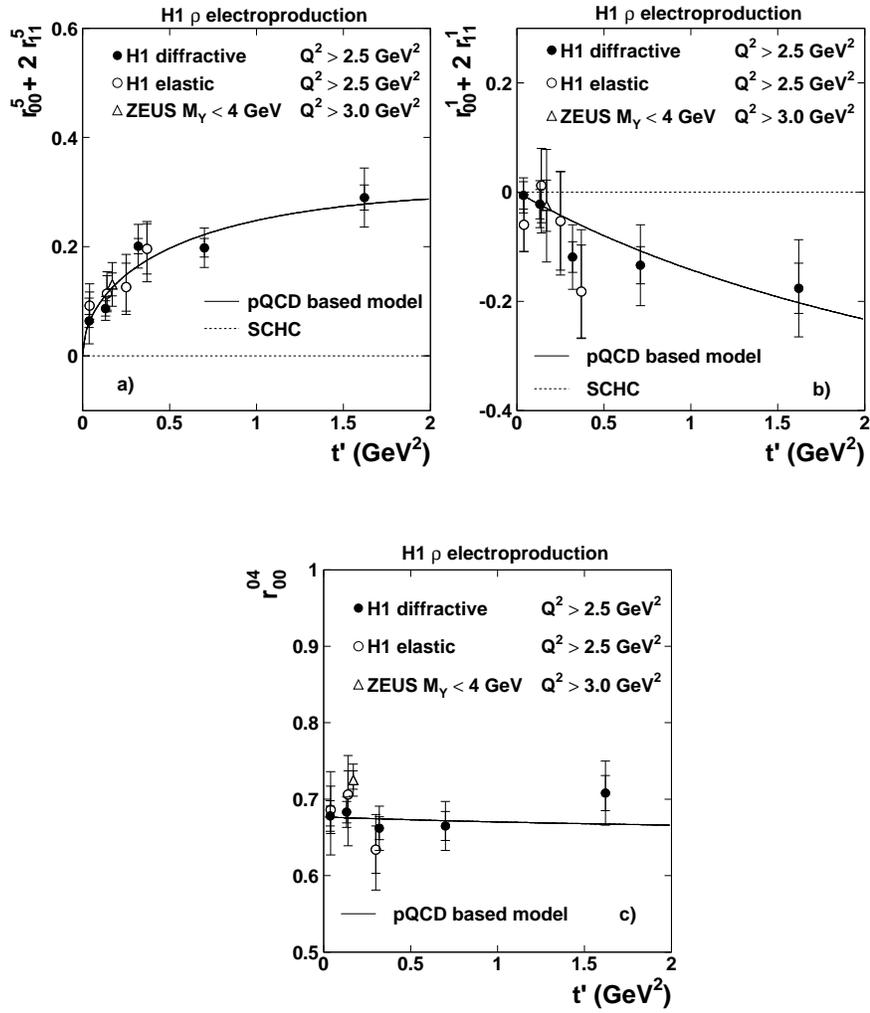

   \centering
   \epsfig{file=H1DISRhoHelicity2000pub_fig4a.eps,width=50mm}
   \hspace*{0.5cm}
   \epsfig{file=H1DISRhoHelicity2000pub_fig4b.eps,width=50mm}
   \hspace*{0.5cm}
   \epsfig{file=H1DISRhoHelicity2000pub_fig4c.eps,width=50mm}
   \caption{\it   The H1 and ZEUS measurement (~\protect\cite{pl:b539:25} and
   references therein) of the
   $t'$-dependence of combinations of
SCHNC spin density matrix elements
$ r^1_{00} + 2r^1_{11}$ and $ r^5_{00} + 2r^5_{11}$ compared with of
the behavior of the SCHC matrix element $r^{04}_{00}$.
The solid curves are from pQCD calculations by Ivanov and Kirschner
\protect\cite{IKspinflip}.}
   \label{fig:H1viol3}
\end{figure}

Among the SCHNC spin density matrix elements the best measured 
ones are the combinations $2r^1_{11} + r^1_{00}$
and $2r^5_{11} + r^5_{00}$, which describe the cross section
single-differential in angle $\Phi_h$ after the integration 
over $\cos\theta_h$: 
\be
\frac{d\sigma}{d\Phi_h} \propto 1 + \sqrt{2 \epsilon
(1+\epsilon)} \cos\Phi_h (r^5_{00}+2r^5_{11}) - \epsilon
\cos2\Phi_h(r^1_{00}+2r^1_{11})\,. 
\label{eq:5.4.2.1}
\ee

Figure~\ref{fig:H1viol1} 
shows also the resent H1 measurement of the
$Q^2$-dependence of these matrix elements for the case of $\rho$ mesons
~\cite{cpaper:ichep2002:989}.
The first combination is significantly non-zero, while the second
is compatible with zero. The results of the direct fit to the
entire angular distribution shown in
Fig.~\ref{fig:ivanov-rho-density} indicate that it is large
non-zero value of $r^5_{00}$ which is a source of departure the
SCHC in the first combination. Since this matrix element is
proportional to the Re(${\cal T}_{00}{\cal T}^*_{01}$), it 
indicates that the
single-flip amplitude ${\cal T}_{01}$  is not zero. The data 
shown in Fig.~\ref{fig:H1viol1} give an estimate of 
the relative strength of this amplitude at an
average value of $|t|\approx 0.15$ GeV$^2$

\be
{|{\cal T}_{01}| \over \sqrt{|{\cal T}_{11}|^2 + |{\cal T}_{00}|^2}} \approx
r^5_{00}\sqrt{{1+R \over 2R}} \approx 8 \pm 3 \%\,.
\label{eq:5.4.2.2} 
\ee

The dominance of $r^5_{00}$ among the helicity violating amplitudes at
large $Q^2$ is in agreement with theoretical 
expectations \cite{KNZspinflip,*KNZspinflip1,IKspinflip,IgorPhD}. All the
other matrix elements do not differ significantly from zero. There
is, however, some indication that the double-flip amplitude is
non-zero at small $Q^2$, see the last row in
Fig.~\ref{fig:ivanov-rho-density}.

The $t$-dependence of the $s$-channel helicity violating
amplitudes has already been shown in Fig.~\ref{fig:H1viol1} 
 within the diffraction cone. In
Fig.~\ref{fig:H1viol3} the data ~\cite{pl:b539:25} on
helicity violating matrix elements $2r^1_{11} + r^1_{00}$ and
$2r^5_{11} + r^5_{00}$ within extended $t$-region, $|t'| < 2$
GeV$^2$ are shown. The behavior of both of these combinations is compatible
with $\propto \sqrt{|t'|}$ and confirms the theoretical
expectations. For comparison, in the same Figure we show the
$t$-behavior of the helicity conserving matrix element
$r^{04}_{00}$ within the same region, which is compatible with
constant.



\subsubsection{Experimental results: helicity properties at large $t$}

\begin{figure}[htbp]
   \centering
   \epsfig{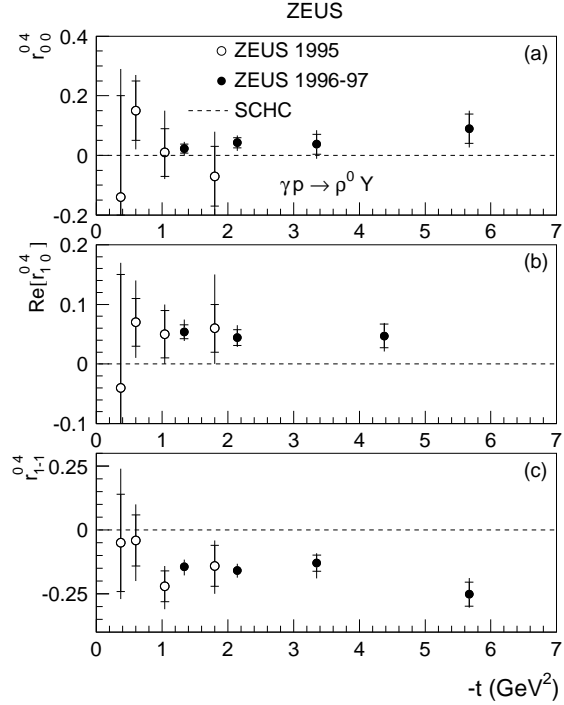}
   \epsfig{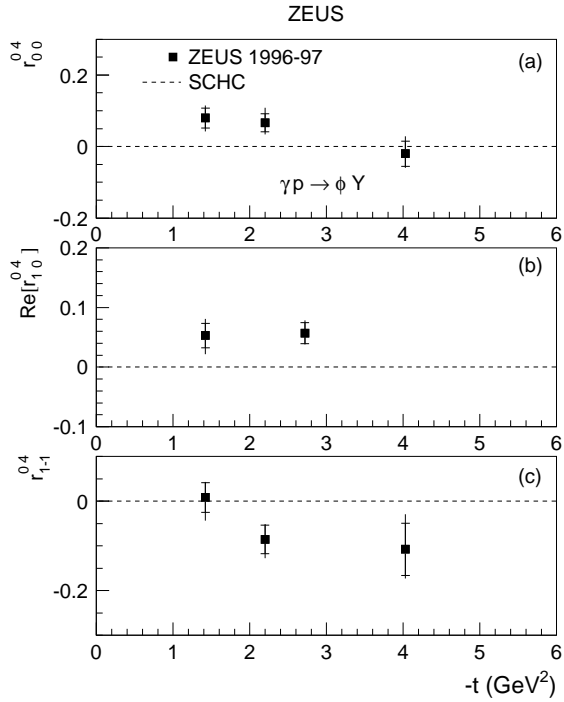}
   \caption{\it   The spin density matrix elements for proton-dissociative
   $\rho$ production (the top panel) and  $\phi$ production 
(the bottom panel) as a function of $-t$ measured by
   ZEUS ~\protect\cite{epj:c26:389}
   (black sumbols) and 
   ~\protect\cite{epj:c14:213} (open symbols). The SCHC prediction is shown as
   the dashed line.}
   \label{fig:highTrhoHelicity}
\end{figure}

\begin{figure}[htbp]
   \centering
   \epsfig{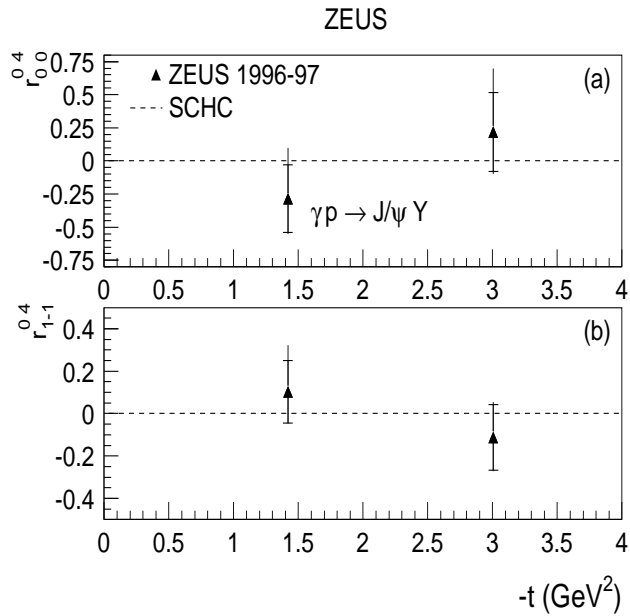}
   \caption{\it   The spin density matrix elements for proton-dissociative
   $J/\psi$ production as a function of $-t$ measured by ZEUS
   ~\protect\cite{epj:c26:389}.
   The SCHC prediction is shown as
   the dashed line.}
   \label{fig:highTjpsiHelicity}
\end{figure}

At large-$t$ only the proton dissociative reaction
$\gamma p \to V Y$ has been studied. 
The experimental data on the spin-density matrix elements
from the recent ZEUS measurement ~\cite{epj:c26:389} of
the large-$t$  $\rho$, $\phi$ and $J/\psi$ photoproduction
are shown in
Figs.~\ref{fig:highTrhoHelicity},
~\ref{fig:highTjpsiHelicity},~\ref{fig:ForshawHelicityRhoLarget} in combination with the
lower-$t$ data ~\cite{epj:c14:213}; the H1 results 
~\cite{pl:b568:205} are shown in 
Fig. \ref{fig:ForshawHelicityJpsiLarget}. 

The dominant feature of the DIS data within the diffraction
cone, $|t| \lsim 1$ GeV$^2$ was large $r^{04}_{00}$ 
driven by the dominant $\sigma_L$ from the transition 
$\gamma_L \to V_L$. Such a transition is 
absent in real photoproduction.
Figures~\ref{fig:highTrhoHelicity},
~\ref{fig:highTjpsiHelicity} show that the
matrix element $r^{04}_{10}$
is positive valued, definitely nonvanishing one.
The results for $r^{04}_{00}$ show that 
the probability to
produce longitudinal $\rho$ or $\phi$ mesons from a
transverse photon increases with $|t|$ up to $4 \div 9 \%$,
but is nowhere close to the SCHNC dominance
of $\sigma_L$ expected in the pQCD two-gluon model of
Ginzburg et al. \cite{GinzburgPanfilSerbo} with the chiral-even
photon-quark-antiquark vertex. The matrix element $r^{04}_{1-1}$ 
is numerically substantial and gives a solid 
evidence for a double-flip contribution. 
All the spin density
matrix elements shown in Figs. 
\ref{fig:highTrhoHelicity},\ref{fig:highTjpsiHelicity}
are small, so that the SCHC is the empirical feature of large-$t$
production.

Within the theoretical approaches reviewed in Section 4.10
the sole source of approximate SCHC is the chiral-odd 
photon-quark-antiquark vertex. The theoretical calculations
\cite{PoludniowskiHight}
are shown in Figs.~\ref{fig:ForshawHelicityJpsiLarget} and
~\ref{fig:ForshawHelicityRhoLarget}. 
The pQCD two-gluon approximation grossly overpredicts 
$r^{04}_{00}$ for all vector mesons.
 The BFKL model
correctly reproduces the gross features of $r^{04}_{00}$ and 
$r^{04}_{1-1}$, but predicts the wrong sign of $r^{04}_{10}$. 
The shown theoretical results are for the 
so-called asymptotic vector meson distribution
amplitude, but the sensitivity to the model wave function is
weak and can not explain this sign conflict with the experiment.

For the heavy quarks in the $J/\Psi$ the Fermi motion is slow
and the BFKL calculations predict much smaller helicity 
flip effects compared to the results for the light vector mesons,
cf. Fig.\ref{fig:ForshawHelicityJpsiLarget}
and \ref{fig:ForshawHelicityRhoLarget}.

\begin{figure}[htbp]
   \centering
   \epsfig{file=ForshawHelicityJpsiLarget.ps,width=70mm,height=80mm,angle=270}
   \caption{\it   The spin density matrix elements for proton-dissociative
   $J/\psi$ production as a function of $-t$ measured by H1
   ~\protect\cite{pl:b568:205}.
The predictions form
the BFKL (solid curves) and pQCD two-gluon approximations
(dashed curves) are shown for a comparison
\protect\cite{PoludniowskiHight}.\bigskip}
   \label{fig:ForshawHelicityJpsiLarget}
\end{figure}

\begin{figure}[htbp]
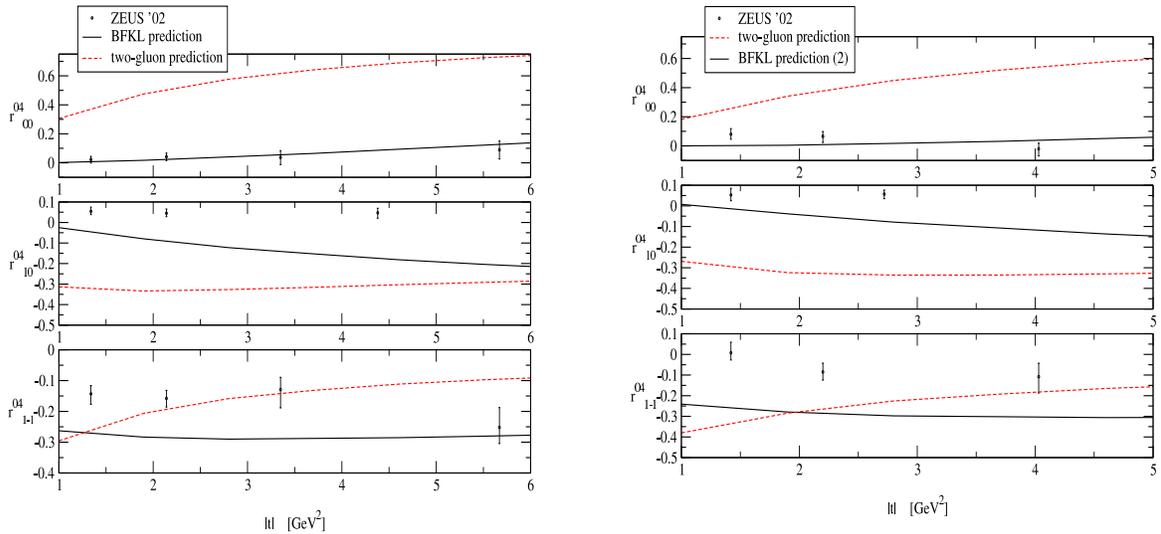

   \centering
   \epsfig{file=ForshawHelicityRhoLarget.ps,width=70mm,height=70mm,angle=270}
\hspace*{1cm}
   \epsfig{file=ForshawHelicityPhiLarget.ps,width=68mm,height=70mm,angle=270}
   \caption{\it   The spin density matrix elements for proton-dissociative
   $\rho$ (left box) and $\phi$ (right box) production as a function of $-t$ measured by 
ZEUS ~\protect\cite{epj:c26:389}. The predictions for from
the BFKL (solid curves) and pQCD two-gluon approximations
(dashed curves) are shown for a comparison
\protect\cite{PoludniowskiHight}.\bigskip}
   \label{fig:ForshawHelicityRhoLarget}
\end{figure}


\newpage


\section{The $Q^2$ dependence }


\subsection{Total cross section $\gamma^* p \rightarrow V p$}


\subsubsection{Theoretical expectations: what is the correct
hard scale for the $Q^2$ dependence?}

As emphasized in Introduction, see Section~\ref{sect1.3}, the vector meson
production at a given $Q^2$ probes the gluon content of a target
at pQCD factorization scale $\Qb^2 \sim (Q^2+m_V^2)/4$, 
see (\ref{eq:1.3.5}).
Theoretical predictions of the $Q^2$-dependence
of the vector meson production cross section in the color dipole
approach were described in detail in Sections~\ref{sect3.4.2}
and 4.9. 
Besides, Section~\ref{sect4.8} shows how the leading log-$Q^2$ approximation 
results are recovered, and improved upon, in the more
consistent $k_\perp$-factorization approach at large $Q^2$.

The theoretical expectation for the 
shape of the $Q^2$-dependence of the transverse amplitude
can be read off Eqs.(\ref{eq:3.4.2.5}),(\ref{eq:3.4.2.6}) and
(\ref{eq:4.9.2}). Performing the $t$-integration in (\ref{eq:4.9.2})
and combining $\sigma_T$ and $\sigma_L$, one finds the non-relativistic master
formula
\bea
\sigma_V =  \sigma_T +\sigma_L 
\approx
{\pi^3  m_V \,\Gamma(V\to e^+e^-)\over  12 \alpha_{em} b_V(\Qb^2)\Qb^8}
\cdot \left(Q^2+ {m_V^2 \over R_{LT}}\right)
\cdot
\left[\alpha_s(\Qb^2)\cdot G(x_g,\Qb^2_{GL})\right]^2\,.
\label{eq:6.1.1.1}
\eea
Although the absolute normalization, the exact value of the pQCD scale
$\Qb^2$ and the departure of $R_{LT}$ from the non-relativistic quark
model expectation $R_{LT}=1$ do depend on the wave function of the vector
meson, and the contribution from the helicity-flip transitions
has not been included, Eq. (\ref{eq:6.1.1.1}) contains all 
the ingredients of the full
pQCD description. 

One can present $\sigma_V$ either as
$\sigma_V= \sigma_T(1+R_V)$ and use the theoretical approximations 
for $\sigma_T$ or substitute $\sigma_T = \sigma_L/R_V$  
and test the theoretical predictions for $\sigma_L$. Brodsky et al.
~\cite{BFGMSvm} 
argued that the end-point, $z\sim 0, z\sim 1$ contributions are 
minimal in, and the pQCD evaluations are more reliable for, the 
$\sigma_L$ and our representation (\ref{eq:6.1.1.1}) corresponds
to the latter, preferred, choice.   

The longitudinal cross section is small at $Q^2 \lsim m_V^2$,
but becomes the dominant feature at larger $Q^2$ and  tames further
the decrease of the total cross section at large $Q^2$.
In view of $R_{LT}\sim 1$ at small $Q^2$, and because the 
$M_V^2$ term can be neglected at large $Q^2$, the factor
$ \left(Q^2+ {m_V^2 \over R_{LT}}\right)$ in the master formula
is not very different from $(Q^2+m_V^2)$. Then the 
 $Q^2$-behavior of the total cross section
would have been $\sim (\Qb^2)^{-n_V}$ with the exponent $n_V=3$ 
modulated by 
the gluon density squared. As we shall see 
below, $R_{LT}<1$ and steadily decreases with $Q^2$, so that 
even without the scaling violations the expected exponent $n_V <3$.

Finally, the $Q^2$ dependence of the diffraction slope can not be ignored.
As we argued in Section 3.4.2, the diffraction slope $b_V(\Qb^2)$
decreases with growing $(Q^2+m_V^2)$, which also enhances
slightly $\sigma_V$ at large $Q^2$. As a matter of fact, the
diffraction slope too is a function of $\Qb^2$ rather than $Q^2$.

To summarize, there are strong theoretical reasons 
\cite{NNPZdipoleVM,NNPZcdVMsyst} for the 
presentation of the experimental data as a function 
of either the scanning radius $r_S$ or $(Q^2+m_V^2)$.
In such a presentation the major flavor dependence is in the 
explicit factor $ m_V \,\Gamma(V\to e^+e^-)$. There is also a hidden
dependence of the absolute normalization, $R_{LT}$ and of the exact
dependence of the hard scale $\Qb^2$ on $(Q^2+m_V^2)$ all of which
depend on the wave 
function of vector mesons \cite{NNZscanVM,Igorhardscale}.


\subsubsection{Theoretical expectations: the impact of 
$x_g$-dependent scaling violations on the $Q^2$ dependence}

The predicted $Q^2$-dependence is driven mainly by two phenomena: 
the shrinkage of the photon
light-cone wave function with $Q^2$ and the resulting decrease of the scanning
radius, and the non-trivial
$Q^2$-dependence of the gluon density at fixed $W$.
The former property leads to the strong decrease of the transverse
cross section $\sim (\Qb^2)^{-4}$ at a sufficiently large $\Qb^2$.
The gluon density, $G(x_g,\Qb_G^2)$, at a fixed $W$, depends on $\Qb^2$ 
also via $x_g \approx \Qb^2/W^2$. At small to moderately large 
$Q^2$, one observes the quick rise 
of the small-$x_g$ gluon density because of the scaling violations.
At larger $Q^2$ the values of $x_g$ are larger, the scaling violations
are weaker, and the decrease of the gluon density towards large $x_g$
takes over, see Fig.~\ref{fig:ivanov2004-g75}. 
Because of such a convex $Q^2$ dependence of $G(x_g,\Qb_G^2)$ 
in the $\log\sigma_V - \log \Qb^2$ plot one 
must see the convex curve.

\begin{figure}[htbp]

   \centering
   \epsfig{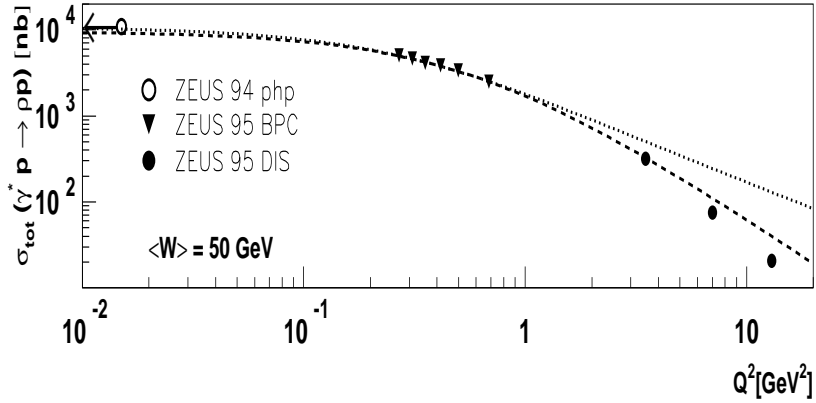}\\
   \epsfig{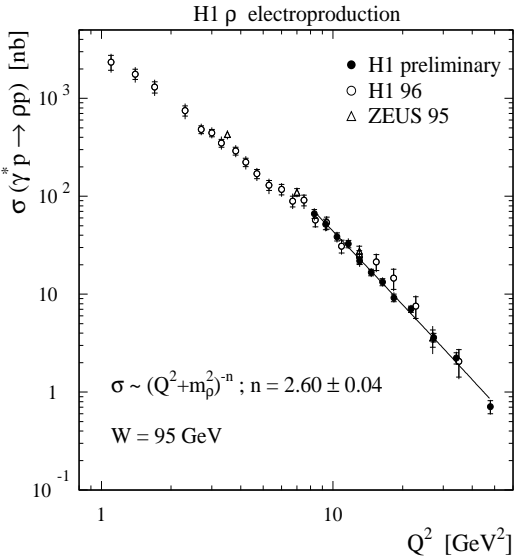}
   \epsfig{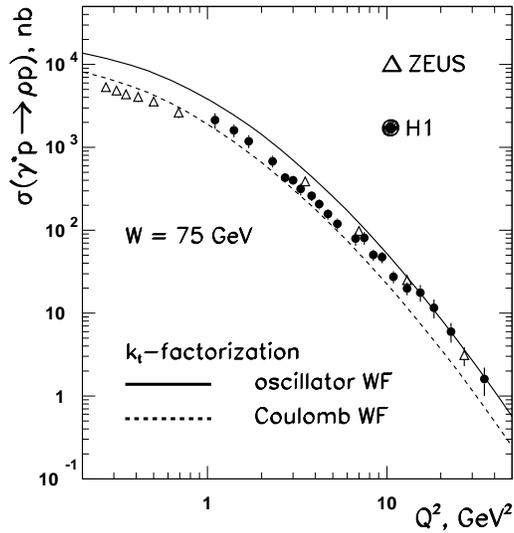}
  \caption{\it   The total cross section of diffractive $\rho$ meson
production as a function of $Q^2$. Upper plot shows ZEUS data from
~\protect\cite{epj:c6:603} fitted together with photoproduction
point ~\protect\cite{epj:c2:247}. The curves represent fits using
the function $\sigma(Q^2) \propto (1 + R(Q^2))/(Q^2 +
m^2_{eff})^2$(dotted line) and function $\sigma(Q^2) \propto {1 /
(Q^2 + m_\rho^2)^n}$(dashed line). The left bottom plot shows
recent H1 data ~\protect\cite{cpaper:ichep2002:989}
 compared with
published data ~\protect\cite{epj:c13:371, epj:c6:603}. The right
bottom plot presents the $k_t$-factorization predictions
\protect\cite{IgorPhD,IgorNumerics} based on
the oscillator (solid line) and Coulomb (dashed line)
wave functions compared with the published H1 and ZEUS points.}
   \label{fig:XsecQ2rho}
\end{figure}


\subsubsection{The $Q^2$-dependence: the experimental data}

The change of the character of the $Q^2$-dependence from small 
to large $Q^2$ is best seen in the data for light vector mesons.
The experimental data on the $Q^2$-dependence of 
the $\rho^0$ production are shown in 
Fig. \ref{fig:XsecQ2rho}. 
The upper plot in Fig. \ref{fig:XsecQ2rho} shows the ZEUS data.
The low-$Q^2$ data, $Q^2< 1~{\rm GeV}^2$, are dominated by
$\sigma_T(Q^2)$ and were fitted to the form 
\be
\sigma_{\rho}(Q^2) \propto {1 + R_{\rho}(Q^2) \over (Q^2 +
m^2_{eff})^{n_{\rho}}}
\label{eq:6.1.3.1}
\ee
where $R_{\rho}(Q^2)=\sigma_L/\sigma_T$ and the exponent 
$n_{\rho}\equiv 2$ as dictated by VDM. The dotted line shows 
the result of the fit including the real photoproduction point.
The fit yields $m_{eff} = 0.66 \pm 0.11$ GeV which is close 
to the mass of the $\rho$-meson as expected in VDM.

\begin{figure}[htbp]
   \centering
   \epsfig{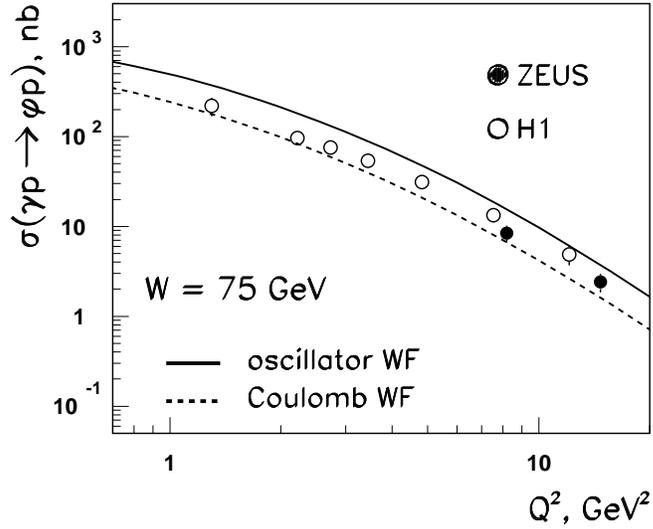}
  \caption{\it    The published H1 ~\protect\cite{pl:b483:360} 
and ZEUS
~\protect\cite{pl:b377:259,pl:b380:220} data  on the
total cross section of diffractive $\phi$ meson
production as a function of $Q^2$. The $k_t$-factorization
predictions \protect\cite{IgorPhD,IgorNumerics}
for the oscillator (solid line) and 
Coulomb (dashed line) wave functions are shown for
comparison.  }
   \label{fig:XsecQ2phi}
\end{figure}

\begin{figure}[htbp]
   \centering
   \epsfig{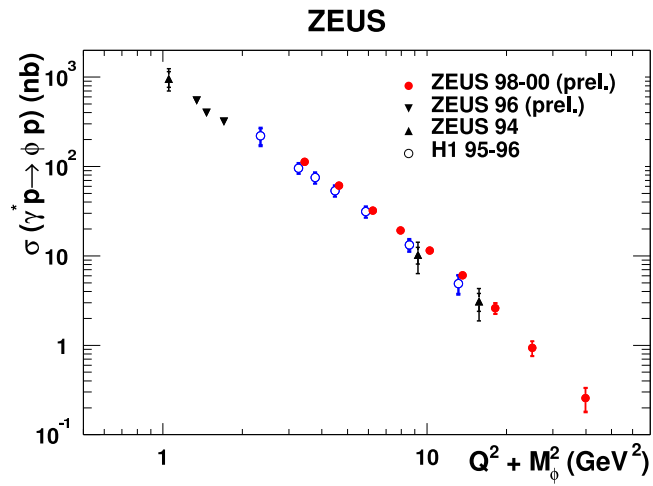}
  \caption{\it    The preliminary  ZEUS data 
~\protect\cite{ZEUS2004_phi_preliminary} on the $\phi$ meson
production cross section are shown together with the
published data shown in Fig.~\ref{fig:XsecQ2phi}.}
   \label{fig:XsecQ2phi2}
\end{figure}

\begin{figure}[htbp]
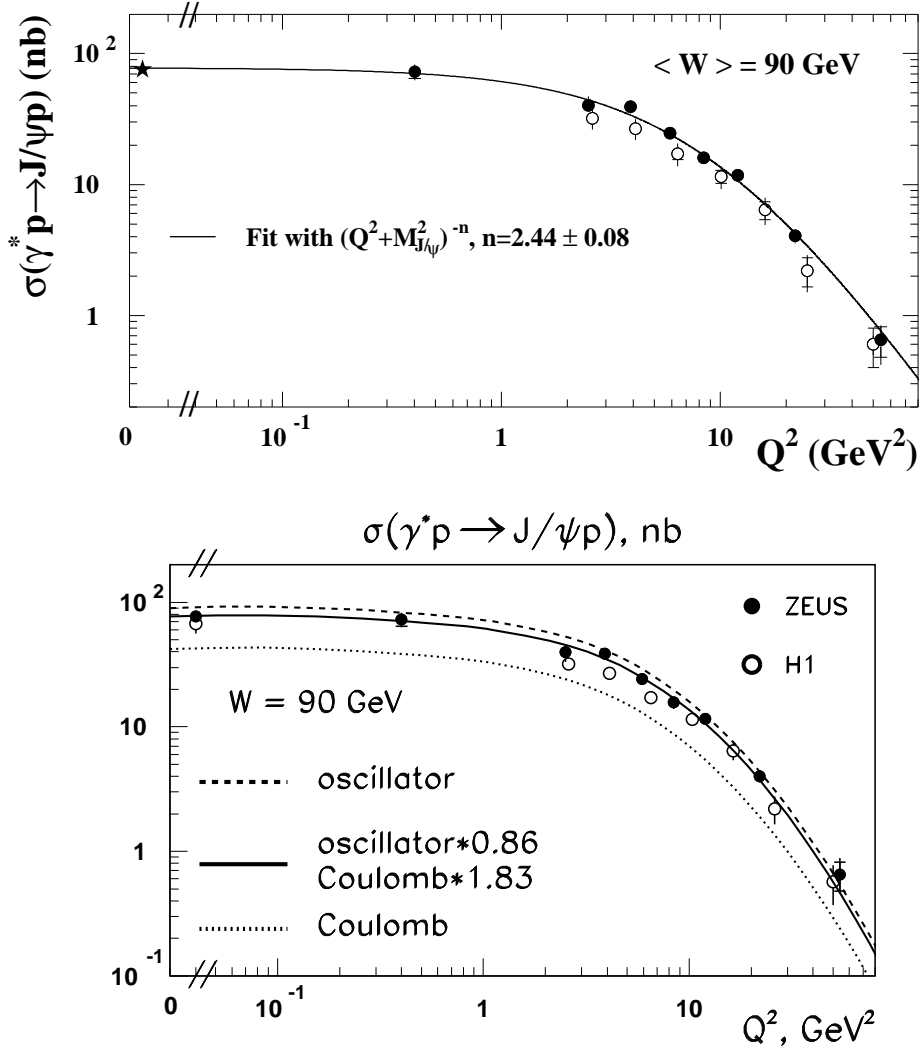


   \centering
   \epsfig{file=JpsiZEUS_H1_Q2dep_Summary.eps,width=125mm}

   \hspace*{1.2cm}
   \epsfig{file=ivanov2004-total-jpsi-q2dep.eps,width=125mm}
  \caption{\it   
Upper plot shows recent ZEUS results (solid symbols) 
on exclusive $J/\psi$ electroproduction cross section 
as a function of $Q^2$ at $\langle W\rangle=90$ GeV
\protect\cite{np:b695:3}.
ZEUS photoproduction
\protect\cite{epj:c24:345} and H1 electroproduction (open symbols)
\protect\cite{epj:c10:373} cross sections 
are also shown.
The full line is a fit to the ZEUS data  
of the form (\protect\ref{eq:6.1.3.2}).
The inner error bars represent the statistical uncertainties,
the outer bars are the statistical and systematic uncertainties added in
quadrature.
The bottom plot shows the $k_t$-factorization
predictions \protect\cite{IgorPhD,IgorNumerics}
based on oscillator (solid lines) and
Coulomb (dashed lines) wave functions compared with the published
data. The solid curve is the $k_{\perp}$-factorization result
upon the normalization to the photoproduction point with
the rescaling factors shown in the figure, the
curves for two wave functions merge within the thickness of lines}
   \label{fig:XsecQ2jpsi}
\end{figure}

The VDM value of the exponent $n_{\rho}=2$ for the low-$Q^2$ data
must be compared to the large-$Q^2$ pQCD expectation, $n=4$, 
without scaling violations, 
see Eq. ~(\ref{eq:3.4.2.8}). Indeed, at higher $Q^2$, the cross
section of $\rho$ production steadily departs from the dotted line.
The parameterization of the form
\be
\sigma_{V}(Q^2) \propto {1 \over (Q^2 +
m_V^2)^{n_{V}}}\,,\label{eq:6.1.3.2}
\ee
yielded $n_{\rho} = 2.32\pm
0.10$ for ZEUS 95 data ~\cite{epj:c6:603}
 at $Q^2>5$ GeV$^2$,
$n_\rho = 2.24\pm 0.09$ for H1 96 data ~\protect\cite{epj:c13:371}
and $n_\rho = 2.60\pm 0.04$ for preliminary H1 data at $Q^2 > 8$
GeV$^2$ ~\cite{cpaper:ichep2002:989}.
The effective exponent $n_\rho$ of the  $Q^2$-decrease 
rises steadily with the region being fitted, from 2 in the soft region to
2.6 at the highest $Q^2$, in good agreement with the theoretical 
expectations of the convex $\sigma_V(Q^2)$. 

Now we comment on the comparison with theoretical predictions from
the  $k_{\perp}$-factorization \cite{IgorPhD,IgorNumerics}.
As we cautioned in Section 1.4, 
the onset of the
truly pQCD hard regime for the light $\rho$-mesons requires 
large $Q^2$, see Eq.~(\ref{eq:1.4.4}). At smaller $x$ the applicability
of hard pQCD description somewhat improves, still the soft-hard 
decomposition of the gluon density
shown in Fig.~\ref{fig:ForshawHelicityJpsiLarget} indicates that the nonperturbative soft component of the
   gluon density remains substantial at gluon momentum squared
   $\kappa^2 \lsim (1\div 2)$ GeV$^2$. According to the discussion
in Section 4.8, see especially Fig. ~\ref{fig:IgorQ2scaleRho}, 
in order to have
the hard scale  $\overline{Q}^2 \gsim
    (1\div 2)$ GeV$^2$ one needs 
   $Q^2 \gsim 10$ GeV$^2$. The $\rho$-production at smaller
$Q^2$ is strongly affected by nonperturbative physics. Still
we notice that the $k_{\perp}$-factorization
predictions \cite{IgorPhD,IgorNumerics} reproduce within
the overall normalization factor $\sim 2$
the measured cross section which drops by
nearly four orders in the magnitude from  the real photoproduction
to the largest value of $Q^2$. One must not be jubilant, though,
since the unintegrated gluon density in the soft region has
been adjusted to reproduce real photoabsorption and DIS at 
small-$Q^2$. 
At $Q^2 \gsim 10$ GeV$^2$ 
there is a very good agreement between  experimentally observed 
and predicted $Q^2$-dependence. In the nonperturbative region 
the Coulomb WF is doing better job at small $Q^2$, but at higher
$Q^2$ the experimental data deviate from the curve for the
Coluomb WF and agree better with the results 
for the oscillator 
WF.

\begin{figure}[htbp]
   \centering
   \epsfig{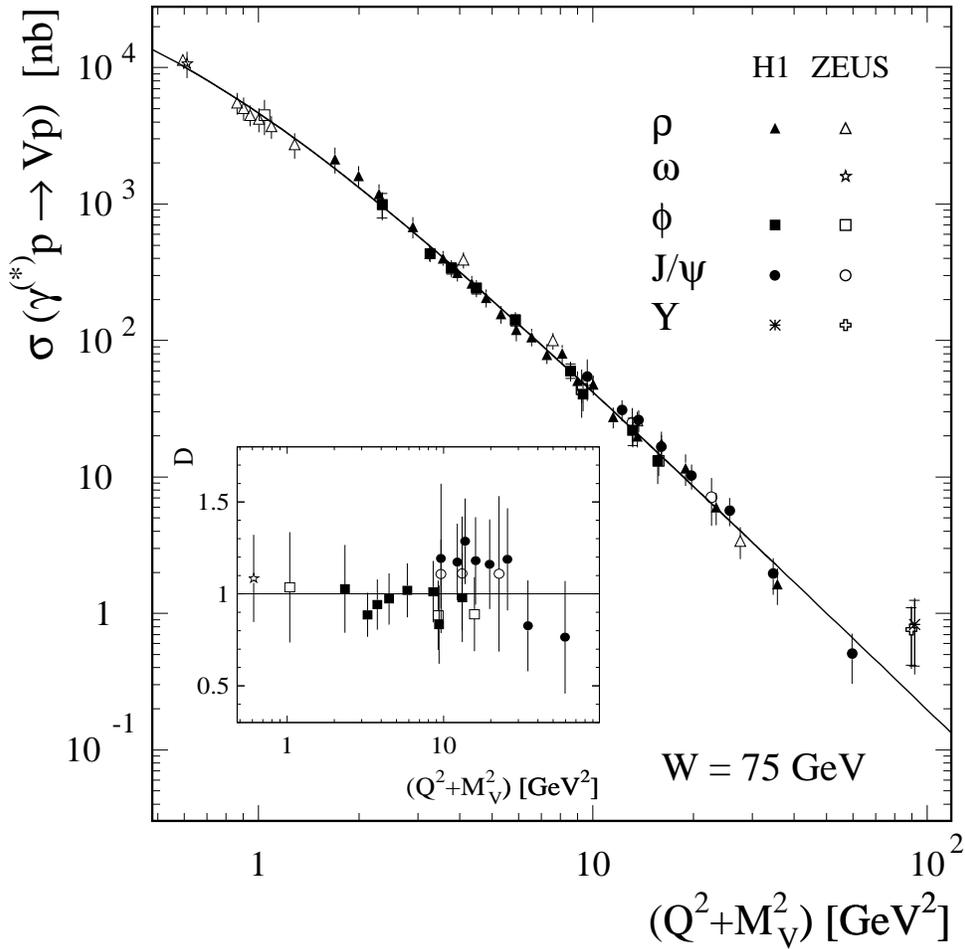}
    \caption{\it   H1 and ZEUS 
     measurements 
     ~\protect\cite{epj:c13:371, epj:c6:603, pl:b338:507, epj:c10:373,
     zfp:c75:215, zfp:c75:607, pl:b377:259, pl:b380:220,epj:c2:247, zfp:c73:73, 
     pl:b483:23, pl:b437:432}
     of the total cross sections
     $\sigma (\gamma^{\star}p \rightarrow V p)$ as a function of $(Q^2 + M^2_V)$
     for elastic $ \rho, \omega, \phi, J/\psi$ and $\Upsilon$ meson production, at
     the fixed value $W$ = 75 GeV. The cross sections 
     were scaled by SU(5) charge-isospin factors.
     The curve is a fit by formula (\protect\ref{eq:6.1.3.3}) to
     the H1 and ZEUS $\rho$ data, and 
     the ratio $D$ of the scaled $\omega$, $\phi$ 
and $J/\psi$ cross sections to this
     parameterization is presented in the insert.}
\label{fig:H1sigmaAllVM}
\end{figure}

The similar pattern is seen in the $\phi$-production shown in
Figs. ~\ref{fig:XsecQ2phi},~\ref{fig:XsecQ2phi2}.
The preliminary high-$\Qb^2$ data shown in Fig.~\ref{fig:XsecQ2phi2}
do agree better with the $k_{\perp}$-factorization results for
the oscillator WF.

Finally, the experimental results for
$J/\psi$, are shown 
in Fig. ~\ref{fig:XsecQ2jpsi}. 
These data correspond to  a  sufficiently large 
hard scale $\Qb^2$. Correspondingly, once the theoretical curves 
are normalized to the photoproduction data as shown in the
bottom plot of Fig.~\ref{fig:XsecQ2jpsi} - here we chose
a normalization to the ZEUS point, - the curves for the
oscillator and Coulomb wave functions become indistinguishable
and the resulting description of the experimentally observed $Q^2$-dependence 
is very good.

A parameterization of the $Q^2$
dependence of the recent ZEUS data 
\cite{np:b695:3} on the $J/\Psi$ cross section in the same form as
(\ref{eq:6.1.3.2}) with an appropriate change of the meson mass yielded
$n_{J/\psi} = 2.44 \pm 0.08$ 
~\cite{cpaper:ichep2002:818}. The impact of the mass term 
in $\Qb^2$ on the exponent $n_V$ is substantial which is
well illustrated by the H1 analysis  ~\cite{pl:b483:360} of 
the combined H1 and ZEUS data on different vector mesons.
A fit performed on the H1 and ZEUS $\rho$ data using the paramtrization 
\be
\sigma_{\rho} (Q^2)\propto {a_1 \over (Q^2+M^2_V+a_2)^{a_3}}
\label{eq:6.1.3.3}
\ee
with the result $a_1 = 10689 \pm 165$~nb, $a_2 = 0.42 \pm  0.09$ GeV$^2$ and $a_3=2.37 \pm 0.10$
is shown in a curve in Fig.~\ref{fig:H1sigmaAllVM}. 
Notice, that the so found exponent $a_3$ is close to 
the recent ZEUS result $n_{J/\psi} = 2.44 \pm 0.08$.
The $Q^2+M^2_V$ dependence of the H1 data on $\phi$ and 
of the combined H1 and ZEUS
data on $J/\psi$ were found to follow the same parameterization
as shown in Fig.~\ref{fig:H1sigmaAllVM}.
 

\begin{figure}[htbp]
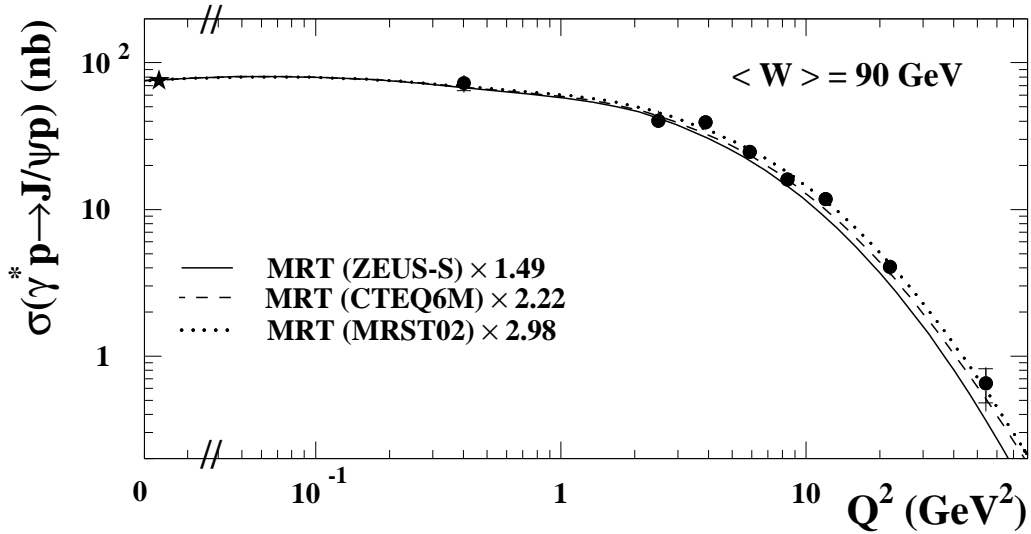

   \centering
   \epsfig{file=Jpsi_MRT_FKS_GLLMN_Q2dep.eps,width=140mm}
\epsfig{file=Jpsi_MRT_Glue_Q2dep.eps,width=140mm}
       \caption{\it   The recent ZEUS data on $Q^2$-dependence of 
$J/\psi$ electroproduction at $\langle W\rangle=90$ GeV
~\protect\cite{np:b695:3}.
The curves in the top box represent the predictions of the QCD models MRT
FKS and GLLMN (see text), the curves in the bottom plot
are the MRT results based on different gluon densities 
in the proton. All theoretical curves are rescaled as indicated 
to fit the ZEUS photoproduction point at $\langle W\rangle=90$ GeV.
The inner error bars represent the statistical uncertainties,
the outer bars are the statistical and systematic uncertainties added in
quadrature. 
An overall normalization uncertainty of $^{+5\%}_{-8\%}$ was not included.
}
\label{fig:ZEUSrecentJpsiQ2}
\end{figure}

\subsubsection{The vector meson production as a probe
of the gluon density in the proton }

The deceptively simple Eq.~(\ref{eq:6.1.1.1})
suggests that vector meson production cross section
discriminating among the different models for the gluon 
density $G(x_g,\Qb_G^2)$. Ryskin was the first to make
this point \cite{RyskinJPsi}, for early
discussion see \cite{LevinKfactVM}. 
As we saw above, in all the cases 
the color dipole/$k_t$-factorization  model 
with the unintegrated glue
adjusted to the proton structure function data is doing
a good job on the $Q^2$-dependence of $\sigma_V(Q^2)$.  
The $k_{\perp}$-factorization results for $\sigma_V(Q^2)$ for the two 
extreme parameterizations of the wave function are shown in 
Figs.~\ref{fig:XsecQ2rho},
~\ref{fig:XsecQ2phi},~\ref{fig:XsecQ2jpsi}. The choice of the wave function 
has a marginal impact on the
predicted $Q^2$-dependence. The two curves do typically envelop the
experimental data points, but a mismatch of the factor $\sim 2$ in the 
overall normalization between the theory and experiment can not be
eliminated at the moment.

Here we illustrate the model dependence with the following
example. Recent ZEUS data on $J/\Psi$ electroproduction
were compared in \cite{np:b695:3} with predictions of the three
pQCD  models by (i)  the extended Bloom-Gilman
duality \cite{GNZlong} based estimates by
Martin et al. (MRT, \cite{martinduality2}) 
for different NLO DGLAP parameterizations of the 
gluon density (making use of the NLO gluon
densities with leading order impact factors is 
somewhat inconsistent, though):
MRST02 \cite{MRST02}, CTEQ6M \cite{CTEQ6M}
and ZEUS-S \cite{ZEUS-S}, (ii) leading $\log Q^2$ estimates
with nonrelativistic $J/\Psi$ wave function by Frankfurt et al. 
(FKS,\cite{Frankfurt1996,Frankfurt1998}) for CTEQ4L 
gluon density \cite{CTEQ} and (iii) the color dipole model 
with unitarity corrections
by Gotsman et al. (GLLMN, \cite{GotsmanJpsi}). 
For the heavy $J/\Psi$ the pQCD scale $\Qb^2$ is
sufficiently large already for real photoproduction.
The results are shown in Fig.~\ref{fig:ZEUSrecentJpsiQ2}.
After the absolute normalization is adjusted 
to the photoproduction
point, the resulting $Q^2$ dependence 
is not much different from the $k_{\perp}$-factorization
results shown in Fig.~\ref{fig:XsecQ2jpsi} and is 
consistent with the experimental data. The renormalization
factors vary from 0.9 for the GLLMN model to 2.98 for
the MRT model with the MRST02 gluon density.
One would conclude that the predictions
for $Q^2$-dependence from pQCD master formula
(\ref{eq:6.1.1.1}) are to a large extent model independent
 ones. Fortuitous models with good reproduction of the 
absolute value
of $\sigma_V(Q^2)$ would not be a wonder,
but one should not rejoice with that and in general 
must be content if 
the theoretical and experimental values of $\sigma_V(Q^2)$
agree within $\sim 50\%$. 
Hereafter we only shall show the $k_{\perp}$-factorization
results \cite{IgorPhD,IgorNumerics} 
for the oscillator parameterization of the wave functions.


\subsection{The flavour dependence: ratios $\sigma_{V}/\sigma_\rho$}

\begin{figure}[htbp]
   \centering
   \epsfig{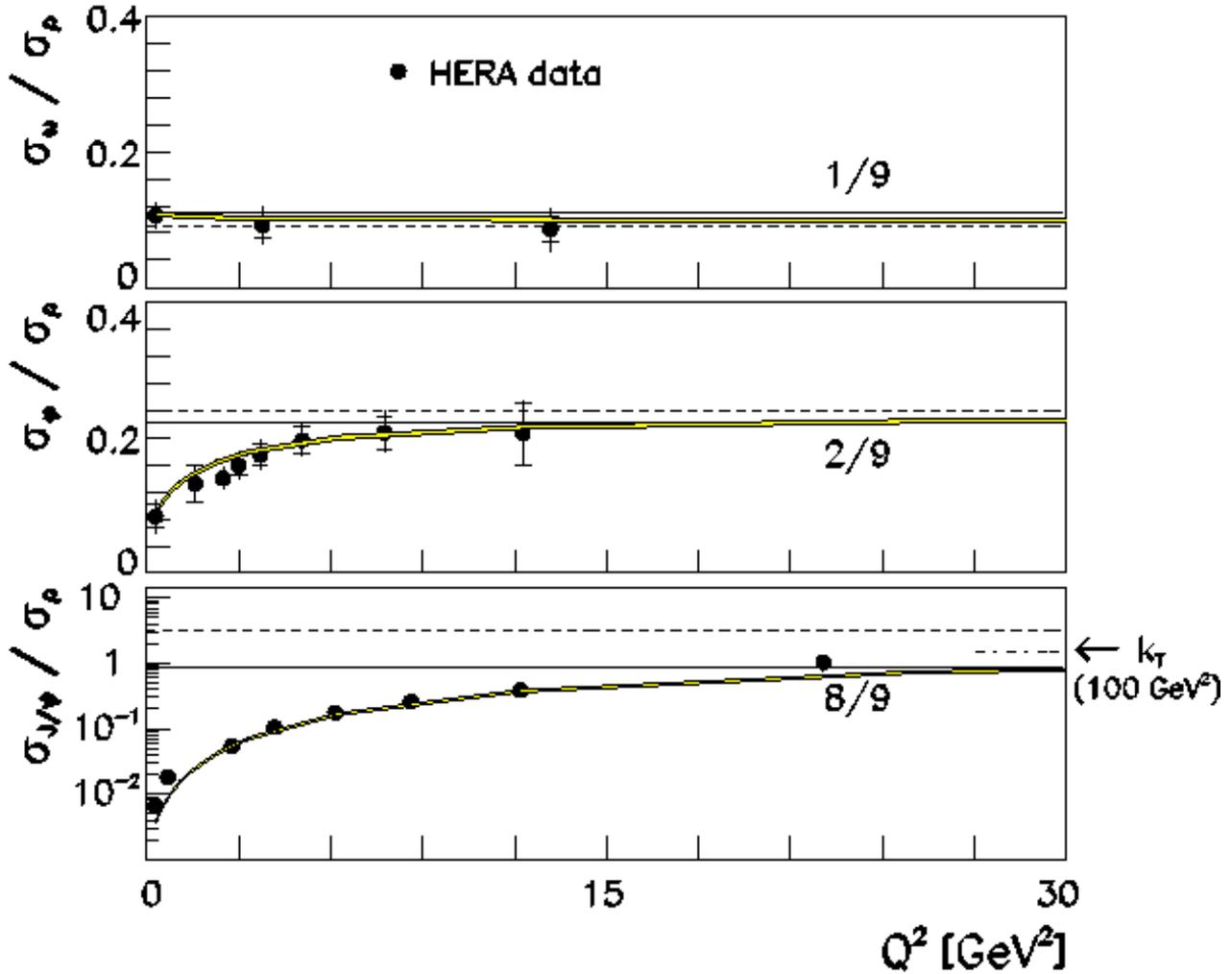}
  \caption{\it   The ratios of the $\omega$ 
  ~\protect\cite{pl:b487:273}, $\phi$ ~\protect\cite{pl:b483:360} (the PHP
  point is calculated using ~\protect\cite{epj:c2:247, pl:b377:259})  and $J/\psi$ 
  (calculated using only recent ZEUS data 
  ~\protect\cite{epj:c24:345,np:b695:3,epj:c2:247,epj:c6:603}, for H1 results see
Figs.~\protect\ref{fig:ivanov2004-tot-all},
~\protect\ref{fig:ZEUSratiosQ2larget}) 
to the $\rho^0$ cross sections as
   a function of $Q^2$. 
    The horizontal solid lines correspond to the SU(4) predictions,
while the horizontal dashed lines correspond to the pQCD
predictions in the non-relativiatic approximation given by
Eq.(\ref{eq:6.2.1.1}).
As shadowed band the corresponding
predictions from the $k_t$-model for the oscillator wave functions
\protect\cite{IgorNumerics} are shown. The width of the band is taken
just arbitrary and does not 
correspond to the theoretical uncertainties. The $k_t$-model prediction
for the $J/\psi / \rho$ ratio at $Q^2$=100~GeV$^2$ is shown separately 
as a dashed-dotted line.}
     \label{fig:ZEUSratiosQ2}
\end{figure}

The mass term in the scanning radius (\ref{eq:1.3.3}) and the
corresponding hard scale $\Qb^2$ of Eq.~(\ref{eq:1.3.5}) change
dramatically from the $\rho,\omega$ to $\phi$ to $J/\Psi$.
For this reason a comparison of $\sigma_V(Q^2)$ for the
different vector mesons as a function of $Q^2$ makes no sense.
This is
clearly seen from Fig.~\ref{fig:ZEUSratiosQ2} which shows
the ratios of cross sections $\sigma_V(Q^2)/\sigma_{\rho}(Q^2)$
as a function of $Q^2$. They exhibit a very steep dependence
on $Q^2$ from real photoproduction to large
$Q^2$: even for the $\phi$ the ratio rises by a factor
of $\approx 3$, whereas for the $J/\Psi$ it rises by more than
two orders of magnitude. The experimentally 
observed $Q^2$-dependence is well reproduced 
within the $k_{\perp}$-factorization approach \cite{IgorNumerics}.

We reiterate the point of Section 6.1 about the approximate
restoration of flavour symmetry if the cross sections are scaled by 
the non-relativistic factor $ m_V \Gamma(V\to e^+e^-)$
and compared at
equal hard scale $\Qb^2$, i.e., at equal $(Q^2+m_V^2)$.
According to the Review of Particle Properties \cite{PDG2002}
\be
{1\over \eta_V^{J/\Psi}} ={ m_V \Gamma(V\to e^+e^-)\over 
m_{J/\Psi}\Gamma(J/\Psi \to e^+e^-)}
= \rho~:~\omega~:~\phi~:~J/\psi~=
0.32~:~0.029~:~ 0.077~: 1\,,
\label{eq:6.2.1.1}
\ee
(the uncertainties in the rhs of (\ref{eq:6.2.1.1}) from the decay widths
vary from several to $\sim 7$ per cent for
light to heavy mesons and are not shown). 

The $J/\psi$ has been chosen as the reference point as it is
the best approximation to a nonrelativistic quarkonium, 
for light vector mesons the non-relativistic approximation is
evidently poor. As a test of theory 
one must rather compare the
predictions for the flavour dependence  at a sufficiently
large $(Q^2+m_V^2) \sim 20$ GeV$^2$. Here the $k_{\perp}$-factorization 
predictions
can be summarized as 
\be
{1\over \eta_V^{J/\Psi}}\Bigr|_{k_{\perp}-{\rm fact}} =
0.68  ~:~0.068~:~ 0.155 ~:~1\,.
\label{eq:6.2.2.1}
\ee
\begin{figure}[htbp]
   \centering
   \epsfig{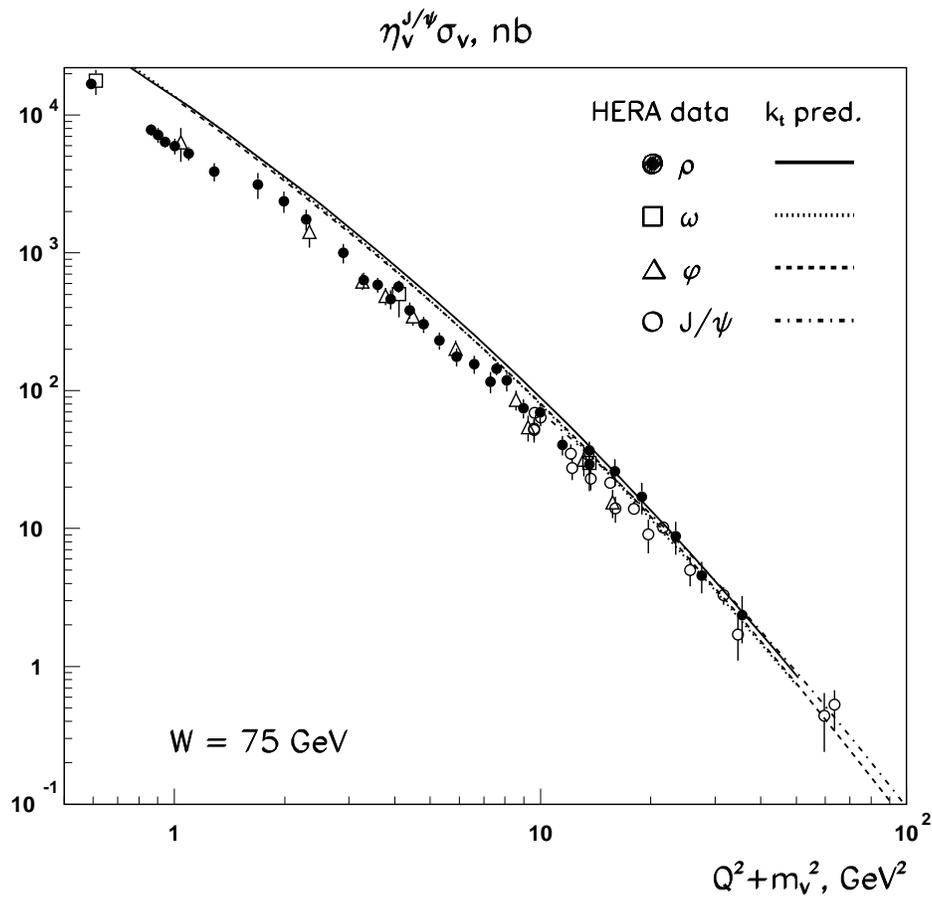}
   \caption{\it   A compilation of the 
flavor-rescaled using $k_{\perp}$-factorization 
Eq.(~\protect\ref{eq:6.2.2.1}) total cross sections
$\eta^{J/\psi}_V\sigma_V(Q^{2})$
for the $\rho,\omega,\phi,J/\Psi$ vector mesons 
~\protect\cite{epj:c6:603, epj:c13:371, epj:c2:247, pl:b487:273, pl:b377:259,
pl:b380:220, pl:b483:360, epj:c24:345, np:b695:3, np:b472:3, epj:c10:373} 
as a function of $Q^2+m_V^2$. Shown are also the corresponding
predictions from the $k_t$-model for the oscillator wave functions.
\protect\cite{IgorNumerics}.}
   \label{fig:ivanov2004-tot-all}
\end{figure}
There is a marginal change from the oscillator to Coulomb 
parameterization for wave functions, the theoretical uncertainty 
in the rhs of (\ref{eq:6.2.2.1}) is about 10 per cent. 
In Fig.~\ref{fig:ivanov2004-tot-all} we plot the flavor-rescaled 
$\sigma_V(Q^2)\eta_V^{J/\Psi}|_{k_{\perp}-{\rm fact}}$ for 
all vector mesons as a function
of $(Q^2+m_V^2)$. The universal $(Q^2+m_V^2)$-dependence is clearly seen
and the overall agreement between the experimentally observed
and theoreticaly predicted flavor dependence is good.

For uncertain reasons, perhaps by misinterpretation of
the discussion in \cite{Frankfurt1996}, one often discusses
the restoration of the $SU(4)$ flavour symmetry,
\be
\rho^0~:~\omega~:~\phi~:~J/\psi = ~1~:~1/9~:~2/9~:~8/9  
= ~1.125~:~0.125~:~0.22(2)~:~1\,.
\label{eq:6.2.1.2}
\ee
Incidentally it is not much different from (\ref{eq:6.2.2.1}), except of $J/\psi$, 
see Fig.~\ref{fig:ZEUSratiosQ2}.
 We caution 
that there are no sound reasons for such an SU(4) ratio of cross
sections even at very large $Q^2$.
The  $SU(4)$-factor weighted empirical cross sections were
shown in Fig.~\ref{fig:H1sigmaAllVM} and can indeed be fitted by the 
flavor-independent universal curve, but  
this agreement with $SU(4)$ rations must be regarded as an accidental 
one.


\subsection{The ratio $R_V=\sigma_L/\sigma_T$}\label{sect6.3}


\subsubsection{Theoretical expectations}

For heavy vector mesons treated as a nonrelativistic quarkonium, 
the pQCD predicts $R_V(Q^2) = \frac{Q^2}{m_V^2} R_{LT}$ with $R_{LT}=1$, 
see (\ref{eq:4.9.3}). Despite all the 
uncertainties with the wave functions, 
important point is that for a meaningful 
evaluation of $R_V(Q^2)$
the transverse and longitudinal
vector meson must be related by the rotation-invariance.
The Fermi motion effects in the longitudinal and transverse amplitudes
are different, though, the latter being more sensitive to 
the end point contributions from $z\sim 0, z\sim 1$. Such
corrections are  automatically incorporated in the color dipole 
and $k_{\perp}$-factorization calculations. 
First, as discussed in Section~\ref{sect4.8}, the hard pQCD scales 
for transverse and longitudinal cross sections do slightly
differ, $\Qb^2_{GT} \lsim \Qb^2_{GL}$, which already leads to a
substantial reduction of $R_{LT}$ at large $Q^2$, as it was 
found in \cite{NNZscanVM}. Second, the 
SCHNC transitions contribute for the most part 
to the transverse cross section, especially
for light vector mesons, further lowering $R_{LT}$. Third, 
as discussed in Section~\ref{sect4.10},
the predictions for $R_V(Q^2)$ are potentially very sensitive to 
the presence of a hard, quasi-pointlike $q\bar q$ component
in the vector meson wave function. As a test of this
sensitivity we evaluate $R_V(Q^2)$ for a squeezed $\rho$-meson
as discussed in Section 4.10. 

One way to circumvent problems with the poorly known 
wave function of vector mesons is  to resort to the 
Bloom-Gilman duality for diffractive DIS \cite{GNZlong}. 
Because of the strikingly different dependence of the
duality integrals for $\sigma_L, \sigma_T$ on the duality interval,
see Eqs. (\ref{eq:3.4.7.6}), (\ref{eq:3.4.7.7}), the
ability \cite{martinduality,martinduality2}
to reproduce both $\sigma_V(Q^2)$ and $R_V(Q^2)$
is not surprising. It is not clear, though, whether such
an evaluation is entirely consistent with the rotation invariance
constraints or not. 

In the color evaporation model discussed in Section 3.4.7
the $J/\Psi$ is a part of the open charm cross section 
and, arguably, the ratio $\sigma_L/\sigma_T$ must be close
to that for the open charm, $R_V \approx R_{DIS} \ll 1$, 
which would conflict the large-$Q^2$ data on $J/\psi$ 
production to be shown below.


\subsubsection{Experimental results}

\begin{figure}[htbp]
   \centering
   \epsfig{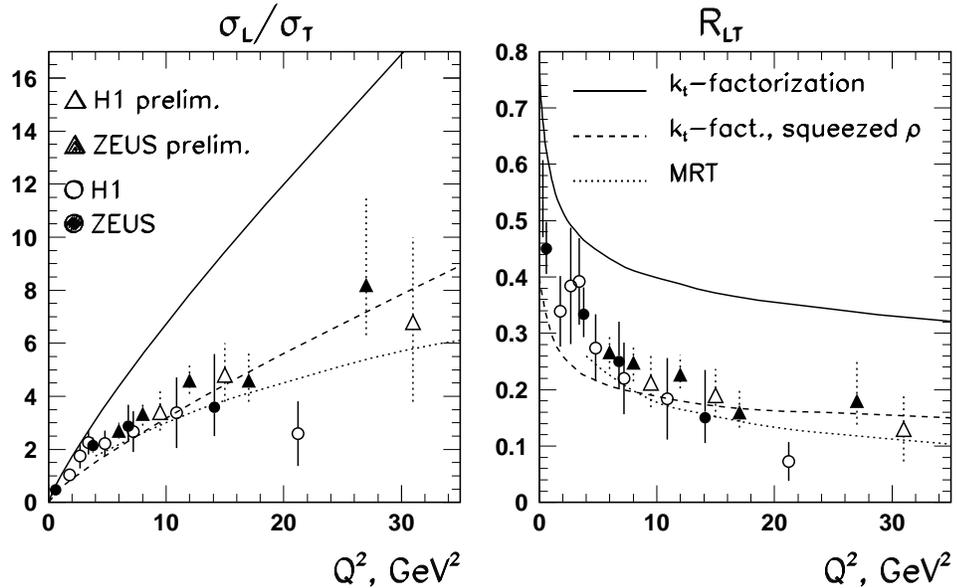}
   \caption{\it    A compilation of the experimental
data on the ratio $R_{\rho} = \sigma_L/\sigma_T$ and $R_{LT}^{\rho}=
R_{\rho}m_{\rho}^2/Q^2$
for the $\rho$ meson production  as a function of $Q^2$. 
The recent preliminary data from H1
~\protect\cite{cpaper:ichep2002:989} and ZEUS
~\protect\cite{cpaper:epc2001:594} measurements are just scanned from the
plots of the correspondent papers and shown together with the
published results
~\protect\cite{epj:c13:371,epj:c6:603,np:b463:3,epj:c2:247}. 
The theoretical predictions are as follows: the dotted curve
shows the estimates by Martin et al. based on the Bloom-Gilman 
duality ~\protect\cite{martinduality}, the solid curve 
and dashed curves 
show the results from the $k_{\perp}$-factorization
for the conservative radius of the $\rho$-meson and
the squeezed $\rho$-meson, respectively 
\protect\cite{IgorPhD,IgorNumerics}.}
   \label{fig:ivanov2004-ratio-rho-lt}
\end{figure}

Figure~\ref{fig:ivanov2004-ratio-rho-lt} shows the ratio
$$
R_{\rho}(Q^2) =
{\sigma_L(\gamma^* p \to \rho p) \over \sigma_T(\gamma^* p \to \rho
p)}\,,
$$
as a function of $Q^2$ for the $\rho$ meson production.
The right plot shows the same data presented in the form of
$$
R_{LT}^{\rho}(Q^2)=
R_{\rho}(Q^2){m_{\rho}^2\over Q^2}\,.
$$
At the small values of $Q^2$ the ratio is an approximately 
linear function
of $Q^2$ and $R_{LT}(0)\sim 1$, but $R_{LT}(Q^2)$ decreases
steadily with rising  $Q^2$ and the growth of $R_{\rho}(Q^2)$ 
slows down.
Whereas some of the published data sets even suggest 
flattening of $R_{\rho}(Q^2)$, the new preliminary data
\cite{cpaper:ichep2002:989,cpaper:epc2001:594} 
indicate the steady large-$Q^2$ rise of $R_{\rho}(Q^2)$

\begin{figure}[!htb]
   \centering
   \epsfig{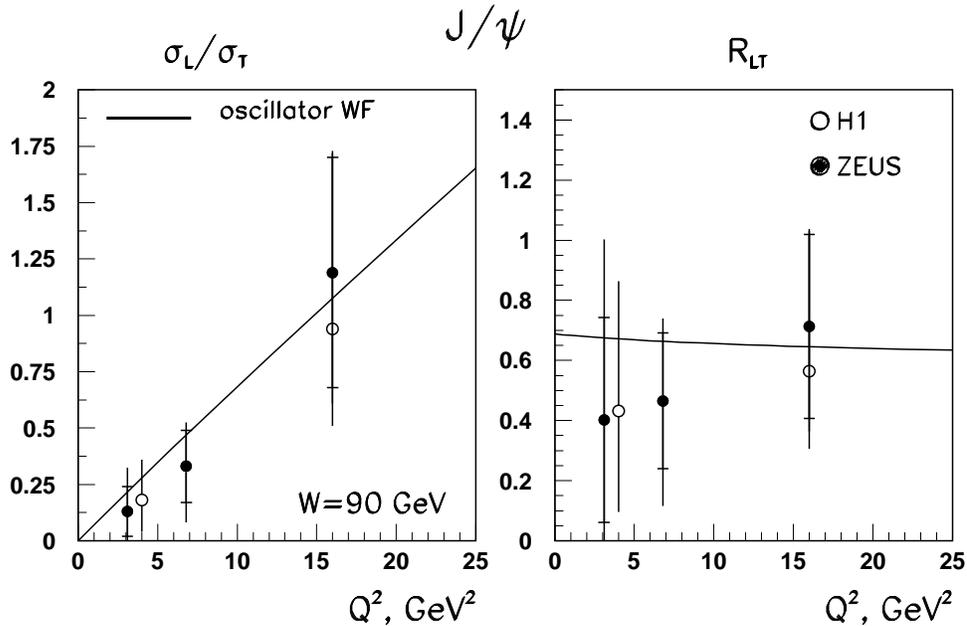}
   \caption{\it    A compilation of the experimental
data on the ratio $R_{J/\psi} = \sigma_L/\sigma_T$ and 
$R_{LT}^{J/\psi}=
R_{J/\psi}m_{J/\psi}^2/Q^2$
for the $J/\psi$ meson production  as a function of $Q^2$.
The open circles
represent the H1 \protect\cite{epj:c10:373}
and solid symbols the ZEUS $J/\psi$ measurements
~\protect\cite{np:b695:3}. 
The solid curve is a prediction
from the $k_{\perp}$-factorization approach \protect\cite{IgorNumerics}.}
   \label{fig:ivanov2004-jpsi_ratiolt}
\end{figure}

The dotted curve shows an evaluation $R_{\rho}(Q^2)$
\cite{martinduality,martinduality2} based on the duality 
approach \cite{GNZlong}. Whereas in the $k_{\perp}$
factorization one starts with the $q\bar{q}$ continuum production
amplitudes, projects it onto the $J^P=1^-$ state and 
averages 
over the masses $M \sim m_V$ of the $q\bar{q}$ state with 
the vector 
meson wave function as the weight function.
In the duality
approach \cite{GNZlong} one rather calculates $|{\cal T}|^2$ for 
a fixed diffractive mass $M$ and then integrates over certain
range of $M$ around $m_V$. For the $\rho$ meson Martin et al.
take the duality interval $M=[M_1,M_2]=[600,1050]$ MeV. 
Because of the different sensitivity of duality integrals for
$\sigma_{L,T}$ to the duality interval, 
see Eqs.~(\ref{eq:3.4.7.5}), (\ref{eq:3.4.7.6})
one readily finds a good description of the $Q^2$ dependence
of $R_{\rho}(Q^2)$ and the expense of a possible conflict
with the rotation invariance. Although the absolute value
of $R_{\rho}(Q^2)$ is essentially adjusted to the data, the
steady rise of $R_{\rho}(Q^2)$ at large $Q^2$ must be regarded
as the genuine pQCD prediction. For the uncertainties with
the duality description of the absolute cross section  
see Section 3.4.7.  

The solid curve shows the
result of the $k_{\perp}$-factorization calculation
with the conservative radius of the $\rho$-meson. Although
at largest $Q^2 \sim 20 {\rm GeV}^2$ the predicted
$R_{LT} \sim 0.3 \ll 1$, as found already in 1994
within the color dipole model \cite{NNZscanVM}, 
there is an obvious disagreement 
with the large-$Q^2$ data. A comparison 
between the data and the $k_t$-factorization
calculations in terms of $\sigma_L$ and $\sigma_T$ separately
shows \cite{IgorPhD} that the source of this discrepancy is the transverse cross
section $\sigma_T$, whose value at high $Q^2$ is significantly
underestimated by the model. A strong sensitivity of 
predictions for $R_{\rho}(Q^2)$ to the
short-distance properties of the vector meson is 
illustrated by the dotted curve which is the 
$k_{\perp}$-factorization result for the squeezed $\rho$-meson.
As anticipated in Section 4.10, the prime effect of higher 
short-distance density in the $\rho$-meson is an
enhancement of $\sigma_T$ and suppression of  $R_{\rho}(Q^2)$.

\begin{figure}[htbp]
   \centering
   \epsfig{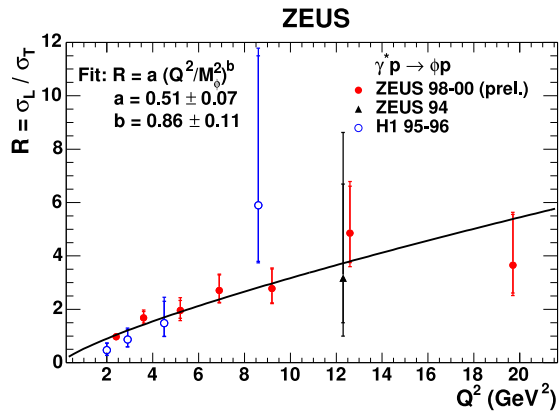}
   \caption{\it    A compilation of the experimental
data on the ratio $R_{\phi} = \sigma_L/\sigma_T$
for the $\phi$ meson production  as a function of $Q^2$. Shown 
are the recent preliminary data from ZEUS
~\protect\cite{ZEUS2004_phi_preliminary}  together 
with the published results from ZEUS ~\protect\cite{pl:b377:259,pl:b380:220}
and H1 ~\protect\cite{pl:b483:360}. The solid curve shows the fit
of the form $R_{\phi} = a(Q^2/m_{\phi}^2)^b$ with the
parameters cited in the figure.}
   \label{fig:Phi_ZEUS2004_RLT}
\end{figure}

Figure~\ref{fig:ivanov2004-jpsi_ratiolt} shows the ratio 
$\sigma_L/\sigma_T$ for the $J/\psi$ mesons
compared with the prediction of the $k_t$-factorization model.
The recent ZEUS result averaged over $Q^2$ is $R_{LT}=0.52\pm 0.16$
\cite{np:b695:3}. The recent high accuracy experimental results 
for large-$Q^2$ $\phi$ production are shown
in Fig.~\ref{fig:Phi_ZEUS2004_RLT}. If compared against 
equal $Q^2/m_V^2$, the theoretical
results and the experimental
data for $R_{LT}$  show similar behavior 
for all three vector mesons: $\rho$, $\phi$ and $J/\psi$.
For instance, the fit to experimental data on $r_{00}^{04}$ for the $\rho$-production
shown in Fig.~\ref{fig:H1ZEUSr0400} can be reinterpreted, via Eq. (\ref{eq:5.3.1.4}) with
$\epsilon \approx 1$, as $R_{\rho}= 
a(Q^2/m_{\rho}^2)^b$ with $a=1/\xi=0.46\pm 0.015$ and $b=k=0.75\pm0.3$ which
agrees perfectly with the recent ZEUS parameterization for the $\phi$
production shown in Fig.~\ref{fig:Phi_ZEUS2004_RLT}: $a=0.51\pm 0.07$ and
$b=0.86\pm 0.11$.

\newpage


\section{The energy dependence and Regge properties of 
diffractive vector meson
production}


\subsection{Theoretical expectations}

The local energy dependence of $\sigma_{\gamma^*p\to Vp}(W^2,Q^2)$
is usually parameterized as 
\be
\sigma_{\gamma^*p\to Vp}(W^2,Q^2) \propto W^{\delta}\, .
\label{eq:7.1.1}
\ee
The exponent $\delta$ for the  
$t$-integrated total cross section is controlled by
the $x$-dependence of the integrated gluon density, 
see Eq.~(\ref{eq:4.8.10}), and the shrinkage of the diffraction cone.
Within the standard exponential approximation for the $t$-dependence 
\be
\sigma_{\gamma^*p\to Vp}(W^2,Q^2)\propto {Im {\cal T}_{LL}^2(W^2,t=0)
\over  {W^4 b_V(W^2,Q^2)}}\, 
\label{eq:7.1.1*}
\ee
and 
\be
\delta = 4[\lambda(\Qb^2_G) - {\alpha^\prime_{\Pom} \over b(\Qb^2_G)}]\, .
\label{eq:7.1.2}
\ee

The microscopic pQCD models of the Pomeron exchange,
see Sects.~\ref{sect3.2}, ~\ref{sect3.3} and Eq.(\ref{eq:3.2.3}),
suggest that the Pomeron is a more complex object than an 
isolated single pole - it is either the branching point or a 
sequence of Regge poles. For this reason the exponent $\delta$ 
can depend on $Q^2$, $m_V$, the energy range, helicities etc. 
Eq.~(\ref{eq:7.1.2}) emphasizes the crucial point \cite{NNZscanVM}
that within the color dipole and $k_{\perp}$-factorization 
approaches $\delta$ only depends on $\Qb^2_G \sim (Q^2+m_V^2)$. 
The corrections to
the $(Q^2+m_V^2)$-scaling are well understood and  marginal:
a slight departure of the hard scale $\Qb^2_G$ from
$\Qb^2$ has been discussed in Sect. 4.8. Specifically, 
 at equal values of $Q^2 + m_V^2$, 
the $\rho$ production is in a somewhat softer regime than the
$J/\psi$ production and one can
expect that $\delta_{J/\psi}$ should be slightly larger than
for the $\delta_\rho$.

\begin{figure}[htbp]
   \centering
   \epsfig{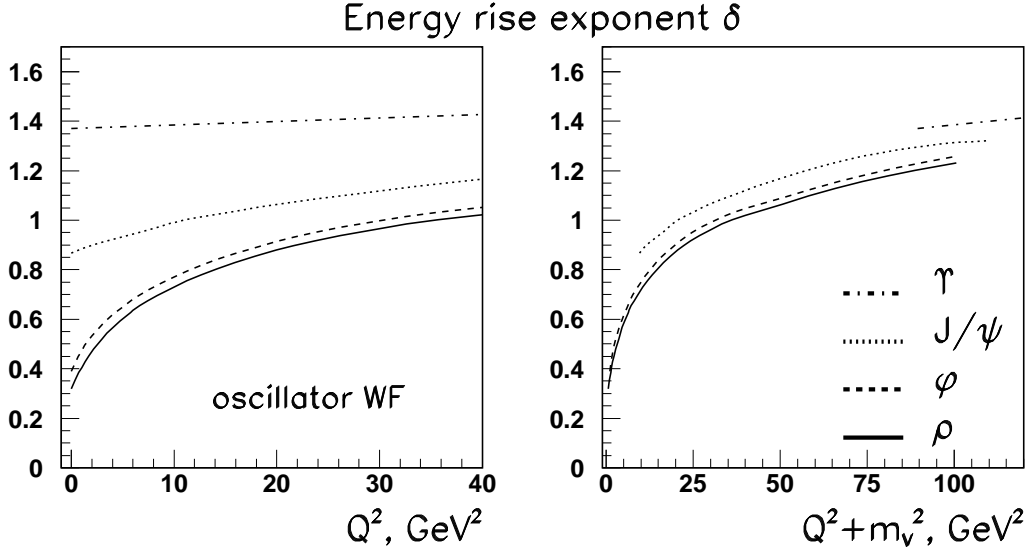}
   \caption{The $k_t$-factorization predictions for 
the exponent $\delta$ of the fits $\sigma \propto W^\delta$
   to the $W$ dependence for different vector mesons
\protect\cite{IgorPhD,IgorNumerics}. Notice a 
strong flavor dependence in the left
box where $\delta$ is plotted as a function of $Q^2$. The 
flavor dependence is much weaker when $\delta$ is plotted 
against $Q^2+M_V^2$ (the right box). }
  \label{fig:VMintercepts}
\end{figure}

The $k_{\perp}$-factorization phenomenology of DIS structure
functions described in Sect.~\ref{sect3.3.3} strongly 
suggests that, for purely numerical reasons, within 
the kinematical range of the HERA 
experiments the vacuum exchange can be approximated by
the two, soft and hard, Pomeron poles with approximately
$Q^2$-independent intercepts. 
The values of the intercepts anticipated in this approximation
were given in Sect.~\ref{sect3.1.4} and Sect.~\ref{sect3.3}.
They translate into $\delta \sim 0.2 \div 0.4$ in the soft region
and much larger $\delta \gsim 1 \div 1.5$ when the interaction becomes
sufficiently hard. Since the pQCD hard scale $\Qb^2$ increases not only
with $Q^2$, but also with $m_V$, the theory predicts
the corresponding hierarchy 
$ \delta_{J/\psi} > \delta_{\phi} > \delta_{\rho}$
at any $Q^2$, including photoproduction; for exact values see 
Fig.~\ref{fig:VMintercepts}.
On the other hand, the variations of $\delta$ from $\rho,\omega$ to
$\phi$ to $J/\Psi$ to $\Upsilon$ are weak against the variable 
$Q^2+m_V^2$.

The energy dependence of the $\psi(2S)$ production
should be discussed separately.
As pointed out in Section~\ref{section2s},
the node of the $2S$ radial wave function 
leads to a partial compensation between contributions from
dipoles with the size
below and above the node. As a result, the typical dipole sizes
that contribute to $\psi(2S)$ production are smaller than
for $J/\psi$ production. This makes the pQCD scale for $\psi(2S)$
production somewhat harder than for $J/\psi$, which results in 
$\delta_{\psi(2S)} > \delta_{J/\psi}$ and the rise of
the cross section ratio $\sigma(\psi(2S))/\sigma(J/\psi)$ with
growing energy is predicted \cite{NNPZZslopeVM,*NNPZZslopeVM1,IgorPhD}.


\subsection{Experimental results: real photoproduction}


\subsubsection{Ground state vector mesons}

\begin{figure}[htbp]
   \centering
   \epsfig{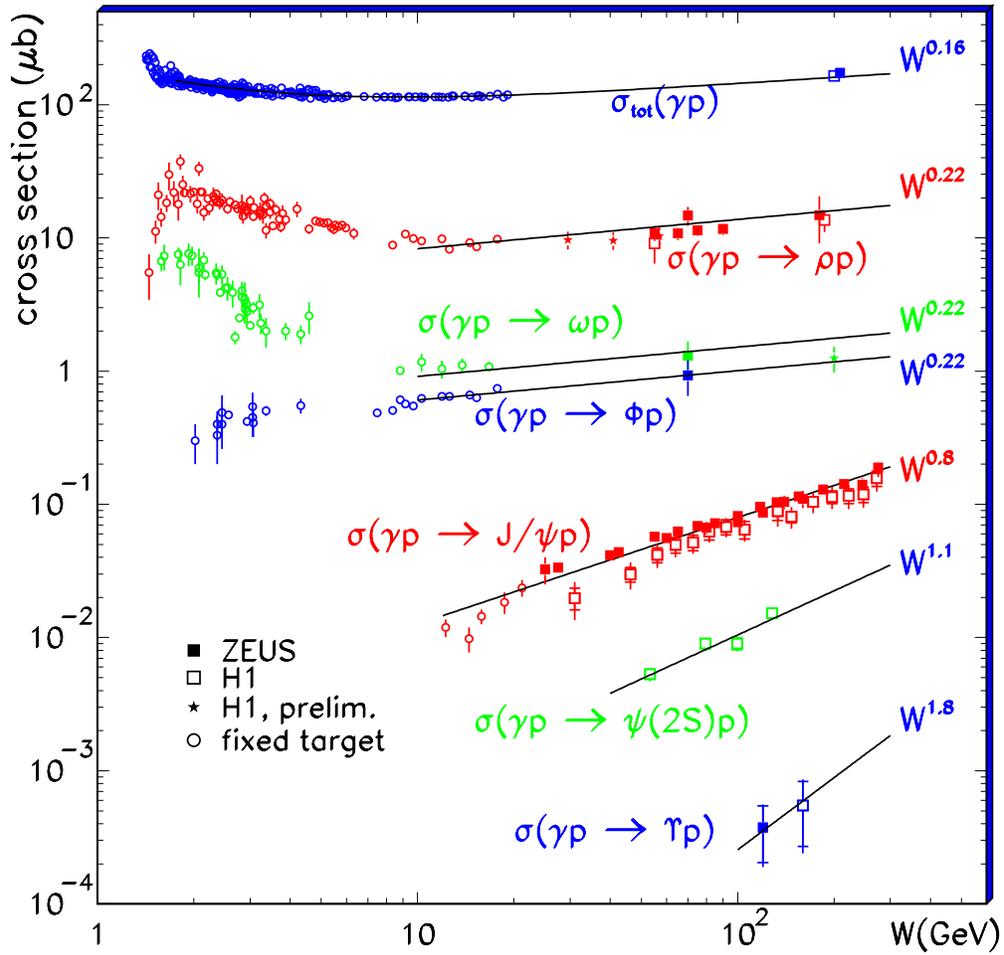}
   \caption{\it The total vector-meson photoproduction 
cross section as a function of $W$
   for all vector mesons measured at HERA shown together with the
   results of fixed target experiments and compared with the total
photoabsorption cross section
   $\sigma_{tot}^{\gamma p}$.}
   \label{fig:WDepAllVM}
\end{figure}

\begin{figure}[htbp]
   \centering
   \epsfig{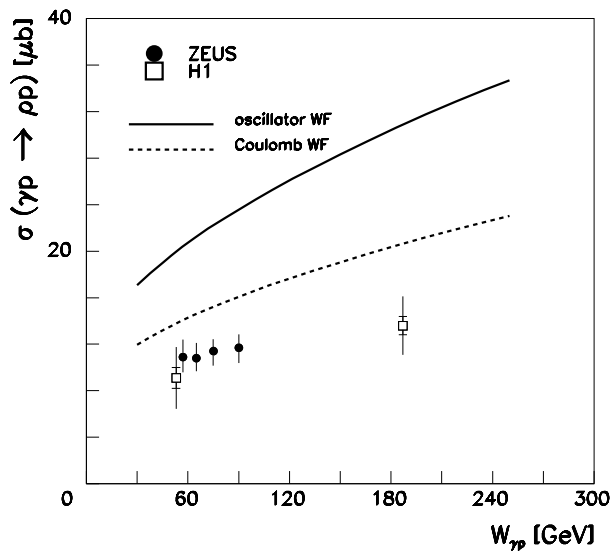}
   \caption{\it The W-dependence of exclusive $\rho$ photoproduction 
cross section measured by the H1~\protect\cite{np:b463:3} 
and ZEUS~\protect\cite{epj:c2:247}
   experiments compared with predictions of the $k_t$-factorization model
for the oscillator and Coulomb wave functions of the $\rho$ meson.}
   \label{fig:rho}
\end{figure}

\begin{figure}[htbp]
  \centering
   \epsfig{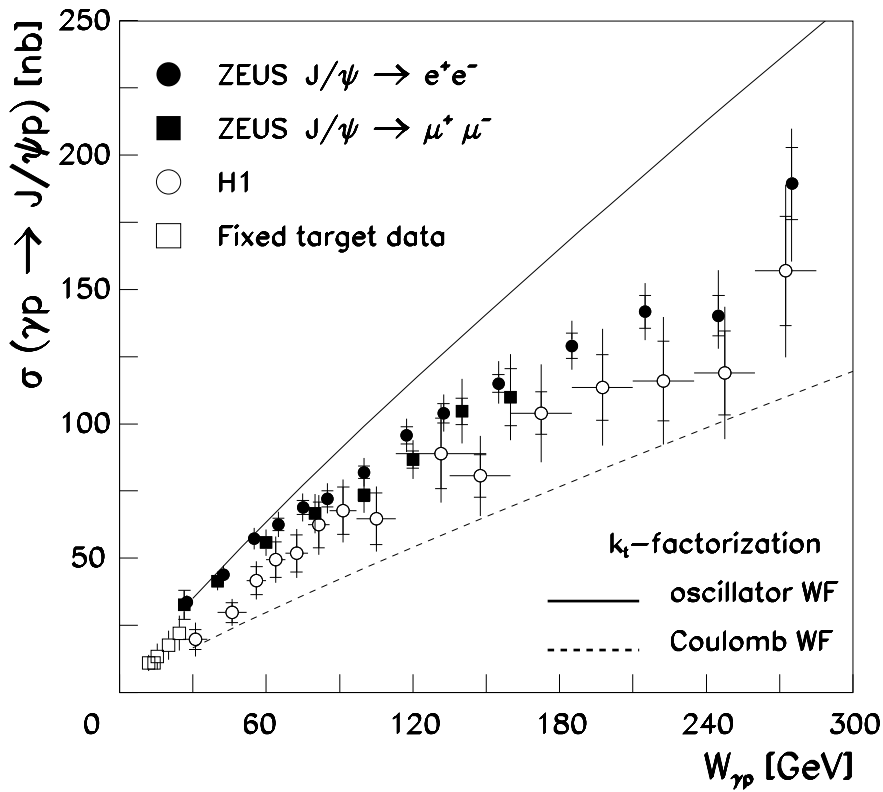}
  \caption{\it A compilation of the experimental data on 
exclusive $J/\psi$ photoproduction
   cross section as a function of   $W$ measured.
The black points show the recent ZEUS data for the
   $J/\psi \rightarrow \mu^+\mu^-$ and $J/\psi \rightarrow e^+e^-$
decay channels    ~\protect\cite{epj:c24:345}.
   The inner bars indicate the statistical uncertainties; the outer bars
   are the statistical and systematic uncertainties added in quadrature.
   Results
   from the H1~\protect\cite{pl:b483:23}
   and fixed target experiments ~\protect\cite{prl:48:73,prl:52:795}
   are shown by open symbols. Recently H1 reported about correction of the
   published cross sections and they are now much closer to the ZEUS ones.
   The $k_t$-factorization 
   predictions ~\protect\cite{IgorNumerics} for the oscillator and Coulomb wave functions
   are shown by the solid and dashed lines, respectively.}
  \label{fig:JpsiPHPW}
\end{figure}

The summary of the experimental data on the energy dependence
 of all vector meson photoproduction cross sections 
measured at HERA is shown in Fig.~\ref{fig:WDepAllVM} 
together with the total cross section
$\sigma_{tot}^{\gamma p}$ and the results from fixed target
experiments. There is a clear pattern of variation of
the exponent $\delta$ from light to heavy mesons:
For the light vector mesons, $\omega$, $\rho$ and
$\phi$, the rise of the cross sections with energy is well described
by $W^{0.22}$ behavior.
Already the first measurement of the $J/\psi$ photoproduction at
HERA showed a much steeper rise with $W$ than that observed for
the light vector mesons. The energy dependence was found to be
$W^{(0.7-0.8)}$, which implies the effective Pomeron intercept around
$\alpha_{\Pom}(0) \approx 1.2$. The still larger $\delta$ has been
found for $\Psi (2S)$, the experimental data on $\Upsilon$ production
are not conclusive yet. Figure~\ref{fig:rho} summarises of 
the experimental data on the $\rho$ photoproduction and 
Fig.~\ref{fig:JpsiPHPW} shows the recent ZEUS
photoproduction measurement \cite{epj:c24:345}
of the $J/\psi$ photoproduction
in a wide $W$ range ($20 \div 300$ GeV) together with earlier data
from H1 and fixed target experiments. The result
of the fit in the form $\sigma \propto W^{\delta}$ to the
ZEUS data with $W>30$ GeV yielded  $\delta = 0.69 \pm
0.02(stat.) \pm 0.03(syst.)$. 

The $k_t$-factorization and color dipole 
calculations, shown in Fig.~\ref{fig:VMintercepts},
predict precisely such a variation of the energy dependence
with the mass of the vector meson.
The real photoproduction of light vector mesons is dominated by 
non-perturbative component of the unintegrated gluon density 
of the proton and/or soft dipoles, and $\delta$ is small,
but the contribution from hard gluons rises from light to heavy 
vector mesons, which is a universal prediction from all
pQCD motivated models. The energy dependence from the
specific $k_{\perp}$-factorization model \cite{IgorPhD,IgorNumerics}
comes out right for both
$\rho$ (Fig.~\ref{fig:rho}) and $J/\Psi$ mesons (Fig.~\ref{fig:JpsiPHPW}).


\subsubsection{Radially excited vector mesons}

\begin{figure}[htbp]
   \centering
   \epsfig{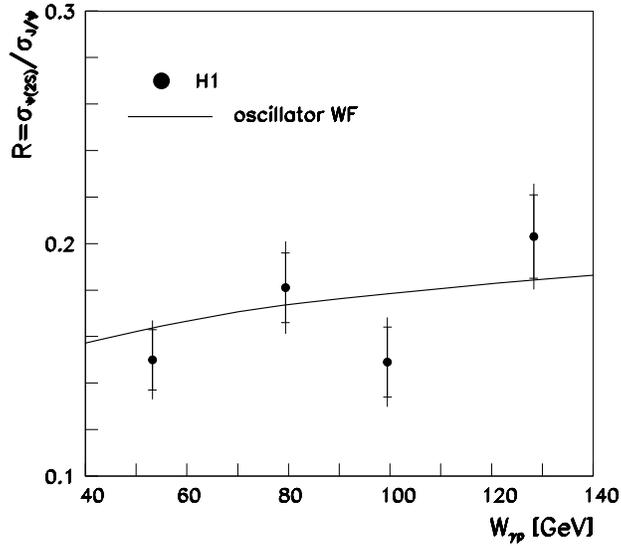}
   \caption{\it The ratio of photoproduction cross sections
   $\Psi (2S)$ over $J/\psi$ as a function of $W$~\protect\cite{pl:b541:251}.
   The $k_t$-factorization predictions 
\protect\cite{IgorNumerics} are shown for comparison.}
   \label{fig:psi2sPHPW}
\end{figure}

The $\Psi (2S)$ photoproduction was studied by
H1~\cite{pl:b541:251} using tagged and untagged data samples of
$\Psi (2S) \rightarrow ll$ and $\Psi (2S) \rightarrow ll \pi \pi$
decay channels, where $ll$ stands for $e^+ e^-$ or $\mu^+ \mu^-$.
The measurement of the ratio $R$ of $\Psi (2S)$ to $J/\psi$
photoproduction cross sections as a function of $W$ is shown in
Fig.~\ref{fig:psi2sPHPW}. The suppression of this ratio,
$R\ll 1$, is due to the node effect, the resulting strong cancellations
between the contribution from the large size and small size 
components of the $\Psi(2S)$ make the predicted ratio
strongly dependent on the model for the wave function. 
The experimental data agree with $k_t$-factorization predictions
based on the oscillator wave function. 
The overall ratio for the data gives the value $R=0.166 \pm
0.007(stat.) \pm 0.008(sys.) \pm 0.007 (BR)$ in a good agreement
with the previous measurements~\cite{pl:b421:385} (for the
branching error calculation see~\cite{pl:b541:251}). 
A fit to this ratio of the form $R
\propto W^{\Delta \delta}$ yields $\Delta \delta = 0.24
\pm 0.17$, which indicates that the energy dependence of the $\Psi (2S)$
photoproduction cross section is slightly steeper than that
for the $J/\psi$ meson. This agrees with the 
color dipole model expectations \cite{NNPZdipoleVM,NNPZZslopeVM,*NNPZZslopeVM1}.

The definitive $\rho'(2S)$ assignment of the excited $\rho$ 
states is still pending and the experimental data on the
$\rho'(2S)$ production are not available yet. Here we 
simply mention that all color dipole 
\cite{NNZrhoprim,DoschRhoRhoprim,NemchikRhoprim}
and/or $k_{\perp}$-factorization \cite{IgorPhD,IgorNumerics}
calculations based on the node effect predict
$\sigma(\rho'(2S))/\sigma(\rho(1S))< 1$ with very
steep $Q^2$ dependence up to $Q^2\gsim (3\div 5)$ GeV$^2$,
whereas the duality relation estimates by Martin, 
Ryskin and Teubner \cite{martinduality} gives 
$\sigma(\rho'(2S))/\sigma(\rho(1S))> 1$. This simply
shows that one must
be careful with application
of duality to radial excitations: because
the mass spectrum of the $q\bar{q}$ pairs which enters
the duality integral does not exhibit any non-monotonous 
$Q^2$-dependence \cite{NZdifDIS,GNZcharm}, the steep
$Q^2$-dependence driven by the node effect can easily
be missed.


\subsubsection{Test of the vector dominance model}

In the color dipole language, the success of the vector-dominance 
model (VDM) for real photoproduction derives from the proximity
of the (quark flavor dependent) color dipole distributions in 
the photon and light vector mesons. The straightforward extension
of the VDM approximation (\ref{eq:3.4.1.5}) to the $\gamma p$
elastic scattering amplitude reads 
\be
{\cal T}(\gamma p\to  \gamma p) = \sum_{V=\rho,\omega,\phi}
\frac{\sqrt{4\pi\alpha_{em}}} {f_V}
{\cal T}(\gamma p\to Vp)\, .
\label{eq:7.2.3.1}
\ee
Assuming the pure imaginary amplitudes, which is a good 
approximation because in real photoproduction ${1\over 4} \delta \ll 1$,
one can extract the photoproduction amplitudes from the 
forward differential cross sections, whereas the $\gamma p$
elastic scattering amplitude is related by optical theorem
to $\sigma_{tot}(\gamma p)$. This leads to the Stodolsky sum rule
\cite{Stodolsky}
\be
\sigma_{tot}(\gamma p) = \sqrt {16 \pi \cdot 
  \left. \frac{\textstyle d \sigma ^{\gamma p \to \gamma p}}
              {\textstyle  d t}             \right|_{t=0} }
 = \sum_{V=\rho^0,\omega,\phi} 
   \sqrt{16 \pi \cdot \frac{4 \pi \alpha}{f^2_V} \cdot 
         \left. \frac{\textstyle d \sigma ^{\gamma p \to Vp}}
                     {\textstyle d t}                      \right|_{t=0} } \, .
\ee
The test of VDM sum rule has been reported by
the ZEUS collaboration \cite{np:b627:3}. 
The VDM analysis of the low-energy data gave 
$f^2_V /4 \pi$ = 2.20, 23.6 and 18.4~\cite{BauerRevModPhys,*BauerRevModPhys1} for
$\rho^0$, $\omega$ and $\phi$, respectively. Then, based 
on the photoproduction data at 70 GeV, the VDM sum rule gives
a value of $ 111 \pm 13 ({\rm exp.})\, \mu {\rm b}$ 
for the photon-proton total cross section at $W_{\gamma p} = 70$ GeV. 
The $\rho ^0$ meson contributes about 85\% of this value.
The interpolation of photon-proton total cross section at a
center-of-mass energy of $W_{\gamma p} = 70$ GeV,
obtained by interpolation between the present measurement 
and the lower energy measurements using the Regge model 
fits, is 
$139\pm 4$ $\mu {\rm b}$. The two numbers are close to each other;
the point that simplified VDM
model does not saturate the sum rule is well known,
for the review see \cite{BauerRevModPhys,*BauerRevModPhys1}.


\subsection{Experimental results: vector mesons in DIS}


\subsubsection{The impact of hard scale on the energy dependence}

\begin{figure}[htbp]
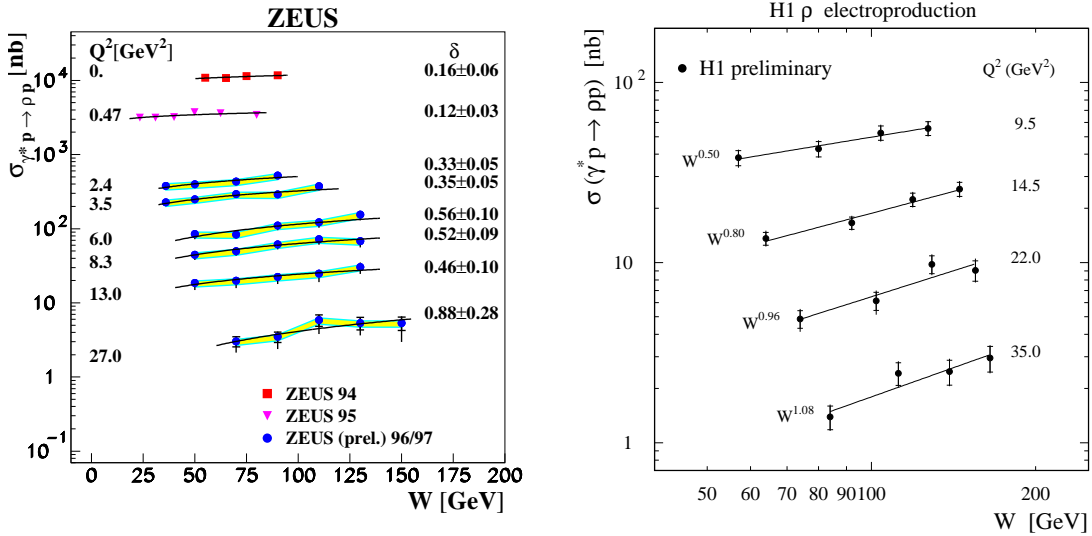

   \centering
   \epsfig{file=ZEUSDISRhoEPS01prelim_fig1.ps,width=80mm}
   \epsfig{file=H1DISRho2002prelim_fig4.eps,width=75mm}
   \caption{\it The $W$ dependence of the $\rho$ electroproduction cross section
   for different values of $Q^2$ as measured by
    ZEUS~\protect\cite{cpaper:epc2001:594}
   and H1~\protect\cite{cpaper:ichep2002:989}. The lines represent the results of fitting
   $\sigma \propto W^{\delta}$ with the $\delta$ values indicated in the
   figures. The shaded area in the ZEUS case indicates additional normalization
   uncertainties due to proton dissociation background.}
   \label{fig:rhoWdepQ2}
\end{figure}

\begin{figure}[p]
\vfill
\begin{center}
\hspace*{-1.1cm}
\epsfig{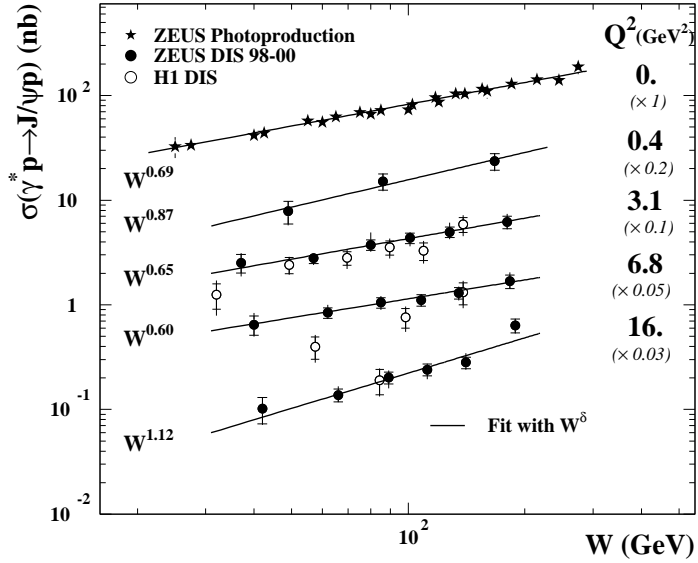}
\epsfig{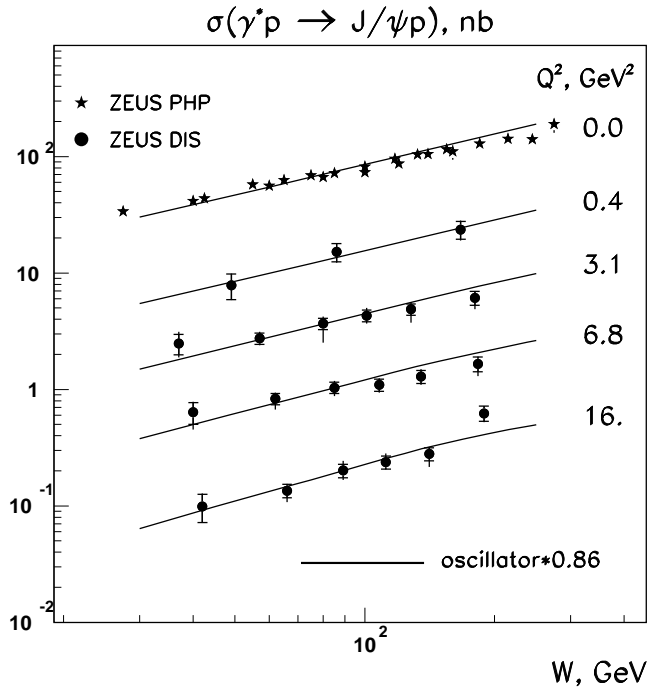}
\end{center}
\caption{\it 
The top plot shows the recent ZEUS results 
on exclusive $J/\psi$ electroproduction cross section 
as a function of W for four values of $Q^2$
\protect\cite{np:b695:3}.
ZEUS photoproduction
\protect\cite{epj:c24:345} and H1 electroproduction 
\protect\cite{epj:c10:373} cross sections 
are also shown.
The full lines are fits to the ZEUS data  
of the form $\sigma \propto W^{\delta}$.
The inner error bars represent the statistical uncertainties,
the outer bars are the statistical and systematic uncertainties added in
quadrature.
An overall normalization uncertainty of $^{+5\%}_{-8\%}$ was not included.
The bottom plot shows the $k_t$-factorization
predictions \protect\cite{IgorPhD,IgorNumerics}
normalized to the ZEUS photoproduction cross section
at $\langle W \rangle =90$ GeV and compared to the recent ZEUS data.}
   \label{fig:JpsiWdepQ2}
\end{figure}

%

The cross section for the exclusive $\rho^0$ electroproduction
measured by ZEUS \cite{cpaper:epc2001:594} and H1
\cite{cpaper:ichep2002:989} as a function of $W$ for different
values of $Q^2$ is presented in Fig.~\ref{fig:rhoWdepQ2}. The
curves represent the result of the $\sigma \propto W^{\delta}$ fits 
to the data. The exponent of the energy dependence increases 
with the $Q^2$ growth from about $0.2$ at low $Q^2$ up to $0.8\div 1.0$ at high
$Q^2$, in agreement with theoretical expectations form the color dipole 
approach shown in 
Fig.~\ref{fig:VMintercepts}.  In contrast to the $\rho$ production, 
the $J/\psi$ production cross-section 
exhibits 
almost the same $W$-dependence for all measured values of $Q^2$, including
the photoproduction limit, see Fig.~\ref{fig:JpsiWdepQ2},
in good agreement with the theoretical prediction from the color 
dipole approach shown
in Fig.~\ref{fig:VMintercepts}.

\begin{figure}[htbp]
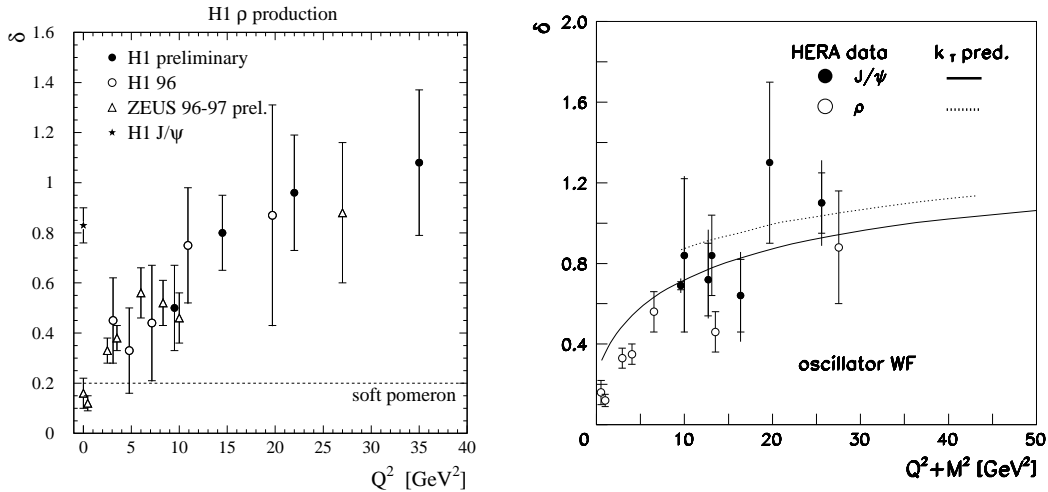

   \centering
   \epsfig{file=H1DISRho2002prelim_fig5.eps,width=68mm}
   \epsfig{file=savin2004-delta-vs-q2bar.eps,width=78mm}
  \caption{\it The exponent $\delta$ of the fits $\sigma \propto W^\delta$
   to the $W$ dependence for the $\rho$ production
as summarized in  ~\protect\cite{cpaper:epc2001:594}
   including the PHP $J/\psi$ point from ~\protect\cite{pl:b483:23} presented as
   a function of $Q^2$ (left plot) and combined $\rho$ and $J/\psi$ data
~\protect\cite{epj:c24:345,np:b695:3,epj:c10:373}
   presented as a function
of $Q^2+M_V^2$ in
the right plot. The
 predictions from the $k_t$-factorization approach 
\protect\cite{IgorPhD,IgorNumerics} are also shown. }
   \label{fig:ExpdeltaQ2}
\end{figure}


The point that the correct hard scale is $\propto (Q^2+m_V^2)$
is clear from Fig.~\ref{fig:ExpdeltaQ2} which 
shows the dependence of the exponent $\delta$ on $Q^2$ (the lhs. box) and
$(Q^2+m_V^2)$ (the rhs box), respectively. For the $\rho$-mesons the both
plots give  a clear indication of the rise of $\delta$
with $Q^2$ and $\Qb^2$. At largest $Q^2$ the exponent $\delta$ 
rises up to $1 \div 1.2$, which implies the rise of the effective
intercept of the Pomeron up to $\alpha_{\Pom}(0) \approx 1.3$.
If it were to be plotted as a function of $Q^2$, the exponent
$\delta$ for the $J/\Psi$ would be completely out of the 
observed trend for the $\rho$-meson, see the lhs box where only the 
PHP point is shown.  
When plotted as a function of $(Q^2+m_V^2)$, the same result
for the $J/\Psi$
is perfectly consistent with the results for the $\rho$ at
the same value of $(Q^2+m_V^2)$, confirming with the theoretical expectation
of the (approximate) flavor symmetry in this variable. 
The theoretical values of $\delta$ shown in Fig.~\ref{fig:ExpdeltaQ2}
were evaluated for the range $W= 50\div 100$~GeV. 
The overall agreement between the experiment and
$k_{\perp}$-factorization approach is
good. The lower box of Fig.~\ref{fig:JpsiWdepQ2} shows shows the $k_t$-factorization
predictions \protect\cite{IgorPhD,IgorNumerics}
normalized to the ZEUS photoproduction cross section
at $\langle W \rangle =90$ GeV; the quality of the theoretical description of
the data points is good.


\subsubsection{Discriminating the models for gluon density}

\begin{figure}[htbp]
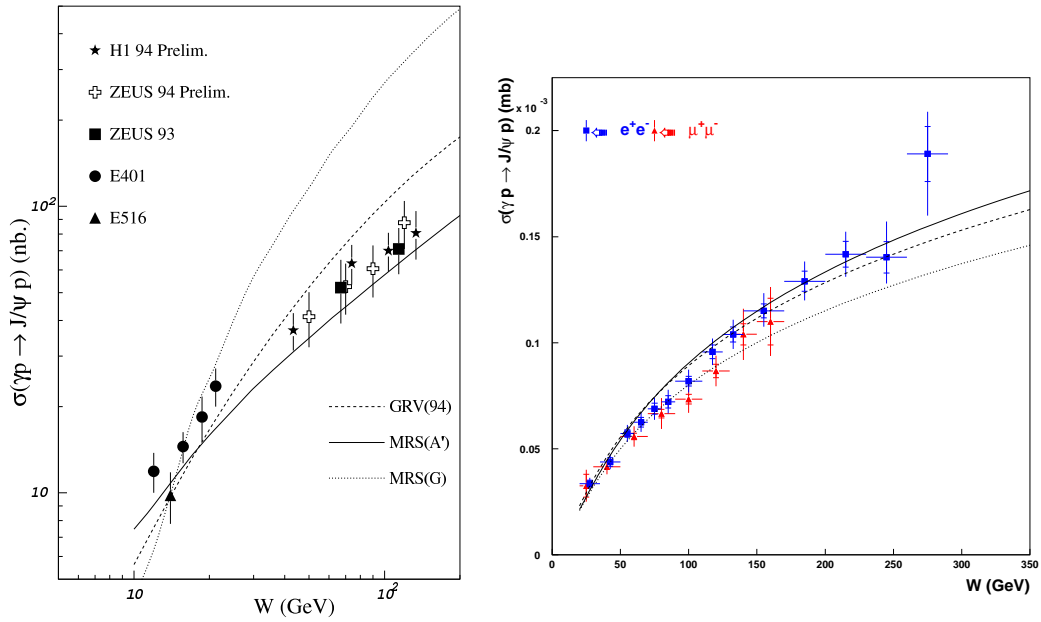

  \centering
   \epsfig{file=RyskinJpsiGlue.ps,width=60mm}
\epsfig{file=FioreJpsiSigma.ps,width=80mm}
  \caption{\it The left box: 
tests of models for the
small-$x$ gluon density using
the early HERA data on 
the energy dependence of $J/\Psi$
photoproduction together with data from fixed target 
experiments. The plot taken from \protect\cite{LevinKfactVM}
shows the pQCD calculations for
the GRV95 \protect\cite{GRV95} and MRS95 \protect\cite{MRS95}
parameterizations of the proton gluon density based on
the pre-1995 data on $F_{2p}(x,Q^2)$ and
illustrates the status of the subject at that time.
The right box: the parameterization of the energy dependence 
of $J/\psi$ production in the Fiore et al. model of
a double-pole Pomeron with intercept $\alpha_{\Pom}(0)=1$ 
\protect\cite{FioreDoublePolePomeron,*Fiore:2001bg}. }
 \label{fig:RyskinJpsiGlue}
\end{figure}

The  early discussion of the
$J/\Psi$ production was centered around the idea \cite{RyskinJPsi}
of stringent tests of models for the gluon density $G(x,Q^2)$.
An example of such a test based on the experimental data which
were available at that time is illustrated by the lhs plot of
Fig.~\ref{fig:RyskinJpsiGlue}. The shown theoretical
curves \cite{LevinKfactVM} are based on the 1995 updates of
parameterizations for the gluon density adjusted
to the HERA data on the small-$x$ growth of $F_{2p}(x,Q^2)$.
The energy behavior of 
each vector meson production can also be reproduced
by a number of simple parameterizations within models 
described in Sects.~\ref{sect3.1.4} and \ref{sect3.4.8}.
An example of a fit to the $J/\Psi$ photoproduction cross section
in a model by Fiore et al. \cite{FioreDoublePolePomeron,*Fiore:2001bg} in
which Pomeron is the double-pole at $\alpha_{\Pom}(0)=1$ 
is shown in the rhs plot of Fig.~\ref{fig:RyskinJpsiGlue}.

\begin{figure}[htbp]
   \centering
   \epsfig{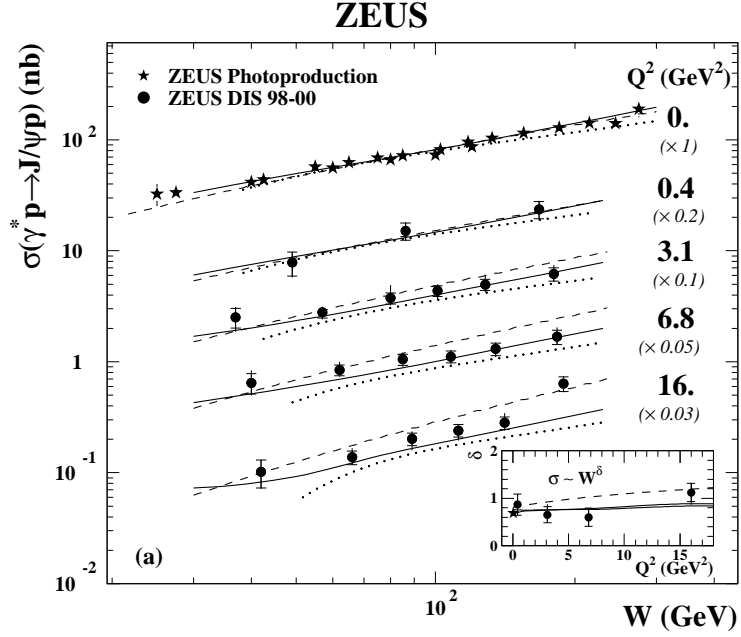}
    \epsfig{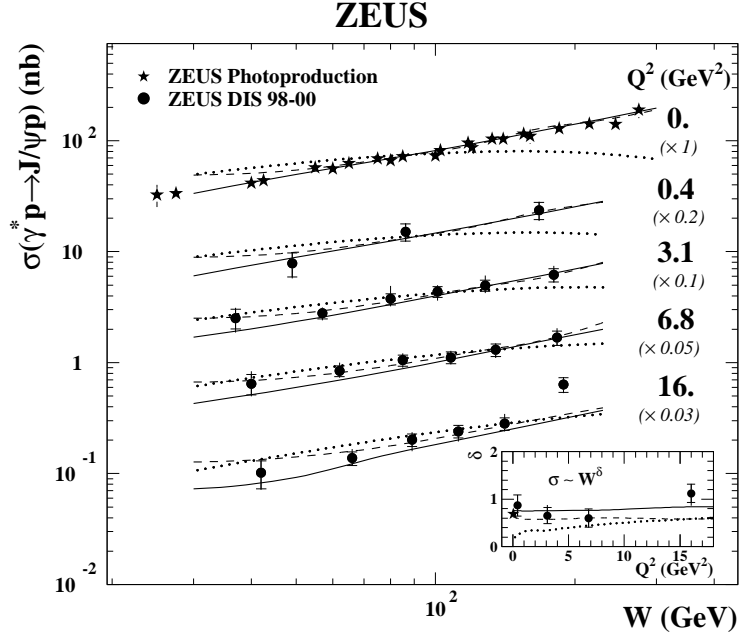}
    \caption{\it The tests of the pQCD predictions
 for various models for the small-$x$
 gluon density of the proton vs. the recent ZEUS data on 
$J/\psi$ production as a function of $W$ for
different values of $Q^2$ \protect\cite{np:b695:3, epj:c24:345}.
The curves in the top box represent 
the predictions of models MRT$\times 1.49$ (solid curve)
FKS$\times 1.7$ (dashed curve) and
 GLLMN$\times 0.9$ (dotted curve), the curves in the bottom plot
are the MRT results based on different gluon densities 
in the proton: the solid curve - ZEUS-S$\times 1.49$,
the dashed curve - CTEQ6M$\times 2.22$, the dotted curve - 
MRST02$\times 2.98$. 
All theoretical curves are rescaled as indicated above
to fit the ZEUS photoproduction point at $\langle W\rangle=90$ GeV.
The inserts show the exponent $\delta$ of the
parameterization $\sigma \propto W^{\delta}$ for
these models.}
\label{fig:ZEUSrecentJpsiW2}
\end{figure}

The past decade brought great improvements in our
understanding of the small-$x$ gluon densities and in the
vector meson productuon data compared to the 1995 situation
shown in Fig.~\ref{fig:RyskinJpsiGlue}.
The discrimination of the models remains weak, though: for the
reasons discussed
in Section 6.4, there are substantial uncertainties in the
absolute normalization which can not be eliminated at the moment.
This point is further illustrated by the recent ZEUS
data on $J/\Psi$ electroproduction shown in
Fig.~\ref{fig:ZEUSrecentJpsiQ2} in comparisons
with the predictions from the more recent
MRT, FKS and GLLMN models for the gluon density
described in Sect. 6.4.
All models need 
large rescaling to adjust the normalization to
the photoproduction at  $\langle W\rangle=90$ GeV. 
At the expense of such rescaling all model calculations
reasonably reproduce the gross features of the energy dependence.
For comparison, the $k_t$-factorization
predictions \protect\cite{IgorPhD,IgorNumerics} shown in the bottom plot of
Fig.~\ref{fig:JpsiWdepQ2} give an equally good description
of the same data using a rescaling factor 0.86. 
One exception is the MRT results for the MRST02
gluon density for small $Q^2$; the likely reason
for the found strong 
discrepancy is that the corresponding
values of $\Qb^2$ are too close to the lower  $Q^2$-boundary
of applicability of this particular set of parton densities
\cite{MRST02}.


\subsubsection{Comparison of vector meson
production and inclusive DIS from Regge model viewpoint} 

\begin{figure}[htbp]
   \centering
   \epsfig{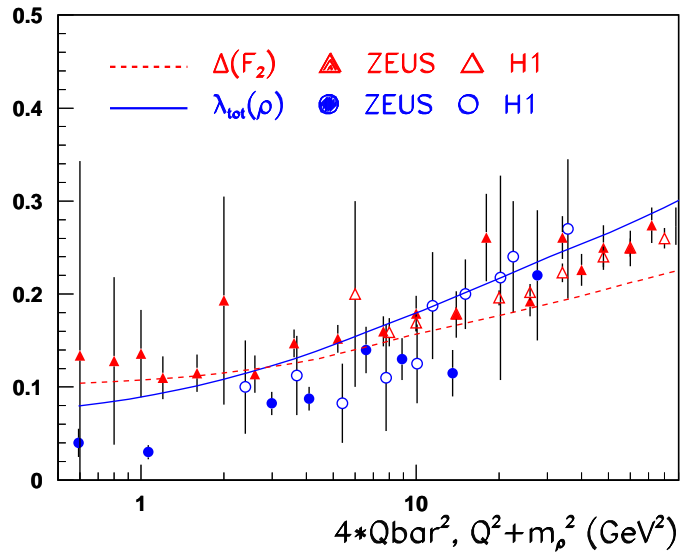}
   \caption{A compilation of the exponents $\lambda_{tot}^{\rho}(Q^2)=
{1\over 4}\delta(Q^2)$ 
 and $\Delta(\Qb^2)$ of Regge fits to the energy dependence of 
the $\rho$ photoproduction cross section, $\sigma_{\rho}(Q^2)
\propto (W^2)^{2\lambda_{tot}^{\rho}(Q^2)}$,
and of the proton structure function \protect\cite{epj:c7:609,pl:b520:183}, 
$F_{2p}(x,\Qb^2={1\over 4}
(Q^2+m_{\rho}^2)) \propto x^{-\Delta(\Qb^2)}$, 
plotted against $(Q^2+m_{\rho}^2)$, in comparison 
with the results from
the $k_{\perp}$-factorization model \protect\cite{IgorPhD,IgorNumerics}.}
   \label{fig:ivanov2004-delta-v-f2}
\end{figure}

We  repeatedly made a point that the  pQCD vacuum exchange is 
not an isolated Regge pole. Section 3.3
and Fig.~\ref{fig:intercepts_grv}
show how the effective exponent of the ${1\over x}$-dependence changes
from $\tau(\bkappa^2)$ for the unintegrated glue ${\cal
F}(x,\bkappa^2)$ to $\lambda(Q^2)$ for the integrated
glue $G(x,Q^2)$ to $\Delta(Q^2)$ for $F_{2}(x,Q^2)$ or
$\sigma^{\gamma^*p}_{tot}(x,Q^2)$. This entails the failure of
naive Regge factorization in a comparison of the energy dependence of
vector meson production and inclusive DIS, which was nicely demonstrated
by ZEUS collaboration \cite{levyDIS2002} in their study of the ratio 
\be
r^V_{tot} = {\sigma_{\gamma^* p \to Vp}(W^2,Q^2) \over
\sigma^{\gamma^*p}_{tot} (W^2,\Qb^2)}\,.
\label{eq:7.3.3.2}
\ee
The argument is as follows. According to Eqs. (\ref{eq:7.1.1}), (\ref{eq:7.1.2}),
\be
\sigma(\gamma^* p \to Vp) \propto (W^2)^{2\lambda_{tot}^{V}(Q^2)},\quad\quad
\lambda_{tot}^{V}(Q^2)= {1\over 4}\delta \approx \lambda(\Qb^2)-{\alpha_{\Pom}'\over b(W^2)}=\alpha_{\Pom}-{\alpha_{\Pom}'\over b(W^2)}-1\, ,
\label{eq:7.3.3.2*}
\ee
where the last form in terms of $\alpha_{\Pom}$ holds if the QCD Pomeron were an 
isolated Regge-pole. Notice, that $\lambda(Q^2)$ as defined in Sect. 3.3.1 
is different from the
exponent $\Delta(Q^2)$ in
\be
\sigma^{\gamma^*p}_{tot} \propto (W^2)^{\Delta(\Qb^2)}
=(W^2)^{\alpha_{\Pom}-1}\, ,
\label{eq:7.3.3.2**}
\ee
see a comparison in Fig.~10.
Such a  difference between the intercepts $\lambda(Q^2)$ and  $\Delta(Q^2)$ and their
substantial dependence on $Q^2$ shown in Fig.~10  already go beyond the rigors of the Regge 
theory. Still one can try to probe the vacuum exchange in the numerator and denominator of
$r_V$ equalizing the relevant hard scales, i.e., evaluating 
the ratio $r^V_{tot}$ with $\sigma^{\gamma^*p}_{tot} (W^2,\Qb^2)$ taken at  
$\Qb^2=(Q^2+m_V^2)/4$.  In Fig. \ref{fig:ivanov2004-delta-v-f2} we present 
$\lambda_{tot}^{V}(Q^2)$ and $\Delta(\Qb^2)$ plotted vs. common hard scale.
Only if one ignored
the difference between  $\lambda(Q^2)$ and  $\Delta(Q^2)$, the $W$-dependence
in (\ref{eq:7.3.3.2*}) and (\ref{eq:7.3.3.2**}) would be controlled by one and
the same $\alpha_{\Pom}$ and one would expect the $W$-dependence of the ratio 
$r^V_{tot}$ of the form 
\be
r^V _{tot}\propto (W^2)^{\eta},  \quad\quad \eta = \alpha_{\Pom}-{\alpha_{\Pom}'\over b(W^2)}-1\, .
\label{eq:7.3.3.3}
\ee

\begin{figure}[htbp]
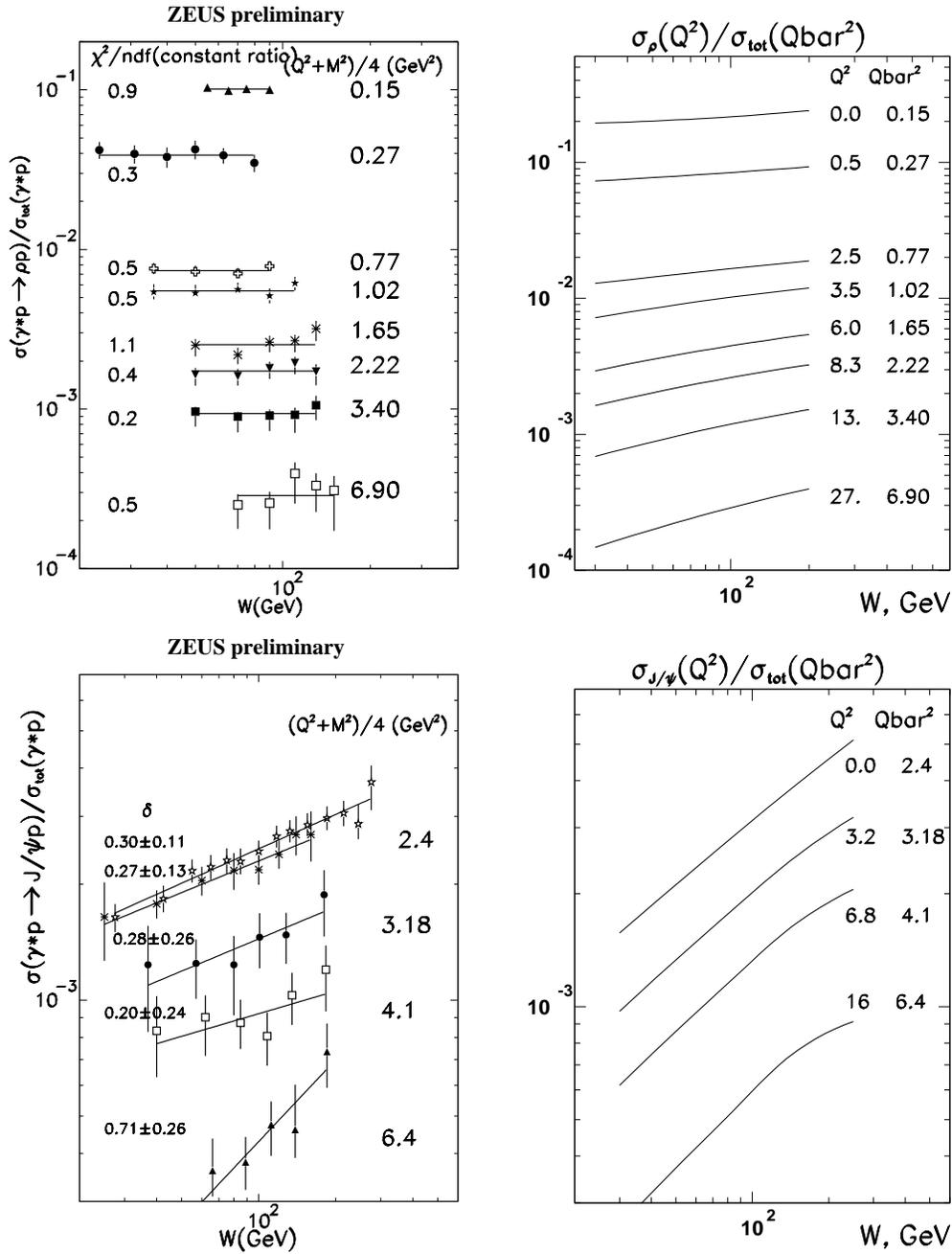

   \centering
   \epsfig{file=levy-rrho-data.eps,width=70mm}
   \epsfig{file=ivanov2004-levy-rho,width=65mm}
   \epsfig{file=levy-jpsi-data.eps,width=70mm}
   \epsfig{file=ivanov2004-levy-jpsi.eps,width=65mm}
   \caption{The ratio $r_V=\sigma_V(Q^2)/\sigma_{tot}(\Qb^2)$
for the $\rho$ production
as function of energy for several values of 
$\Qb^2={1\over 4}(Q^2+m_V^2)$: the lhs plots show 
(a) the experimental data on the
$\rho$ and $J/\psi$ production  \protect\cite{levyDIS2002},
the rhs plots show the corresponding theoretical expectations
for the energy dependence of $r_V$ \protect\cite{IgorNumerics}.
}
   \label{fig:LevyRhoRatio}
\end{figure}

The experimental results on $r^V_{tot}$ for the $\rho$ and $J/\Psi$ are
shown in the lhs plots of Fig.~\ref{fig:LevyRhoRatio}. They are consistent with 
little or no $W$-dependence for the $\rho$ production. Notice, that an
approximate constancy of $r_V$ for the $\rho$-production at small $Q^2$ is 
very much reminiscent of the familiar very weak 
energy dependence of the ratio $\sigma_{el}/\sigma_{tot}$
in $\pi N, KN, NN$ interaction, see plots in \cite{PDG2002}. Here the smallness
of the exponent $\eta$ is to a large extent due to the term
${\alpha_{\Pom}'/ b(W^2)}$ from the shrinkage of
the diffraction cone. The $W$-dependence of $r^V_{tot}$ for the $J/\Psi$ production
is substantial. 
This hard-scale and process-dependence of $\eta$ has been
interpreted as an evidence for the breaking of the Regge factorization.

The energy dependence of $r^V_{tot}$ expected from $k_t$-factorization is 
shown in the rhs  boxes of Fig.~\ref{fig:LevyRhoRatio}. It includes
the effect of shrinkage of the diffraction cone. The theoretical
results do correctly reproduces the trend of the experimental data shown in 
lhs boxes of Fig.~\ref{fig:LevyRhoRatio}. Although the ZEUS experimental
results for the $\rho$ mesons are consistent with $r^V_{tot}=const$, 
within the error bars they do not exclude the theoretically expected 
weak energy dependence shown in the rhs plot. The $k_{\perp}$-factorization
correctly describes the change of the energy dependence of 
$r^V_{tot}$ from the light $\rho$ to heavy $J/\Psi$.

\newpage


\section{The $t$-dependence and properties
of diffractive cone}


\subsection{Low-$t$: proton-elastic and
proton-dissociative events}

\begin{figure}[htbp]
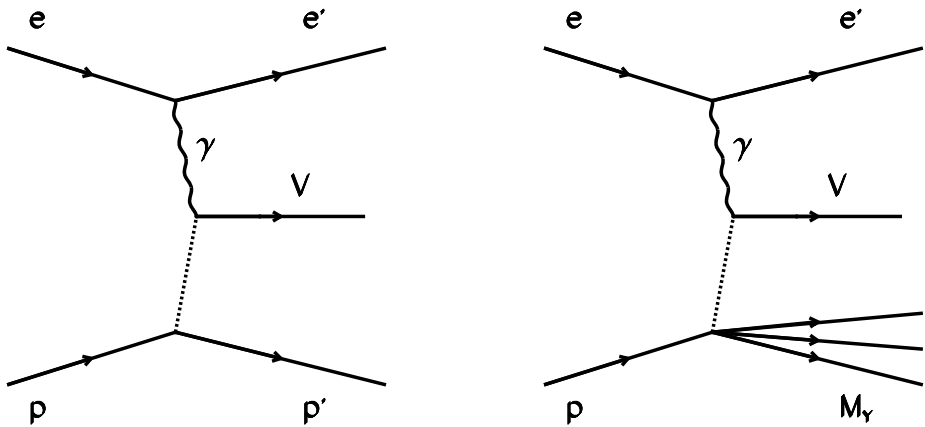

   \centering
   \epsfig{file=feyn_el.eps,width=70mm}
   \epsfig{file=feyn_pdiss.eps,width=70mm}
   \caption{\it Schematic diagram of proton-elastic
   (left) and proton-dissociative (right) vector-meson production
   in $ep$ interaction: $ep \rightarrow eVp$ and $ep \rightarrow eVY$ .}
   \label{fig:elpdissFd}
\end{figure}

The vector meson production 
differential cross section
$d\sigma / dt$ exhibits  a pronounced forward diffraction cone, which
spans up to $|t| \lsim 1.0$ GeV$^2$. Such a diffraction cone is 
familiar from hadronic scattering processes.  
Within diffraction cone, the differential cross section falls off
with $|t|$ approximately exponentially, see Sect,~\ref{sect3.1.3}
and Sect.~\ref{sect3.1.4}. In close similarity to hadronic
scattering, the dominant process here is the proton-elastic 
(hereafter just "elastic") production 
of ground state vector mesons $\gamma p \rightarrow V p$, 
see Fig.~\ref{fig:elpdissFd} left, where the label "elastic" 
is a reminder of the VDM relationship between the photoproduction
and $VP$ elastic scattering amplitude, recall Section 3.4.1.

At larger $t$ the elastic production dies out and the 
proton-dissociative reaction  $\gamma p \rightarrow V Y$ 
takes over. One can argue \cite{HoltmannDD} that the relative 
importance of the elastic and 
proton-dissociative reaction is precisely the same as
in proton-nucleus and proton-proton scattering described in 
Section 3.1.5. At small $t$ within the diffraction cone
the proton-dissociative production will be smaller than
the elastic production but still the biggest background contribution 
to the elastic vector meson production.

Experimentally, the direct separation of the two processes 
is only possible if the leading proton is measured in the Leading (Forward) Proton
Spectrometer (LPS, FPS) or if the hadrons from the proton-dissociative 
system $Y$
with a sufficiently high mass ($>3 \div 4$ GeV) are observed in the
forward part of the detector. Because the forward part of each
detector has a beam-pipe hole the smaller mass states $Y$ will 
just escape undetected.  
In the case when the proton or its excitation escapes through
the beam-pipe hole undetected, one needs to estimate the
portion of the proton-dissociative events
and subtract it from the visible cross section
based on the Monte Carlo modeling. Such a procedure
leads to sizeable systematic uncertainties.

\begin{figure}[htbp]
   \centering
   \epsfig{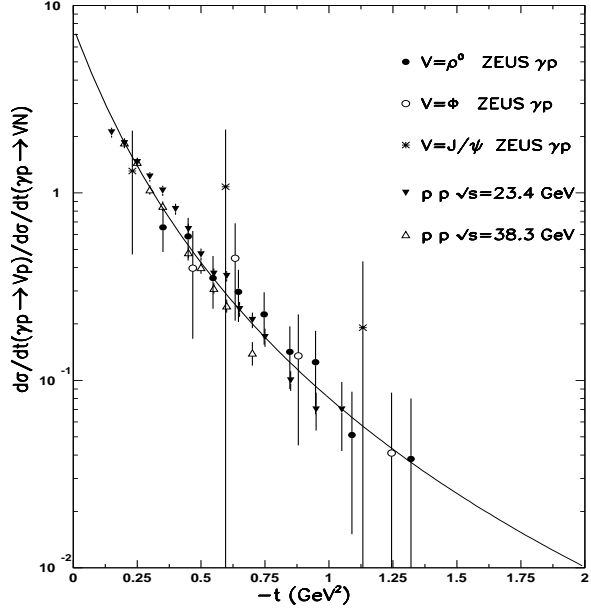}   \epsfig{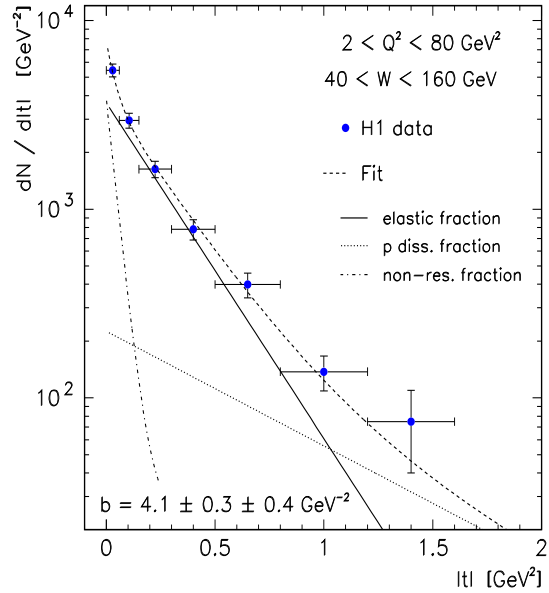}

   \caption{\it 
Upper plot shows 
the ratio of elastic to proton-dissociative differential cross
   sections as a function of $-t$ for vector-meson photoproduction
   ~\protect\cite{epj:c14:213}, together with data from $pp$
   reactions~\protect\cite{np:b108:1} at $\sqrt{s}=23.4$ and $38.3$~GeV. The
   curve is result of combined fit to all the data.
Bottom plot presents the 
$|t|$ distribution for the $J/\psi$ sample as measured by
   H1 ~\protect\cite{epj:c10:373}. The dashed line is result
   of a fit taking the background contributions into account. The full line
   corresponds to the elastic contribution assuming an exponential distribution.
   The contributions from proton dissociation and non-resonant background are
   shown separately.
}
   \label{fig:H1jpsi-tdep-decomp}
\end{figure}

If the Regge factorization decomposition (\ref{eq:3.1.3.2})
were exact, then the $t$-dependence from the $\gamma\to V$
transition vertex and from the $t$-channel exchange will 
cancel in the ratio
\be
Ratio(el/diss)=\frac{d\sigma ({\gamma p \rightarrow V p})}{dt} \Bigg/
\frac{d\sigma ({\gamma p \rightarrow VN})}{dt}\, ,
\label{eq:8.1.1}
\ee
which must be flavor independent and controlled by the change 
of the $t$-dependence from $p\to p$ to $p\to Y$ transition
\cite{HoltmannDD}. 
Furthermore, the Regge factorization predicts that $Ratio(el/diss)$
will not change from photoproduction of vector mesons to 
$pp$ scattering, 
\be
\left.Ratio(el/diss)\right|_{pp} =
\frac{d\sigma ({p p \rightarrow p p})}{dt} \Bigg/
\frac{d\sigma ({p p \rightarrow p N})}{dt}
\approx \left.Ratio(el/diss)\right|_{\gamma V}\, .
\label{eq:8.1.2}
\ee

In Fig.~\ref{fig:H1jpsi-tdep-decomp} the experimentally
measured ratio $Ratio(el/diss)$
is shown as a function of $t$ for $\rho$, $\phi$ and $J/\psi$
mesons photoproduced at HERA. 
The same ratio for the $pp$
reaction is shown for a comparison. All ratios coincide
within the errors, which is consistent with the hypothesis of
Regge factorization and lends a support to the separation
procedure. The elastic reaction production
dominates at $|t| \lsim 0.4-0.6 ~{\rm GeV^2}$, the proton-dissociative
production takes over at larger $|t|$. 
Notice a close similarity to a comparison of the elastic and 
nucleus-dissociative $pA$ scattering shown in 
Fig.~\ref{fig:pC12elasticdissociative}.
Figure~\ref{fig:H1jpsi-tdep-decomp} shows the typical interplay
of the $t$ dependence of the elastic and proton-dissociative 
production in electroproduction of  $J/\psi$
as measured by H1~\cite{epj:c10:373}.
One sees that the relative contribution of the
proton-dissociative events is increasing rapidly with $t$ making
the elastic measurement possible only at rather low $t$ values
(typically $|t| < 0.6 \div 1.0$ GeV$^2$). 

At still larger momentum transfers, $|t| \gg 1 ~{\rm GeV^2}$, 
the vector meson
production is dominated by proton-dissociative reaction. Besides
that, the production mechanism changes and the exponential
fall-off of the differential cross section is superseded by
the inverse powers of $|t|$ or $(m_V^2+|t|)$.



\subsection{Diffraction cone at low-$t$}


\subsubsection{The diffraction slope versus hard scale: theoretical expectations}

The $t$-dependence of the differential cross section as small-$t$
is usually parameterized in terms of the diffraction slope $b_V$, see Eq.
(\ref{eq:3.1.3.1}). The fitted values of $b_V$ depend slightly on
the range of $t$, the more refined parameterization 
$${d\sigma \over dt} \propto 
\exp(-b_V |t| - c|t|^2)
$$ with the curvature parameter $c$ allows one to
extend the 
fits of the experimental data up to $|t| \sim 1$ GeV$^2$.
The allowance for the curvature $c$ does not 
shift the value of fitted $b_V$ significantly. Several other
definitions of the effective slope can be encountered in
literature, e.g., the derivative of the logarithm of 
the differential cross section at $t=0$, 
$$
b_V={1\over \sigma}\cdot{d\sigma\over dt }\Big|_{t=0}\,,
$$
or, less often, the definition in terms 
of the average momentum transfer squared, $\langle |t|\rangle$,
$$ b_V={1 \over \langle|t|\rangle}\,,$$ 
they can differ from
$b_V$ defined by fits to Eq. (\ref{eq:3.1.3.1}) by $1 \div 2$ GeV$^{-2}$.

The decomposition (\ref{eq:3.1.3.2}) of the diffraction slope into
the target transition, beam transition and the $t$-channel
exchange components is exact in the simple Regge model. Similar 
decomposition holds also in the $k_{\perp}$-factorization and
color dipole approaches despite breaking of the strict
Regge factorization. In the photoproduction the key point is 
the shrinkage of the $q\bar{q}$ state of the photon from light to 
heavy quarks accompanied by the related decrease of 
the radius of the ground state vector meson, see Section 3.4.2, 
which entail the hierarchy of 
diffraction slopes
\be
 b_{J/\psi}^{el} < b_\phi^{el} < b_\rho^{el}\,.
\label{eq:8.2.1.1}
\ee

The $Q^2$-dependence of the diffraction slope is driven
by the decrease of the scanning radius $r_S$ with $Q^2$, see
Section~\ref{sect3.4.2}, Eq.~(\ref{eq:3.4.2.7}), and the $Q^2$-dependence
of the Regge shrinkage term through
the Regge parameter $W^2/(Q^2+m_V^2)$, see Section~\ref{sect4.5}, 
Eq.(\ref{eq:4.5.2}):
\be
b(Q^2) = b_0 + 2\alpha'_{\Pom}\log\left({x_0 W^2 \over Q^2 + m_V^2}\right) 
+ {A \over Q^2 + m_V^2}
\label{eq:8.2.1.2} 
\ee
For $x_0=8.3\cdot 10^{-4}$ as defined in Section 4.5, $b_0 \approx B_N$ 
is the approximately flavor-independent 
contribution from the proton. 
The shrinkage of the diffraction cone is for the
most part a property of the $x_g$-dependent skewed gluon
density, see Section 4.5. Besides that, because the
scaling violations do depend on $x_g$, the scanning radius
and hard scale $\Qb^2$ will exhibit slight energy dependence,
which shall affect the energy dependence of the differential
cross section \cite{NNPZZslopeVM,*NNPZZslopeVM1,IgorShrinkage}. The net
effect of such non-Regge corrections is a substantial reduction
of the observed $\alphappom$ from the input
$\alpha'_{BFKL}=0.25$ GeV$^{-2}$ in the 
parameterization (\ref{eq:4.5.3}).

According to Eq.~(\ref{eq:3.4.2.7}),
the diffraction slopes for different vector mesons should be 
close if taken at equal values of the scanning radius $r_S$ 
or equal values of $(Q^2 + m_V^2)$ \cite{NZZslope}. More
detailed analysis in \cite{NNPZZslopeVM,*NNPZZslopeVM1} has shown that
for the $J/\psi$ the diffraction slope is slightly smaller,
by $\sim 0.5$ GeV$^{-2}$, 
than for the $\rho$ at the same $(Q^2 + m_V^2)$. 
In addition, the arguments presented in Section~\ref{sect4.8}
suggest a somewhat larger scanning radius and larger diffraction
slope for the transverse amplitude ${\cal T}_{11}$ compared
with the longitudinal amplitude ${\cal T}_{00}$. The effect has
been suggested and evaluated for the first time in 
\cite{NNPZZslopeVM,*NNPZZslopeVM1}. Even for light vector mesons
at small $Q^2$ the expected difference is small, $ \sim 0.5\div 1$
GeV$^{-2}$, for heavy vector mesons the contribution
to the diffraction slope from the scanning radius
is small and the effect is negligible one.

The arguments of Section 3.1.5 are fully applicable to vector
meson production. For the proton dissociative photoproduction
one expects a substantially smaller diffraction slope
\be
 b^{diss}_V (Q^2)\approx  b_V^{el}(Q^2) -  B_N\,.
\label{eq:8.2.1.3}
\ee       
The principal point is that the difference of diffraction slopes
for the elastic 
and proton-dissociative reaction must be approximately
$Q^2$ and flavor independent.


\subsubsection{Experimental results: real photoproduction}

The experimental results on the diffraction slope $b$ measured
in photoproduction at HERA are summarized in Tab.~\ref{tab:bslope}.

\begin{table}[htbp!]

\begin{sideways}
\begin{minipage}[b]
{\textheight} \vspace{-0.5cm}

\begin{center}
\begin{tabular}{||c|c|c|c|c|c|c||}
\hline

Process & Value of slope~$b$,~GeV$^{-2}$ & $W$, GeV & $|t|, GeV^{2}$ & Exp. & Ref. \\
\hline\hline
 $\gamma p \rightarrow \rho p$ & $10.31 \pm 0.77(stat.) \pm 0.52(sys.)$ &
  25-70 & 0.073-0.45 & H1 & ~\protect\cite{cpaper:ichep2002:991} \\
 $ $ & $10.9 \pm 0.3(stat.) ^{+1.0}_{-0.5}(sys.) $ &
  50-100 & 0.-0.5 & ZEUS & ~\protect\cite{epj:c2:247} \\
 $ $ & $6.0 \pm 0.3(stat.) ^{+0.6}_{-0.3}(sys.) \pm 0.4(mod.)$ &
  85-105 & 0.4-1.2 & ZEUS & ~\protect\cite{epj:c14:213} \\
\hline
 $\gamma p \rightarrow \rho Y$ & $5.8 \pm 0.3(stat.) \pm 0.5(sys.) $ &
  50-100 & 0.025-0.5 & ZEUS & ~\protect\cite{epj:c2:247} \\
 $ $ & $2.4 \pm 0.2(stat.) ^{+0.2}_{-0.1}(sys.) \pm 0.3(mod.)$ &
  85-105 & 0.4-1.2 & ZEUS & ~\protect\cite{epj:c14:213} \\
\hline\hline
 $\gamma p \rightarrow \phi p$ & $7.3 \pm 1.0(stat.) \pm 0.8(syst.)$ &
  60-80 & 0.1-0.5 & ZEUS & ~\protect\cite{pl:b377:259} \\
 $ $ & $6.3 \pm 0.7(stat.) \pm 0.6(sys.) \pm 0.3(mod.)$ &
  85-105 & 0.4-1.2 & ZEUS & ~\protect\cite{epj:c14:213} \\
\hline
 $\gamma p \rightarrow \phi Y$ & $2.1 \pm 0.5(stat.) \pm 0.3(sys.) \pm 0.4(mod.)$ &
  85-105 & 0.4-1.2 & ZEUS & ~\protect\cite{epj:c14:213} \\
\hline\hline
 $\gamma p \rightarrow J/\psi p$ & $4.73 \pm 0.25(stat.) ^{+0.30}_{-0.39}(sys.)$
  & 40-150 & 0-1.2 & H1 & ~\protect\cite{pl:b483:23} \\
$ $& $4.99 \pm 0.13(stat.) ^{\pm 0.45}(sys.)$
  & 40-150 & 0.07-0.9 & H1 & ~\protect\cite{pl:b541:251} \\
 $ $ & $4.15 \pm 0.05(stat.) ^{+0.30}_{-0.18}(sys) $ &
  20-290 & 0-1.2 & ZEUS & ~\protect\cite{epj:c24:345} \\
 $ $ & $4.0 \pm 1.2(stat.) ^{+0.7}_{-1.1}(sys) ^{+0.4}_{-0.6}(mod.)$ &
  85-105 & 0.4-1.2 & ZEUS & ~\protect\cite{epj:c14:213} \\
\hline
 $\gamma p \rightarrow J/\psi Y$ & $0.7 \pm 0.4(stat.) \pm 0.2(sys.)^{+0.5}_{-0.3}(mod.)$ &
  85-105 & 0.4-1.2 & ZEUS & ~\protect\cite{epj:c14:213} \\
$ $ & $1.07 \pm 0.03(stat.) \pm 0.11(sys.)$ &
  40-150 & 0.15-3 & H1 & ~\protect\cite{pl:b541:251}\\
\hline \hline
 $\gamma p \rightarrow \Psi(2S) p$ & $4.69 \pm 0.57(stat.) \pm 0.46 $
  & 40-150 & 0.07-0.9  & H1 & ~\protect\cite{pl:b541:251} \\
\hline
 $\gamma p \rightarrow \Psi(2S) Y$ & $0.59 \pm 0.13(stat.) \pm 0.12(sys.)$ &
  40-150 & 0.15-3 & H1 & ~\protect\cite{pl:b541:251}\\
\hline \hline
\end{tabular}
\caption{\it  Recent HERA measurements of the $b$ slope for PHP
elastic and proton-dissociative vector-meson production.}
\label{tab:bslope}
\end{center}

\end{minipage}
\end{sideways}

\end{table}

The values of diffraction slopes depend on the $t$-region, where the 
fits are performed, which is familiar from $\pi p, pp$ scattering
\cite{Burq,Schiz}. The flavor dependence is consistent with the
theoretical expectations described in Section 8.2.1: the diffraction 
slope rises with the size of the meson with the exception of 
the $\Psi(2S)$. The charmonium models \cite{NovikovPhysRep,QuiggPhysRep}
suggest the radius of the $\Psi(2S)$ as large as the radius of the $\phi$, 
so that the diffraction slope of elastic $\Psi(2S)p$ scattering 
would be the same as in $\phi p$ elastic scattering. Then in the
naive VDM one would expect $b_{\psi(2S)}\approx b_{\phi}$. In
contrast to that, the counterintuitive 
\be
b_{\Psi(2S)} - b_{J/\Psi} \approx -0.5~{\rm GeV^{-2}}
\label{eq:8.2.2.1}
\ee
was
predicted in \cite{NNPZZslopeVM,*NNPZZslopeVM1}  on the basis of node
effect \cite{NNNcomments}. The H1 results for both elastic
and proton-dissociative  $\Psi(2S)$ production
\cite{pl:b541:251} are consistent with this prediction.
Within the
experimental error bars $\Delta b  =
b^{el} - b^{diss} \approx (4.5\div 5) ~{\rm GeV}^{-2}$ 
is flavor independent which 
agrees perfectly with the theoretical
expectation (\ref{eq:3.1.5.6}), (\ref{eq:8.2.1.3}) and
must be contrasted to a strong
flavor dependence of $b_{el}$. 

Notice a very small $b_{diss}(\gamma p \rightarrow J/\psi Y)$,
which is consistent with the equally small slope for
the double-dissociative hadronic diffraction $pp\to XY$, for the
compilation of the hadronic data from the fixed target to
ISR energies see \cite{C75,C80}.


\subsubsection{Experimental results: $Q^2$-dependence of the 
diffraction slope}

\begin{figure}[htbp]
   \centering
   \epsfig{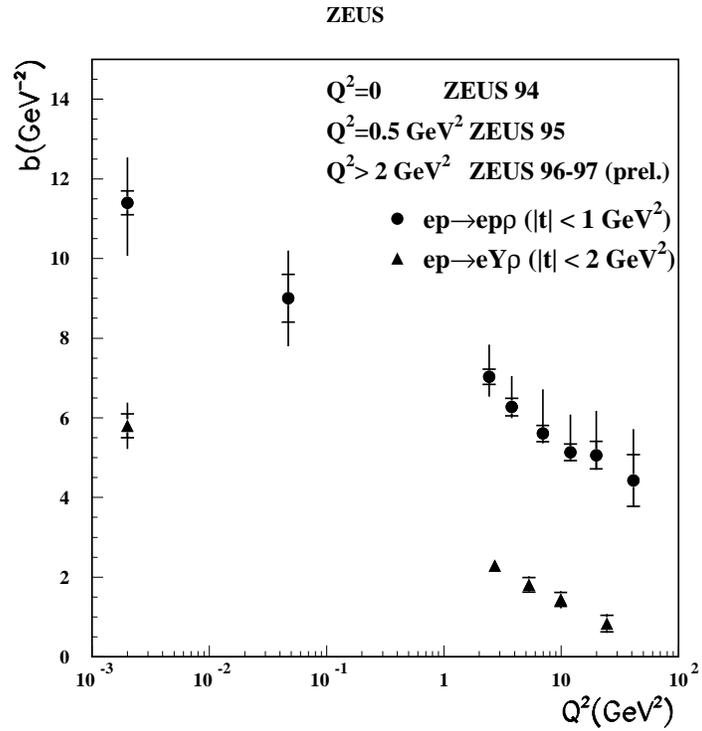}
   \hspace*{-0.5cm}
   \epsfig{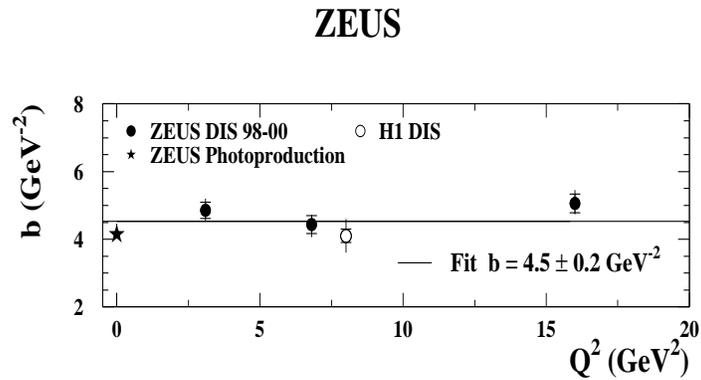}
   \caption{\it The preliminary ZEUS data on diffraction slope $b$ for 
elastic and proton-dissociative production of
the   $\rho$ ~\protect\cite{cpaper:ichep2002:818}) upper plot. Bottom plot shows
the recent ZEUS \protect\cite{np:b695:3} and H1
 \protect\cite{epj:c10:373} data for   elastic production of the $J/\psi$ 
 mesons as a function
   of $Q^2$.}
   \label{fig:ZEUSbslopeQ2}
\end{figure}

 Figure~\ref{fig:ZEUSbslopeQ2} shows the $Q^2$-dependence of
the diffraction slopes $b_\rho(Q^2)$ and  $b_{J/\psi}(Q^2)$. 
In the case of 
the $\rho$-meson, a strong $Q^2$ dependence of the $b_\rho$ is observed:
$b \approx 11$ GeV$^{-2}$ in the photoproduction limit and
drops down to $4\div 5$ GeV$^{-2}$ in the hard
electroproduction. In striking contrast, in the case of $J/\psi$
production the $Q^2$-dependence is very weak or absent, 
$b_{J/\psi} \approx 4.5$ GeV$^{-2}$ was found throughout
the entire $Q^2$ range.

Figure~\ref{fig:ZEUSbslopeQ2} shows also the comparison between the
elastic and proton-dissociative slopes in $\rho$ meson production.
One clearly sees an approximately $Q^2$-independent 
difference $\Delta b = 
b_\rho^{el}(Q^2) - b^{diss}_\rho
\approx 4\div 5$ GeV$^{-2}$ between the two data sets, which 
coincides with the
photoproduction value shown in Tab.~\ref{tab:bslope}. 
This is an important confirmation of the theoretical expectation
(\ref{eq:8.2.1.3}). 

\begin{figure}[htbp]
   \centering
   \epsfig{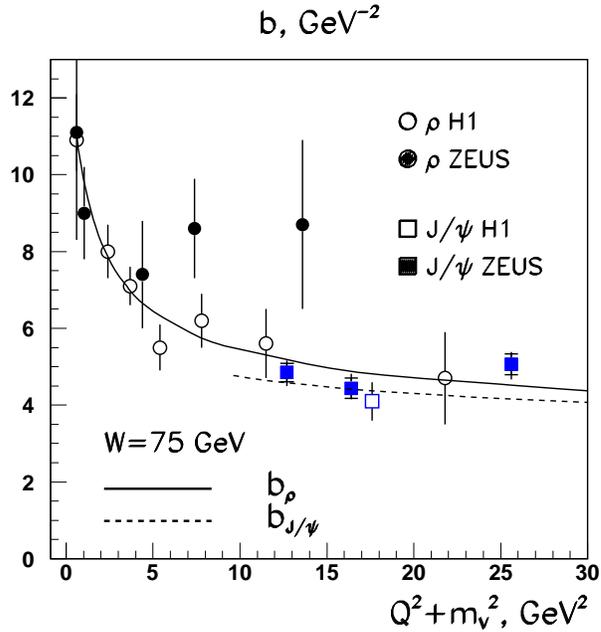}
   \caption{\it The diffraction slope $b$ of elastic production of
  the  $\rho$~(H1:\protect\cite{epj:c13:371},~
ZEUS: \protect\cite{epj:c2:247,epj:c6:603}, 
the data from ~\protect\cite{cpaper:ichep2002:818} are not shown) and
   $J/\psi$~(H1:\protect\cite{epj:c10:373}, 
ZEUS:\protect\cite{np:b695:3})  mesons as a function
   of $Q^2+M_V^2$. The predictions of from the $k_t$-factorization 
approach \protect\cite{IgorPhD,IgorNumerics}
   are shown.}
   \label{fig:ivanov-slope-rhojpsi}
\end{figure}

Figure~\ref{fig:ivanov-slope-rhojpsi} shows the above results for
elastic slopes $b_\rho^{el}(Q^2)$ and $b_{J/\psi}^{el}(Q^2)$ 
plotted as a function of $(Q^2+m_V^2)$. The strong flavor
dependence obvious if one compares the photoproduction values
$b_\rho^{el}(Q^2=0)$ at $b_{J/\psi}^{el}(Q^2=0)$ is dramatically
reduced in the variable  $(Q^2+m_V^2)$. The theoretical curves
are from the $k_t$-factorization model \cite{IgorNumerics}.
Because of flavor dependent departure of the hard scale $\Qb^2$
from the simple estimate (\ref{eq:1.3.5}), the theoretical
predictions for the $J/\psi$ are slightly lower than for the 
$\rho$. Similar results were found earlier in the color dipole
model \cite{NNPZZslopeVM,*NNPZZslopeVM1,NemchikSlopeVM}. The origin of the weak $Q^2$-dependence
for the $J/\Psi$ is in a very small contribution to 
$b_{J/\psi}^{el}(Q^2)$
from $b_{\gamma^* V}(Q^2) \propto r_S^2 \propto 1/(Q^2+m_V^2)$,
see Eqs. (\ref{eq:3.4.2.7}), (\ref{eq:8.2.1.2}).

The similar approximate $(Q^2+m_V^2)$ scaling holds for the
proton-dissociative reaction: the results for
$b_{\rho}^{diss}$ for the largest values of $Q^2$ in
Fig.~\ref{fig:ZEUSbslopeQ2} are perfectly consistent
with the photoproduction value for $b_{J/\Psi}^{diss}$
from Tab.~\ref{tab:bslope}.




\subsection{Shrinkage of the diffraction cone and
the Pomeron trajectory}


\subsubsection{The $W$-dependence in photoproduction}

\begin{figure}[htbp]

   \centering

   \epsfig{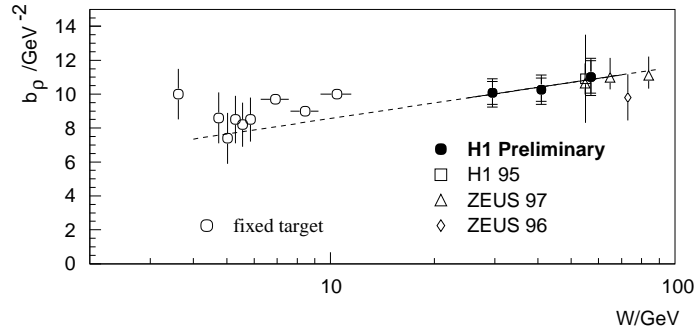}
   \caption{\it The diffraction slope $b_{\rho}$ as a function of $W$ for
   $\rho$ photoproduction ~\protect\cite{cpaper:ichep2002:991}.
   The continuous line shows the result of the fit of the form
   (\protect\ref{eq:3.1.3.4}) to the recent H1 measurement,
   other HERA and fixed target measurements are shown for comparison;
   the extrapolation of the fit
   to the low $W$ region is indicated by the dashed line. 
  }
   \label{fig:slopePHPrho}
\end{figure}

\begin{figure}[htbp]
   \centering
   \epsfig{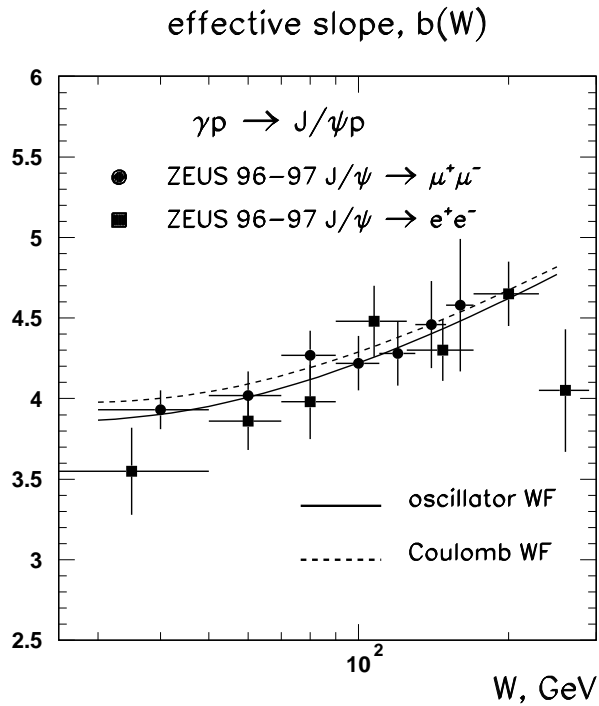}
   \caption{\it The slope $b_{J/\psi}$ as a function of $W$ for $J/\psi$
   photoproduction as measured by ZEUS ~\protect\cite{epj:c24:345}.
   The results of the $k_t$-factorization calculations  
   \protect\cite{IgorPhD,IgorNumerics}
   based on the oscillator
   (solid line) and Coulomb (dashed line) wave functions
   are compared with the data.}
   \label{fig:slopePHPJpsi}
\end{figure}

Figures ~\ref{fig:slopePHPrho} and ~\ref{fig:slopePHPJpsi} show the
energy dependence of the diffraction slope $b$ for the $\rho$ and
$J/\psi$ photoproduction, respectively. One sees a steady growth
of the diffraction slope, i.e., the shrinkage of diffractive cone.
When parameterized in terms of (\ref{eq:3.1.3.4}), the $\rho$ 
photoproduction data
yield $\alphappom = 0.23 \pm
0.15(stat.)+^{0.10}_{0.07}(syst.)$~GeV$^{-2}$ ~\cite{epj:c2:247}.
This result for the soft reaction is consistent with $\alpha_{\Pom}'
=0.25$ GeV$^{-2}$ found in the simple Regge pole description
of elastic $pp$ scattering. The $J/\Psi$ photoproduction must be
regarded as a hard reaction, here the evidence for nonvanishing
$\alphappom$ is even stronger. Fitting to the form
$b=b_0(90 GeV) + 4 \alphappom ln(\frac{W}{90})$ gave $\alphappom = 0.116 \pm
0.026(stat.)^{+0.010}_{-0.025}(syst.)$~GeV$^{-2}$
~\cite{epj:c24:345}.

\subsubsection{The Pomeron trajectory}

In addition to fitting the differential cross section $d\sigma/dt$ at
given $W$ and evaluating the diffraction slope $b$ as a function of $W$, 
one can study the $W$-dependence of the differential cross sections at 
fixed  $t$. According to Eq.~(\ref{eq:3.1.1.2}) the $W$-dependence
of the differential cross section is $\propto W^{4(\alphapom(t)-1)}$,
so that one can extract $\alphapom$ versus $t$ , i.e. measure the effective
trajectory $\alphapom(t)$ of the $t$-channel vacuum exchange.

\begin{figure}[htbp]
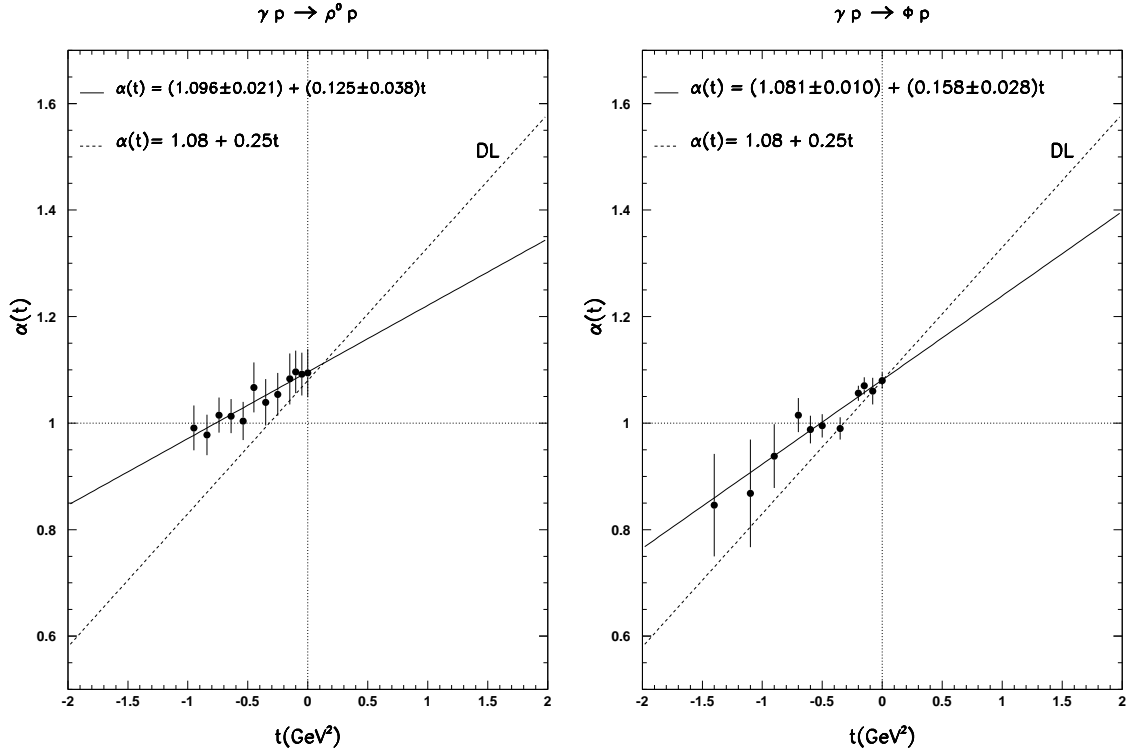

   \centering
   \epsfig{file=ZEUSHighTRhoPhiJpsi1999pub_fig22.eps,width=75mm}
   \epsfig{file=ZEUSHighTRhoPhiJpsi1999pub_fig24.eps,width=75mm}
   \caption{\it Determination of the Pomeron trajectory from the reactions
   $\gamma p \rightarrow \rho^0 p$ and $\gamma p \rightarrow \phi p$.
   The dots are results of ZEUS measurements ~\protect\cite{epj:c14:213}.
   The solid lines are results of the linear fit. The DL 
parameterization for the soft Pomeron trajectory 
\protect\cite{DLsigtot} is shown for
   comparison as a dashed line.}
   \label{fig:trajectoryPHP1}
\end{figure}

\begin{figure}[htbp]
   \centering
   \epsfig{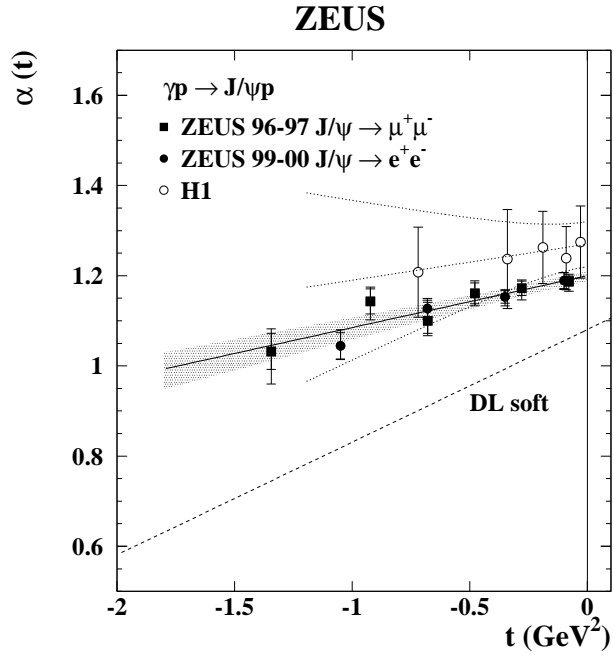}
   \epsfig{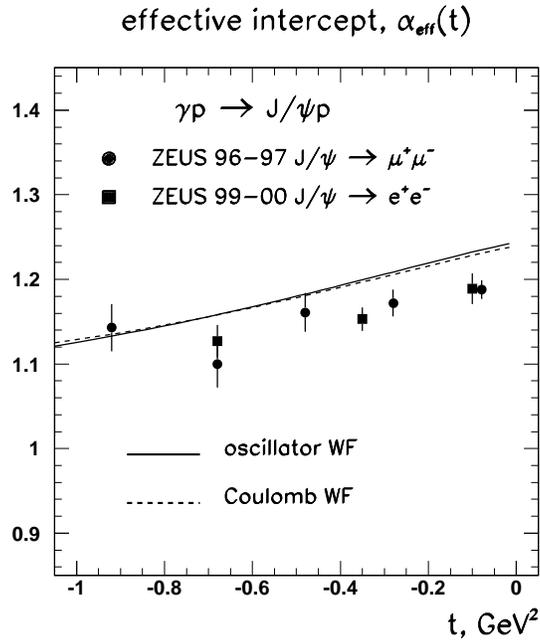}
   \caption{\it Pomeron trajectory as obtained from $J/\psi$
   photoproduction measurements by ZEUS ~\protect\cite{epj:c24:345}.
   The results of H1 measurements ~\protect\cite{pl:b483:23} are
   shown for comparison.
   Linear fits to the ZEUS and H1 data are shown and compared with the
   DL parameterization for the soft Pomeron trajectory 
\protect\cite{DLsigtot}. Bottom. The predictions from the $k_t$-factorization 
approach
    \protect\cite{IgorPhD,IgorNumerics} are compared with the ZEUS data.}
   \label{fig:trajectoryJpsi}
\end{figure}

\begin{figure}[htbp]
   \centering
   \epsfig{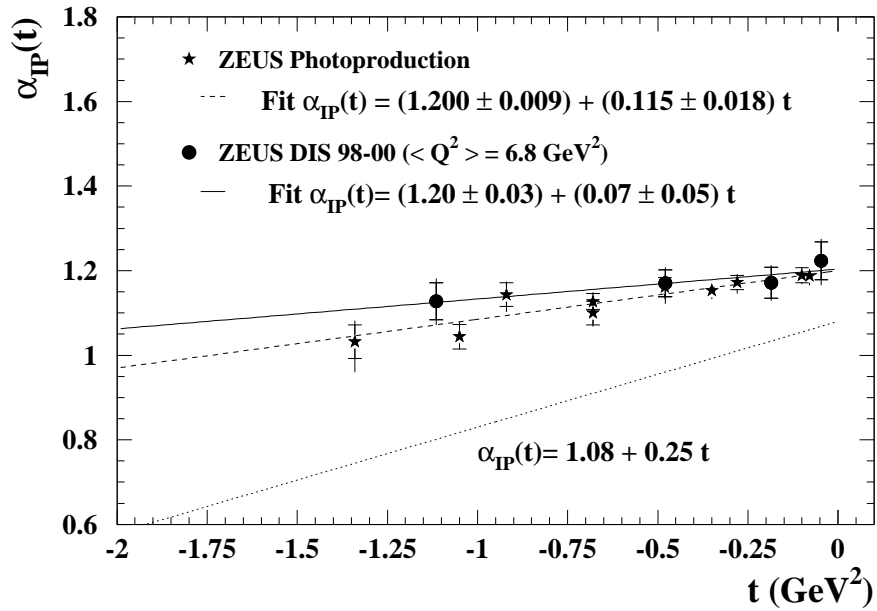}
   \caption{\it A comparison of the Pomeron trajectory extracted form
the ZEUS data on real photoproduction and electroproduction of the 
$J/\Psi$~\protect\cite{epj:c24:345,np:b695:3}. The solid lines are results of the linear fit. The 
Donnachie-Landshoff parameterization \protect\cite{DLsigtot} of the
soft Pomeron trajectory is shown for
   comparison as a dashed line.}
   \label{fig:ZEUSPomerontrajectory}
\end{figure}

Figures ~\ref{fig:trajectoryPHP1}, ~\ref{fig:trajectoryJpsi}
demonstrate the $\alpha_{\Pom}(t)$ measured by ZEUS in the
photoproduction of the $\rho$, $\phi$ and $J/\psi$ mesons. The
obtained parameters of the effective Pomeron trajectories are:
\bea
&&\alpha_{\Pom}(\rho;t) = (1.096 \pm 0.021) + (0.125 \pm 0.038)\cdot t\,;
\nonumber\\
&&\alpha_{\Pom}(\phi;t) = (1.081 \pm 0.010) + (0.158 \pm 0.028)\cdot t\,;\\
&&\alpha_{\Pom}(J/\Psi; t) = (1.200 \pm 0.009 \pm {}^{+0.004}_{-0.010}) + (0.115
\pm 0.018 \pm {}^{+0.008}_{-0.015})\cdot t\quad\,.\nonumber 
\label{eq:8.3.2.1} 
\eea 
which must be compared with the DL parameterization 
(\ref{eq:3.1.4.11}).
For all the vector mesons the determined slope of 
the vacuum trajectory is significantly non-zero. Within the 
error bars, the value of $\alphappom$ appears 
to be insensitive to the particular type of the vector meson.

\begin{figure}[htbp]
   \centering
   \epsfig{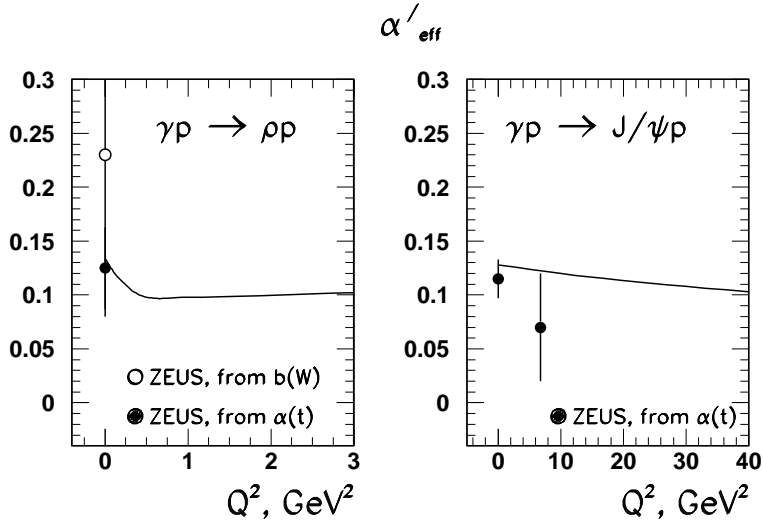}

   \vspace*{-1cm}
   \caption{\it The $k_t$-factorization model predictions  
\protect\cite{IgorPhD,IgorNumerics}
for
the $Q^2$-dependence of the slope of the 
effective Pomeron trajectory, $\alphappom$, for the $\rho$~
\protect\cite{epj:c2:247,epj:c14:213}
and $J/\psi$ ~\protect\cite{epj:c24:345,np:b695:3} production. 
The results for $\alphappom$ obtained from the study of cone shrinkage $b(W)$
and from the trajectory analysis $\alpha_{\Pom}(t)$ are shown separately by
open and filled dots, correspondingly.}
   \label{fig:ivanov-shrinkageQ2}
\end{figure}

The dependence of the extracted Pomeron trajectory on $Q^2$
has been studied by ZEUS for the case of $J/\Psi$ production. 
Figure~\ref{fig:ZEUSPomerontrajectory} shows comparison between the
Pomeron trajectory at $Q^2 = 0$ and $Q^2 = 6.8$ GeV$^2$. The 
$k_{\perp}$-factorization predicts a slight increase of
the effective intercept with $Q^2$, see Fig.~\ref{fig:VMintercepts}.
 Even if $\alpha'_{BFKL}$ in the 
parameterization (\ref{eq:4.5.3}) were a constant --- the
solutions of the color dipole BFKL equation \cite{NZZslopePisma}
for the 
diffraction slope exhibit weak $\bkappa^2$-dependence of 
$\alpha_{BFKL}'$, though \cite{NZZslope}, --- the above 
described extraction  will yield a
weakly $Q^2$- and flavor-dependence values of 
$\alphappom$. Such a  non-Regge effects in
the extracted $\alphappom$ have been
evaluated in \cite{IgorShrinkage,IgorNumerics} and
demonstrated in Fig.~\ref{fig:ivanov-shrinkageQ2}.
Because of the same non-Regge effects, the theoretical 
results for the effective vacuum trajectory are sensitive
to the wave function of the vector meson. The effect 
can be seen from at the bottom plot of Fig.~\ref{fig:trajectoryJpsi},
it is negligible for all the practical purposes.

\subsection{The gluon-probed radius of the proton
and the Pomeron exchange radius}

The above presented experimental data on the diffraction
slope for the $J/\Psi$ production give
\be
B_N = b_{2G} \approx (4\div 4.5)~{\rm GeV^{-2}}\, .
\label{eq:8.4.1}
\ee
It is the non-perturbative parameter which defines the
form factor of the proton probed by color-singlet
two-gluon current, $G_{2G}(t)\approx \exp({1\over 2}b_{2G}t)$.
It must be compared to the electromagnetic form factor 
$G_{em}(t)$, the familiar dipole parameterization for
which, $G_{em}(t) \propto 1/(\Lambda^2-t)^2$ with 
$\Lambda^2 \approx 0.7$ GeV$^2$ amounts to
\be
B_{em} \approx {4\over \Lambda^2} \approx (5.5\div 6) ~{\rm GeV^{-2}}\, .
\label{eq:8.4.2}
\ee
The departure of $b_{2G}$ from $B_{em}$ must be regarded 
as substantial one. The photon only couples to charged 
partons in the proton, whereas the gluonic form factor is sensitive
to the distribution of all color charges - the (anti)quarks and gluons
-  in the proton. Should one interpret the
observed inequality $b_{2G} < B_{em}$ as an evidence for the gluonic lump
in the center of the proton?  

The same problem is present in $\pi p,pp$ scattering. For
instance, Schiz
et al. find that at 200 GeV the observed $t$-dependence of scattering
amplitudes is very well reproduced by the product of the
charge form factors of the beam and target particles \cite{Schiz}. One would
interpret that as an equality of the strong interaction and electromagnetic
radii, $b_{2G}\approx B_{em}$, at this particular energy. However, 
in hadronic scattering the shrinkage of the diffraction cone is
rapid, and at lower energies $\sim 10$ GeV, still
in the applicability domain of the Regge model, both $b_{\pi p}$
and $b_{pp}$ are substantially lower than at 200 GeV, and one 
runs into the same problem with $b_{2G} < B_{em}$.

A very small value of the gluon radius of the proton, as a matter of
fact, $b_{2G} \ll B_{em}$, has been
found by Braun et al. from the QCD sum rules \cite{Mankiewicz2G}.
The QCD sum rule results depend on the interpolating field for
the nucleon, though. Braun et al. take the $qqqg$ operator, whether
the radius evaluated for the hybrid higher Fock component of the 
proton applies to the whole proton or not remains an open issue.

Another small parameter which emerges in small-$t$ diffraction 
is the Pomeron exchange radius which is probed either in the 
double double proton-dissociative $pp$ scattering or in 
proton-dissociative vector meson production at small scanning
radius $r_S$ such as $\gamma p \to J/\Psi Y$ of the 
$\rho$ production in large-$Q^2$ DIS, $\gamma^* p \to \rho Y$: 
\be
b_{\Pom} \approx  b(pp \to XY) \sim b(\gamma p \to J/\Psi Y)
\sim b(\gamma^* p \to \rho Y) \sim 1~{\rm GeV^{-2}}\, . 
\label{eq:8.4.3}
\ee
The latter two reactions have a hard scale on the vector meson 
side, and one may link the anomalously small $b_{\Pom}$ to the
small propagation radius of perturbative gluons $R_c \sim 1$ GeV$^{-1}$.
The $pp$ double diffraction reaction probes the soft Pomeron,
why do the soft and hard Pomeron exchanges have equally small
exchange radius is an open issue.


\subsection{Beyond the diffraction cone: large $|t|$ as a hard scale}


\subsubsection{Large-$t$ vector mesons as a Mueller-Navelet
isolation of the hard BFKL exchange}

The basics of the pQCD treatment, and helicity properties, 
of large-$t$ vector meson production were reviewed in Sections 4.10
and 5.4. Here we focus on the $t$- and $W^2$-dependence of 
the large-$t$ cross section.

On the theoretical side the large-$|t|$ production of vector
mesons is a very promising testing
ground for ideas on the BFKL Pomeron, because the large momentum
transfer $\bDelta$ flows along the BFKL Pomeron from the 
target to the projectile. There is a close analogy to the
long sought Mueller-Navelet isolation of the hard BFKL
exchange by selecting DIS events with one hard jet in the
target fragmentation region and the second hard jet in the 
photon fragmentation region \cite{MuellerNavelet}. 
Within the effective parton 
description of the proton-dissociative $\gamma p \rightarrow VY$
in terms of the elastic scattering $\gamma q \rightarrow Vq'$
the recoil quark (antiquark, gluon) with large transverse
momentum $\bDelta$ gives rise to the target hard jet of Mueller
\& Navelet, whereas the large-$t$ vector meson is a substitute
for the Mueller-Navelet hard jet in the photon fragmentation
region. The vector meson production is even more advantageous
because one has an access to larger rapidity gaps between the
vector meson and the system $Y$ than it would be possible
in the Mueller-Navelet two-jet process. 

The pQCD two-gluon approximation misses the dependence on
the rapidity gap and the total normalization must be adjusted 
to the experimental cross section. Otherwise, as we shall
see below, it is doing a reasonable job on the $t$-dependence.
The state of the art BFKL based calculations use Lipatov's
leading order fixed-$\alpha_S^{BFKL}$ approximation for the 
unintegrated off-forward gluon density. By the logics of the 
calculation one may expect
\be
\alpha_S^{BFKL}\sim \alpha_S(|t|)
\label{eq:8.5.1.1}
\ee
and expect the Regge energy dependence (\ref{eq:4.11.6})
with the BFKL 
trajectory given by Eq.~(\ref{eq:3.2.2}).


\subsubsection{Theoretical expectations for flavor 
dependence at large-$t$}

\begin{figure}[htbp]
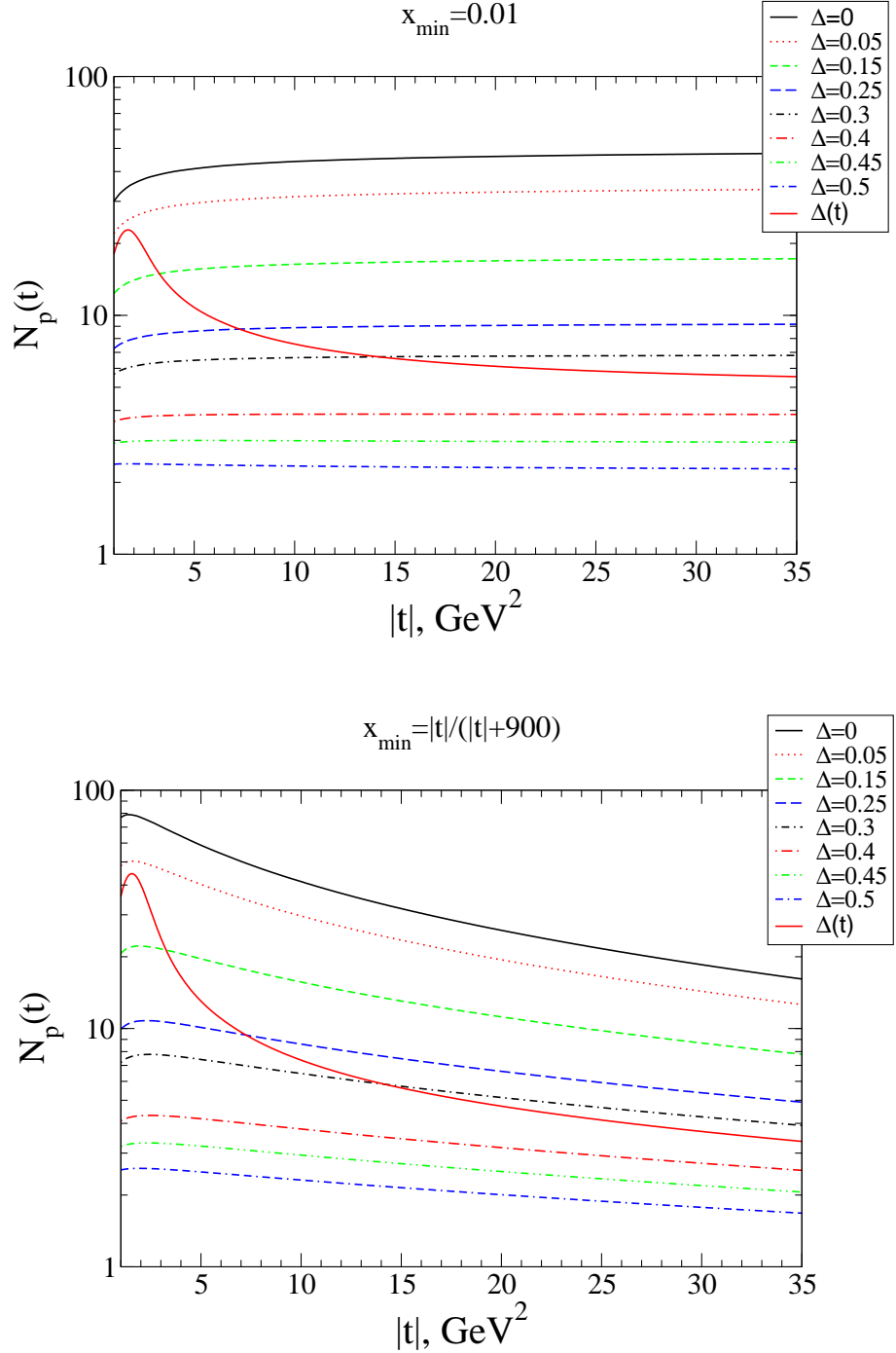

   \centering
 \epsfig{file=EffectivePartons1.eps,width=120mm} \vspace{1cm}\\
\epsfig{file=EffectivePartons2.eps,width=120mm}
  \caption{\it The $t$-dependence of the effective number of 
partons $N_{p}(t)$ for the ZEUS (top box)
and the H1 (bottom box) kinematical cuts for different values
of the intercept $\alpha = 1+\Delta$ \protect\cite{Fedya}. Shown also are
the curves for the $t$-dependent trajectory 
$\alpha_{\Pom}(t)$ parameterized by Eq.~\protect\ref{eq:8.5.3.2}
and shown in Fig.~\protect\ref{fig:EffectiveInterceptLarget}. }
\label{fig:EffectivePartons}
\end{figure}

A crude reinterpretation of very involved theoretical calculations 
\cite{ForshawRyskinLarget,GinzburgIvanovHight,BartelsHight,ForshawHight,PoludniowskiHight}
starting with the pQCD subprocess $\gamma^* q \to V q'$ is as follows:  

First, the relationship between the cross section of the 
theoretical partonic subprocess 
$\gamma q \to V q'$ and the experimentally observed
$\gamma p \to V Y$ involves the effective number of partons
in the proton,
\be
 N_p(t) = \int^{1}_{x_{min}} dx' 
(x')^{2(\alpha_{\Pom}(t)-1)}\left({81 \over 16}g(x',t) + 
\sum_{f} [q(x',t)+\bar{q}(x',|t|)]\right)\, ,
\label{eq:8.5.2.01}
\ee  
where the Regge dependence on $x'$ is reabsorbed into the
flux of equivalent partons. After $N_p(t)$ is factored out,
one obtains the cross section of the partonic subprocess
at a fixed energy $W_{\gamma q}$:
\be
\left. {d\sigma(\gamma q \to V q') \over dt}
\right|_{W_{\gamma q}=W_{\gamma p}} =
{1\over N_p(t)}\cdot {d\sigma(\gamma p \to V Y) \over dt}.
\label{eq:8.5.2.02}
\ee  
The
cut $x_{min}=0.01$ used by ZEUS collaboration \cite{epj:c26:389}
gives $N_p(t)$ shown in Fig.~\ref{fig:EffectivePartons}. H1 
collaboration \cite{pl:b568:205} imposes
the cut $M_Y^2 < M_{max}^2 = 900~{\rm GeV^2}$, which translates
into 
\be
x_{min} = {|t| \over M_{max}^2+|t|}\, .
\label{eq:8.5.2.03}
\ee
Evidently, for $\alpha_{\Pom}(t)=1$ the number of partons diverges
as $x_{\min}\to 0$ and at $x_{min}=0.01$ it exhibits rise with
$|t|$ because of the scaling violations in the gluon density, 
see Fig. \ref{fig:EffectivePartons}. 
In the opposite to that, for the H1 cut (\ref{eq:8.5.2.03})
the decrease of $N_{p}$ with $|t|$ is driven by the rise of
$x_{min}$, see Fig.~ \ref{fig:EffectivePartons}. The sensitivity
of $N_{p}$ to the trajectory $\alpha_{\Pom}(t)$ is very strong:
for $\Delta_{\Pom}=\alpha_{\Pom}(t)-1=0$ it is a true number of
partons, is very large and is a steep function of $x_{min}$,
for $\Delta_{\Pom}=0.5$ the integral
(\ref{eq:8.5.2.01}) is reminiscent of the momentum sum rule
integral and yields weakly $t$-dependent $N_{p} \approx 2.5$. see
Fig. \ref{fig:EffectivePartons}.

\begin{figure}[htbp]
   \centering
   \epsfig{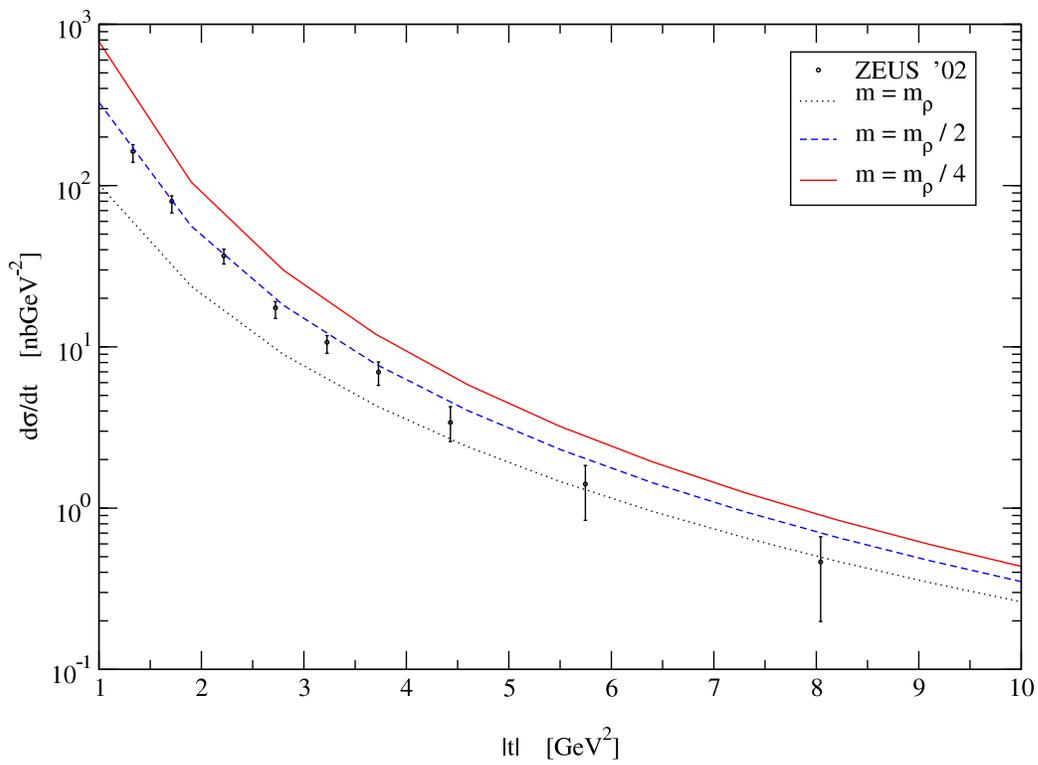}
   \caption{\it An example of the sensitivity of predictions
for the $\rho$ production cross section to the mass $m$ of the quark in
the vector meson 
\protect\cite{PoludniowskiHight}.}
   \label{fig:ForshawMassDepLarget}
\end{figure}

Second, in view of an approximate SCHC with dominant $\sigma_T$
the cross section must be proportional to $m_V^3 \Gamma(V\to e^+e^-)$.
That does not exhaust the flavor dependence because 
the onset of the hard regime does obviously depend on the 
mass of the heavy quark. The
pQCD two-gluon exchange approximation suggests that
for slow Fermi motion in vector mesons the appropriate 
hard scale is
\be
\Qb_t^2 \approx (m_V^2+|t|)
\label{eq:8.5.2.1}
\ee 
(\cite{ForshawHight,PoludniowskiHight} and references
therein). Although the Fermi motion can change the coefficient 
in front of $m_V^2$, one must conclude that for the 
$J/\Psi$ the large $t$ means $|t| \gg m_{J/\psi}^2 \sim 
10$ GeV$^2$. The numerical studies by 
Poludniowski et al.
\cite{PoludniowskiHight} show that even for light mesons
the variation of the constituent quark mass from
$m_{\rho}/4$ to $m_{\rho}/2$ to  $m_{\rho}$ changes
the predicted cross section by a factor of $\sim 2$ 
even at $|t|$ as large as 10 GeV$^2$. see
Fig.~\ref{fig:ForshawMassDepLarget}.

Third, in the pQCD two-gluon approximation the target quark
is regarded as pointlike one and the $t$ dependence is entirely due
to the $\gamma \to V$ transition vertex with the hard 
scale $\Qb_t^2$. Then, upon the $x$-integration in (\ref{eq:4.11.1}), 
\be
{d\sigma_{2G}(\gamma q \to V q') \over dt}
\propto 
\alpha_S^2(\Qb_t^2)\alpha_S^2(|t|)
{m_V^3 \Gamma(V\to e^+e^-) \over {(\Qb_t^2)}^4}\, ,
\label{eq:8.5.2.2}
\ee
where we indicated the natural choice of the running couplings of gluons 
to the target quark, $\alpha_S(|t|)$, and to quarks in the vector 
meson, $\alpha_S(\Qb_t^2)$.
Based on the experimental data on vector meson decays
\cite{PDG2002} the prediction for flavor dependence of
$d\sigma_V^{diss}/N_p(t)$  at identical values of
the hard scale $\Qb_t^2$ is
\be
\rho~:~\omega~:~\phi~:~J/\psi ~ = ~1~:~0.8\times {1\over 9}~:~
2.1 \times {2\over 9}~:~56\times {8\over 9}\,.
\label{eq:8.5.2.3}
\ee
Recall that in the studies of the
$Q^2$ dependence one had to compare the cross sections at
at identical  $(Q^2+m_V^2)$. 

Fourth, in the BFKL approximation the target quark becomes
effectively non-pointlike one
and introduces the approximately flavor independent 
factor $\propto 1/|t|$ to the cross section, whereas the 
$t$-dependence from the $\gamma \to V$ transition vertex
will be weaker:
\be
\left. {d\sigma_{BFKL}
(\gamma q \to V q') \over dt}\right|_{W_{\gamma q}=W}
\propto
\alpha_S^2(\Qb_t^2)\alpha_S^2(|t|){m_V^3 \Gamma(V\to e^+e^-) 
\over |t| {(\Qb_t^2)}^3} \times 
\left({W^2 \over \Qb_t^2}\right)^{2\Delta_{BFKL}}\, .
\label{eq:8.5.2.4}
\ee
Consequently, the flavor dependence (\ref{eq:8.5.2.3})
at identical values of the hard scale $\Qb_t^2$
must be tested for $|t|N_{p}^{-1}(t)(d\sigma_V^{diss}/dt)$ rather   
than $d\sigma_V^{diss}/dt$. 
Notice an extra suppression 
$\propto 1/(\Qb_t^2)^{2\Delta_{BFKL}}$ coming from the Regge
parameter.  To run the strong couplings 
in (\ref{eq:8.5.2.4}) is to go beyond the accuracy of the 
scaling BFKL approximation.


\subsubsection{The experimental results: 
measuring the trajectory of the hard BFKL Pomeron}

The $W$-dependence of the proton-dissociative vector meson 
production has been measured by both ZEUS and H1 collaborations.

\begin{figure}[htbp]
   \centering
   \epsfig{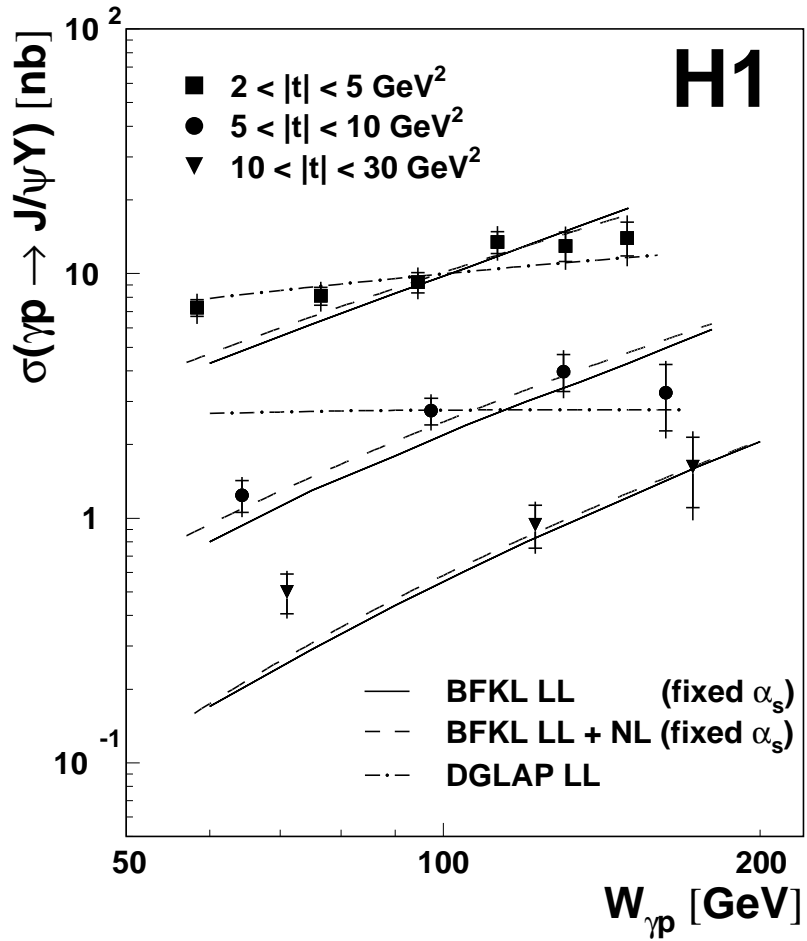}
   \caption{\it The H1 results for the $J/\Psi$ photoproduction cross section 
as a function of $W_{\gamma p}$ in 
three bins of $|t|$ \protect\cite{pl:b568:205}.
The inner error bars correspond to the statistical error and
the outer error bars are the statistical and systematic
errors added in quadrature.
The solid lines show the predictions from the BFKL model
\protect\cite{Enberg:2002zy},the dashed-dotted curve is the result 
from the DGLAP model \protect\cite{Gotsman:2001ne}.}
\label{fig:H1JpsiWdepLarget}
\end{figure}

\begin{table}
\begin{center}
\begin{tabular}{|c|c|c|c|}
\hline
$|t|$ range     & $ \langle |t| \rangle $  &     &  \\   
($\rm GeV^2$) &  (${\rm \ GeV^2}$)  &   
\raisebox{1.5ex}[-1.5ex]{$\delta$}  &  \raisebox{1.5ex}[-1.5ex]{$\alpha(t)$} \\   
\hline \hline
$2-5$     &      
$ 3.06$   &
$0.77   \pm 0.14  \pm 0.10 $ &  
$1.193   \pm 0.035  \pm 0.025   $  \\ \hline
$5-10$  &      
$ 6.93   $   &  
$1.29   \pm 0.23  \pm 0.16  $ &
$1.322   \pm 0.057  \pm 0.040  $ \\ \hline
$10-30$ &      
$ 16.5  $   &
$1.28   \pm 0.39  \pm 0.36  $ &  
$1.322   \pm 0.097  \pm 0.090  $  \\ \hline
\end{tabular}
\caption{\it The value of $\delta$ obtained when applying 
a fit to the data of the form
$\sigma (W) \propto {W}^{\delta}$ for each $|t|$ range,
together with the corresponding vacuum trajectory $\alpha(t)$ obtained 
from $\alpha (t) = (\delta + 4)/4$.
The first uncertainty is statistical and the second is systematic.}
\label{tab:H1interceptsLarget}
\end{center}
\end{table}

\begin{figure}[htbp]
   \centering
   \epsfig{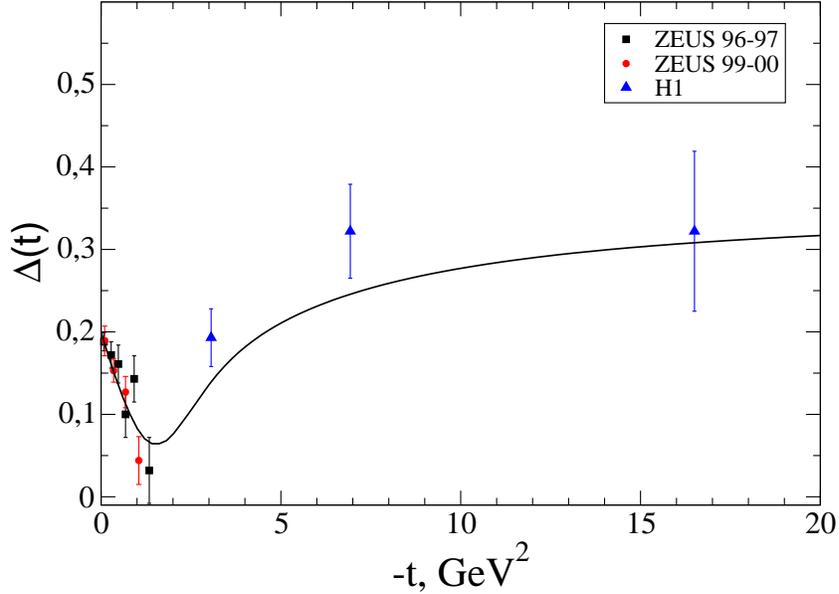}
   \caption{\it A compilation of the ZEUS and H1 results on
the determination of the Pomeron trajectory $\alpha_{\Pom}(t)$
($ \Delta (t) = \alphapom(t)-1$) from the $J/\Psi$ photoproduction. The solid curve shows
a possible interpolation from the regime of shrinking 
diffraction cone at small $t$ to the hard BFKL regime
at large $t$. }
\label{fig:EffectiveInterceptLarget}
\end{figure}

The absolute $W$ dependence of the $J/\psi$ production
has been measured by
H1 Collab. \cite{pl:b568:205}. The experimental data are
shown in Fig.~\ref{fig:H1JpsiWdepLarget}, the results
of the Regge fits are presented in Tab.~\ref{tab:H1interceptsLarget}. 
These results from H1 
give a solid evidence for $\alpha_{\Pom}(t) >1$ at large $|t|$.
The found values of $\Delta_{\Pom}=\alpha_{\Pom}(t)-1$ are close
to the leading order BFKL prediction with $\alpha_S^{BFKL} 
\sim (0.1\div 0.15)$, a comparison with the NLO intercept is
unwarranted at the moment.
On the other hand, the large-$t$ extrapolation of the ZEUS
results (\ref{eq:8.3.2.1}) suggests $\alpha_{\Pom}(t) < 1$ for
$|t| \gsim 2$ GeV$^2$. (For the $\rho$ and $\phi$ mesons 
the similar crossover takes place at $|t|\sim 1$ GeV$^2$,
one would readily attribute that to the process being still
soft.) Now recall that the shrinkage at small-$t$ is driven
by the infrared growth of $\alpha_S$ by which the low-$t$ 
BFKL evolution becomes sensitive to the infrared region
around the finite propagation
radius $R_c$ for perturbative gluons. In contrast to that,
in the Mueller-Navelet large-$t$ regime the large 
momentum transfer $\sim \bDelta$ flows through propagators of
all $t$-channel gluons, the infrared contribution will 
be suppressed and gross features of the fixed
$\alpha_S^{BFKL} \sim \alpha_S(t)$, leading log${1\over x}$,
BFKL evolution will be recovered. One can fancy the 
nonlinear $|t|$ dependence of the vacuum trajectory of
the form 
\be
\alpha(t)=1.2 + (0.16~{\rm GeV^{-2}}) t\cdot {\Lambda^6 \over |t|^3 + \Lambda^6}
+0.16{t^4 \over t^4 + \Lambda^8}
\label{eq:8.5.3.1}
\ee
shown in Fig.~\ref{fig:EffectiveInterceptLarget} for $\Lambda^2 = 2.5$~GeV$^2$.

\begin{figure}[htbp]
   \centering
   \epsfig{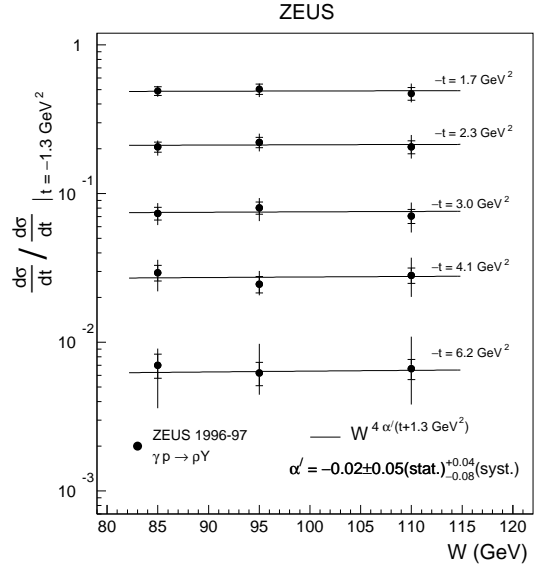}
   \epsfig{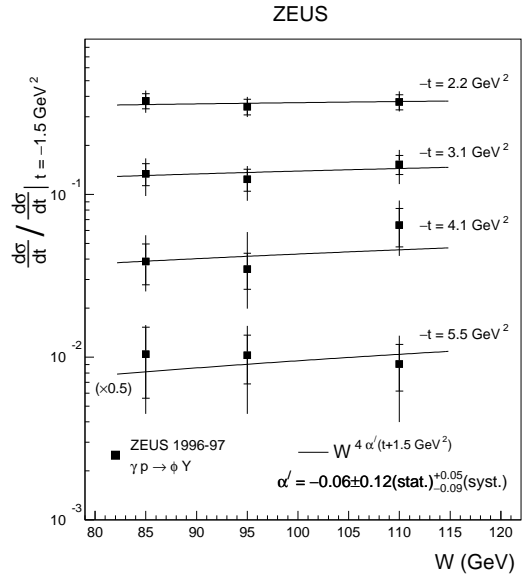}
   \caption{\it The $d \sigma / dt$ ratios for $\rho$(left) 
   and $\phi$(right) production
   cross sections as a function of $W$ in five(four) $t$ intervals.
   The lines represent the result of the fit with Eq.~\ref{eq:8.5.3.1}.}
   \label{fig:ZEUSshrinkageLarget}
\end{figure}

\begin{figure}[htbp]
   \centering
   \epsfig{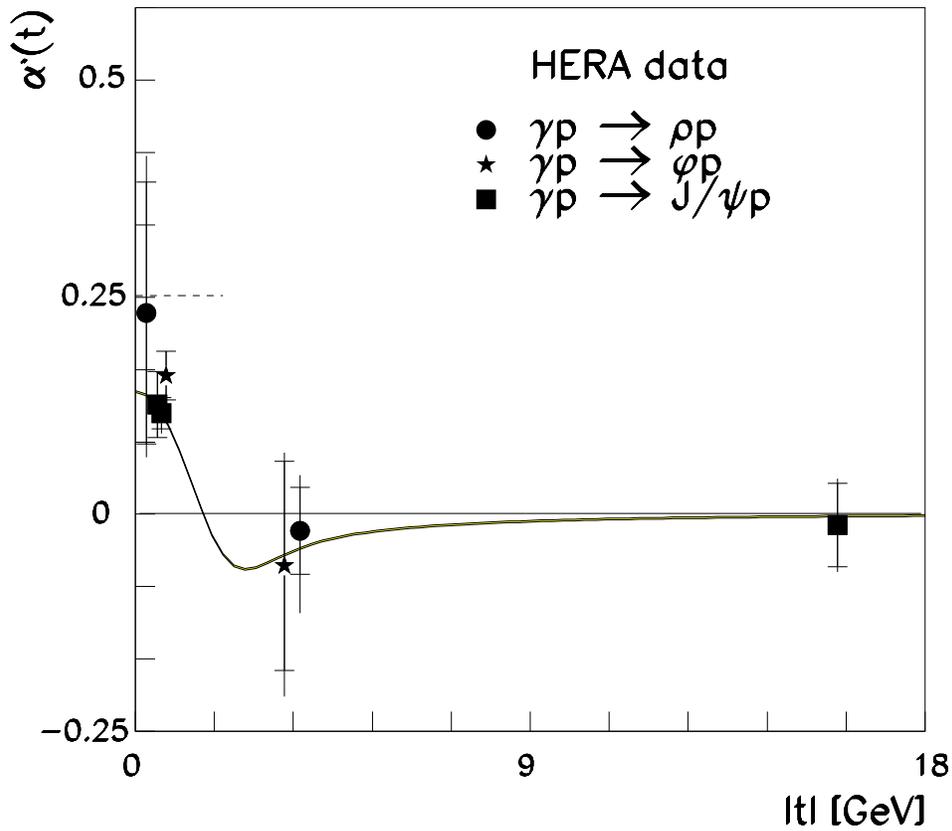}
   \caption{\it HERA results on $\alpha^\prime$ for 
the elastic and proton-dissociative (at $|t| > 1.3 ~\rm GeV^{2}$) 
vector meson production  
~\protect\cite{epj:c2:247,epj:c14:213,epj:c24:345,epj:c26:389,pl:b568:205}
compared with the $\alpha'(t)$ from the parameterization
(\protect\ref{eq:8.5.3.1}). The reference value for
soft hadronic interactions
$\alpha^\prime_{soft}=0.25~{\rm GeV^{-2}}$ is shown as a dashed line.
  The points are put in the center of $|t|$ range
   in which $\alpha^\prime$ is measured.
  The vertical inner bars indicate
  the statistical uncertainty and the outer bars represent 
  the statistical and
  systematic uncertainties added in quadrature.
}
   \label{fig:AlphaPrimSummary}
\end{figure}

The Ansatz (\ref{eq:8.5.3.1}) for the Pomeron
trajectory turns over at $|t| \sim 1\div 2$ GeV$^2$, which
is close to the natural scale $R_c^{-2} \sim (0.5\div 1)$ GeV$^2$.
The change of the sign of the derivative $\alpha'(t)$ from 
small to large $t$ is supported by the ZEUS experimental data
shown in Figs.~\ref{fig:ZEUSshrinkageLarget} and ~\ref{fig:AlphaPrimSummary}.
Although the $W$-dependence of the efficiency of the 
photoproduction tagger hinders the direct measurement of
the absolute $W$-dependence 
and the determination of $\alpha_{\Pom}(t)$ 
with the present ZEUS data \cite{epj:c26:389},
the large-$t$ slope of the vacuum exchange trajectory can
be measured in the tagger independent manner,
\be
{(d\sigma(W)/dt)\over {\left.(d\sigma(W)/dt)\right|_{t=t_0}}}
\propto W^{4\alpha'(t-t_0)}
\label{eq:8.5.3.2}
\ee 
The experimental data from ZEUS for this ratio and the 
found values of the slope of the vacuum trajectory $\alpha'$
are shown in Fig.~\ref{fig:ZEUSshrinkageLarget}. The summary of
the low-$t$ and high-$t$ results for $\alpha'$ is presented in
Fig.~\ref{fig:AlphaPrimSummary}, where we also show $\alpha'(t)$
for the parameterization (\ref{eq:8.5.3.1}).

\subsubsection{The experimental results: 
the $t$-dependence for a nucleon target }

\begin{figure}[htbp]
   \centering
   \epsfig{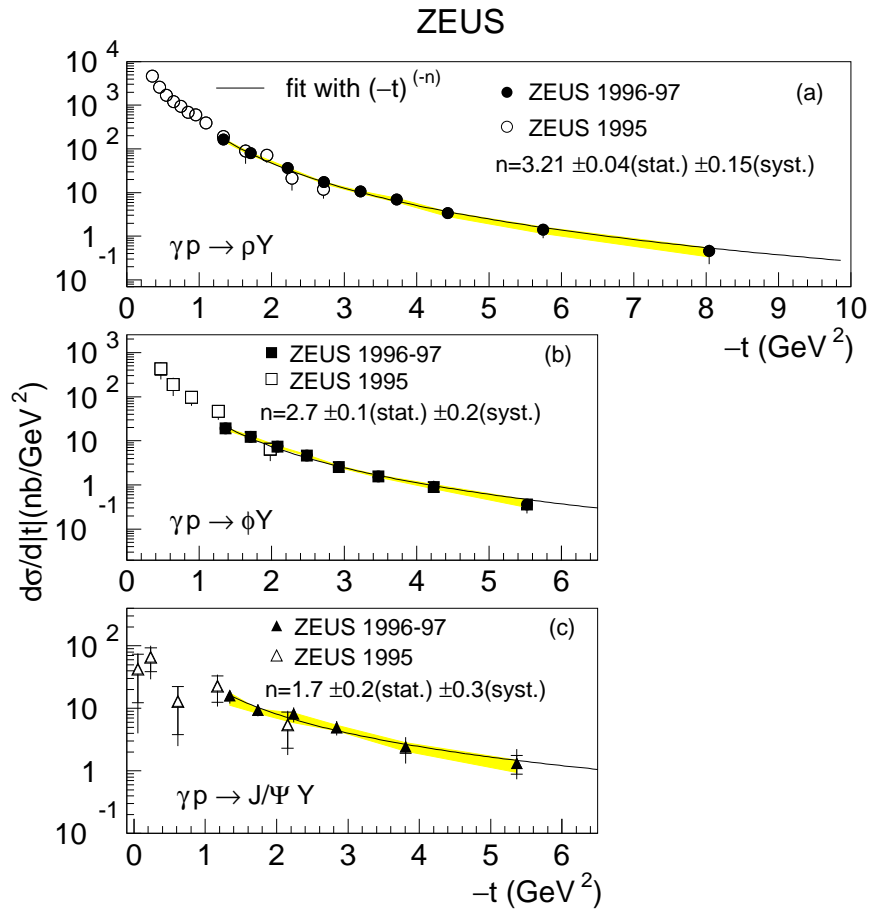}
  \caption{\it The ZEUS results for the differential cross sections
    $d \sigma_{\gamma p \rightarrow VY}/d|t|$ in the range
    $80<W<120$~GeV and $x>0.01$ for $\rho^0$, $\phi$ and $J/\psi$
    ~\protect\cite{epj:c26:389} production.
    The lines are results of the fit to the data with the function
    $A(-t)^{-n}$. The shaded bands represent the correlated uncertainties due
    to the modeling of the hadronic system $Y$.}
   \label{fig:ZEUSrhoY}
\end{figure}

\begin{figure}[htbp]
   \centering
   \epsfig{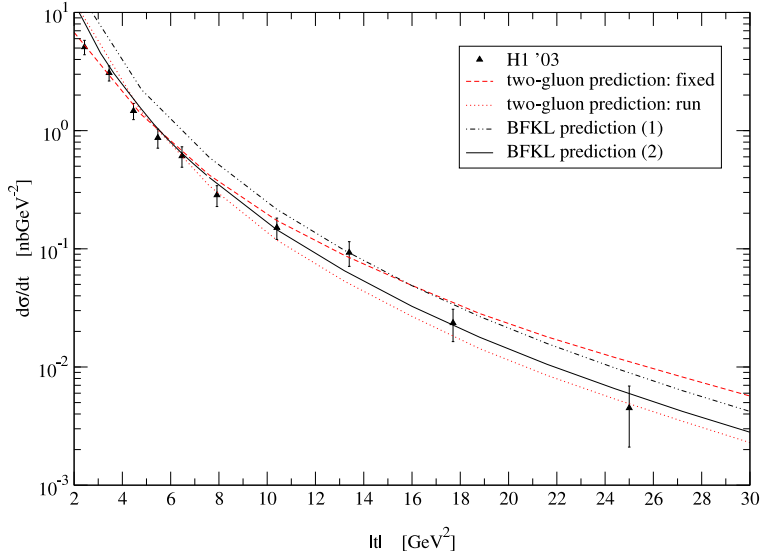}
   \caption{\it The differential cross section $d\sigma(\gamma p \to V Y)/dt$
for the $J/\psi$ photoproduction at large $t$ in the range 
$80 < W < 120$ GeV and $x'>|t|/(900({\rm GeV^2})+|t|) $ from H1 collaboration
\protect\cite{pl:b568:205}. The theoretical results for the
pQCD two-gluon exchange with fixed and running $\alpha_S$
and scaling BFKL approximations are from Poludniowski et al.
\protect\cite{PoludniowskiHight}. }
   \label{fig:ForshawSigmaJpsiLarget}
\end{figure}

The $t$-dependence of the $\rho$, $\phi$ and $J/\psi$ meson
proton-dissociative production cross section at high-$t$ is shown
in Figs.
~\ref{fig:ZEUSrhoY} and ~\ref{fig:ForshawSigmaJpsiLarget}. 
It is much slower than the
exponential one typical of the diffraction cone and is in broad
agreement with the inverse power law as discussed in Section 8.2.

The quantitative interpretation of the experimental data
taken at the moderately large $t$ depends on the choice of the 
hard scale. A fit to the observed $t$-dependence
in the form $d\sigma/d|t| \propto |t|^{-n_V}$ yielded for the
ZEUS data \cite{epj:c26:389} the exponents
\bea
n_\rho &=& 3.21 \pm 0.04(stat.) \pm 0.15 (syst.) \quad 
({\rm ZEUS, \quad 1.2 < |t| < 10~GeV^2})\,;\label{eq:8.5.4.1}\\
n_\phi &=& 2.7 \pm 0.1(stat.) \pm 0.2 (syst.) \quad 
({\rm ZEUS, \quad 1.2 < |t| < 6.5~GeV^2})\,;\label{eq:8.5.4.2}\\
n_{J/\psi} &=& 1.7 \pm 0.2(stat.) \pm 0.3 (syst.) \quad 
({\rm ZEUS, \quad 1.2 < |t| < 6.5~GeV^2})\,.
\label{eq:8.5.4.3}
\eea
which must be compared to $n_V\approx 4$ expected from theory.
The values of $|t| < 6.5$ GeV$^2$ in the ZEUS data on the $J/\Psi$ 
production are arguably too small for the onset of the
true large-$t$ behaviour. The $J/\Psi$ production
data from H1 \cite{pl:b568:205} extend
to $|t| < 30$ GeV$^2$, see fig.~\ref{fig:ForshawSigmaJpsiLarget},
 and give the exponent
\bea
n_{J/\psi} &=& 3.00 \pm 0.08(stat.) \pm 0.05 (syst.) \quad 
(H1, \quad {\rm 4.46 < |t| < 30~GeV^2})\,.
\label{eq:8.5.4.4}
\eea

\vspace*{1cm}

\begin{figure}[htbp]
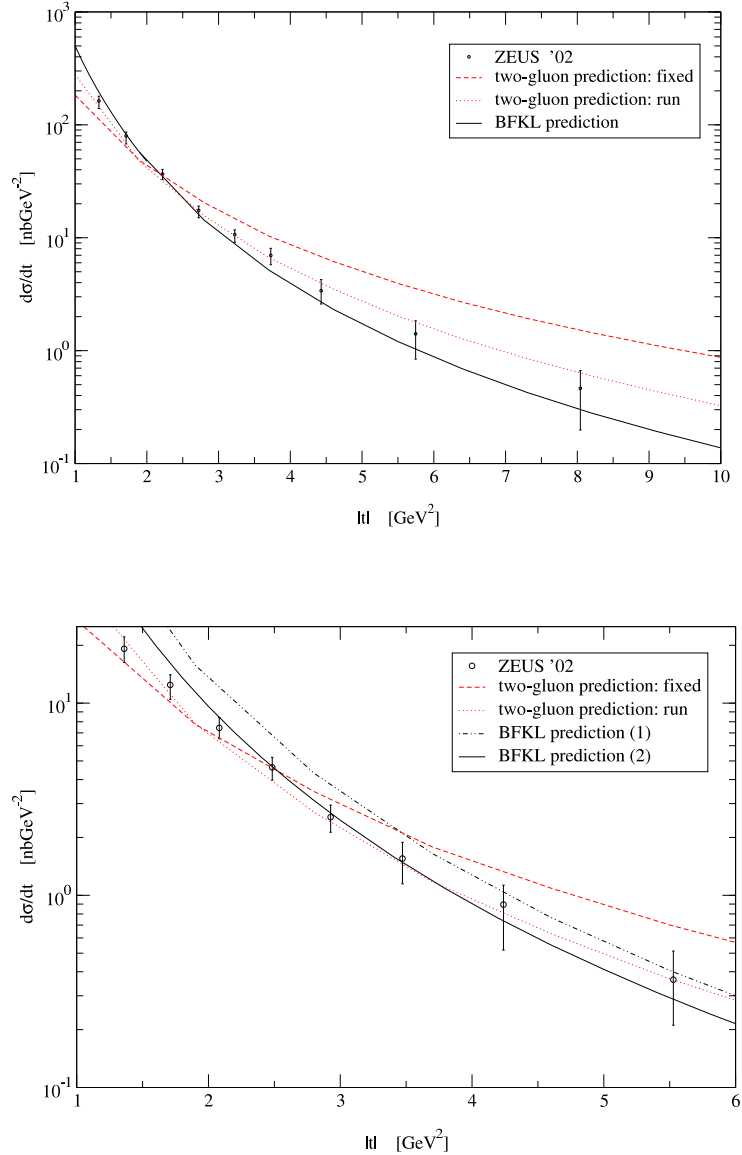

   \centering
   \epsfig{file=ForshawSigmaRhoLarget.ps,width=70mm,angle=270}
\bigskip\\bigskip\\
   \epsfig{file=ForshawSigmaPhiLarget.ps,width=70mm,angle=270}
   \caption{\it The differential cross section $d\sigma(\gamma p \to V Y)/dt$
for the $\rho$ (upper plot) and $\phi$ (lower plot)
photoproduction at large $t$ in the range 
$80 < W < 120$ GeV and $x'>0.01$ from ZEUS collaboration
\protect\cite{epj:c26:389}. The theoretical results for the
pQCD two-gluon exchange with fixed and running $\alpha_S$
and scaling BFKL approximations are from Poludniowski et al.
\protect\cite{PoludniowskiHight}. }
   \label{fig:ForshawSigmaRhoLarget}
\end{figure}

A comparison of the results (\ref{eq:8.5.4.3})
and (\ref{eq:8.5.4.4}) shows an importance of the finite
mass effects in the $t$-dependence, see also 
Fig.~\ref{fig:ForshawMassDepLarget}. The theoretical
calculations by Poludniowski et al. \cite{PoludniowskiHight}
within the scaling BFKL approximation are shown in 
Figs. ~\ref{fig:ForshawSigmaJpsiLarget}.\ref{fig:ForshawSigmaRhoLarget},
 and clearly show an
improvement from the pQCD two-gluon to BFKL approximation.
The impact of the running strong coupling on the pQCD two-gluon
results for the $t$-dependence is substantial, the BFKL 
calculations are for the fixed coupling. The results for
the $\phi$ and $J/\Psi$ are based on the parameters of
the model which were adjusted to the $\rho$ photoproduction.

\subsubsection{The experimental results: 
the flavor and $t$-dependence for a partonic subprocess
$\gamma q \to q' Y$ }

The most direct test of the BFKL approach to large-$t$ vector
mesons is provided by the reanalysis \cite{Fedya2} 
of the experimental data in
terms of the cross section of partonic subprocess
\be
\left. |t|{d\sigma(\gamma q \to V q') \over dt}
\right|_{W_{\gamma q}=W_{\gamma p}} =
{|t|\over N_p(t)}\cdot {d\sigma(\gamma p \to V Y) \over dt}
 \propto {1 \over (|t|+m_V^2)^{n_V}}\, ,
\label{eq:8.5.4.5}
\ee
Here the residual $t$-dependence in the rhs 
probes the true dynamics of hard $\gamma V$ transition,
see  Eq.~(\ref{eq:8.5.2.4}). 
In such a representation the scale-invariant BFKL approximation
predicts the flavor independent
\be
n_V=3+2\Delta_{BFKL} \approx (3.2\div 3.6)\,.
\label{eq:8.5.4.6}
\ee
The major problem with the extraction of the partonic
cross section is that the absolute value and 
$t$-dependence of the number of effective partons $N_p(t)$ 
exhibits a strong sensitivity the the pomeron trajectory, see 
Fig.~\ref{fig:EffectivePartons}. The resulting 
uncertainty propagates into the magnitude and
$t$-dependence of the cross section of $ |t|{d\sigma(\gamma q \to V q')/ dt}$
in (\ref{eq:8.5.4.5}) and, consequently, into
the determination of the exponents $n_V$ from the fit of the
partonic cross section to the parameterization (\ref{eq:8.5.4.5}).

For the consistency with direct experimental measurements 
of the pomeron trajectory one must use the 
parameterization (\ref{eq:8.5.3.1}) 
which correctly reproduces all the
features of the H1 and ZEUS data shown in 
Figs.~\protect\ref{fig:EffectiveInterceptLarget}
and \protect\ref{fig:AlphaPrimSummary}.  
The resulting fits to the ZEUS data \protect\cite{epj:c26:389} on the $\rho,\phi$ and 
$J/\Psi$ production at $\langle W_{\gamma p}\rangle = 100 GeV$
yield $n_{\rho} = 2.08 \pm 0.06$, $n_{\phi} = 1.83 \pm 0.13$,
$n_{J/\psi}(ZEUS) = 0.78 \pm 0.64 $. In the H1 data \protect\cite{pl:b568:205}
on the $J/\Psi$ production the energy $\langle W_{\gamma p}(t)\rangle$
slightly rises with $t$ which introduces a certain bias into the
$t$-dependence and enhances $n_{J/\psi}$. Neglecting that bias 
and  excluding the point at lowest $t$ yields  $n_{J/\psi}(H1) = 2.55 \pm 0.2 $. 
   
The above cited error bars do not include the theoretical
uncertainties of $N_p(t)$ connected to the parameterization of the
$t$-dependence of $\alpha_{\Pom}(t)$.  
For the sake of illustration, we cite here the results found if
$N_p(t)$ is evaluated for fixed $\alpha_{\Pom}(t) = 1+\Delta_{BFKL}
\approx 1.25$, although such a flat $\alpha_{\Pom}(t)$ is inconsistent
with the H1 and ZEUS data shown in Figs.~\protect\ref{fig:EffectiveInterceptLarget}
and \protect\ref{fig:AlphaPrimSummary}. In 
this case the $t$-dependence of $N_p(t)$ will be 
much weaker for both the ZEUS and H1 cuts, see 
Fig.~\ref{fig:EffectivePartons}, it doesn't change 
substantially with the further increase of $\Delta_{BFKL}$. 
The partonic cross sections
extracted from the same data will have much steeper $t$-dependence and
the fitted exponents $n_V$ will be substantially larger than for the
pomeron trajectory of Eq.~(\ref{eq:8.5.3.1}): $n_{\rho}=
2.86 \pm 0.05,~~n_{\phi} = 2.66 \pm 0.12,~~n_{J/\psi}(ZEUS)=
3.88 \pm 0.62\sim 3$ for the ZEUS data and 
$n_{J/\psi}(H1) \ 3.86 \pm 0.26$ for the H1 data. 
The statistical error bars in fitted values of $n_{V}$ for the two 
choices of $N_p(t)$  are misleading because the model dependence 
in the extraction of the exponent $n_V$ of 
the $t$-dependence is much larger than the statistical error bars. 
In their scale-invariant BFKL calculations shown in 
Fig.~\ref{fig:ForshawSigmaJpsiLarget},
Poludniowski et al. \cite{PoludniowskiHight} use fixed $\alpha_S=0.25$,
which amounts to even larger $\Delta_{BFKL}=0.66$ and $n_{V}\approx 4.3$.
Within those uncertainties, the data on different vector mesons
do not exclude the flavor independent $n_V$ and 
the observed $t$-dependence does not conflict the 
BFKL expectation (\ref{eq:8.5.4.6}). 

A very large $\alpha_{\Pom}(t) = 1+\Delta_{BFKL}=
1.66$ used in the theoretical calculations  \cite{PoludniowskiHight}
conflicts the experimental data on the Pomeron trajectory shown
in Figs.~\protect\ref{fig:EffectiveInterceptLarget}
and \protect\ref{fig:AlphaPrimSummary}. Furthermore, the effective number
of partons $N_p(t)$ evaluated with $\Delta_{BFKL}=0.66$ 
is by the factor $\sim $(5-10) smaller
than for the experimentally suggested trajectory shown in 
Fig.~\protect\ref{fig:EffectiveInterceptLarget}. This uncertainty 
is not discussed in ~\cite{PoludniowskiHight} and casts a 
shadow on the agreement between the theory and experiment
in the magnitude of the cross section.  
Before drawing firm conclusions on the status of the 
BFKL approach one needs much better understanding of the Pomeron 
trajectory and incorporation of the realistic Pomeron trajectory 
into the theoretical formalism.

\begin{figure}[htbp]
   \centering
   \epsfig{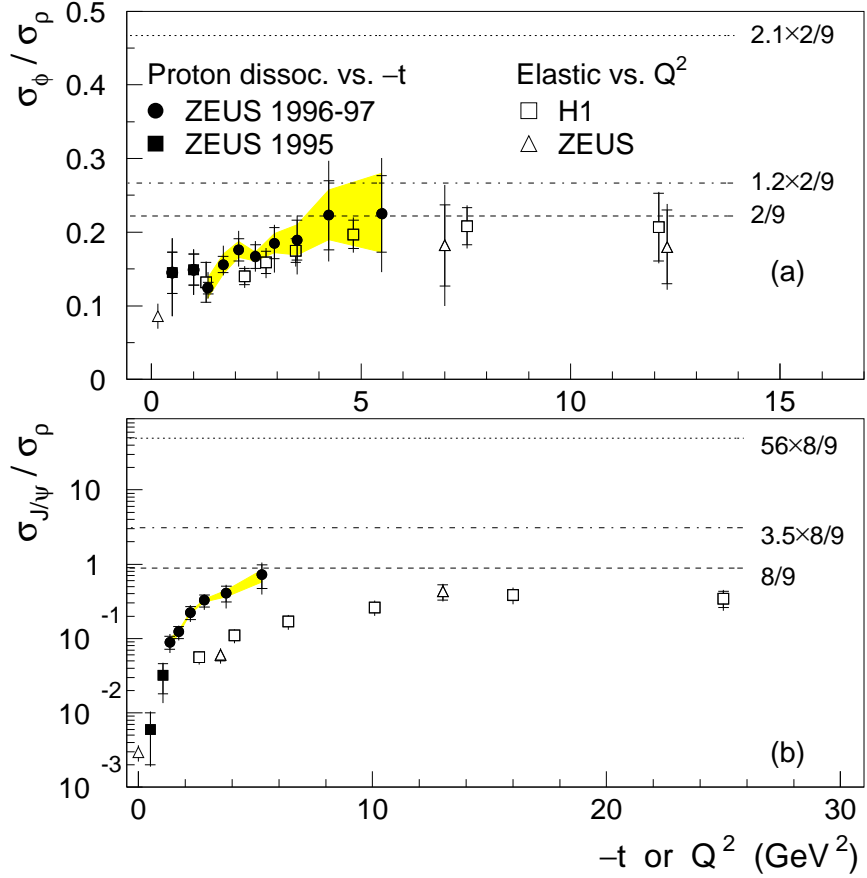}
  \caption{(a) The ratio of the $\phi$ to $\rho^0$ cross sections as
   a function of $-t$ or $Q^2$. The $\phi/\rho^0$ results as a function 
   of $-t$ for proton-dissociative photoproduction from this analysis 
   are shown with solid circles and those from the 
   ZEUS 1995~\protect\cite{epj:c14:213} measurement with the solid squares.
  The
  shaded bands represent the size of the correlated uncertainties due 
  to the  modeling of the dissociative system, $Y$.
   Open triangles at $Q^2\approx 0 ~{\rm GeV^2}$~\protect\cite{pl:b377:259},
   $Q^2 =  7 {\rm GeV^2}$~\protect\cite{pl:b487:273} and 
   $Q^2 =  12.3 {\rm GeV^2}$~\protect\cite{pl:b380:220} represent the 
   $\phi/\rho^0$ ratio of the elastic cross sections as a function of 
   $Q^2$ from ZEUS, while the open squares represent those from 
   H1~\protect\cite{pl:b483:360}. 
(b) The ratio of the $J/\psi$ to $\rho^0$ cross sections as
   a function of $-t$ or $Q^2$. 
  The same convention for symbols as for $\phi/\rho^0$ ratio is used.
  Open triangles at $Q^2\approx 0 {\rm GeV^2}$~\protect\cite{zfp:c75:215}
  and $Q^2 = 3.5, 13 {\rm GeV^2}$~\protect\cite{epj:c6:603} 
represent the ZEUS
  measurements,
  while the open squares represent 
those of H1~\protect\cite{epj:c13:371,epj:c10:373}.     
  The dashed lines correspond to the SU(4) predictions, while the 
  dotted and dashed-dotted correspond to the pQCD values given by 
  Eqs.~\ref{eq:8.5.2.3} and~\ref{eq:6.2.1.1}, respectively.
}
     \label{fig:ZEUSratiosQ2larget}
\end{figure}

The same model-dependence of extraction of the partonic observable
$|t|d\sigma(\gamma q \to V q')/dt=|t|( d\sigma(\gamma p \to V Y)/dt)/N_p(t)$ 
affects the discussion of the flavor-dependence of large $t$
cross sections. First, the point that neither $Q^2$ nor $|t|$ are 
the correct hard scales to compare different vector mesons is illustrated
by Fig.~\ref{fig:ZEUSratiosQ2larget}. Whereas for light $\phi$
meson the $t$-dependence, as well as the $Q^2$-dependence 
of the ratio $\sigma_\phi/\sigma_\rho$ is weak, 
the ratio $\sigma_{J/\psi}/\sigma_\rho$
changes by more than two orders of magnitude. The apparent approach to the
SU(4) ratios at largest measured value $|t|$ is misleading - no true asymptotics
can be reached at $|t| < m_{J/\psi}^2$.

\begin{figure}[htbp]
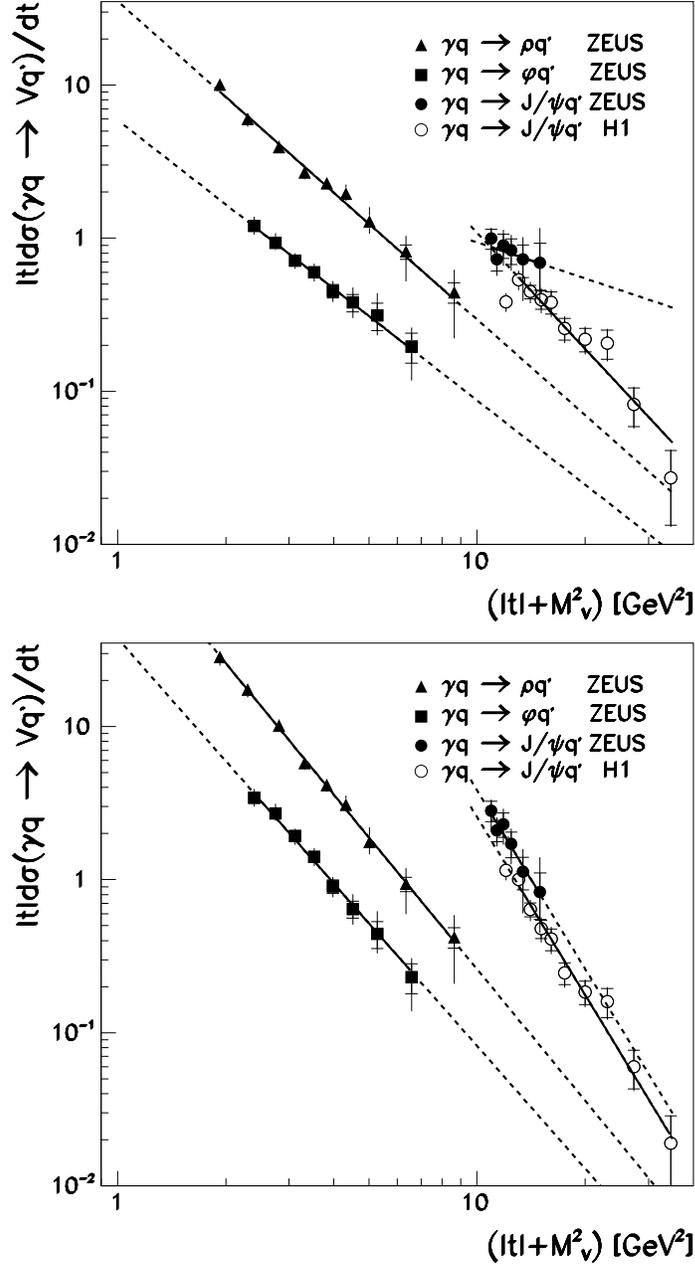

   \centering

   \vspace*{-2cm}
   \epsfig{file=RhoPhiJpsiComparisonLarget.eps,width=105mm}

   \vspace*{-2cm}
   \epsfig{file=RhoPhiJpsiComparisonLargetFixedPom.eps,width=105mm}
  \caption{The $t$ dependence of the large-$t$ vector meson
production measured by ZEUS \protect\cite{epj:c26:389} and H1 
\protect\cite{pl:b568:205}
presented as a differential cross section of the partonic subprocess
$|t|{d\sigma(\gamma q \to V q')/dt}=|t|( d\sigma(\gamma p \to V Y)/ dt)/N_p(t)$. 
The 
straight lines are results of the fit by Eq.~\protect\ref{eq:8.5.4.5},
the lowest $|t|$ data point of H1 has been excluded from the fit.
The left box shows the results for the number of effective partons
 $N_p(t)$ evaluated for the
Pomeron trajectory $\alpha_{\Pom}(t)$ of Eq.~(\protect\ref{eq:8.5.3.1})
shown in Fig.~\protect\ref{fig:EffectiveInterceptLarget}, the
right box is for $N_p(t)$ evaluated for the fixed $\Delta_{\Pom}=0.25$.}
     \label{fig:RhoPhiJpsiComparisonLarget}
\end{figure}

Strong departure from the $SU(4)$ ratios is evident
from Fig.~\ref{fig:RhoPhiJpsiComparisonLarget}, where we show
the ZEUS and H1 data in the form of the partonic subprocess
observable $|t|{d\sigma(\gamma q \to V q')/dt}=
|t|( d\sigma(\gamma p \to V Y)/dt)/N_p(t)$ plotted as a function 
of $(|t|+m_V^2)$. Although the experimental data on the $\rho,\phi$ 
and $J/\Psi$ vector mesons don't have an overlap in $(|t|+m_V^2)$,
it seems safe to extrapolate the ZEUS data on the 
$\rho$ and $\phi$ production to $(|t|+m_{J/\psi}^2)=12.5$ GeV$^2$
typical of the ZEUS data on the $J/\psi$ production. The so
extrapolated $\rho$ and $\phi$ cross sections 
have the factor of $\sim 2$ uncertainty.

We recite from (\ref{eq:8.5.2.3}) the flavor dependence of 
$(1/N_p(t))(d\sigma_V^{diss}/dt)$  at identical  $(Q^2+m_V^2)$
based on the vector meson 
decay properties:
\be
\rho~:~\omega~:~\phi~:~J/\psi ~ = ~1~:~0.8\times {1\over 9}~:~
2.1 \times {2\over 9}~:~56\times {8\over 9}\,.
\label{eq:8.5.4.6*}
\ee
If the number of effective partons $ N_p(t)$ is 
evaluated with the pomeron trajectory (\ref{eq:8.5.3.1}) 
which correctly reproduces the H1 and ZEUS data shown in 
Figs.~\protect\ref{fig:EffectiveInterceptLarget}
and \protect\ref{fig:AlphaPrimSummary}, then the extrapolation  
of the $\rho$ and $\phi$ cross sections to $(|t|+m_{J/\psi}^2)=12.5$ GeV$^2$ 
gives the cross section ratios (within the factor $\sim 2$ extrapolation
uncertainty)
\be
\rho~:~\phi~:~J/\psi ~ = 1~:~
{1\over 2} \times 2.1\times {2\over 9}~:~{1\over 15} \times 56\times {8\over 9}\, .
\label{eq:8.5.6.1}
\ee
If $N_p(t)$ is evaluated for $\Delta_{BFKL}=0.25 =$ const, then the same 
extrapolation gives slightly different
cross section ratios 
\be
\rho~:~\phi~:~J/\psi ~ \approx 1:~
{2\over 3}\times 2.1\times {2\over 9}~:~{1\over 7}\times 56\times {8\over 9}
\label{eq:8.5.6.2}
\ee
For the both choices of the $N_{p}(t)$ 
the principal effect is an enhancement of the light vector meson 
production with respect to the $J/\Psi$ production. Such an
enhancement due to the chiral-odd $\gamma q\bar{q}$ transitions  
\cite{IvanovChiralOdd}
is present in calculations of Poludniowski et al \cite{PoludniowskiHight}.

\newpage
\section{Summary and conclusions}

Slightly more than a decade ago, quite unexpectedly, HERA has become 
a unique facility for exploring the diffractive physics in an
entirely new domain of many different hard scales. 
 Compared to the fixed target
data, the center of mass energy $W$, the explored regions
of $Q^2$ and $t$ were extended by one order of magnitude. 
By now the principal features
of the flavor-, $t$-, $Q^2$- and $W^2$-dependences of the
observed cross sections, polarization properties of
produced vector mesons are well established. 
The high statistics of the data from HERA
allowed for the first ovservation
of SCHNC in high energy diffraction.   
Regarding the experimental situation, there is an
overall consistency between the experimental data from
H1 and ZEUS Collaborations and the early data from
fixed target experiments.

What did change in our understanding 
of high-energy
diffractive scattering after that decade of amassing
high-precision experimental data by H1 and ZEUS Collaborations
at HERA? What was the impact of these data on theoretical
ideas on high-energy vacuum exchange in the $t$-channel 
- the Pomeron? How strongly is the post-HERA pQCD Pomeron
different from the pre-HERA Pomeron
approximated by an isolated Regge pole with an
intercept $\alpha_{soft}\approx 1.1$? Is our theoretical
understanding of diffraction sufficient to get all the 
information we can, and would like to, from the available
experimental data?

The scaling violation in inclusive DIS is a classical 
short-distance phenomenon dominated by scales $\sim 1/Q$
\cite{DGLAP1,*DGLAP2,DGLAP3,*DGLAP4,DGLAP5}. 
Thanks to the 1970's groundbreaking works by Fadin, Kuraev, 
Lipatov and Balitsky 
\cite{FKL75,KLF77,*KLF771,BL} the theory was well prepared to the 
observed departure from the DGLAP evolution and the 
BFKL Pomeron reinterpretation of a steep small-$x$ rise 
of structure functions discovered at HERA. 

The principal virtue of diffractive vector meson 
production in DIS at HERA is that it is a short distance 
dominated process. At small to moderate $t$ within the
diffraction cone, the short-distance property is 
quantified by a scanning radius $r_S$
\cite{KZcharmonium,NNNcomments,KNNZrs,KNNZct}, at large-$t$
the short scale is set by $1/\sqrt{|t|}$. The early
discussion of importance of large-$t$ photoproduction of
vector mesons as a testing ground of ideas on the pQCD
Pomeron goes back to mid-80's work by Ginzburg, Panfil 
and Serbo \cite{GinzburgPanfilSerbo}. The idea of unified color dipole
description of inclusive DIS and diffractive vector mesons, 
the concept of the scanning radius 
\cite{KZcharmonium,NNNcomments,KNNZrs,KNNZct} and
the importance of $\Qb^2 = {1\over 4}(Q^2+m_V^2)$ as 
a hard scale at small $t$ \cite{RyskinJPsi,NNZscanVM}, 
and the possibility of
vector meson production as a testing ground for models
of gluon density in the proton  emerged in early 90's 
in works by Kopeliovich, Nemchik, Zakharov et al. 
\cite{KNNZct} and Ryskin \cite{RyskinJPsi}
and Brodsky et al. \cite{BFGMSvm}. Ever since then the collective effort 
by many groups has led to a fairly refined treatment
of vector meson production at small to large $|t|$
and to an understanding of an accuracy and limitations of
the leading log${1\over x}$ pQCD approaches.
The wealth of the experimental data collected by H1 and
ZEUS confirmed all the gross features of the pQCD
based description of the process.

An approximate 
$(Q^2+m_V^2)$-scaling of all observables - total
cross section \cite{NNZscanVM}, the diffraction slope
\cite{NZZslope}, the exponent
of the energy dependence \cite{NNZscanVM} --- alias the Pomeron intercept --- 
has been the recurrent theme
in our discussion of the experimental data, its 
experimental confirmation at HERA must be regarded as a 
major discovery and an undoubted 
success of the pQCD approaches. The pQCD
dictated relationship between the energy dependence 
of vector meson production 
and inclusive DIS encoded
in terms of the energy dependence of the gluon density
\cite{KNNZct,RyskinJPsi,BFGMSvm}
has been confirmed experimentally beyond reasonable
doubt and is still another major discovery at HERA. 

Success with theoretical predictions of the absolute
cross sections is modest one. 
The $(Q^2+m_V^2)$ as a hard scale for small-$t$
diffractive vector mesons is entirely analogous to $Q^2$
as a hard scale in inclusive DIS. The pQCD only predicts
the dependence on those hard scales starting from certain
soft input.
Within the color dipole and
$k_{\perp}$-factorization approaches this input is
universal for vector meson production and inclusive
DIS, still it does not come from first principles of
pQCD. Only one leg of the pomeron, which couples to 
the $\gamma^*V$ transition, rests on the hard pQCD
ground, the second leg which couples to the proton
is always in the soft regime. 
Once the $k_{\perp}$-factorization 
and other related pQCD model predictions
are normalized to the $J/\Psi$ photoproduction data,
the description of the observed $Q^2$- and energy 
dependence of $\sigma_{J/\Psi}(Q^2)$ is close to a perfect 
one, which must be regarded as a real success of pQCD
in the domain of hard diffractive scattering.

Nonetheless a factor of $\sim 2$ sensitivity of pQCD model 
predictions to the wave function of vector mesons is a well 
established limitation of the leading order $\log{1\over x}$
formalism and can not be eliminated at the moment. 
The ratio of longitudinal and transverse cross sections,
$R_V=\sigma_L/\sigma_T$, is an example of an observable
which exhibits especially strong sensitivity to the wave 
function of light vector vector mesons
\cite{KolyaCracow}. A very good
demonstration of the sensitivity to the wave function
is a node effect in the $\Psi(2S)$ production
\cite{KZcharmonium,NNNcomments,KNNZrs} which
suppresses the cross section, makes it grow with
energy faster than the $J/\Psi$ production and
leads to counterintuitive inequality of diffraction slopes 
for the $\Psi(2S)$ and $J/\Psi$ production \cite{NNPZZslopeVM,*NNPZZslopeVM1}.   
The further tests of pQCD predictions for diffractive
vector meson production call upon the development
of NLO $k_{\perp}$-factorization formalism. This includes
the theoretical understanding of the effect of higher
Fock states in vector mesons and the derivation of
the $k_{\perp}$-factorization impact factors for
the $\gamma^* V$ transition to NLO in log${1\over x}$.

The works by D.Ivanov and Kirschner \cite{IKspinflip}
and Kuraev et al. \cite{KNZspinflip,*KNZspinflip1},
in conjunction with B.Zakharov's \cite{SlavaSpinFlip}
early pQCD motivated 
discussion
of helicity flip in hadronic scattering,
have led to understanding of SCHNC as a generic property 
of high-energy scattering the origin of which does
not require an applicability of pQCD. The emerging
phenomenology of spin properties of diffractive vector
mesons has been very successful at small $t$, the
SCHNC in high energy small-$t$ diffractive scattering
is sill another major discovery at HERA. On the
theoretical side the chapter
has not been closed - further studies of the impact
of chiral-odd $q\bar{q}$ components in the photon 
suggested by D.Ivanov et al. \cite{IvanovChiralOdd,*IvanovChiralOdd1}
are in order. They were
found to be crucial \cite{PoludniowskiHight}
for theoretical explanation of
an approximate SCHC in the large-$t$ photoproduction data. 
Here the BFKL based phenomenology of the absolute normalization
and $t$-dependence is in a reasonably good shape, a fly in the ointment 
is a sign discrepancy in the helicity-flip amplitude
found by Poludniowski et al \cite{PoludniowskiHight}.

What new did we learn about the hard Pomeron trajectory?
In inclusive DIS the pQCD Pomeron can only be probed
at $t=0$ and one can not tell a difference between
the fixed branching point and moving pole options
for the Pomeron. In diffractive vector meson production
the full $t$-dependence of the Pomeron exchange
can be probed.
The experimental observation by ZEUS Collab.
of Gribov's shrinkage
of the diffraction cone in $J/\Psi$ production is 
an important evidence for the hard Pomeron being a
moving $j$-plane singularity --- this is definitely
the first important new finding on the hard
Pomeron trajectory beyond the 
reach of inclusive DIS. 
One option is that the pQCD Pomeron is 
a sequence of isolated
moving Regge poles as advocated in the pioneering 
Fadin-Kuraev-Lipatov work on the BFKL Pomeron
\cite{FKL75}.
Unfortunately, the accuracy of the combined 
set of the proton structure function and vector 
meson production data does not yet allow one to resolve 
the fine structure of those poles. The theoretical
discussion has been confined to the evaluation of
the slope of trajectories of these poles 
\cite{NZZslope}, the
large-$t$ behavior of trajectories remains an open issue.
The color dipole approach predicts an approximately
$Q^2$-independent shrinkage of the diffraction cone,
there is a weak evidence for that from ZEUS measurements, but
the experimental situation is not conclusive yet. 

Large $t$ as a hard scale brings in new opportunities.
In this case the pQCD Pomeron is expected to be in hard regime all
the way through from the target to $\gamma^*V$
transition. The experimental information on large-$t$
vector mesons is very exciting. 
The H1 data on large-$t$
$J/\Psi$ mesons gave a very interesting evidence,
supported also by ZEUS data, for the antishrinkage,
$\alpha'_{\Pom} <0$, and emergence of
the hard BFKL Pomeron exchange with intercept
$\alpha_{\Pom}\sim 1.3$ at $|t| \gsim 3$ GeV$^2$.
This is the second important new finding on
the hard Pomeron trajectory beyond the 
reach of inclusive DIS. Such a transition 
from the shrinkage to the antishrinkage is 
plausible, but has not yet been explored 
theoretically. Here important issues for future
theoretical studies are 
the sensitivity of the turn over from shrinkage
to antishrinakge to the infrared regularization
of pQCD and its $(Q^2+m_V^2)$-dependence. To this
end, an experimental 
study of the interplay 
of two hard scales --- $(Q^2+m_V^2)$ and $|t|$ --- would
be most interesting.  The flavour and $t$-dependence
of large-$t$ cross sections does not conflict the 
estimates based on the leading order BFKL approach
\cite{PoludniowskiHight},
but higher precision data are needed for more
definitive conclusions.

To summarize, the program of diffractive vector meson
studies at HERA was exceptionally fruitful one. The
matching theoretical development followed, still 
more work is needed: the pressing issues include
the $t$-dependence of the Pomeron trajectory from
small to large $t$, understaning the r\^ole of higher 
Fock states in vector mesons and derivation of NLO 
$k_{\perp}$-factorization, the further studies of
helicity properties of large-$t$ vector mesons. On
the experimental side, new results on vector mesons
are expected form several more years of run of HERA.

\section*{Acknowledgements}

The authors are grateful to Fyodor F.Pavlov for assistance
with interpretation of the high-$t$ data and Jeff Forshaw for the
useful correspondence on high-$t$ vector meson production. 
The work of one of the authors (I.P.I) has been partially
supported by grants  INTAS-00-00366, RFBR-02-02-17884,
NSh-2339.2003.2, ''Universities of Russia'' UR-02.01.005 and
DESY.

\section*{Note added}

After the main body of this review has been completed, 
D.Yu. Ivanov and his 
collaborators reported a long waited  
 NLO analysis of $\gamma^* V$ production
\cite{DimaNLO,DimaNLO2,DimaNLO31,*DimaNLO32}. It is a very involved calculation
and not yet a full fledged
NLO $k_{\perp}$-factorization analysis because the vector mesons have still
been treated in the collinear approximation. The major expectation was
that the NLO calculations would fix more reliably the magnitude of the
production amplitude, specifically, the NLO amplitudes must have a 
stability window as a function of the so-called factorization scale. 
To this end, the NLO results exhibit a discouraging
instability of the pQCD expansion. First, the NLO corrections 
are twice as large in the magnitude, and of the 
opposite sign, than the LO amplitude of photoproduction of the $J/\Psi$
\cite{DimaNLO2}.
Second, the NLO amplitude for the electroproduction of the $\rho$
lacks an expected stability window vs. the factorization scale
\cite{DimaNLO31,*DimaNLO32}. A further analysis of NLO correction, 
for instance, studies of the sensitivity of the stability window
to models for skewed parton densities,
and an independent rederivation are called upon to clarify this important
issue.

\newpage
\providecommand{\etal}{et al.\xspace}
\providecommand{\coll}{Coll.\xspace}
\catcode`\@=11
\def\@bibitem#1{%
\ifmc@bstsupport
  \mc@iftail{#1}%
    {;\newline\ignorespaces}%
    {\ifmc@first\else.\fi\orig@bibitem{#1}}
  \mc@firstfalse
\else
  \mc@iftail{#1}%
    {\ignorespaces}%
    {\orig@bibitem{#1}}%
\fi}%
\catcode`\@=12

\end{document}